# Iron Oxide Surfaces

Gareth S. Parkinson

Institute of Applied Physics, TU Vienna, Austria

## Abstract

The current status of knowledge regarding the surfaces of the iron oxides, magnetite ($Fe_3O_4$), maghemite ($\gamma$-$Fe_2O_3$), hematite ($\alpha$-$Fe_2O_3$), and wüstite ($Fe_{1-x}O$) is reviewed. The paper starts with a summary of applications where iron oxide surfaces play a major role, including corrosion, catalysis, spintronics, magnetic nanoparticles (MNPs), biomedicine, photoelectrochemical water splitting and groundwater remediation. The bulk structure and properties are then briefly presented; each compound is based on a close-packed anion lattice, with a different distribution and oxidation state of the Fe cations in interstitial sites. The bulk defect chemistry is dominated by cation vacancies and interstitials (not oxygen vacancies) and this provides the context to understand iron oxide surfaces, which represent the front line in reduction and oxidation processes. Fe diffuses in and out from the bulk in response to the $O_2$ chemical potential, forming sometimes complex intermediate phases at the surface. For example, $\alpha$-$Fe_2O_3$ adopts $Fe_3O_4$-like surfaces in reducing conditions, and similarly $Fe_3O_4$ adopts $Fe_{1-x}O$-like structures in further reducing conditions still. It is argued that known bulk defect structures are an excellent starting point in building models for iron oxide surfaces.

The atomic-scale structure of the low-index surfaces of iron oxides is the major focus of this review. $Fe_3O_4$ is the most studied iron oxide in surface science, primarily because its stability range corresponds nicely to the ultra-high vacuum environment, and because it is an electrical conductor, which makes it straightforward to study with the most commonly used surface science methods such as photoemission spectroscopies (XPS, UPS) and scanning tunneling microscopy (STM). The impact of the surfaces on the measurement of bulk properties such as magnetism, the Verwey transition and the (predicted) half-metallicity is discussed.

The best understood iron oxide surface at present is probably $Fe_3O_4$(100); the structure is known with a high degree of precision and the major defects and properties are well characterised. A major factor in this is that a termination at the $Fe_{oct}$-O plane can be reproducibly prepared by a variety of methods, as long as the surface is annealed in $10^{-7}$-$10^{-5}$ mbar $O_2$ in the final stage of preparation. Such straightforward preparation of a monophase termination is generally not the case for other iron oxide surfaces. All available evidence suggests the oft-studied ($\sqrt{2}\times\sqrt{2}$)R45° reconstruction results from a rearrangement of the cation lattice in the outermost unit cell in which two octahedral cations are replaced by one tetrahedral interstitial, a motif conceptually similar to well-known Koch-Cohen defects in $Fe_{1-x}O$. The cation deficiency results in $Fe_{11}O_{16}$ stoichiometry, which is in line with the chemical potential in ultra-high vacuum (UHV), which is close to the border between the $Fe_3O_4$ and $Fe_2O_3$ phases. The $Fe_3O_4$(111) surface is also well studied, but two different surface terminations exist close in energy and can coexist, which makes sample preparation and data interpretation somewhat tricky. Both surfaces exhibit Fe-rich surfaces as the sample selvedge becomes reduced. The $Fe_3O_4$(110) surface forms a one-dimensional (3×1) reconstruction linked to nanofaceting to expose the more stable $Fe_3O_4$(111) surface. $\alpha$-$Fe_2O_3$(0001) is the most studied hematite surface, but difficulties preparing stoichiometric surfaces under UHV conditions have hampered a definitive determination of the structure. There is evidence for at least three terminations: a bulk-like termination at the oxygen plane, a termination with half of the cation layer, and a termination with ferryl groups. When the surface is reduced the so-called "bi-phase" structure is formed, which eventually transforms to a $Fe_3O_4(111)$-like termination. The structure of the bi-phase surface is controversial; a largely accepted model of coexisting $Fe_{1-x}O$ and $\alpha\text{-}Fe_2O_3(0001)$ islands was recently challenged and a new structure based on a thin film of $Fe_3O_4(111)$ on $\alpha\text{-}Fe_2O_3(0001)$ was proposed. The merits of the competing models are discussed. The $\alpha\text{-}Fe_2O_3(012)$ "R-cut" surface is recommended as an excellent prospect for future study given its apparent ease of preparation and its prevalence in nanomaterial.

In the latter sections the literature regarding adsorption on iron oxides is reviewed. First, the adsorption studies of molecules ($H_2$, $H_2O$, CO, $CO_2$, $O_2$, HCOOH, $CH_3OH$, $CCl_4$, $CH_3I$, $C_6H_6$, $SO_2$, $H_2S$, ethylbenzene, styrene, and $Alq_3$) is discussed, and an attempt is made to relate this information to the reactions in which iron oxides are utilized as catalysts (water-gas shift, Fischer-Tropsch, dehydrogenation of ethylbenzene to styrene) or catalyst supports (CO oxidation). The known interactions of iron oxide surfaces with metals are described, and it is shown that the behaviour is determined by whether the metal forms a stable ternary phase with the iron oxide. Those that do not, (e.g. Au, Pt, Ag, Pd) prefer to form three-dimensional particles, while the remainder (Ni, Co, Mn, Cr, V, Cu, Ti, Zr, Sn, Li, K, Na, Ca, Rb, Cs, Mg, Ca) incorporate within in the oxide lattice. The incorporation temperature scales with the heat of formation of the most stable metal oxide. A particular effort is made to underline the mechanisms responsible for the extraordinary thermal stability of isolated metal adatoms on $Fe_3O_4$ surfaces, and the potential application of this model system to understand single atom catalysis and sub-nano cluster catalysis is discussed. The review ends with a brief summary, and a perspective is offered including exciting lines of future research.

Contents

Abstract

List of Abbreviations Used

# List of Abbreviations Used

| | |
|---|---|
| AES | Auger electron spectroscopy |
| AFM | Atomic force microscopy |
| APDBs | Antiphase domain boundaries |
| CEMS | Conversion electron Mössbauer spectroscopy |
| CTR | Crystal truncation rod |
| DBT | Distorted bulk truncation |
| DFT | Density Functional Theory |
| DFT+U | Density Functional Theory with additional Hubbard $U$ |
| DOS | Density of states |
| ESRF | European Synchrotron Radiation Facility |
| FCC | Face centred cubic |
| $Fe_{oct}$ | Octahedrally coordinated Fe cation |
| $Fe_{tet}$ | Tetrahedrally coordinated Fe cation |
| FWHM | Full width at half maximum |
| GGA | Generalized gradient approximation |
| GW | Greens function (G) with screened Coulomb interaction (W) |
| HCP | Hexagonal close packed |
| HREELS | High-resolution electron energy loss spectroscopy |
| HSE | Heyd-Scuseria-Ernzerhof |
| IPES | Inverse photoemission spectroscopy |
| IRAS | Infrared reflection absorption spectroscopy |
| L | Langmuir (1 L = $1.33 \times 10^{-6}$ mbar.s) |
| LEED | Low energy electron diffraction |
| LEED *IV* | Quantitative Low energy electron diffraction |
| LEEM | Low energy electron microscopy |
| LEIS | Low energy ion scattering |
| MEIS | Medium energy ion scattering |
| MBE | Molecular beam epitaxy |
| MDS | Metastable deexcitation spectroscopy |
| ML | Monolayer |
| MRI | Magnetic Resonance Imaging |
| nc-AFM | Non-contact atomic force microscopy |
| NEXAFS | Near edge x-ray absorption fine structure |
| OH | Hydroxyl group |
| PBE | Perdew–Burke–Ernzerhof functional |
| PBE+U | Perdew–Burke–Ernzerhof functional with additional Hubbard $U$ |
| PDOS | Partial density of states |
| PEEM | Photoemission electron microscopy |
| PEC | Photo-electro-chemical |
| PES | Photoemission spectroscopy |
| PLD | Pulsed laser deposition |
| RF | Radio frequency |
| RHE | Reference hydrogen electrode |

| | | |
|---|---|---|
| RHEED | Reflection high energy electron diffraction | |
| RMS | Root mean square | |
| SCV | Subsurface cation vacancy | |
| SMSI | Strong metal support interaction | |
| SP-LEEM | Spin polarized Low energy electron microscopy | |
| STM | Scanning tunnelling microscopy | |
| STS | Scanning tunnelling spectroscopy | |
| SXRD | Surface x-ray diffraction | |
| TEM | Transmission electron microscopy | |
| TPD | Temperature programmed desorption | |
| TPR | Temperature programmed reduction | |
| TSMR | Transition metal surface resonance | |
| $T_C$ | Curie temperature | |
| $T_V$ | Verwey transition temperature ($T_V$ = 125 K) | |
| UHV | Ultra-high vacuum | |
| UPS | Ultraviolet photoemission spectroscopy | |
| $V_O$ | Oxygen vacancy | |
| WGS | Water-gas shift | |
| XAS | X-ray absorption spectroscopy | |
| XES | X-ray emission spectroscopy | |
| XMCD | X-ray magnetic circular dichroism | |
| XPS | X-ray photoelectron spectroscopy | |
| XRD | X-ray diffraction | |
| XSW | X-ray standing waves | |

# 1. Introduction

## 1.1 Motivation

Iron and oxygen are two of the four most common elements in the Earth's crust, and iron oxides form naturally through the weathering of Fe-containing rocks both on land and in the oceans. The natural abundance of iron oxides in rocks, soils, and dust in the atmosphere ensures that they play an important role in geochemistry [1]. The Curiosity rover recently confirmed the existence of hematite (α-$Fe_2O_3$) on Mars, which is partly responsible for the red hue, and its analysis offers clues to the history of liquid water in the planet's environment [2]. Here on Earth, iron oxides have integrated in many biological systems; magnetite ($Fe_3O_4$), for example, aids navigation of magnetotactic bacteria [3], is thought to play a similar role in the beaks of homing pigeons [4], and is even found (in an as yet unknown capacity) in the human brain and other body tissue [5].

Iron oxides have proven invaluable materials to mankind over the millennia, starting with the pre-historic use of iron oxide containing Ochre pigments to decorate cave walls. Our first experiences with magnetism came through $Fe_3O_4$ containing rocks, and compass-like instruments based on $Fe_3O_4$ were already used for religious purposes in China as early as 200 BC. The development of $Fe_3O_4$-based compasses for navigation occurred in Europe as early as 850 AD. Through the 20$^{th}$ century the iron oxides were at the forefront of discovery in science. For example, $Fe_3O_4$ was one of the first mineral structures solved by Bragg in 1915 [6], and Verwey [7] discovered one of the first metal-insulator transitions in $Fe_3O_4$ in 1939. Néel took $Fe_3O_4$ as the prototypical example of his theory of ferrimagnetism [8].

Today, by far the single most important use (by volume) of iron oxides is as a source of Fe, which is subsequently processed to make steel. Rocks containing high amounts of hematite and magnetite are mined from the ground, and these can be easily reduced by carbo-thermal reduction ($Fe_2O_3$ + 3CO → 2Fe + 3$CO_2$). This process has been known for a long time, but is of course ultimately a very complicated surface mediated reaction. Other common uses of iron oxides include corrosion protective coatings ($Fe_3O_4$, or "black rust" thin films are prepared by a process known as bluing of steel), in recording media, and catalysis (the water-gas shift reaction and styrene production utilize iron-oxide-based catalysts).

Looking to the future, there has been a resurgence of research into iron oxide materials recently. In their aptly titled perspective article "The Iron Oxides Strike Back:…", Tartaj et al. [9] describe how the exciting properties of iron oxides, coupled to their low toxicity, stability, and economic viability, make them ideal for application in a wide range of emerging fields. For example, α-$Fe_2O_3$ has almost the ideal band gap for PEC water splitting, performed via the reaction: $H_2O$ → ½ $O_2$+$H_2$ ($E_0$ = 1.23 V), which is a way to convert solar energy into useful chemical energy whilst providing a major source of $H_2$. Current research aims to improve the efficiency with which charge carriers can be separated in α-$Fe_2O_3$ via control of the morphology, and significant effort is made to reduce the required over-potential through surface engineering. The deposition of small amounts of cobalt, for example, reduces the over-potential by 0.1 V [10]. Interestingly, although the iron oxide catalyst for the water-gas shift reaction is well-established in industry, the toxicity of the chromium stabilizing it has led to new work to develop alternatives.

A second emerging field with huge potential is biomedicine [11-13]. Here, the interest stems largely from the ability to manipulate the location of non-toxic iron-oxide nanoparticles inside the body by the application of an external magnetic field. For example, $Fe_3O_4$ nanoparticles are used as a contrast agent in MRI scanners, and much effort is made to coat magnetic nanoparticles with drugs, which can then be directed and held in the desired location to deliver a targeted treatment [14; 15]. In hyperthermia treatment, tumours can be killed by the local temperature increase that occurs when $Fe_3O_4$ nanoparticles are placed in a rapidly varying magnetic field. In general, these biomedical applications are based on the attachment of ligands to the nanoparticles, which is clearly a surface issue.

Tartaj et al. [9] also highlight promising applications in energy storage (in particular the utilization of α-$Fe_2O_3$ as an anode material for lithium ion batteries) and heterogeneous catalysis. They discuss the use of iron oxide-based catalysts in the Fenton reaction [16], ethylbenzene dehydrogenation [17], and Fischer-Tropsch synthesis [18]. It is interesting to note, however, that while iron oxides are reactive enough to catalyse certain reactions, they are often used as the inexpensive support for metal nanoparticles, and there is mounting evidence that support effects can play a significant role in such systems [19], particularly when nanoparticles enter the sub-nano regime. Many research papers are published about the ability of iron oxide nanoparticles ($Fe_3O_4$ and γ-$Fe_2O_3$) to remove heavy metals from contaminated water [20; 21]. Again, the ability to remove the particles post treatment with a magnet makes this solution appealing [22]. Zero-valent iron particles, oxidised at the surface in the aqueous environment, are used in the degradation of chlorinated hydrocarbons in wastewater [23].

One final application worth highlighting is novel electronic devices. Already in the 1980's DFT-based calculations predicted $Fe_3O_4$ to be a half-metallic ferromagnet [24] i.e. a metallic conductor for one spin channel, but an insulator in the other. This led to much interest in using $Fe_3O_4$ as a source of spin-polarized current for spintronics devices [25; 26] but, unfortunately, the performance of prototype devices never reached the heights expected (e.g. [27-29]), which was often attributed to a "magnetic dead layer" at the interface [30]. As will be shown here, iron oxide surfaces frequently reconstruct, and the resulting surfaces can exhibit markedly different electronic and magnetic structure to the bulk compound. Interface engineering offers an interesting opportunity to improve performance. More recently, researchers have thought to use the Verwey transition as a switch in the emerging field of "Mottronics" [31].

In this review I hope to provide an overview of what has been learned about iron oxide surfaces over the 20-30 years of metal-oxide surface science. I also hope to provide the interested reader with an introduction to the iron oxides in general and provide a resource for useful data and constants, although this cannot hope to compare to the exhaustive books "Iron oxides in the laboratory: Preparation and characterization" [32] and "The Iron Oxides: structure, properties, reactions, occurrences and uses", by Cornell and Schwertmann [1]. As we go through the many experiments, I hope to convey the importance of high-quality stoichiometric samples in research, although this point is made very strongly by Walz in his excellent paper "The Verwey Transition – A Topical Review" [33]. More generally, I want to show how the most interesting surface phenomena are linked to the ease with which iron oxides switch between different structures and stoichiometry in response to the external environment, and that well known defect structures from bulk studies can guide the construction of models of iron oxide surfaces. I hope to clarify some confusing results, and highlight instances where consensus remains some way off.

In many ways the knowledge of iron oxide surfaces today is similar to that of $TiO_2$ in 2003, when Diebold published her seminal Surface Science Report "The Surface Science of Titanium Dioxide" [34]. Given the inexorable year-on-year rise in the number of publications regarding $TiO_2$ and the vast number of citations this review has received, it is clear that a well-timed review can act as a springboard to encourage new researchers to join the field. The $TiO_2$ review made the vital point that the $TiO_2$(110) surface could be reproducibly prepared via the familiar surface science procedure of sputter/anneal cycles, resulting in a surface of precisely known structure and properties; the crucial first step to any successful surface science project. Here, I hope to demonstrate that our understanding of the $Fe_3O_4$(100) surface has now reached a similar level, and crucially, that a monophase $Fe_{oct}$-O layer can be easily prepared on both natural and synthetic single crystals, as well as thin films. I will demonstrate that all the available experimental and theoretical evidence are consistent with the subsurface cation vacancy (SCV) model of the surface [35] (see section 3.3.3).

A central point in this review is that the bulk defect chemistry of the iron oxides makes them intrinsically different to other well-known metal oxide surfaces. In most systems studied to date, oxygen vacancies ($V_O$s) dominate in the bulk when the materials are reduced in the ultrahigh vacuum environment, and an equilibrium concentration of these defects forms at the surface. One of the major lessons of metal oxide surface science to date is that $V_O$s dope electrons into the lattice, underlie surface reconstructions, and provide the active sites for surface chemistry. A similar treatment of $Fe_3O_4$, however, leads to Fe interstitials in the bulk; the $O^{2-}$ lattice remains intact. This in turn leads to excess Fe at the surface, and eventually Fe-rich surface terminations. I will show that the defect structures present in the bulk of iron oxide materials often occur at the surfaces, and that these motifs can serve as a useful guide in the development of new structural models of the surfaces. Note that this behaviour is not unique to iron oxides, the oxides of Ni, Mn, and Co exhibit similar bulk phases and defect chemistry, but as wide-gap insulators, these materials have not been extensively studied. I believe that in the $Fe_3O_4$(100) surface, we have developed a prototypical system to study this class of metal oxide compounds.

Iron oxide surface science has been reviewed before by leading groups in the field. Weiss and Ranke [17] provided an excellent overview of results up to 2002, with a particular focus on their studies of ethylbenzene-styrene conversion. Freund and co-workers have discussed iron oxides as part of several broader reviews of metal oxide surface science. Most recently, Kuhlenbeck, Shaikhutdinov and Freund [36] discussed $Fe_3O_4$ and α-$Fe_2O_3$ surfaces as part of their 2013 Chemical Reviews article, part of a special issue devoted to metal oxide surface science. Both of these prior reviews focus somewhat on iron-oxide thin films grown epitaxially on metal supports, in part because this is how both groups prepared their samples. Here, an attempt is made where possible to emphasize work that has been performed on iron-oxide single crystals, and to highlight where the existence of a semi-infinite sink of Fe in the bulk has a significant impact on the surface behaviour.

## 2 The Iron Oxides

## 2.1 General Overview

Before looking the iron oxides individually, it is instructive to first consider them as a class. The iron oxides are all based on a close packed $O^{2-}$ anion lattice, with the smaller Fe cations occupying octahedrally and tetrahedrally coordinated interstices in between (Figure 1). Under the most reducing conditions, wüstite ($Fe_{1-x}O$) is formed. It crystallizes in the rocksalt structure, contains $Fe^{2+}$ in octahedral sites, and is often non stoichiometric with a cation deficiency. Under oxidising conditions, hematite (α-$Fe_2O_3$) is formed. α-$Fe_2O_3$ crystallizes in the corundum structure and contains $Fe^{3+}$ in octahedral sites. In between, there is magnetite ($Fe_3O_4$), an (inverse) spinel with $Fe^{3+}$ in tetrahedral sites and a 50:50 mixture of $Fe^{2+}$ and $Fe^{3+}$ in octahedral sites. Finally, when $Fe_3O_4$ is oxidised directly, $Fe^{2+}$ is converted to $Fe^{3+}$ within the spinel structure, and compensating iron vacancies ($V_{Fe}$) appear in the octahedral sublattice. The defective spinel structure is remarkably robust, and can accommodate the full range of stoichiometry between $Fe_3O_4$ and $Fe_2O_3$. In the extreme case, all Fe is oxidised to $Fe^{3+}$, and maghemite (γ-$Fe_2O_3$) is formed. γ-$Fe_2O_3$ is metastable against transformation to α-$Fe_2O_3$, but exists partly because the conversion from the spinel to the corundum structure requires the $O^{2-}$ lattice to be converted from fcc to hcp. In contrast, switching between γ-$Fe_2O_3$, $Fe_3O_4$ and $Fe_{1-x}O$ is remarkably fluid because this only requires a rearrangement of the cations within a fcc oxygen lattice. A cursory glance at the Fe-$O_2$ phase diagram (Figure 2) reveals that γ-$Fe_2O_3$, $Fe_3O_4$ and $Fe_{1-x}O$ can all be the stable bulk phase at $O_2$ pressures and temperatures accessible in a UHV surface science experiment. Reduction [37; 38] and oxidation [39-43] of the various compounds have been extensively studied and found to be driven mostly by cation diffusion to and from the surface, so it should not be surprising that the surface terminations are strongly dependent on the preparation conditions.

The properties of the different iron-oxide phases are summarized in Table 1. The data are taken from references [1], unless otherwise stated.

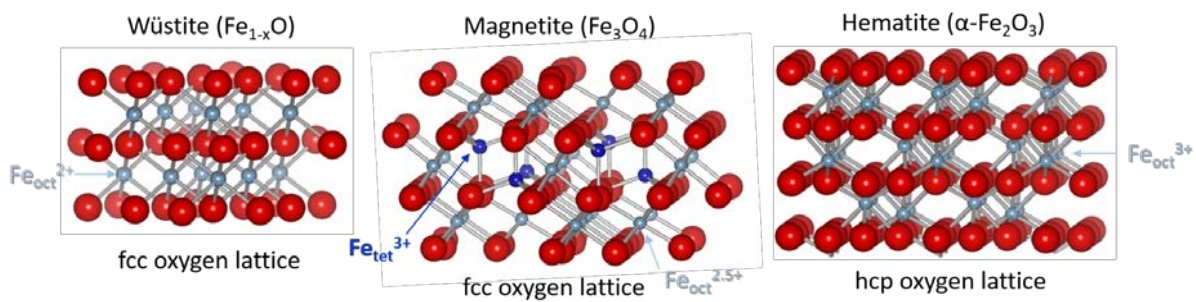

Figure 1: The iron oxides are based on a close packed $O^{2-}$ anion lattice with metal cations in octahedral and tetrahedral coordinated interstitial sites. Maghemite (not shown) is isostructural with magnetite, but with Fe vacancies on the octahedral sublattice.

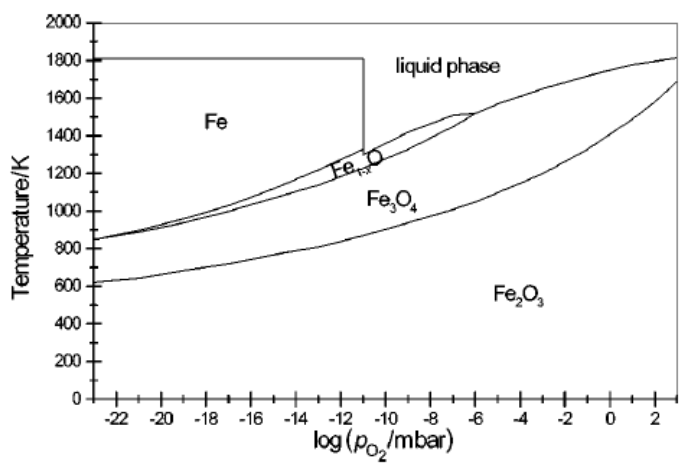

Figure 2: Phase diagram for the Fe-O$_2$ system calculated using the equi-therm program. Reproduced from Ref. [44] by permission of the PCCP Owner Societies

Table 1: Properties of the Iron Oxides (data taken from refs. [1])

<u>Atomic radii</u>

Fe (metallic) = 0.126 nm, O (covalent) = 0.066 nm

<u>Ionic Radii</u>

$Fe^{2+}$ = 0.082 nm, $Fe^{3+}$ = 0.065 nm, $O^{2-}$ = 0.14 nm

| *Mineral* | *Wüstite* | *Magnetite* | *Maghemite* | *Hematite* |
|---|---|---|---|---|
| *Formula* | $Fe_{1-x}O$ | $Fe_3O_4$ | $\gamma$-$Fe_2O_3$ | $\alpha$-$Fe_2O_3$ |
| *Cation* | $Fe^{2+}$ | $Fe^{2+}$/$Fe^{3+}$ | $Fe^{3+}$ | $Fe^{3+}$ |
| *Structure type* | Defect Rocksalt | Inverse Spinel | Defect Spinel | Corundum |
| *Crystallographic system* | Cubic | Cubic | Cubic / Tetragonal | Hexagonal |
| *Space group* | Fm3m | Fd3m | P4$_3$32 (see section 2.3) | R$\bar{3}$c |
| *Anion stacking* | FCC (111) | FCC (111) | FCC (111) | HCP (001) |
| *Lattice parameters (nm)* | a = 0.4302–0.4275 | a = 0.8396 | a = 0.83474 | a = 0.50436  c = 1.37489 |
| *Formula units / unit cell* | 4 | 8 | 8 | 6 |
| *Colour* | Black | Black | Reddish brown | Red |
| *Density (g.m$^{-3}$)* | 5.9–5.99 | 5.18 | 4.87 | 5.26 |
| *Mohs Hardness* | 5 | 5 | 5.5 | 6.5 |
| *Melting point °C* | 1377 | 1583-1597 | - | 1350 |
| *Boiling point °C* | 2512 | 2623 | | |
| *Magnetism* | Antiferro- | Ferri- | Ferri- | Weak Ferro- / Anti- |
| *Neèl (Curie) Temperature °C* | 203-211 | 850 | 820-986 | 956 |

| | Heat of Formation (kJ.mol$^{-1}$) | -251 | -1012.6 | -711.1 | -742.7 |

70 years of intense research on the Verwey transition in Fe$_3$O$_4$ has seen this compound become one of the best characterized metal oxides. As such, there is much data in the literature tracing the evolution of bulk properties with temperature, particularly close to the transition temperature (125 K). In what follows, several old but potentially very useful plots for the surface scientist are reproduced (in some instances not with ideal figure quality), with comparison to other iron oxides, where possible. In Figure 3 the original conductivity versus temperature (more specifically 1000/T) plot of Verwey is reproduced. Note the two order of magnitude decrease at the Verwey transition. Figure 4 shows the specific heat capacity of Fe$_3$O$_4$ in the range of 5—350 K (from ref.[45]), whereas high temperature data 300—1000K can be found in ref. [46]. Whether the spike in $C_P$ at the Verwey transition consists of one or two peaks depends on stress in the sample [47]. Also provided is specific heat capacity data for α-Fe$_2$O$_3$ from ref. [48]. No specific heat anomaly is reported to occur with the magnetic transitions $T_C$ = 858 K in Fe$_3$O$_4$ and $T_N$ = 250 K in α-Fe$_2$O$_3$.

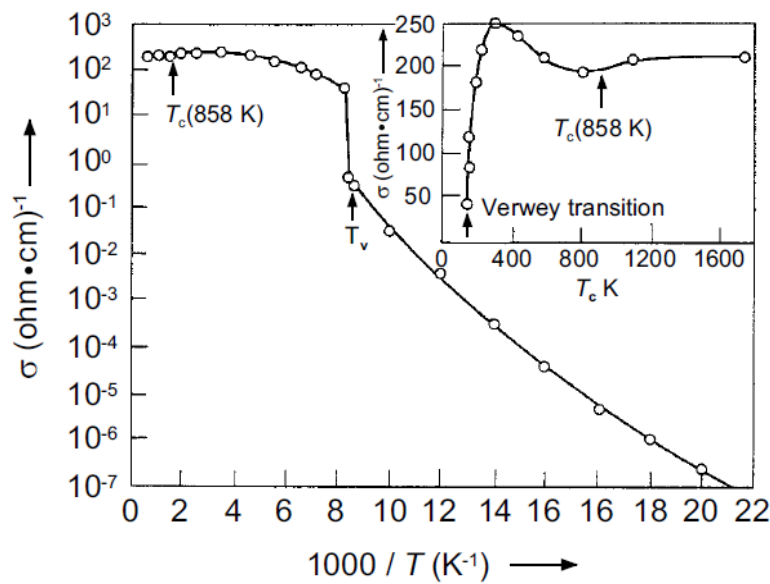

Figure 3: Electrical conductivity σ of Fe$_3$O$_4$ as a function of temperature. Note the Verwey transition at 125 K for a stoichiometric sample. Figure reproduced from the review of Walz [33]. Data originally from ref. [49].

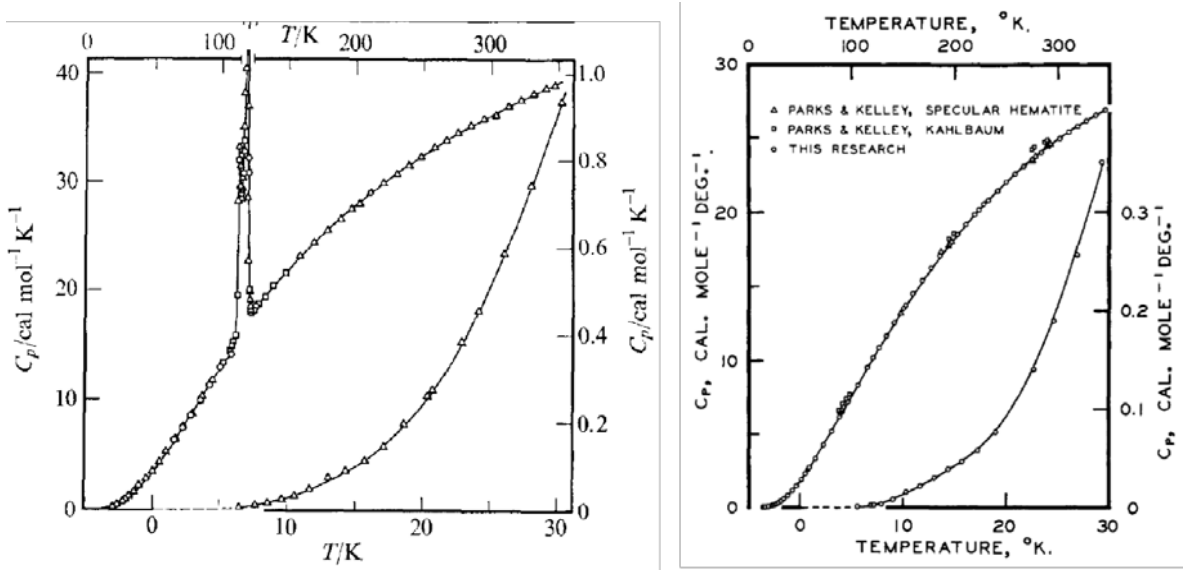

Figure 4: Heat capacity $C_P$ for $Fe_3O_4$ (left) and α-$Fe_2O_3$ (right). Low temperatures are shown using the lower curves and axes, whereas higher temperatures are shown using the upper curves and axes. Magnetite data reproduced from ref. [45; 46]. $Fe_2O_3$ figure reprinted with permission from Ref. [48]. Copyright 1959 American Chemical Society.

The thermal conductivity of $Fe_3O_4$ and α-$Fe_2O_3$, as determined in ref. [50], are shown in Figure 5. A mean regression of the data from several samples of different thickness is shown as solid lines in Figure 5, which leads to the following equations for thermal conductivity (K):

$$K(\text{magnetite}) = 0.0423 - 1.37 \times 10^{-5}\, T \qquad (1)$$

$$K(\text{hematite}) = 0.0839 - 6.63 \times 10^{-5}\, T \qquad (2)$$

Data for the thermal expansion coefficient of $Fe_3O_4$ from ref. [51] are included as Table 2. The optical properties shown in Figure 6 are particularly important for those wishing to perform IRAS experiments on $Fe_3O_4$ single crystals. The reflectivity of $Fe_3O_4$ is sufficient to acquire good quality experimental data in reflection, but the similarity of the real and imaginary component of the dielectric function can lead to unusual line shapes. This topic is discussed in detail in section 4.7.

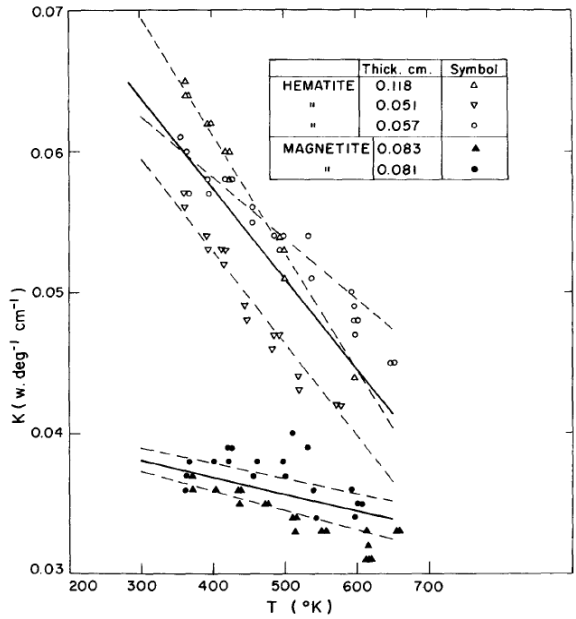

Figure 5: Thermal conductivity of $Fe_3O_4$ and $Fe_2O_3$ single crystals of different thickness as a function of temperature. Figure reproduced from ref. [50].

Table 2: Linear thermal expansion coefficient of $Fe_3O_4$ [51].

| T / K | α (×$10^{-6}$ $K^{-1}$) | T / K | α (×10-6 $K^{-1}$) |
|---|---|---|---|
| 298 | 10.41 | 773 | 16.54 |
| 373 | 11.38 | 973 | 19.05 |
| 473 | 12.68 | 1073 | 20.35 |
| 573 | 13.97 | 1173 | 21.61 |
| 673 | 15.26 | 1273 | 22.85 |

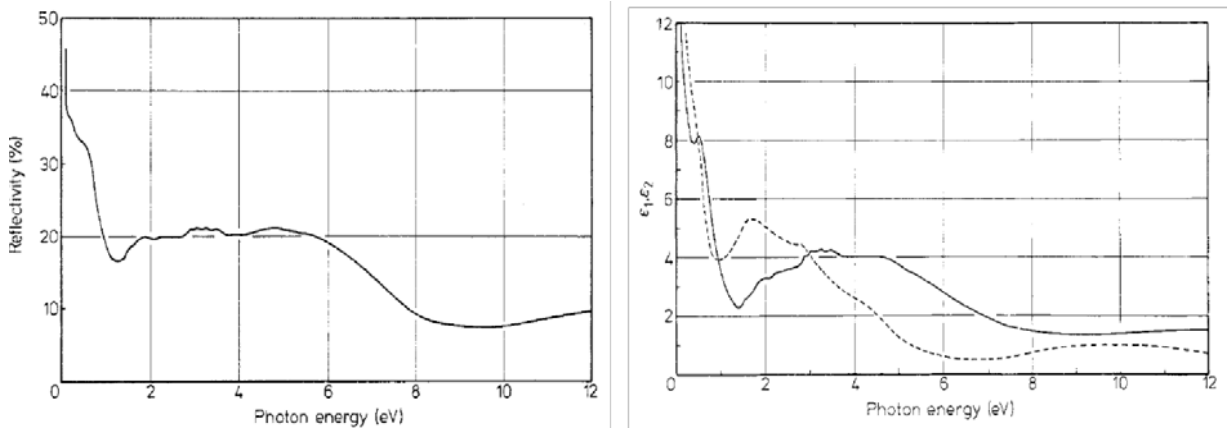

Figure 6: Optical properties on $Fe_3O_4$. (left) Reflectivity as a function of photon energy. (right) Real (dashed) and imaginary (solid) parts of the dielectric function of $Fe_3O_4$ at 300 K. Figures reproduced from Ref. [52].

Given that much of the interest in the iron oxides is due to their magnetic properties, it would be remiss to not include standard values for the bulk here. In Table 3 data are tabulated for magnetic transition temperatures and saturization magnetitaion of the iron oxides. Also included are the anisotropy constant (corresponding to the relevant easy axis), and the magnetostriction constant $\lambda$, which is the (saturation) change in volume in response to an applied magnetic field. All data were obtained from ref. [1]. Surface magnetism is discussed in section 2.2.5.

Table 3: Magnetic Properties of the Iron Oxides. Data from ref. [1].

|  | Transition Temperatures in K | Saturation Magnetization $\sigma_s$ (300 K) $JT^{-1}kg^{-1}$ | Anisotropy Constant $K_{eff}$ $Jm^{-3}$ | Magneto-striction Constant $\lambda$ | Magnetic Hyperfine Field $B_{hf}$ 295 K | Magnetic Hyperfine Field $B_{hf}$ 4 K |
|---|---|---|---|---|---|---|
| **$Fe_3O_4$** | $T_C$ = 850 $T_V$ = 120 | 92-100 | $10^4$-$10^5$ | $35 \times 10^{-6}$ |  |  |
| **$\gamma$-$Fe_2O_3$** | $T_C$ = 820-956 | 60-80 | $10^5$ | $35 \times 10^{-6}$ | 50 | 54.6 |
| **$\alpha$-$Fe_2O_3$** | $T_C$ = 956 $T_M$ = 260 | 0.3 | 1-6$\times 10^5$ | $8 \times 10^{-6}$ | 51.8 | 54.2 |

In what follows the structure and properties of the different iron oxides are described in detail, particularly where these properties pertain to applications for which the surface/interface properties play an important role.

## 2.2 Magnetite ($Fe_3O_4$)

### 2.2.1 Bulk Structure

$Fe_3O_4$ is just one of more than 150 metal oxide materials to crystallize in the spinel structure. Spinels have the general formula $AB_2O_4$, where A and B designate either different cation species, as in the case of $MgAl_2O_4$ and $CuFe_2O_4$, or different oxidation states of the same cation, as in $Fe^{2+}Fe^{3+}_2O_4$ and $Co^{2+}Co^{3+}_2O_4$. The spinel structure is based on an fcc $O^{2-}$ anion lattice, in which 1/8 of the tetrahedral and 1/2 of the octahedral interstices are occupied (see Figure 7) [53]. In a "normal" spinel (e.g. $MgAl_2O_4$ and $Co_3O_4$) the $A^{2+}$ cations occupy the tetrahedral interstitial sites, while the $B^{3+}$ cations occupy the octahedral sites. However, if the $A^{2+}$ cation has a large crystal field stabilization energy [53], these atoms occupy half of the octahedral sites, and the displaced $B^{3+}$ cations take up the tetrahedral coordination; such materials are known as "inverse spinel". Note there is a large literature regarding the site preference in spinel compounds, which is beyond the scope of the current review. The interested reader is referred to ref. [53]. The topic of this section, $Fe_3O_4$, is an inverse spinel, and the coexistence of $Fe^{2+}$ and $Fe^{3+}$ in the octahedral sublattice leads directly to many of its interesting material properties. In this review, a slightly different nomenclature will be used, with cations in the octahedral and tetrahedral sites given the subscript "oct" and "tet", respectively. This allows easier comparison between the different iron oxides.

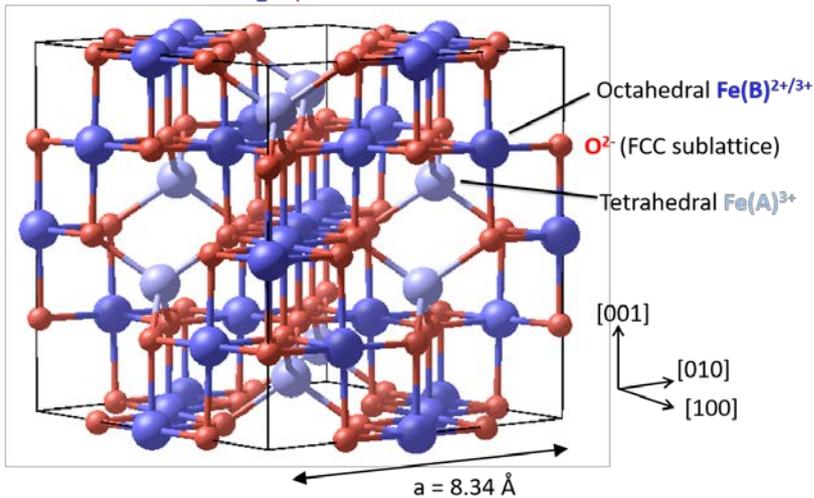

Figure 7: The inverse spinel structure of $Fe_3O_4$ is based on a fcc arrangement of $O^{2-}$ cations in which $Fe^{3+}$ cations occupy 1/2 of the tetrahedral interstices, and a 50:50 mix of $Fe^{3+}$ and $Fe^{2+}$ cations occupy 1/8 of the octahedral interstices. At room temperature, rapid hopping of electrons on the $Fe_{oct}$ sublattice results in the conductivity of a poor metal ($10^3 - 2.5 \times 10^4$ $\Omega^{-1}m^{-1}$ [33]).

## 2.2.2 $Fe_3O_4$ Magnetic and Electronic Structure

In his landmark paper of 1948 [8], Néel used $Fe_3O_4$ as an example to illustrate his theory of ferrimagnetism. Prior to Néel's work, $Fe_3O_4$ had been classified as a ferromagnet, and a magnetization of 4 $\mu_B$ per formula unit had been measured. Neel proposed that the tetrahedral and octahedral sublattices are antiferromagnetically aligned in $Fe_3O_4$, such that the $Fe^{3+}$ cations on each sublattice cancel each other, and the magnetic moment arises entirely from the moment of the remaining $Fe^{2+}$ cation.

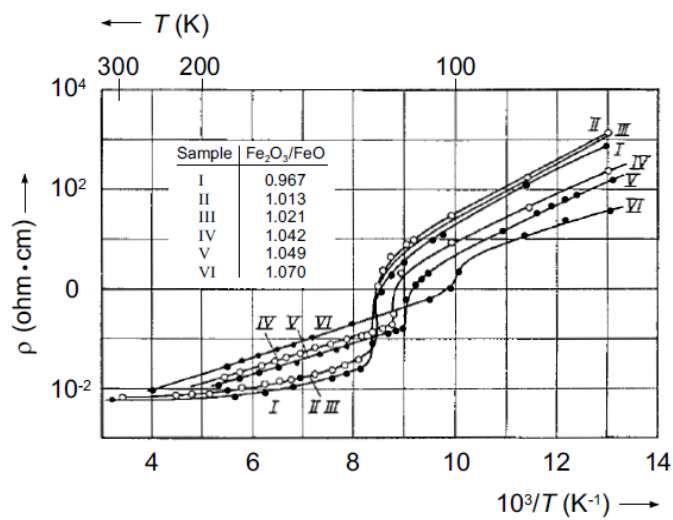

Figure 8: The original resistivity versus temperature plot of Verwey from his 1939 article [7] showing the dependence of the transition on sample stoichiometry. Note that an $Fe_2O_3$/FeO ratio of 1 corresponds to stoichiometric $Fe_3O_4$. Figure reproduced from ref. [33].

The DC conductivity of a stoichiometric $Fe_3O_4$ single crystal is plotted as a function of temperature in Figure 3. At room temperature the conductivity is similar to the poor metals (Ga, In, Sn…), which is more than sufficient for surface science studies using the usual electron spectroscopies and STM, and partly underlies the use of $Fe_3O_4$ as a prototype compound for the spinel class. On cooling, the conductivity decreases appreciably with temperature until 125 K, where it drops by a factor of 100 in a sharp first-order transition (the Verwey transition [7]). Nevertheless, LEED, photoemission, and STM experiments have been successfully conducted on $Fe_3O_4$ single crystals at liquid nitrogen temperature, although no surface science experiments have been reported at liquid helium temperatures as yet.

The discovery of the Verwey transition [7] prompted numerous attempts to understand and model the conduction mechanism in $Fe_3O_4$. For full details the reader is referred to an excellent summary by Walz [33], which covered developments up to 2002. Verwey [7] interpreted his original data (Figure 8) in an ionic picture, postulating that conductivity occurred via the hopping of electrons on the $Fe_{oct}$ sublattice ($Fe^{2+}$ - $e^-$ ⇆ $Fe^{3+}$). Below the transition, this hopping was supposed to be frozen out, with long-range charge order established amongst the $Fe^{2+}$ and $Fe^{3+}$ cations. Subsequently, most conduction models have been based on either band conduction and/or small-polaron hopping. In the band picture [54] the two $Fe_{oct}$ atoms per formula unit distribute their 11 d-electrons across two distinct bands, with ten spin-down electrons occupying a lower energy band, and one electron the higher energy band. This upper band crosses the Fermi level, and is responsible for the metallic conduction. However, the increase in conductivity up to room temperature appears inconsistent with this model. The alternative approach has been to treat the system in the framework of small-polarons [55]. However, models based on non-interacting polaron predict a maximum in the conductivity at 600 K, rather than 305 K, as was observed experimentally. The best fit to the data achieved to date comes from Ihle and Lorenz [56; 57], who proposed that conductivity arises from a superposition of small-polaron band and small-polaron hopping mechanisms. Their insight was to include polaron-polaron interactions, which leads to a short range ordering. In such a picture, band conduction *does* lead to an increase in conductivity with temperature up to ≈ 250 K because destruction of the short-range order is a thermally activated process. Small-polaron hopping increases with temperature and becomes comparable with band conduction around 300 K, and eventually dominates at high temperature.

A simple schematic representation of the *d* levels of the $Fe_{oct}^{2+}$ cation in $Fe_3O_4$ is shown in Figure 9. An exchange splitting of ≈ 3.5 eV causes the spin-up and spin-down levels to straddle the Fermi level, and the $t_{2g}$ and $e_g$ levels are further split by 2 eV by the crystal field. The $Fe_{oct}^{2+}$ cation has six *d* electrons, the first five of which occupy the majority (spin up) states, while the sixth occupies the minority spin band at $E_F$. In an ionic view, it is easy to see why this atom thus has a net moment of 4 $\mu_B$ (low spin state). An $Fe_{oct}^{3+}$ cation, on the other hand, has five electrons that occupy the majority states only, and has a net magnetic moment of 5 $\mu_B$ (high-spin state).

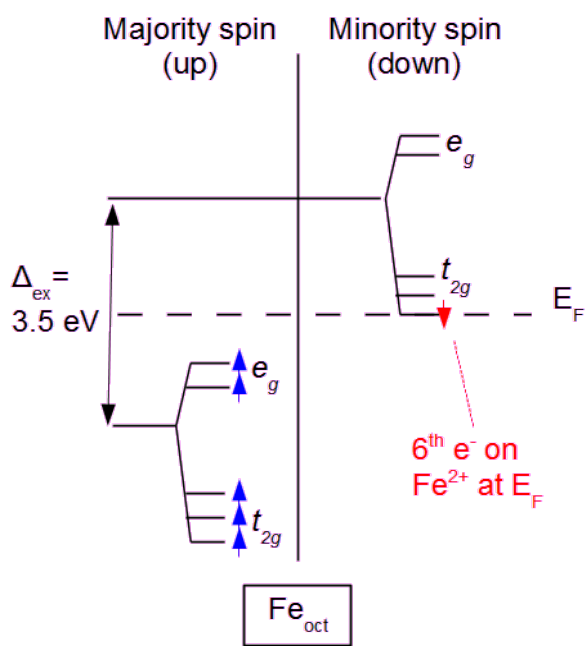

Figure 9: Sketch of the d-levels for the $Fe_{oct}^{2+}$ cations in $Fe_3O_4$. Five *d* electrons occupy the majority spin up states, and the sixth occupies a band at $E_F$ giving rise to conductivity. A schematic for the $Fe_{oct}^{3+}$ cations would be the same without the 6th electron at $E_F$. Sketch redrawn after ref. [58]. Copyrighted by the American Physical Society.

Early density functional theory based calculations for $Fe_3O_4$ were performed as early as 1984 [24; 59-61]. A representative band structure calculated using the local spin density approximation [61] is shown in Figure 10. A crucial aspect is that a band associated with the minority $t_{2g}$ electrons crosses $E_F$, but a band gap exists in the majority spin-up channel. This property, known as half metallicity, makes $Fe_3O_4$ potentially useful as a source of spin polarized current in spintronics applications [25]. Note that, because states derived from the O atoms only begin to appear ≈ 4 eV either side of $E_F$, STM measurements of $Fe_3O_4$ are dominated by the Fe cations. Over the last decade much theoretical effort has focussed on modelling bulk $Fe_3O_4$ (not to mention the surface). Today, most authors use a DFT+U method to account for strong electron correlations in this material. This U parameter effectively penalises the partial orbital occupancy found by DFT, and thus leads to a disproportionation of the equivalent $Fe_{oct}^{2.5+}$-like atoms into distinct cations that are interpreted as $Fe^{2+}$-like and $Fe^{3+}$-like. The charge-ordered structure that results from a U parameter of ≈ 4 eV is typically seen as representative of the low temperature Verwey phase as it reproduces the small bandgap (≈ 90 meV) that has been observed in experiments. Further details of the Verwey transition are discussed in Section 2.2.4, and an excellent summary of the influence of the Hubbard U parameter and structural distortions on DFT calculations is given in ref. [62]. Recent hybrid functional DFT calculations also suggest that the minority-spin electrons occupy a band at $E_F$ [63].

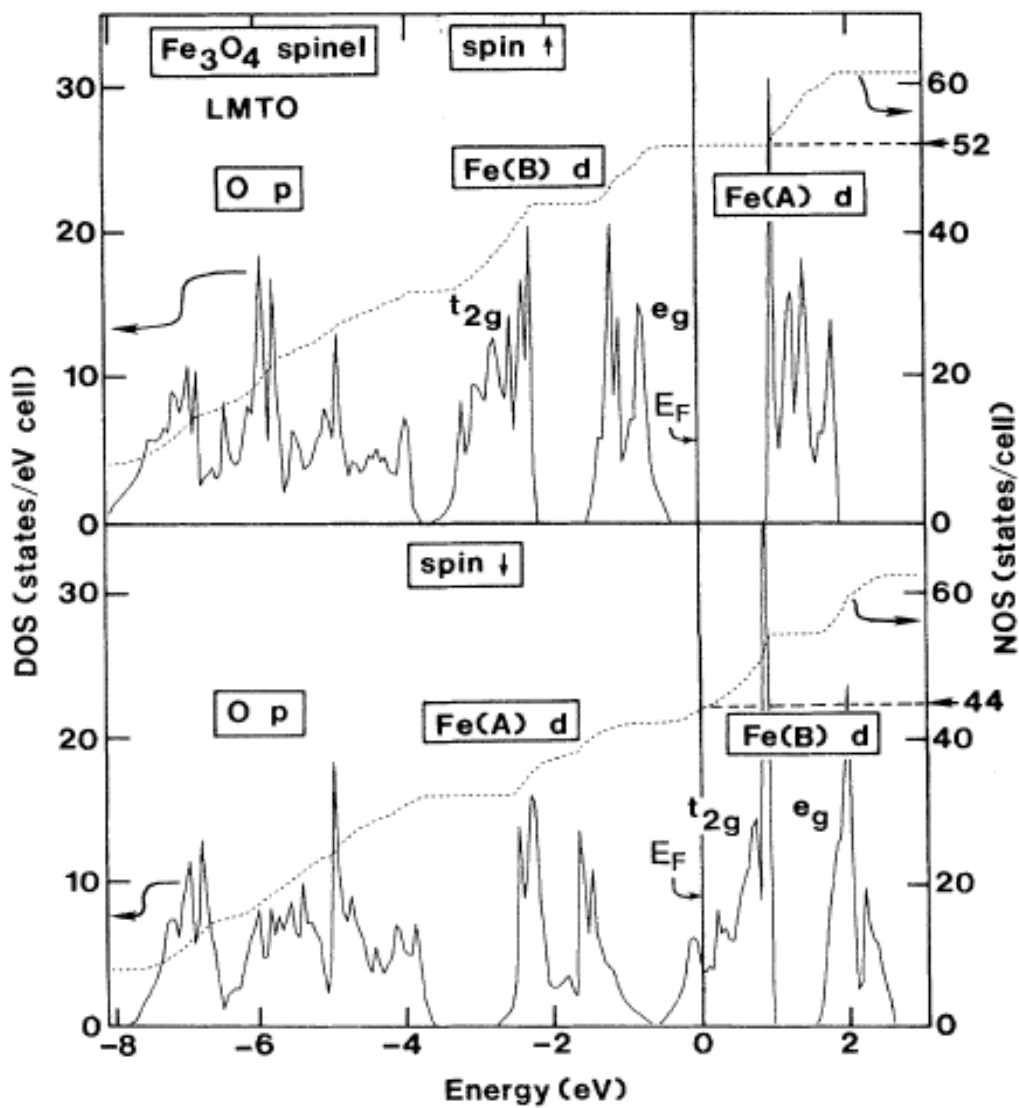

Figure 9: Spin projected one electron densities of states for $Fe_3O_4$ calculated using the local spin density approximation, reproduced from ref. [61]. Interestingly, $t_{2g}$ states associated with the $Fe_{oct}^{2+}$ cations cross $E_F$ in the minority spin channel, whereas there is a band gap in the majority spin channel.

Over the last decade, most DFT-based investigations of $Fe_3O_4$ surfaces utilize a Hubbard U of ≈4 eV for the $d$ states on the Fe cations, e.g. refs. [35; 64-70]. Note that this choice ensures that the "bulk" of the slab is similar to the charge/orbital ordered Verwey phase expected at 0 K. However, the majority of experiments are performed above 125 K, where the $Fe_{oct}$ cations are equivalent. The extent to whether this influences surface properties such as adsorption is not much discussed, although theoretical work has shown that water adsorption modifies the charge-order present the subsurface layers of the slab [71].

The electronic structure of $Fe_3O_4$ has been studied experimentally using photoemission on many occasions. Before discussing recent results it is important to emphasize that the use of $Fe^{2+}$ and $Fe^{3+}$ to describe the different cations in $Fe_3O_4$ is only reasonable in a formal sense. The true extent of charge disproportionation between the different cations is undoubtedly much smaller, and there is some debate whether it exists at all [72]. Recent work has shown that below $T_V$, where charge disproportionation should be "frozen in", the oxidation states of Fe cations vary between 2.4+ and 2.9+ [73; 74]. In the room temperature phase all $Fe_{oct}$ cations are crystallographically equivalent, and Mössbauer spectroscopy, for example, suggests that all $Fe_{oct}$ are equivalent, although this is likely a time average. A thorough topical review from 2004 summarizes experimental data from a range of techniques, and calls the ionic model into question [72].

A particularly interesting issue for the surface scientist is how to interpret Fe *2p* XPS spectra. Typically, the ionic model is invoked, a state at 708.5 eV is assigned to $Fe^{2+}$-like cations, and a peak closer to 710.5 eV is linked to $Fe^{3+}$-like cations [75]. Very recently, Taguchi et al. [76] performed an XPS study using bulk-sensitive hard x-rays, and suggested that the peak usually associated with the $Fe^{2+}$-like cations is too large to account for with an ionic model. Rather, this peak results from charge transfer between the Fe *3d* states on all the (equivalent) $Fe_{oct}$ cations and a state at the top of the valence band. Excellent agreement between the intensity of the peak in the bulk-sensitive experimental data and an Anderson Impurity Model was demonstrated. Interestingly, the authors showed that this peak is very weak in data acquired near a (100) surface. As will be shown in section 3.3.3, this surface reconstructs to form an outermost unit cell with significantly different electronic structure.

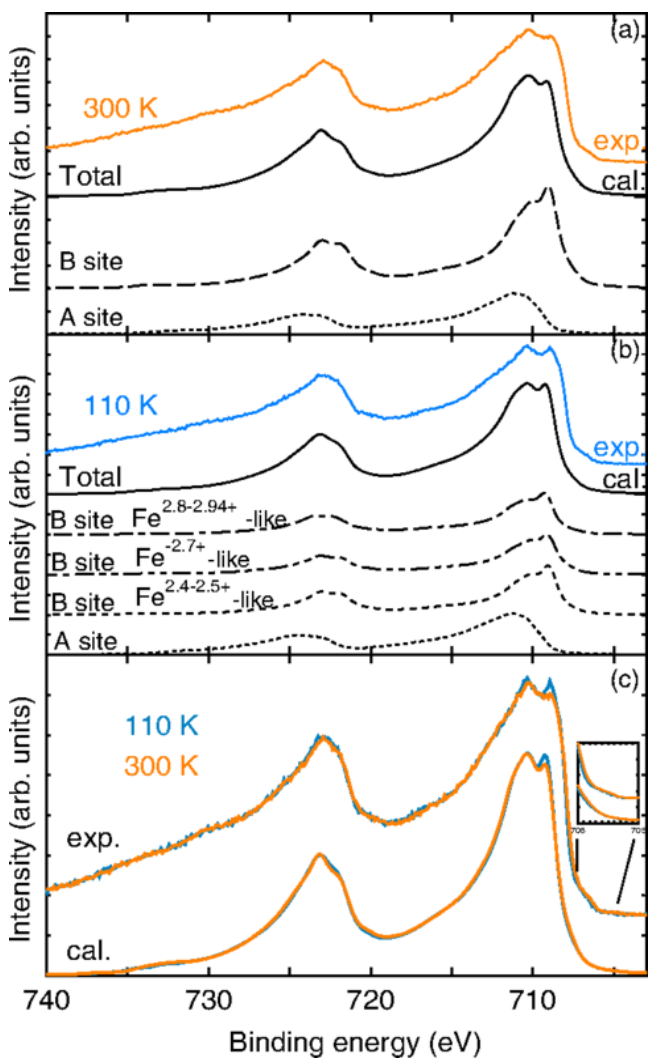

Figure 11: Calculated temperature dependent Fe *2p* spectra of $Fe_3O_4$ based on an Anderson Impurity model, and experimental XPS data acquired using hard x-rays (7.94 keV). (a) Theory-experiment comparison at 300 K. Two contributions based on the $Fe_{tet}$ (or A site) and $Fe_{oct}$ (or B site) cations are required to reproduce the experimental data. Note that the $Fe_{oct}$ cations produce a double peak, including the peak at 708.5 eV. (b) Below the Verwey transition (110 K) the Fe *2p* XPS spectra are fit using three distinct charge states for the $Fe_{oct}$ (B site) cations, in accordance with recent x-ray diffraction results [73; 74]. (c) Direct comparison of experimental and calculated XPS spectra at 300 K and 110 K. Reprinted with permission from ref. [76]. Copyright 2015 by the American Physical Society.

The valance band of $Fe_3O_4$ has also been measured on many occasions using photoemission [59; 77-81]. The spectra from the (100) and (111) facets exhibit similar features, but the relative intensities differ somewhat, particularly for low photon energies (see Figure 12). The authors attribute this difference to a contribution from the different surfaces [77], although it should be noted that the cross section of different peaks can also vary with geometry and incident electron energy. At higher photon energies, where the probing depth is greater and the bulk contribution to the overall signal higher, the spectra are more similar. The small peak closest to $E_F$ (at ≈0.6 eV) is assigned to the $d^6 \rightarrow d^5$ transition on the $Fe_{oct}^{2+}$ cations (in the ionic picture). This peak has been found to exhibit weak band dispersion [78; 80], predicted theoretically [24; 60; 61] and is thus consistent with a band conduction model. The next peak, located at ≈ -2.5 eV, is usually associated with photoemission from $Fe^{3+}$-like cations. Taking these assignments we see that the (100) surface is significantly enriched in $Fe^{3+}$ with respect to (111), which fits well with the SCV model for the (100) surface where the outermost 4 layers are predicted to contain only $Fe^{3+}$-like cations (see Section 3.3.3) [35]. Taguchi et al. [76] also studied the valance band in their studies using bulk sensitive hard x-rays, and concluded that there is finite DOS at $E_F$ at room temperature and that their results were consistent with a model of equivalent $Fe^{2.5+}$-like cations.

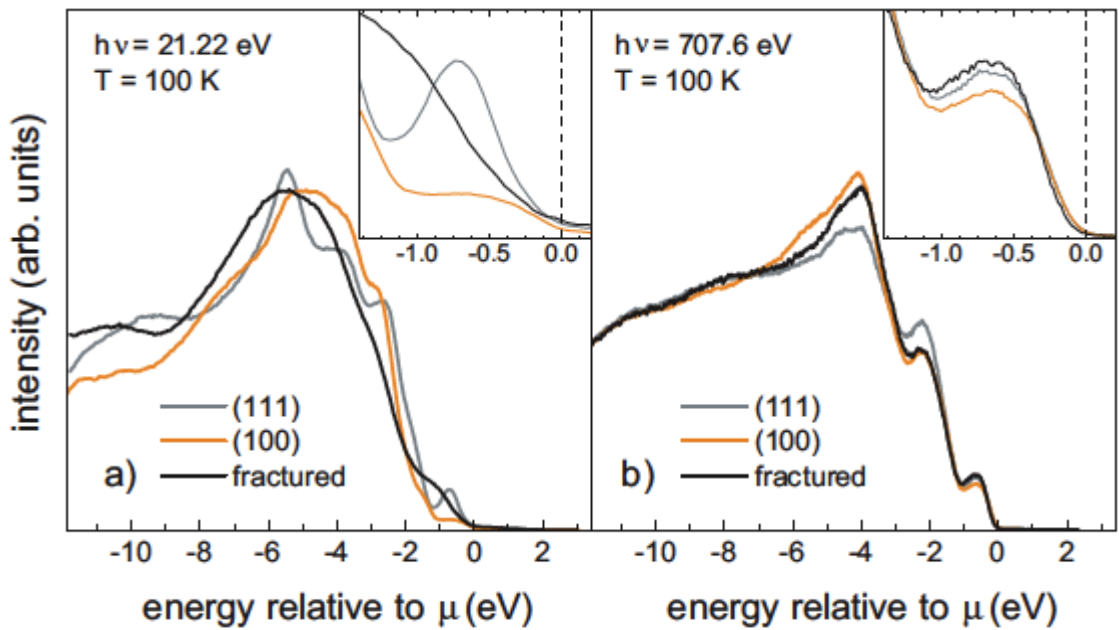

Figure 12: Photoelectron spectroscopy spectra showing the valance band region of synthetic $Fe_3O_4$ single crystal samples exposing different facets using 21.22 eV photons (a) and 707.6 eV photons (b).

The (111) and (100) surfaces were prepared by sputter/anneal cycles with post annealing in oxygen, while the "fractured" surfaces were prepared at 100 K. The authors report the fractured surface was rough and faceted. The same peaks are present in each spectrum, but their relative intensity differ significantly, particularly for the more surface sensitive, lower photon-energy measurements. Figure reproduced from ref. [77].

### 2.2.3 $Fe_3O_4$ - Half-Metallicity

A key result of band structure calculations shown in Figure 10 is that the $Fe_3O_4$ is predicted to be a half-metallic ferrimagnet [24; 60; 61]. This has led to much interest in utilizing $Fe_3O_4$ as a source of spin polarized current in spintronics devices [25]. However, the magnetoresistance of $Fe_3O_4$-based device prototypes has continually disappointed, with device prototypes not exhibiting the performance expected for a 100 % spin polarized electrode [28]. Although the poor performance is most frequently attributed to spin flips occurring at the material interface, Alvardo and Bagus [82], Huang et al. [83], and later Chambers and co-workers [84], have argued that electron correlation effects limit the spin polarization of bulk $Fe_3O_4$ to a maximum of 66 %. As of yet there is no unambiguous experimental confirmation, and the matter remains controversial.

The key issue is that the normal go-to technique to establish half metallicity is Andreev reflection [85], but this method requires temperatures well below the Verwey transition temperature, and thus cannot be applied to the cubic phase of $Fe_3O_4$. To circumvent this, several groups have performed spin-resolved photoemission experiments of the room temperature phase [83; 84; 86-89]. However, photoemission is an intrinsically surface sensitive technique (see Figure 11), and the experiment inevitably probes a mixture of the surface and bulk properties. Looking at the literature it seems that the lowest measured spin polarizations were obtained from samples exposing the (100) surface, which is now known to undergo a major reconstruction stretching four layers deep into the subsurface [35]. Consequently, the 40-55 % spin polarization values measured from $Fe_3O_4$(100) likely include a strong influence from the surface. Indeed, Chambers and co-workers [84] estimated that a magnetic dead layer would need to stretch approximately 1 unit cell into the bulk to reconcile their experimental value of 40 % with the band structure predictions. On the (111) surface, where reconstruction appears to be less severe, a spin polarization of - 80% has been measured, which is sufficiently larger than the -66.7 % that emerges from the ionic model.

The most recent publication to address the half metallicity issue using photoemission measured a spin polarization of -72 % for $Fe_3O_4$(100) thin films grown on MgO(100), again using spin polarized photoemission. It is not clear why the measured spin polarization is so far in excess of that measured for other $Fe_3O_4$(100) samples. However, it must be noted that the photon energies were very low and all aspects of the data, including the band dispersion of the states close to $E_F$, were found to agree with the results of DFT+U calculations. Unfortunately, the calculations were performed prior to the emergence of the SCV structure of the (√2×√2)R45° reconstruction (see section 3.3.3); it would be interesting to see whether the conclusions still hold with the modified surface stoichiometry. It must also be noted that the electronic structure of the $Fe_3O_4$(100) surface depends quite strongly on the Hubbard *U* parameter employed in DFT+U calculations.

In recent years Pratt, Yamauchi, and co-workers have published a series of papers [90-94] using metastable helium atom scattering as a probe of the Fermi-level spin polarization. For further details of this fascinating technique the reader is referred to refs. [92; 95; 96], but essentially, an incident beam of He* (He($2^3$S)) atoms decay via resonance ionization followed by an Auger neutralization processes. The electron that fills the He$^+$1s hole must have a spin opposite to the He$^+$ ion, so electron emission is a function of the spin-dependent DOS of the sample. The cross-section for the process is large, which makes the method extremely surface sensitive. In the experiment, the spin moment of the incident He* atoms is aligned parallel and anti-parallel to the sample magnetization, and the sample current induced by electron emission, $I(V_S)$, is measured as a function of sample voltage (see Figure 13a). The kinetic energy of the ejected electrons is given by $E_{kin}=E_{eff}-E_1-E_2-2\phi$, where $E_{eff}$ is the effective ionization energy of He, $E_1$ and $E_2$ are the binding energy of two electrons involved, and $\phi$ is the surface work function. At $V_{max}$ (the highest sample bias where electrons are emitted), $E_{kin}$ is maximised and both electrons originate from $E_F$. The spin asymmetry at $E_F$ is then determined by:

$$A(V_S) = \frac{I_\uparrow(V_S) - I_\downarrow(V_S)}{I_\uparrow(V_S) + I_\downarrow(V_S)}$$

.

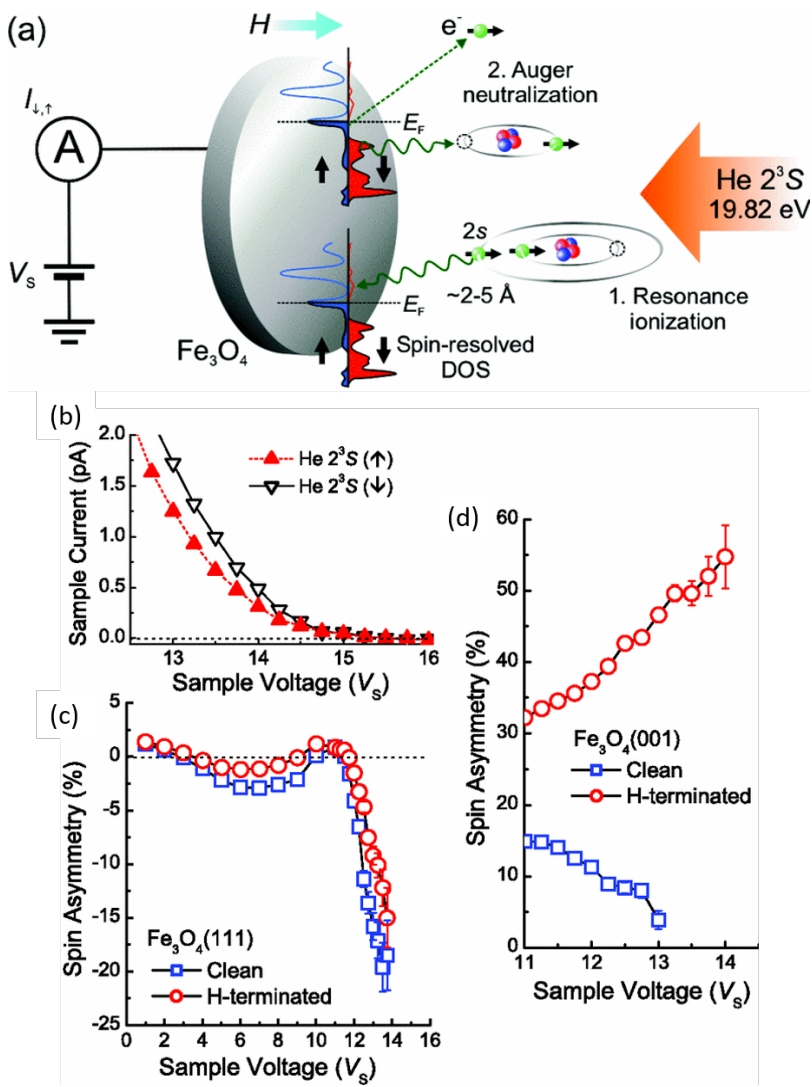

Figure 13: (a) Schematic of the metastable Helium atom scattering experiment. Incident He* atoms undergo a spin dependent Auger neutralisation process such that the induced sample current is a function of the spin dependent sample DOS. At high sample voltage the electrons involved originate from close to $E_F$. (b) Sample current as a function of sample voltage for parallel and anti-parallel spin

alignment with respect to the Fe$_3$O$_4$ single crystal. At $V_{max}$, the electrons involved in the Auger electron process originate close to $E_F$, and the observed asymmetry is related to the different DOS in the surface layer. (c) The corresponding spin asymmetry for Fe$_3$O$_4$(111) together with that of the hydrogen-terminated surface. (d) The spin asymmetry is much smaller at the Fe$_3$O$_4$(100) surface, but hydrogen adsorption induces a drastic enhancement. Figure 13 adapted with permission from Ref. [92]. Copyrighted by the American Physical Society.

The (111) surface is found to exhibit a strong *positive* spin polarization [92] by this approach (Figure 13b), rather than the negative spin polarization expected for the bulk. The authors claim that this surprising result can be consistent with the *negative* 80 % measured by photoemission due to the very different probing depths of the methods. The clean (100) surface (Figure 13c), on the other hand, is found to exhibit a small, negative spin polarization. Again, the authors conclude that the electronic structure of the outermost layer differs significantly from the bulk, and therefore that photoemission is not a suitable technique to measure the bulk spin polarization of Fe$_3$O$_4$. Subsequent work has focused on the impact of adsorption on the spin polarization at the (100) surface. A significant increase in spin polarization, from 5% to > 50%, was observed when the surface was treated with atomic hydrogen [97]. (A large increase in the DOS at the $E_F$ was also observed in photoemission experiments in this author's group [98], but no spin polarized measurements were made), and similar effects have been reported for benzene [90] and boron [94], while theoretical calculations from this group predict similar behaviour for group IV elements [93]. Again, it must be stressed that theoretical calculations for the Fe$_3$O$_4$(100) surface conducted prior to 2014 are based on the (now superseded) DBT model (see section 3.3.2) of the Fe$_3$O$_4$(100) surface [35]. However, at least in the case of boron, the similar calculations were performed for the current SCV model (see section 3.3.3) of the surface using both DFT and DFT+U, and a similar enhancement of the surface spin polarization at $E_F$ was found [94].

In summary, although the matter is not conclusively settled, the available evidence suggests that Fe$_3$O$_4$ may be half metallic in the bulk, as predicted by theoretical calculations. However, surface effects hamper both experimental verification and implementation in devices. An exciting avenue for future research is surface engineering through adsorption, as pioneered by Pratt and co-workers [90-94; 99], which may allow tailoring of the interfacial spin polarization in Fe$_3$O$_4$-based spintronics devices.

## 2.2.4 Surface Verwey Transition

The Verwey transition, as mentioned above, is characterized by a first-order drop in the conductivity by two orders of magnitude at ≈ 125 K (See Figure 2). However, this is not the only change; the crystal symmetry distorts and changes from cubic to monoclinic and the magnetic easy axis rotates from <111> in the cubic phase to the <001> monoclinic axis. Despite intense research spanning several decades the finer details of the transition remain controversial, particularly regarding the existence, or lack, of charge order. The latest results, performed on a single monoclinic domain of Fe$_3$O$_4$ suggest that the crystal exhibits a monoclinic *Cc* symmetry below the transition, and the charge order hypothesis is essentially correct, albeit with smaller charge disproportionation, in the range 2.45-2.55 e$^-$ [74]. For more information regarding the Verwey transition the reader is directed to an extensive review by Walz [33] and the latest work by the Attfield group [73; 74]. Interestingly, the transition temperature is known to depend on a variety of factors, including stoichiometry [100], purity and strain. In this author's research group, the Verwey transition temperature of new Fe$_3$O$_4$ single crystals is measured to assess the stoichiometry and purity of samples prior to using them for surface science studies. If a sample is contaminated often no Verwey transition is observed at all.

The earliest photoemission experiments typically utilized cleaved single crystal samples, and reported the opening of a small band gap ≈ 50-70 meV at the Verwey transition temperature [77; 81; 101; 102]. Later, with the advent of metal-oxide surface science, experiments were conducted on (100) and (111) single crystal surfaces cleaned by sputtering/anneal cycles. In this case, no first order transition was observed, rather a slow monotonic increase in onset energy with increasing temperature [77], or sometimes, no change at all [78]. It should be noted though that sometimes only two measurements were made, e.g. at room temperature and liquid $N_2$ temperature, so it is impossible to distinguish whether a first order transition occurred at $T_V$. Interestingly for the surface scientist, these data clearly suggest that the surfaces formed by cleaving and sputter/anneal cycles are different, but since there is precious little in the surface science literature regarding cleaved samples, so the focus here must be on the surfaces prepared *in-situ*.

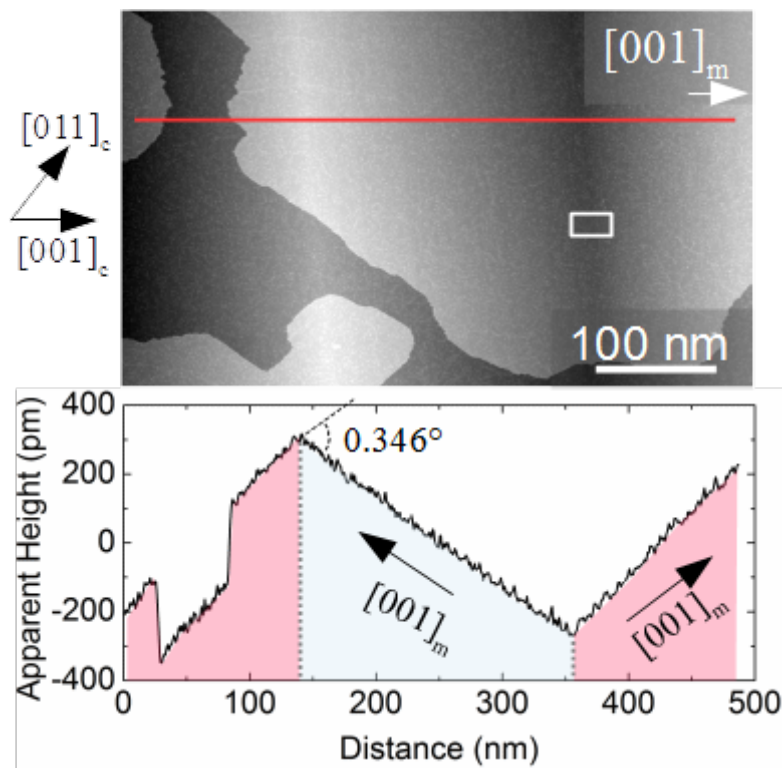

Figure 14: Large scale STM image ($V_{sample}$ = +1V, $I_{tunnel}$ = 0.3 nA) of the $Fe_3O_4$(100) surface acquired at 78 K. The surface exhibits a roof-like distortion due to the formation of the distorted monoclinic twins in the underlying bulk. Figure adapted from ref. [103].

All available evidence suggests that the $Fe_3O_4$(100)-($\sqrt{2}\times\sqrt{2}$)R45° surface does not undergo the Verwey transition with the bulk at 125 K. LEED *IV* measurements reveal that the atomic-scale structure of the $Fe_3O_4$(100)-($\sqrt{2}\times\sqrt{2}$)R45° surface is unchanged [64], although STM and LEEM images show that a long range (500 nm) roof-like structure emerges because the surface layers are buckled slightly by the emergence of the monoclinic twins in the underlying bulk [103] (see Figure 14). SP-LEEM results are consistent with the magnetic easy axis rotating from [110] to [100] in the low temperature phase [103]. Low-energy ion scattering experiments reveal an increase in backscattered $Ar^+$ ions below $T_V$, which is attributed to differences in the neutralization probability and the reduced transparency of the sample in the Verwey phase [104-106]. In scanning tunnelling spectroscopy, Jordan et al. observe a 200 meV gap at both room temperature and 95 K [58]. In 2007, Lodziana described the (√2×√2)R45° reconstruction of $Fe_3O_4$ as a "surface Verwey transition" [69], and suggested that the surface remains in the insulating state well above the bulk Verwey transition temperature. This conclusion was based on DFT+U calculations, which found a stoichiometric bulk termination to exhibit charge and orbital order coupled to lattice strains, which leads to the opening of a band gap in the surface layers. Although the (√2×√2)R45° reconstruction observed in UHV experiments has subsequently been shown to be based on a non-stoichiometric reconstruction unrelated to the Verwey transition [35], a bulk termination, if it could be stabilized by cleaving a sample in UHV for example, could provide an interesting avenue to study charge and orbital ordering with high-resolution surface science methods.

On the (111) and (110) surfaces, there are no comparisons of the atomic-scale structure above and below the Verwey transition, and scanning tunnelling spectroscopy results are contradictory. Jordan et al. [58] observed a small band gap at room temperature and 100 K, as they did for the (100) surface, but Shimizu et al. [107] report metallic behaviour at room temperature, and a small gap at 78 K. In the latter study, the authors propose that the discrepancy arises from surface defects, and that the Jordan et al. result is less reliable because many different surface sites (including defects) were averaged to create the STS I$V$ spectra. For further details regarding the Verwey transition and charge ordering in low dimensions, the reader is referred to the very recent topical review by Bernal-Villamil and Gallego [108].

### 2.2.5 Surface Magnetism

In the simple ferrimagnetic picture $Fe_3O_4$ exhibits a magnetic moment of + 4 $\mu_B$ per formula unit solely due to the $Fe_{oct}^{2+}$ cations. The $Fe_{tet}^{3+}$ (-5 $\mu_B$) and the $Fe_{oct}^{3+}$ (+5 $\mu_B$) cations cancel each other. This compares well with the experimental value of 4.07 $\mu_B$ [109], and provides support for the proposed half-metallic conduction. Since the $Fe^{3+}$ cations are in a high-spin $3d^5$ configuration, one would not expect an orbital moment from these cations, whereas the $Fe_{oct}^{2+}$ cations are in a $3d^6$ configuration with the additional minority spin electron in a band at the Fermi level, and a significant orbital moment is expected [110; 111]. Over recent years experiments have reported large differences in the magnetic moment, and/or no orbital moment at all [109; 112-114]. In 2011 Goering [115] revisited the topic via XMCD and XAS experiments, and concluded that the simple picture of integer $Fe^{3+}$ and $Fe^{2+}$ cations is not reasonable, and that the situation is more band-like. Consequently, the "$Fe^{3+}$" cations will possess a significant orbital moment, although this is not usually measured due to cancelation between the $Fe_{oct}$ and $Fe_{tet}$ sublattices. Note that the idea of non-integer charge states is in line with the prevailing view regarding the nature of the Verwey transition, discussed in Section 2.2.4. It is important to note however, that the surface sensitivity can be significant when the x-ray absorption is measured in the specific case of partial electron yield. Therefore as in the case of half-metallicity, one has to be cautious using surface sensitive techniques to measure bulk properties. This point was illustrated very clearly by the work of Martin-Garcia et al. [116], who used a similar approach to analyse XMCD data acquired from a well-characterised $Fe_3O_4$(100) surface exhibiting the SCV reconstruction [35] (discussed in Section 3.3.3). A reduced magnetic moment of 3.4 $\mu_B$ was measured because all Fe within the outermost four layers are $Fe^{3+}$-like, and the amount of octahedral and tetrahedral cations differs from the bulk. A small orbital moment of 0.3 $\mu_B$ was measured, the origin of which may lie in the "hidden" orbital moment of $Fe^{3+}$ cations.

Magnetism at the $Fe_3O_4$(100) surface has been investigated in detail in recent years by Figuera and co-workers [103; 116-118]. Micrometer-sized domains, imaged in real space using spin-polarized LEEM (see Figure 15) [103; 117], were found to have the easy axes aligned with the <110> directions in the surface plane. This contrasts to the bulk, where the easy axis is <111> [1; 119]. It was concluded that the magnetocrystalline anisotropy is relatively weak in the cubic room-temperature phase, and that the dipolar forces (shape anisotropy) dominate, allowing the easy axes to bend towards the surface and reduce the stray field. This is also thought to explain the curved domain walls observed in Figure 15. Cooling through the Verwey transition, the easy axes align with the local *c* axis of the monoclinic Verwey phase [103].

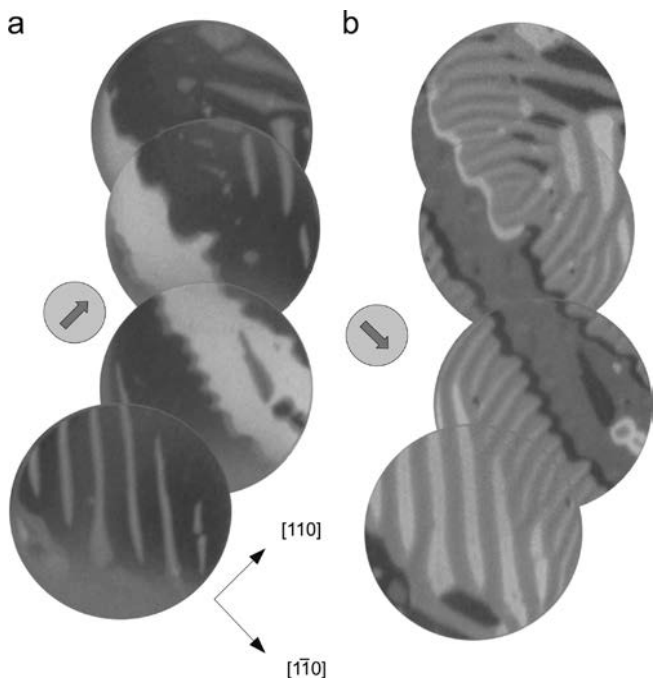

Figure 15: Spin-polarized low energy electron microscopy images of the magnetic structure at the $Fe_3O_4$(100) surface. Each circular image has a field of view of 12 μm. These images reveal domains aligned with the <110> directions. Light/dark contrast reveal magnetic domains aligned with the spin-polarization of the incident electron beam (arrow in figure), which is rotated by 90° between (a) and (b). Reprinted from ref. [117], with permission from Elsevier.

At the (111) surface the situation is conceptually similar to that described above for (100) in that the out of plane easy axes rotate to lie in the surface <112> type directions [120-122]. Interestingly, Monti et al. [123] have shown that thin films of $Fe_3O_4$(111) retain ferrimagentic character down to a thickness of two unit cells, which is important for application of $Fe_3O_4$ in spintronics devices, and subsequently the same group have imaged the magnetic domains using SPLEEM [124]. On the $Fe_3O_4$(110) surface, the bulk easy axes lie in the surface plane [122], and the magnetic domains appear to be terminations of the bulk magnetic structure [125; 126].

## 2.2.6 Bulk Defects

The bulk defect chemistry of iron oxides is crucial to understand the behaviour of the surfaces. Dieckmann and co-workers have studied the bulk defects in $Fe_3O_4$, and $Fe_3O_4$-based solid solutions over many years [127-129], and established that an excess of Fe cations is formed at low oxygen activities, whereas cation vacancies form at high oxygen pressures (see Figure 16 for pressure dependence at 1273 K). Note that the anion lattice remains unchanged throughout.

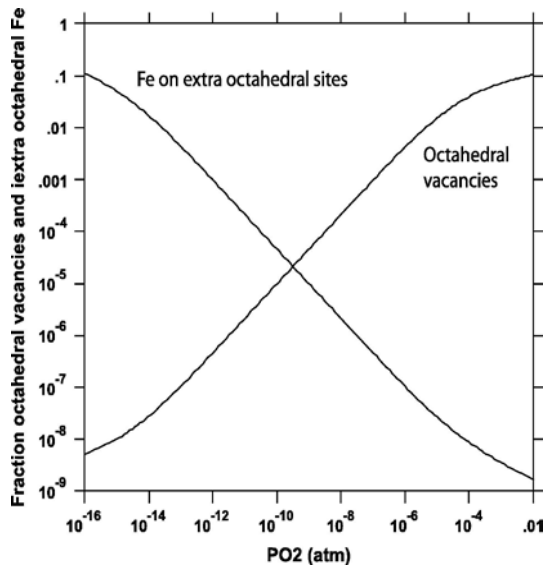

Figure 16: Fraction of octahedral vacancies and Fe interstitials in $Fe_3O_4$ as a function of $O_2$ pressure at 1273 K calculated using Thermo-calc [130] and the thermodynamic data of Sundman [131]. Reprinted from ref. [132], with permission from Elsevier.

Tracer diffusion experiments [133; 134] and Mössbauer spectroscopy [135] reveal that diffusion occurs through vacancy jumps on the normally occupied $Fe_{oct}$ sublattice in oxidising conditions, and through diffusion of between interstitial sites, possibly involving the concerted motion of two atoms, in a reducing environment. Interestingly the diffusion constant goes through a minimum as the dominant mechanism changes (see Figure 17). From such plots, the enthalpy of diffusion is estimated as +630 kJ/mol for the interstitial diffusion, and -140 kJ/mol for the vacancy diffusion. The negative sign appears because the enthalpy of formation of cation vacancies is negative, and thus vacancy diffusion increases with decreasing temperature. Using the barrier for vacancy hopping determined by Dieckmann (137 kJ/mol) and a prefactor of $10^{13}$/s, one can estimate that appreciable bulk diffusion (taken as 1 hop per second) begins at around 550 K. This is in line with STM experiments, where excess Fe atoms deposited on the $Fe_3O_4$(100) surface diffuse into the bulk at around 473-523 K (see section 3.3.4). $^{18}$O tracer diffusion experiments [136] suggest the existence of two types of minority defects; isolated $V_O$s in the presence of Fe interstitials, and $V_O$ clusters in the presence of Fe vacancies. Diffusion in $Fe_3O_4$ is described in some detail a recent book chapter by van Orman and Crispin [137].

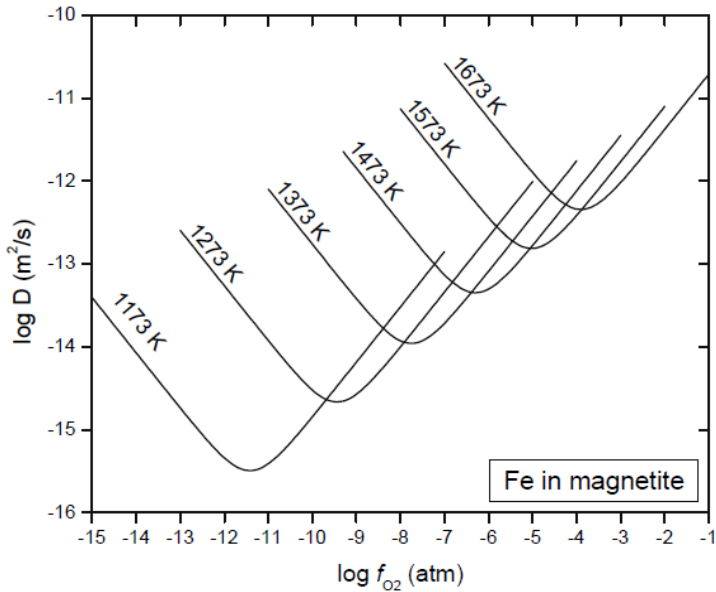

Figure 17: Diffusion coefficient of Fe cations in $Fe_3O_4$ at different temperatures as a function of $O_2$ fugacity. The minimum in each case corresponds to a switch in the diffusion mechanism from vacancy mediated (higher $f_{O2}$) to interstitial mediated (lower $f_{O2}$). Figure reproduced from ref. [137]. Copyright the Mineralogical Society Of America.

Changes to the physical and electronic structure near cation defects in $Fe_3O_4$ have been calculated using a DFT+U approach [138]. Cation vacancies in the octahedral sublattice are preferred, and nearby Fe cations are found to relax toward the defect, while O anions diffuse away. In the iron rich case, occupation of interstitial $Fe_{tet}$ and $Fe_{oct}$ sites are close in energy. The authors conclude that such defects reduce the saturation magnetization of $Fe_3O_4$, but do not significantly modify the density of states in the vicinity of the Fermi level. Similar conclusions were made for $Fe_3O_4$ containing $V_O$s [139]. Consequently, it seems defective $Fe_3O_4$ remains half-metallic, and cannot explain the poor performance of $Fe_3O_4$ based spintronics devices.

## 2.3 Maghemite (γ-$Fe_2O_3$) - Bulk Structure and Properties

γ-$Fe_2O_3$ is most often formed by the oxidation of $Fe_3O_4$, (although it can also be formed by dehydration of Fe hydroxides) and contains only $Fe^{3+}$ cations within the spinel structure. Fe vacancies are thus required to ensure charge neutrality, and the formula can be written $(Fe^{3+}_8)_{tet}[Fe^{3+}_{40/3}V_{Fe\ 8/3}]_{oct}O_{32}$ to illustrate that Fe vacancies occur only on the octahedral sublattice. The lattice parameter of 0.83474 Å is slightly smaller than that of $Fe_3O_4$ (0.8396 Å), and like $Fe_3O_4$, γ-$Fe_2O_3$ is ferrimagnetic. Given the antiferromagnetic alignment of the octahedral and tetrahedral sublattices remains, γ-$Fe_2O_3$ has a net magnetic moment of 2.5 $\mu_B$ per formula unit. The high Curie temperature of ≈ 950 K has seen γ-$Fe_2O_3$ utilized in recording media. The extent of order amongst the cation vacancies has been investigated (e.g. [140-143]) and there is evidence that three different vacancy distributions can occur depending on the size of the crystals [144] and preparation conditions:

(1) cubic structure with random distribution of the vacancies (space group $Fd3m$);
(2) vacancies distributed as the Li cations in $LiFe_5O_8$ (space group $P4_332$) [140]
(3) an ordered distribution of the vacancies with tetragonal symmetry and a three-fold doubling along the $c$-axis (space group $P4_12_12$) [145].

A recent DFT+U study [146] found that ordered Fe vacancies represent the lowest energy configuration, and reproduce the experimentally observed band gap of 2.07 eV (Figure 18). The size of the band gap is different for different spin orientations, which makes γ-Fe$_2$O$_3$ potentially useful as a spin filter in spintronics devices. The top of the valence band is primarily of O 2$p$ character, while the occupied 3$d$ levels of Fe lie around 6–7 eV below the Fermi level. Note this is quite different to the electronic structure of isostructural Fe$_3$O$_4$. The bottom of the conduction band is mainly composed of unoccupied Fe$_{oct}$ states. The authors therefore suggest that maghemite is a charge-transfer type of insulator, and that the first excitation term should correspond to the transfer of electrons from the O$^{2-}$ anions to the octahedral Fe$^{3+}$ cations.

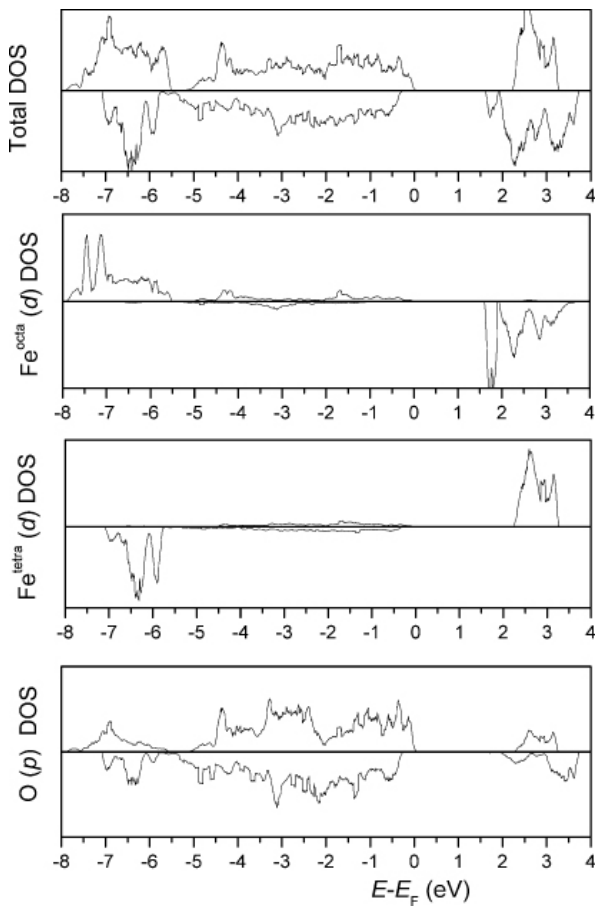

Figure 18: Electronic density of states for the vacancy-ordered structure of γ-Fe$_2$O$_3$, and its projection over Fe 3$d$ and O 2$p$ orbitals. Figure reproduced from ref. [147].

## 2.4  Hematite (α-Fe$_2$O$_3$) - Bulk Structure and Properties

α-Fe$_2$O$_3$ crystallizes in the corundum structure [148], and is isostructural with Al$_2$O$_3$. The unit cell is hexagonal, with a = 0.5034 nm and c = 1.375 nm, and contains 6 formula units of Fe$_2$O$_3$. The structure is most easily seen as a slightly distorted hcp stacking of O$^{2-}$ anions in the c direction (2.29 Å between the planes), with Fe$^{3+}$ cations in two thirds of the octahedral interstices (see Figure 19). Iron cations are each in a high-spin $d^5$ configuration, with spins coupled ferromagnetically in (001) basal planes and antiferromagnetically along the [001] direction. The magnetic moment of per atom is determined to be 4.6 μ$_B$ per atom [149].

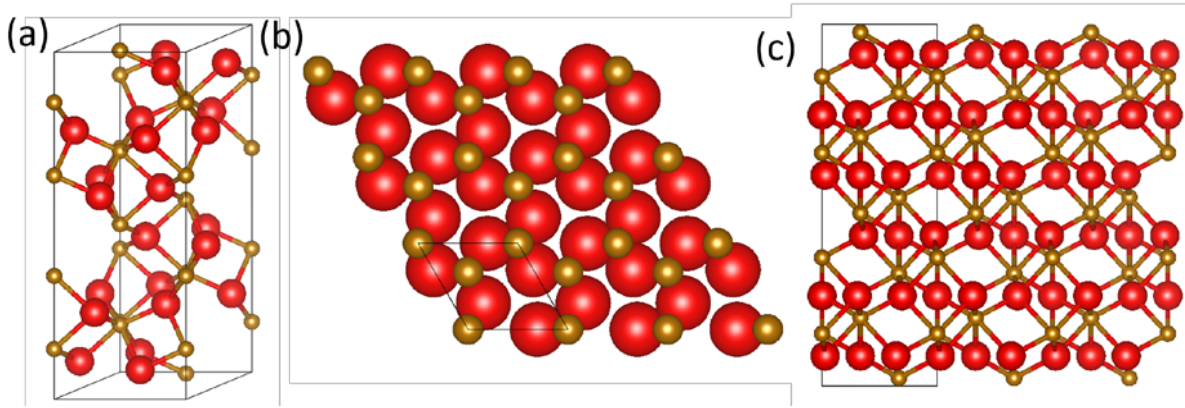

Figure 19: Hematite. (a) The hexagonal unit cell of α-Fe$_2$O$_3$ contains 6 formula units. O$^{2-}$ anions are red, Fe$_{oct}^{3+}$ cations are brown. (b) Fe$_{oct}^{3+}$ cations occupy two thirds of the octahedral interstitial sites between hexagonal close packed O$^{2-}$ planes (drawn in space filling style with unit cell indicated). (c) Side view of the α-Fe$_2$O$_3$ structure that shows how the Fe$_{oct}^{3+}$ cations are not coplanar. Note that α-Fe$_2$O$_3$ is based on a hcp (i.e. ABAB stacking) anion sublattice.

Below the Néel temperature of 955 K hematite exhibits antiferromagnetic ordering, but above the Morin temperature ($T_M$ = 260 K) spin canting between the two Fe$_{oct}^{3+}$ sublattices leads to a weak spontaneous magnetisation [150-152]. Below $T_M$ the magnetic moments align with the [111] direction, and the magnetization vanishes. Conductivity in α-Fe$_2$O$_3$ is poor in the absence of impurities [153], but occurs via a small-polaron hopping mechanism between Fe cations in between the close packed O planes [154; 155]. Consequently conduction in α-Fe$_2$O$_3$ is highly anisotropic, some four times larger in the basal plane than along [0001] [156]. The dielectric response is also anisotropic, with measured dielectric constants ε of 24.1 and 20.6 in the basal plane and [0001] direction, respectively [157].

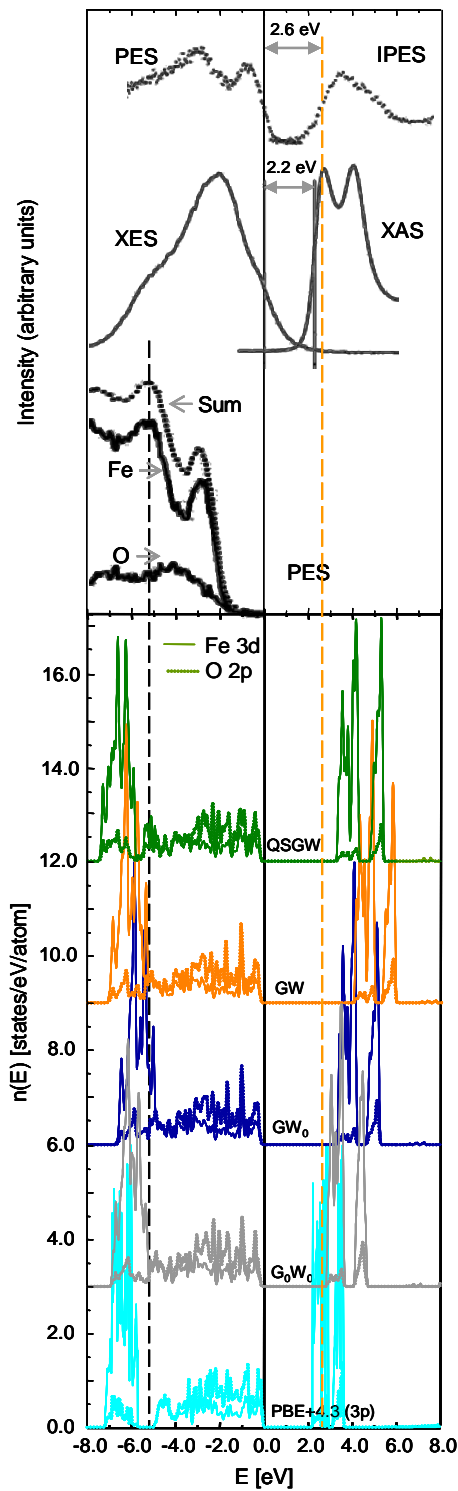

Figure 20: Experimental spectra and PDOS from PBE+U (denoted PBE+4.3) and various GW calculations using the PBE+4.3 input wavefunction. The topmost PES/IPES data is reproduced from ref. [158], the XES/XAS data from ref. [159], and the site-specific PES data from ref. [160]. The PDOS from different calculations are shown, with the Fe 3$d$ states plotted as solid lines, while the O 2$p$ states are plotted as dotted lines. The Fermi energy is set to be zero, highlighted by a black solid vertical line. The black vertical dashed line is used to reference positions of Fe 3$d$ states. The orange vertical dashed line is positioned at 2.6 eV, referenced to the measured PES/IPES gap. Reproduced from Ref [161] with permission of The Royal Society of Chemistry.

Hematite has an indirect band gap of ≈2.2 eV that is of *d-d* origin [159; 162; 163], but a direct optical gap of 2.7 eV [164]. Standard DFT calculations underestimate the band gap by 75 % and predict a magnetic moment per atom of just 3.4 $\mu_B$ per atom, but this can be corrected using a Hubbard *U* [165; 166].  Hybrid DFT can also be used, giving good results with the inclusion of 12 % Hartree-Fock exchange at short ranges [167]. The best agreement with experimental properties was recently found using a many-body Green's function method known as the GW approximation (GWA). In their paper, Liao and Carter [161] plotted the results of experimental measurements of the band gap against the results of several calculations. This plot is shown in Figure 20.

## 2.4.1 Defects and Diffusion

Knowledge of point defects and diffusion in bulk $\alpha$-Fe$_2$O$_{3-\epsilon}$ was reviewed in 1993 by Dieckmann [168], who concluded that oxygen vacancies are the dominant defect between 1373 and 1573 K. The topic has been studied on several occasions since, both experimentally [169] and theoretically [170; 171]. Simultaneous measurements of $^{57}$Fe and $^{18}$O tracer diffusion reveal that oxygen diffusion is faster than that of Fe [169]. In general it seems that both $V_O$ and interstitial Fe can form, but Catlow et al. [172] proposed that electronic disorder far outweighs ionic diffusion and is responsible for electric conductivity. Defective $\alpha$-Fe$_2$O$_3$ prepared under O$_2$-poor conditions has been proposed to possess $V_O$s, which leads to enhanced carrier concentration [173]. However, a recent DFT+U study predicts that $V_{Fe}$ is the most stable defect in oxygen-rich conditions, and that Fe interstitials and $V_O$s are equally probable under oxygen-poor conditions. Since Fe interstitials and vacancies form donor and acceptor states, respectively [170], the authors conclude  that these defects are most likely responsible for charge carriers under ambient conditions.

## 2.4.2  Photoelectrochemical Water Splitting

In recent years much of the interest in $\alpha$-Fe$_2$O$_3$ has been focused on utilizing it as a photoanode material for photoelectrochemical cells [174; 175], so it is useful to discuss the relevant properties here. $\alpha$-Fe$_2$O$_3$ has several advantages for the task of water splitting; it is highly stable in water, cheap, easy to synthesize, and the band gap allows the absorption of visible light. The conduction band edge is too low to allow for the water splitting reaction (2H$_2$O + 2e$^-$ → H$_2$ + 2OH$^-$), so a voltage has to be applied to observe the reaction that is over and above the half-reaction's thermodynamically determined reduction potential (this is known as the over-potential, and is related to the efficiency of the cell) [176]. Consequently there have been many attempts to dope the material with metals to both engineer the band gap and reduce the required over-potential [167; 177-179]. In many cases this improves performance, but in the case of Nb it does not [180]. The valence band edge is more than 1 V lower than the water oxidation potential, but the kinetics of the oxygen evolution reaction (4OH$^-$ → O$_2$ + 4e$^-$ + 2H$_2$O) at the surface are slow, possibly due to the presence of surface states at the hematite/electrolyte interface [175; 181]. A common approach to solve this problem is to introduce catalysts at the interface to speed up the reaction [182-184]. For example, Grätzel and co-workers [185] recently improved the photocurrent to 3 mA/cm$^2$ by introducing IrO$_2$ nanoparticles at the Fe$_2$O$_3$ surface. A second problem is the short hole diffusion length (2-4 nm [186]) and hole mobility of ≈ $10^{-2}$ cm$^2$V$^{-1}$s$^{-1}$, which means that only holes created near the surface participate in the process. However, the penetration depth of light is high, so many of the incident photons create electron-hole pairs that simply recombine. The most common solution to this problem is to produce micro-structured materials (e.g. [187]) with a large surface area. A recent prototype combining Pt doping of the $\alpha$-

Fe$_2$O$_3$, a wormlike morphology, and a Cobalt phosphate co-catalyst achieved a PEC water oxidation activity of 4.32 mA/cm$^2$ at 1.23 V vs. RHE under simulated 1-sun irradiation, 100 mW/cm$^2$, which is 34 % of the maximum theoretical activity [188]. Several materials issues are well known; for example, reduction of the surface to Fe$_3$O$_4$ in a reducing environment results in rapid performance degradation [186].

Clearly though, improving the performance of hematite in water splitting is very much a surface issue, and gaining a better understanding of the structure/function relationship at α-Fe$_2$O$_3$ surfaces is of the upmost importance.

## 2.5 Wüstite (Fe$_{1-x}$O)

Wüstite (Fe$_{1-x}$O) takes the cubic rocksalt structure, with Fe$^{2+}$ cations in octahedral sites. In the (111) direction, the material is made up of alternating planes of fcc O anions and metal cations (see Figure 1). In practice, this compound is always defective, with (1-$x$)-values ranging from 0.83 to 0.95. The defects are known to be Fe$^{3+}$ cations on tetrahedral sites linked to four Fe$_{oct}$ vacancies (so-called Koch-Cohen defects [189]), which further agglomerate into clusters of four Fe$_{tet}$ interstitials linked to 13 Fe$_{oct}$ vacancies. Exactly how the clusters join together is still debated [190; 191], but it is thought to be either along the <100> or <110> type directions [191]. It is interesting to note that a similar motif underlies the SCV reconstruction at the Fe$_3$O$_4$(100) surface [35] (see section 3.3.3). The most common occurrences of this compound are in the Earth's lower mantle [192], but it also forms as an intermediate in the reduction of Fe$_3$O$_4$ and Fe$_2$O$_3$ to Fe. Fe$_{1-x}$O can be stable above 840 K in reducing conditions (see Figure 2), but disproportionates into Fe and Fe$_3$O$_4$ if cooled slowly below 840 K. If the material is quenched rapidly from 840 K, Fe$_{1-x}$O can exist as a metastable phase. There is an ongoing debate if FeO forms in the non-equilibrium low-temperature reduction of iron oxides [193].

Fe$_{1-x}$O exhibits antiferromagnetism below the Néel temperature of 200 K, with the Fe$^{2+}$ magnetic moments aligned parallel to the close packed (111) planes, but in opposite directions from one plane to the next. The Fe defect clusters discussed above are thought to affect the magnetic properties [194; 195].

## 3. Iron Oxide Surfaces

## 3.1 Surface Energies

Natural and synthetic crystals of Fe$_3$O$_4$ are often octahedrally shaped, and enclosed by (111) planes [1], although various different shapes and sizes can be grown, for example, by changing the concentration of KOH and the polyols used when synthesizing Fe$_3$O$_4$ from FeSO$_4$ in solution [196] (see Figure 21). The authors suggest that high levels of KOH lead to the adsorption of OH anions on certain planes.

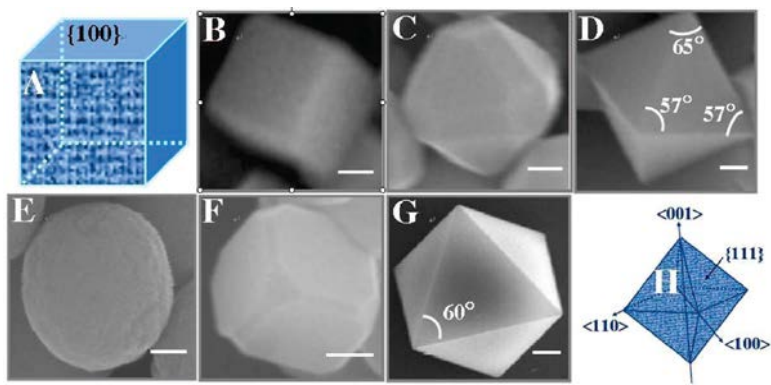

Figure 21: Fe$_3$O$_4$ particles range in shape from cubes enclosed in (100) planes to octahedra with (111) surfaces (B-D), and from spheres to truncated octahedra (E-G) depending on the growth conditions. Reprinted with permission from ref. [196]. Copyright 2008 American Chemical Society

Density functional theory based calculations have been used several times to compare the surface energies of the low-index facets of Fe$_3$O$_4$ [146; 197; 198]. Most recently, Santos-Carballal et al [146] using a DFT+U approach, calculated a Wulff construction enclosed by (001) and (111) facets, which have surface energies of 0.96 and 1.10 Jm$^{-2}$, respectively (see Figure 22). It must be noted, however, that the energies depend on the termination, and that the Fe$_{tet}$ termination of (001) considered does not appear to form on the basis of experimental work. Yang et al. [197], on the other hand, find that the (111) surface is less favourable than the (110) and (100) in their DFT+U calculations, and should be kinetically hindered.

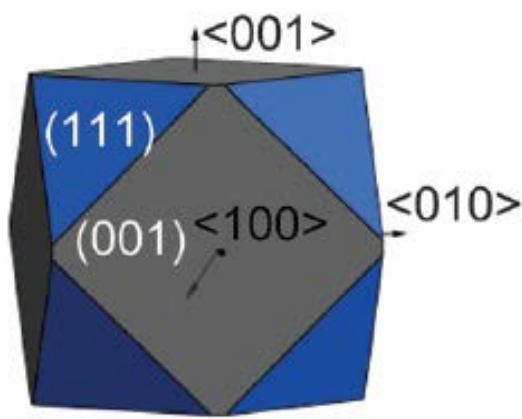

Figure 22: Wulff construction for Fe$_3$O$_4$ based on surface energies derived from DFT+U calculations. Figure adapted from ref. [146]. Published by the PCCP Owner Societies.

α-Fe$_2$O$_3$ particles grown in hydrothermal conditions are found to be either rhombohedral, plate-like or rounded [199; 200]. As with Fe$_3$O$_4$, the shape can be modified by the pH of the respective solution during growth [200]. Calculation of the surface energies (Table 4) of different facets suggest that (012) surfaces (also known as (1$\bar{1}$02) in hexagonal notation) should be commonplace, and that the optimum thermodynamic shape of hematite nanoparticles is a pseudocube (shape (d) in Figure 23) enclosed entirely by (012) surfaces. Such pseudocubes are often observed in nature.

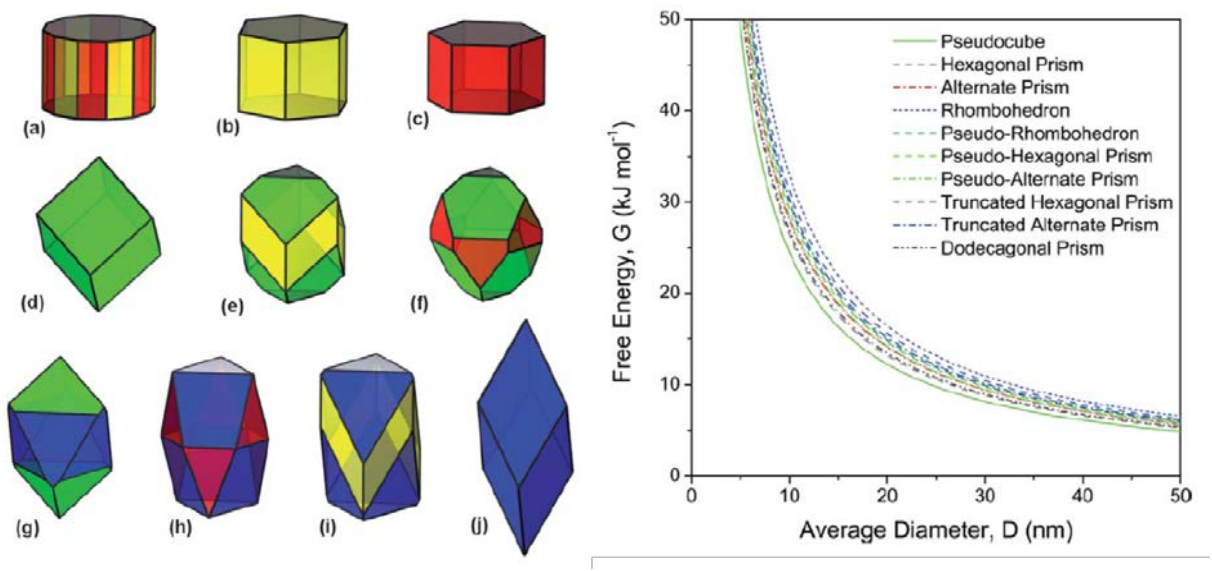

Figure 23: Potential equilibrium shapes of α-$Fe_2O_3$. The surface planes are colour coded such that green is (012), (110) is red, (100) is yellow and blue is (101). (right) Shape (d), the pseudocube, is found to be the most stable shape for all sizes considered using PBE+U calculations [201], followed by the hexagonal prism (shapes a, b, and c). These shapes are commonly observed in experiment [1]. Reproduced from Ref. [201] with permission from the PCCP Owner Societies

Table 4: Calculated surface energies (J/m$^2$) of low-index planes of α-$Fe_2O_3$ for PBE+U [201] and interatomic potential [202] methods.

| Surface | PBE+U [201] | Interatomic potential [202] |
|---|---|---|
| 001-Fe | 1.145 | 2.5 |
| 001-OH | 0.782 | - |
| 012 | 1.056 | 1.956 |
| 100 | 1.369 | 2.251 |
| 101 | 1.306 | 2.844 |
| 110 | 1.230 | 2.329 |

## 3.2   $Fe_3O_4$ - Samples and Preparation

### 3.2.1  Single Crystals

The first surface-science investigations of $Fe_3O_4$ began in the early 1990's. Wiesendanger et al. [203-207] published a series of papers working with mechanically polished natural $Fe_3O_4$(100) single crystals rinsed in ethanol. In-situ cleaving was reported to result in clean, but rough surfaces, but no further details of the cleaved surfaces were published. The surface termination of the polished crystals was found to vary depending on the method of in-situ preparation. Annealing to 810 K resulted in a termination at the $Fe_{oct}$-O plane, evidenced by the observation of rows of protrusions linked to the surface $Fe_{oct}$ rows that are separated by 6 Å and rotate by 90° with each 2.1 Å step in STM images. Note that surface O atoms are not imaged because the O 2$p$ states are not in the vicinity of the Fermi level [24; 60; 208] (despite one claim to the contrary [209]). When the same sample was prepared by sputter/anneal cycles a different termination was observed (which we can now identify as the so-called "Fe-dimer" termination (see section 3.3.4) [68; 210; 211]. Fe-rich terminations can occur after sputter/anneal cycles because O anions are preferentially removed during $Ar^+$ sputtering [212]. Studies using single crystals in this author's laboratory have found that the $Fe_{oct}$-O termination is recovered if a stoichiometric $Fe_3O_4$(100) single crystal, synthetic or natural, is annealed in $10^{-7} – 10^{-5}$ mbar $O_2$ at approximately 900 K in the final stage of sample preparation. Figuera and co-workers [213] have shown that during such an annealing cycle $O_2$ reacts with Fe supplied from the bulk, resulting in hundreds of layers of virgin $Fe_3O_4$(100) surface. The resulting surface is thus clean, flat and well ordered. The price for this reproducibility is that repeated oxidation of an $Fe_3O_4$(100) sample over many months results in $Fe_2O_3$ inclusions, which grow along the (110) directions (see Figure 24) [214]. Furthermore, an initially flat $Fe_3O_4$(100) surface subjected to many sputter/anneal cycles in UHV shows many small pyramids. These features can be found in STM, and appear as many steps close together, with a large (100) plateau on top. These features appear to grow as $Fe_3O_4$ layers grow around a screw dislocation when annealing in $O_2$. A stunning LEEM movie of this process was published as supplementary information with ref. [213], two frames of which are shown in Figure 25. An important consequence of this growth process is that annealing the surface in $^{18}O$ should result in a completely isotopically labelled surface, which can be very useful in IRAS experiments [70].

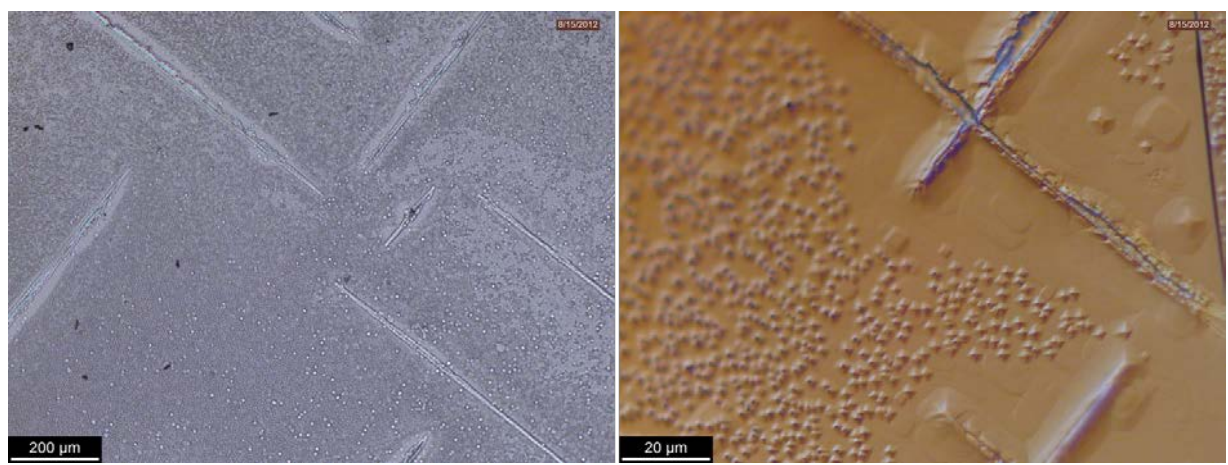

Figure 24: Optical microscopy images of a synthetic $Fe_3O_4$(100) sample that was visibly matte after being exposed to several hundred $Ar^+$ sputter/ $O_2$ anneal cycles over the course of a two year period. The lines are hematite inclusions (also visible to the eye as a hatched appearance on the sample surface) aligned with the <110> directions. Images provided by Dr. Juan de la Figuera, Madrid.

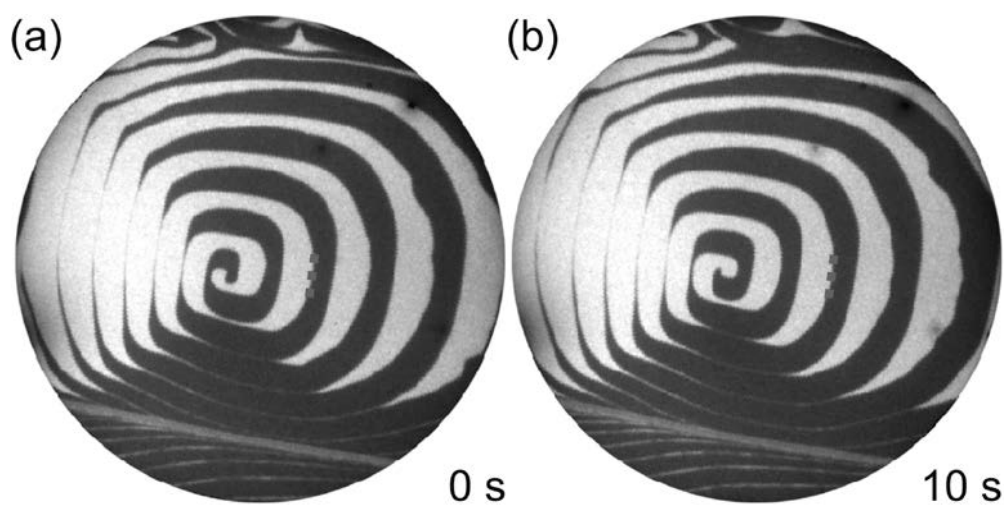

Figure 25: Dark field LEEM images extracted from a movie showing the growth of $Fe_3O_4$(100) layers around a screw dislocation during oxidation of a single crystal sample in $10^{-6}$ mbar $O_2$ at 900 K. Black and white contrast are neighbouring $Fe_3O_4$(001) terraces. Field of view in the images is 20 μm. Figure adapted from ref. [213].

Work has also been performed on $Fe_3O_4$(111) single crystals, but reproducible preparation seems more difficult because two different terminations are close in energy in the chemical potential regime accessed in UHV conditions. After many cleaning cycles the surface can become reduced, and oxygen annealing is required to recover a stoichiometric surface. Most work on this surface has been performed on thin film samples grown on Pt(111) (see next section). A small number of studies have been performed on $Fe_3O_4$(110) single crystals, and the resulting surface is similar to thin films grown on MgO(110) (see section 3.5).

High-quality natural $Fe_3O_4$ single crystals are available from several vendors such as Surface Preparation Labs, CrysTec, MaTeck and Surfacenet in Europe, and Princeton Scientific, Commercial Crystal Labs, and MTI in the USA. Such samples can exhibit trace contamination by elements such as K and Ca, but these segregate and can be removed by sputter/anneal cycles [215; 216]. Note that $Fe_3O_4$ single crystals can be quite brittle, and sometimes crack during mounting, particularly if cracks and/or grain boundaries are already present in the sample. Undoubtedly the best option for single crystal work is to use a stoichiometric synthetic sample. Fortunately, because studies of the Verwey transition phenomena requires stoichiometric high-quality samples, growth recipes are extremely well developed [33; 217]. Synthetic single crystals used in this author's work were grown by the group of Prof. Mao at Tulane University using the floating-zone approach.

### 3.2.2 Thin Film Growth

An alternative way to prepare iron oxide surfaces is to grow a thin film on a lattice-matched substrate [218]. This approach was used extensively by Weiss and Ranke [17] to compare the structure and reactivity of FeO(111), $Fe_3O_4$(111) and α-$Fe_2O_3$(0001) films on Pt(111). The major benefit of this approach is that a fresh iron oxide film is grown for each experiment, so the surface is less likely to depend on sample history, although Fe does diffuse into the Pt substrate over time, and eventually an FeO(111) thin film can be prepared simply by oxidising the Pt(111).

Thin films have also been utilized to study the $Fe_3O_4$(100) surface. The best procedures for growing $Fe_3O_4$(100) thin films using MBE, and results up to the year 2000 were reviewed by Chambers [219]. Most studies have utilized MgO(100) [220] as a substrate because the lattice constant is almost exactly half that of $Fe_3O_4$. However, because $Fe_3O_4$ islands nucleate with equal probability with two distinct registries with respect to this substrate, many bulk domain boundaries form, and these are known to dominate properties such as magnetoresistance. Typically the growth is performed by depositing Fe in an $O_2$ environment, often using a plasma source to activate the $O_2$ [210; 221-227]. It was noticed early on that Mg tends to diffuse into the film even at the moderate growth temperature of 450 °C [221; 222; 228; 229], as might be expected given that $MgFe_2O_4$ is a well-known spinel compound. The Chambers group solved this issue by first growing an intentionally inter-diffused $Mg_xFe_{3-x}O_4$ film as a buffer, and then growing a pure $Fe_3O_4$(100) film at 250 °C [224; 225]. An alternative is to first grow a 200 Å buffer layer of Fe(100) directly on the MgO substrate, which blocks Mg out-diffusion [210], but there are reports of poor structural quality [230] and the surfaces often exhibit Fe-rich terminations [210]. Utilizing a Pt(100) substrate, Davis et al. [211] found a series of Fe-rich terminations that could be tuned via the thickness of the Fe buffer layer. These studies serve to illustrate that the termination of $Fe_3O_4$ surfaces strongly depend on the stoichiometry of the sample, as well as the chemical potential during preparation.

Recently, $Fe_3O_4$(100) thin films have been grown using PLD on a variety of substrates including $SrTiO_3$(100), Si(100), MgO(100), and ZnO(0001) [118; 231-235], as well as by RF magnetron sputtering [236].

### 3.2.3 Oxidation of Fe Single Crystals

The oxidation of Fe has been studied for many years from a surface-science perspective. Fromhold and Cook [237] proposed that the oxidation of metals proceeds via two coupled currents; metal ions flow from the metal to the surface creating a field that aids Schottky emission of electrons. The validity of this approach was confirmed for the oxidation of Fe(100) [238], and makes sense particularly given the defect chemistry of the iron oxides discussed above. However, two recent theory papers [239; 240] suggest that the initial oxidation proceeds through oxygen diffusion to the iron/iron-oxide interface, before iron diffusion takes over once $Fe_3O_4$ or $Fe_2O_3$-like phases are formed. Rather than attempt to review the whole field, which is well beyond the scope of this review, I will highlight a few recent studies on the low-index surfaces, and the interested reader is directed to references therein. However, it is important to note that the oxidation of Fe is somewhat complex, and the surface phases formed strongly depend on the conditions. Consequently, this approach is not generally used to generate a model system for surface science studies of iron oxides.

The initial adsorption of $O_2$ on Fe results in ordered phases on Fe(110) [241; 242] and Fe(100) [243; 244] at room temperature, but oxide formation occurs rapidly on the more open Fe(111) surface [245]. On Fe(100), recent studies show that deposition of several Langmuir of oxygen at room temperature induces the formation of two-ML thick FeO-like islands occupying about 30-40 % of the surface area. Sample annealing induces surface flattening and the formation of a well-ordered p(1×1) structure, in which a complete ML of oxygen atoms resides in hollow sites.

On Fe(110) the most recent computational results suggest O atoms adsorb in pseudo-threefold hollow sites [242]. At low temperatures, the oxide surfaces produced through $O_2$ exposure are not well ordered, but at 673 K several different phases have been observed as a function of $O_2$ exposure.

The oxidation of Fe(110) has recently been studied using a host of complementary surface science techniques (LEED, XPS, NEXAFS, STM, and STS) [246; 247]. At the lowest coverage a zigzag structure was observed in STM, which transformed into $Fe_{1-x}O(111)$-like layers, as observed previously [248]. On further oxidation the growth mode changed and 3D triangular islands of $Fe_3O_4(111)$ nucleated and grew. Such behaviour is reminiscent of thin films of Fe oxides grown by oxidation of Fe deposited on other metal supports [17; 249].

The oxidation of Fe(111) was studied by Qin et al. [245], and recently revisited [250]. In the former study, oxidation was rapid, with both $Fe_3O_4$ and $Fe_2O_3$ forming at 300 K but predominantly $Fe_3O_4$ at 500 K. The latter study was a comprehensive surface science investigation utilizing AES, LEED, XPS, ISS and STM. The oxidation was also found to be very fast, even at 200 K, and a thick film formed in $50 \times 10^{-6}$ mbar·s $O_2$. At high pressures, the oxidised surface is somewhat disordered and appeared to be $Fe_2O_3$-like. Annealing this surface at 773 K resulted in the desorption of much oxygen, and a significant change in the morphology, with large islands residing over the film with (6×6) periodicity. At present, the structure of the (6×6) phase is not known.

## 3.3 $Fe_3O_4(100)$ - The $Fe_{oct}$-O (B-Layer) Termination

### 3.3.1 Early Models

The earliest structural model of the $Fe_3O_4(100)$ surface was proposed by Wiesendanger and co-workers based on STM experiments using a magnetic Cr tip. Every second pair of Fe atoms along the row direction were observed to be brighter, which was interpreted in terms of a magnetic contrast and a static bimodal ($Fe^{2+}$, $Fe^{2+}$, $Fe^{3+}$ $Fe^{3+}$...) charge order in the surface layer [205]. Subsequent work by the Shvets group using antiferromagnetic tips saw similar contrast [251; 252], and also proposed the (√2×√2)R45° reconstruction results from the charge order present within an ideal bulk terminated surface. However, this model does not appear to be under active consideration today, partly because such a termination is polar and should be unstable. In the (100) direction, $Fe_3O_4$ consists of evenly spaced planes of $Fe_{tet}$ and $Fe_{oct}$-O atoms, and each plane has a net charge of +/- 6 electrons (Figure 26). This means the surface region must possess a net charge of +/- 3 electrons to compensate the internal field [253; 254], either through reconstruction and/or electronic reconfiguration (i.e. changing the charge on the atoms from those of the equivalent bulk atoms).

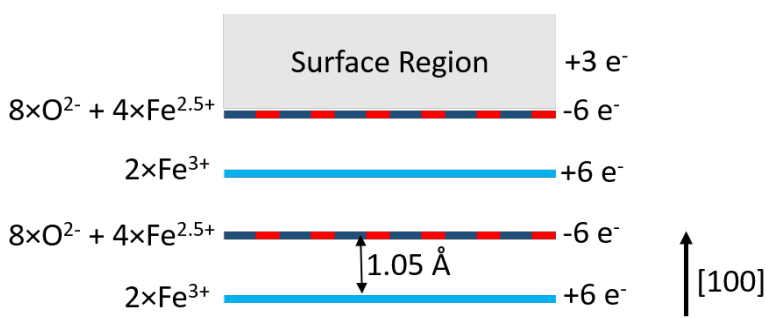

Figure 26: Polarity compensation at the $Fe_3O_4(100)$ surface. In the (100) direction, $Fe_3O_4$ is made up of alternating layers containing two $Fe_{tet}^{3+}$ cations (A layer) and a plane comprising eight $O^{2-}$ anions, two $Fe_{oct}^{3+}$ cations and $Fe_{oct}^{2+}$ cations (B layer). Thus, the A and B layers have a net charge of +/-6 electrons, and $Fe_3O_4$ is a polar surface. Compensation requires that the surface region has a net charge of +3 electrons.

The simplest realisation of the (√2×√2)R45° reconstruction observed in LEED (Figure 27a) satisfying charge compensation would be a termination containing one $Fe_{tet}^{3+}$ atom per unit cell, but this is not observed (Figure 27b). An alternative [255] [226] would be a termination containing 1 $V_O$ per unit cell, together with a slight adjustment of the charge on the surface $Fe_{oct}$ cations to create a surface layer with a net charge of -3 electrons (Figure 27c). Such a surface would be expected to be highly reactive to water [256], and this has not observed in experiment [257]. In what follows the recent models of the $Fe_3O_4$(100) surface are discussed.

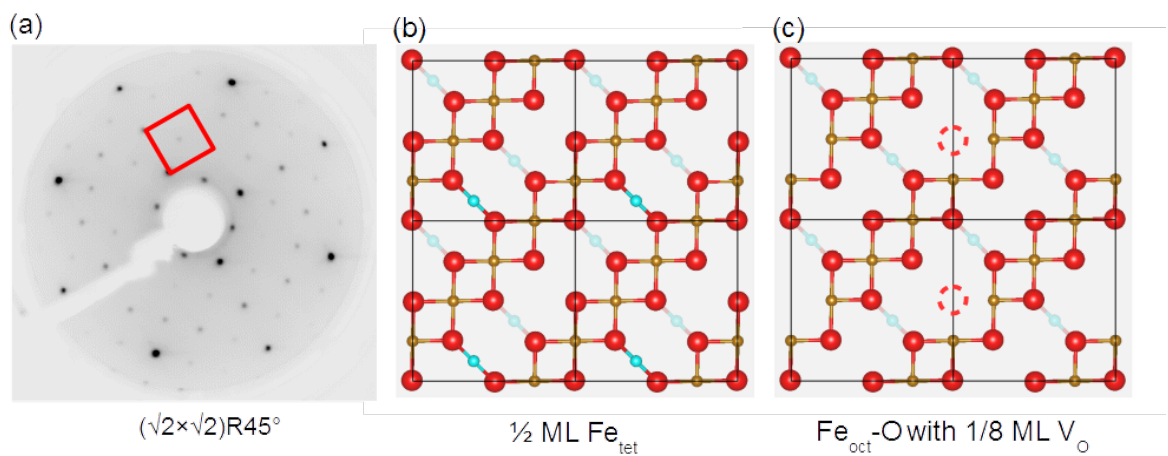

Figure 27: (a) LEED pattern from the $Fe_3O_4$(001)-(√2×√2)R45° surface with an electron energy of 90 eV. The (1×1) unit cell is marked in red. (b) The ½ ML $Fe_{tet}$ termination and (c) the 1/8 ML $V_O$ termination. $Fe_{oct}$ atoms are gold, $Fe_{tet}$ cyan, and $O^{2-}$ anions are red. $V_O$s are marked by red dashed spheres. These early models have been superseded by the SCV reconstruction (Section 3.3.3).

### 3.3.2 The Distorted Bulk Truncation (DBT)

Starting in 2005, Pentcheva and co-workers studied the $Fe_3O_4$(100) surface using a variety of methods and proposed the (√2╳√2)R45° reconstruction to result from a lattice distortion coupled to subsurface charge order [64; 71; 258] (see Figure 28a). Lodziana [69] came to a similar conclusion using DFT+U calculations, and proposed that the reconstruction could be understood in terms of a surface Verwey transition. The proposed charge order opens a small (0.2 eV) band gap in DFT+U calculations, consistent with STS measurements [58]. However, again it was not clear how such a bulk terminated model can be reconciled with polarity compensation criteria, but the structure was widely accepted because it reproduced qualitatively the undulating $Fe_{oct}$ rows observed in STM images, and because both LEED *IV* and SXRD measurements were performed and seemed to confirm the proposed structure. However, the Pendry R-factor for the LEED investigation was poor ($R_P$ = 0.34) [64], and the SXRD results merely demonstrated that the bulk termination was a better fit to the data than several other candidate structures [258].

### 3.3.3 The Subsurface Cation Vacancy (SCV) Structure

The first sign of a severe problem with the DBT model came with the study of metal adsorption on the $Fe_3O_4$(001) surface. Au adatoms, deposited at room temperature, were observed to bind exclusively in just one of the two available bulk continuation $Fe_{tet}$ sites per (√2×√2)R45° unit cell [259]. This observation is highly surprising given the subtlety of the lattice relaxations in the DBT model, and was initially linked to the proposed subsurface charge order [259]. However, DFT+U calculations could not reproduce the observed site preference with a bulk terminated $Fe_3O_4$(001) structure [35]. A major breakthrough came with the study of Co adsorption (see section 5.2.1) [260]; Co adsorbs as an adatom at room temperature with the same site preference as Au, but rapidly incorporates within the surface lattice. Clearly, such facile incorporation suggests the presence of vacancies in the cation lattice. Since all the surface cations are observed in STM images, it was concluded that the vacancies must be in the subsurface layers. Moreover, when 1 Co atom per unit cell was deposited on the surface, the (√2×√2)R45° reconstruction spots disappear and a (1×1) LEED pattern is observed. This strongly suggests that Co fills cation vacancies present in the subsurface layers leading to the recovery of a bulk-truncation. Based on these observations, the SCV model of the (√2×√2)R45° reconstruction was developed.

The SCV model of the (√2×√2)R45° reconstruction [35], published in 2014, is consistent with all experimental observations to date. Crucially, the Pendry R-factor ($R_P$) achieved in LEED experiments is 0.125. This level of agreement between experimental and theoretical curves is on par with that achieved for clean metal surfaces, and provides a high level of confidence in the proposed structure. For comparison, the prior LEED *IV* study reported a $R_P$ = 0.36 for the DBT structure, citing a high number of defects for the poor agreement. However, when this same experimental data is used with the SCV model, an $R_P$ of 0.13 is achieved [35]. Consequently, we can be sure that the same surface was measured in both experiments, despite the different samples, preparation conditions and measurement temperatures employed. In the SCV model the surface $Fe_{oct}$-O layer remains stoichiometric, but is distorted by a rearrangement of the cations in the subsurface layers. Specifically, an additional interstitial $Fe_{tet}$ atom ($Fe_{int}$) in the second layer replaces two $Fe_{oct}$ atoms from the third layer Figure 28b), a net reduction of one Fe atom per unit cell. The presence of $Fe_{int}$ in the (S — 1) layer blocks adsorption of metal adatoms in one of the bulk continuation $Fe_{tet}$ sites, consistent with experimental observations for metal adsorption mentioned above [35; 68; 98; 212; 257; 259; 261; 262]. Interestingly, the replacement of $Fe_{oct}^{2+}$ cations by an $Fe_{tet}^{3+}$ cation is the hallmark of Koch-Cohen defects in bulk $Fe_{1-x}O$ [189]. DFT+U calculations [35] also reveal that the SCV structure is thermodynamically more stable than the DBT structure over the entire range of oxygen chemical potentials accessible in UHV. A subsequent, as yet unpublished, SXRD study confirms the LEED structure. The structure can be obtained in .cif format from ref. [263].

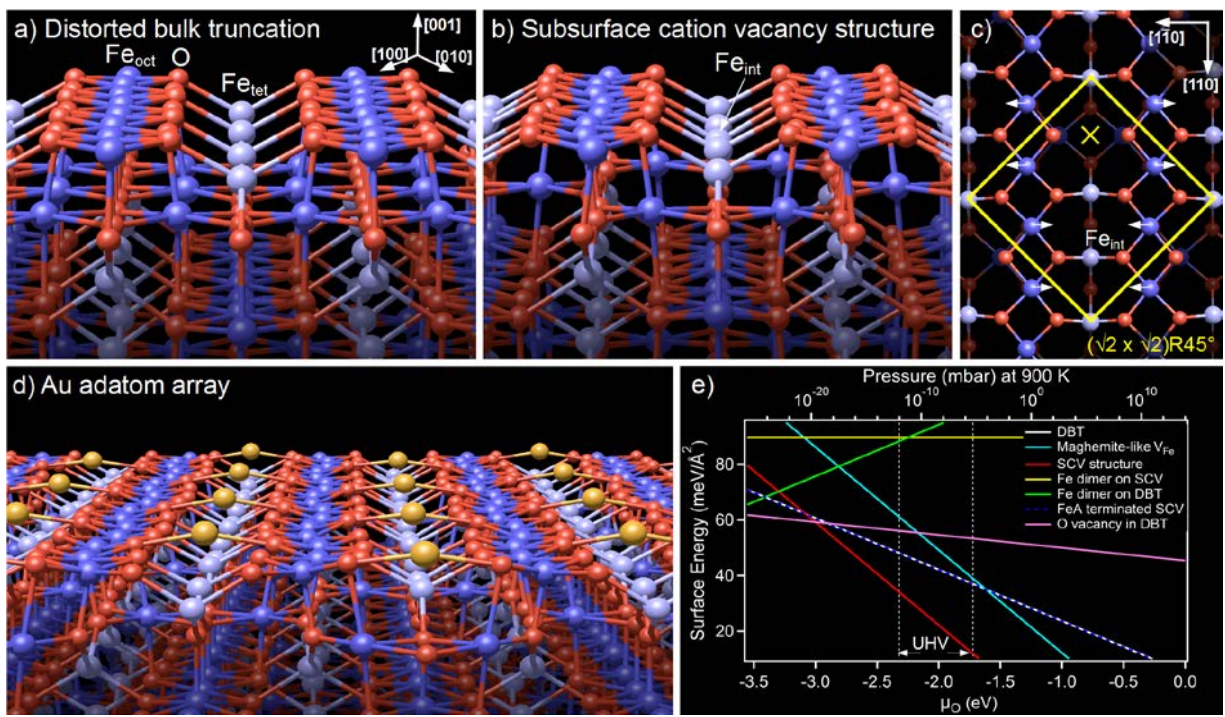

Figure 28: The DBT (a) and SCV (b) models of $Fe_3O_4$(100) determined by DFT+U calculations. (c) Top view of the SCV reconstruction. The × shows the preferred "not-blocked" adsorption site for metals, as calculated for Au in (d). (e) The theoretical surface phase diagram has been updated in comparison to that published in ref. [35] to include various reduced surfaces. The SCV reconstruction is red, and a "maghemite-like" surface with a subsurface $Fe_{oct}$ vacancy per unit cell is cyan. The DBT (white) and $Fe_{tet}$ terminated SCV reconstruction (blue) have the same stoichiometry and are indistinguishable in energy (dashed line). Finally, a surface with one $V_O$ per unit cell (pink) is predicted to be favourable in reducing conditions, but has yet to be realised experimentally. Two configurations for the Fe-dimer surfaces were calculated with two additional Fe atoms on the SCV (yellow) and DBT reconstruction (green). Redrawn after ref. [35]. Reprinted with permission from AAAS.

The SCV reconstruction can be rationalized as an oxidised skin that forms because the oxygen chemical potential in UHV experiments is close to the border of stability of the bulk $Fe_3O_4$ and $Fe_2O_3$ phases [35]. The outermost unit cell has an intermediate stoichiometry, $Fe_{11}O_{16}$ and DFT+U calculations predict that the outermost unit cell contains $Fe^{3+}$ cations only, consistent with XPS measurements [35]. The replacement of $Fe_{oct}$ by $Fe_{tet}$ cations underlies the well-known Koch-Cohen defects in $Fe_{1-x}O$, which result from oxidation of the compound. Interestingly, the DFT+U calculations also predict that polarity compensation is ultimately achieved in the SCV structure when two O atoms in the surface layer have a magnetic moment, which would be consistent with a -1 oxidation state. The two O atoms are those without a subsurface $Fe_{tet}$ neighbour, i.e. those to which metal atoms strongly bind in adsorption experiments [35]. All other O atoms in the surface region are $O^{2-}$. If we take these calculated oxidation states for the atoms we find that the outermost unit cell possesses a net charge of -3 electrons, exactly what is required to compensate the polarity in the [001] direction. It will be fascinating if this prediction is verified experimentally as $O^{1-}$ is exceedingly rare in highly stable oxide compounds.

STM images of the $Fe_3O_4$(100) surface depend strongly on the applied bias, and the tip condition (Figure 29). For tunneling voltages below ≈1.5 V empty states, tunneling is dominated by distorted $t_{2g}$ orbitals, and the protrusions imaged in STM are not centred above the atom cores. At higher bias (>1.5 V) tunneling from $e_g$ states contributes and the rows become much straighter, and at 3 V the contrast reflects the topography. It must be noted however, that tunneling at high bias typically results in tip instability, and can influence the surface in unwanted ways (e.g. desorb OH groups). STM images acquired at negative sample bias (imaging filled states) also show the surface $Fe_{oct}$ atoms, and appear similar to those acquired in empty states. However, this group's experience is that the STM tip is generally less stable in this imaging condition. Figure 30 shows recently acquired (as yet unpublished) images acquired using a low temperature STM/AFM based on the qPlus sensor. This allows simultaneous acquisition of images based on the force and tunnelling current. Clearly, the STM image (0.4 V empty states) exhibits the typical undulating protrusions, while the AFM channel shows the $Fe_{oct}$ rows to be much straighter.

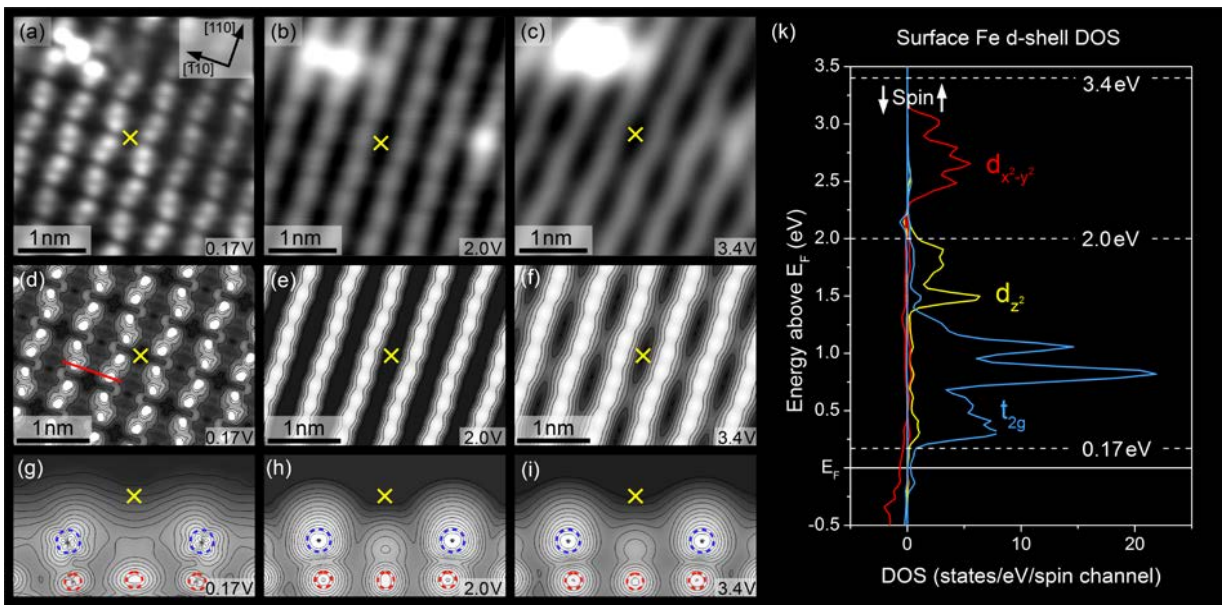

Figure 29: (a-c) Empty states STM images of the $Fe_3O_4$(100) surface strongly depend on the applied bias voltage. STM simulations (d-f) and electron density plots reveal that images below 1.5 V are dominated by tunnelling into distorted $t_{2g}$ orbitals, see DOS plot in (k) while images above 2V are more representative of the location of the atom cores. Figure adapted from [35]. Reprinted with permission from AAAS.

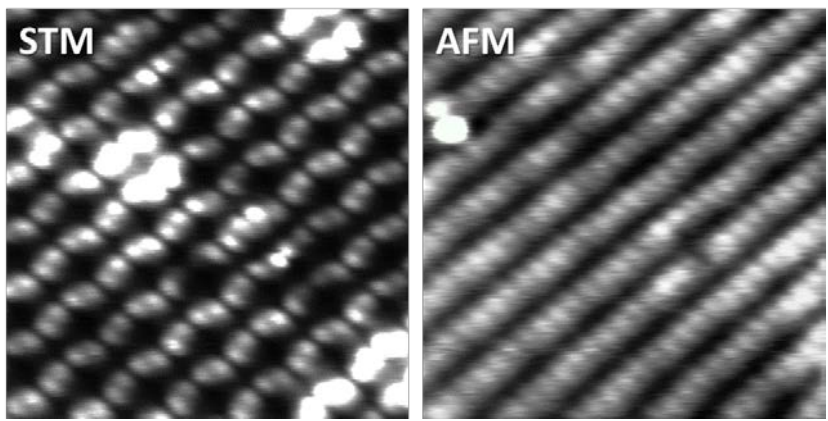

Figure 30: Simultaneously acquired STM (left) and AFM (right) images of the $Fe_3O_4$(100)-($\sqrt{2}\times\sqrt{2}$)R45° surface at 78 K. The images were acquired in constant height mode with $V_{sample}$ = 0.4 V, quartz tuning

fork k=1800 N/m, and an oscillation amplitude A = 0.22 nm. Images provided prior to publication by Dr. Martin Setvin, Institute for Applied Physics, TU Vienna.

In a recent STM study, Spiridis et al. [264] claimed that imaging at very low bias allows to see through the surface layer and image the $Fe_{oct}$ atoms in the second layer directly. It is not clear why this phenomenon should occur, and no justification for the layer-selective tunnelling was provided. Such an imaging mode could not be reproduced in this author's laboratory, but we note that very low tunnelling bias leads to a highly unstable tip due to the lack of DOS near $E_F$. Indeed, the image shown in Figure 29a (0.17 V bias) was acquired in constant-height mode because constant-current was highly unstable. An alternative explanation for the images of Spiridis and co-workers could be that changes to the STM tip lead to imaging of a neighbouring terrace, where the $Fe_{oct}$ rows are rotated by 90 °.

The discovery of the SCV structure requires the re-evaluation of previous studies where conclusions were drawn on the basis of the distorted bulk truncation model. In what follows potential problems will be highlighted, and possible re-evaluation offered where possible.

### 3.3.4  High-Temperature (1×1) Transition

Bartelt et al. [265] observed that the (√2×√2)R45° reconstruction is reversibly lifted when a natural $Fe_3O_4$(100) sample is heated above ≈ 450 °C, and found similar results for a thin film grown on $SrTiO_3$. The amplitude of the reconstruction spots decreased in LEED and the spots broadened significantly through the transition until a (1×1) pattern was visible above 488 °C (Figure 31). The authors report that the transition temperature was independent of $O_2$ partial pressure, and no significant change in the surface morphology was observed in LEEM. Both findings suggest no change in surface stoichiometry. Based on their observations a 2[nd] order Ising-like transition was proposed, with disorder in the subsurface charge-order pattern as the order parameter. With the SCV model of the (√2×√2)R45° reconstruction, however, an analogous transition to (1×1) requires rearrangement of the subsurface cations such that the vacancy/interstitial combination becomes disordered over the long range, and at present the mechanism of the transition is not known.

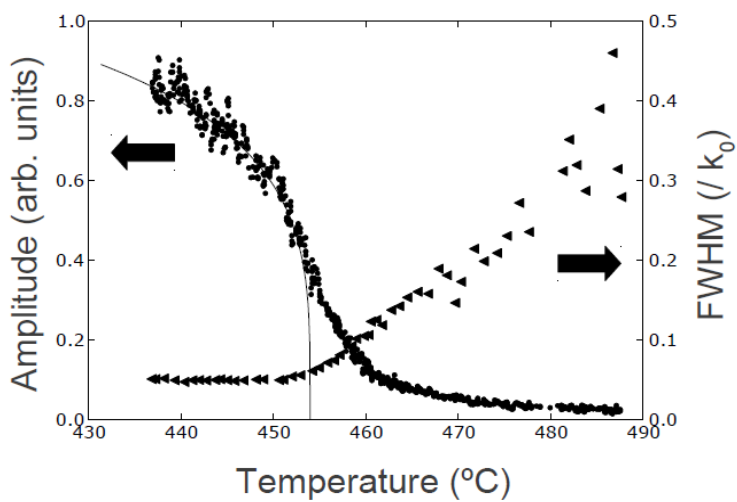

Figure 31: Amplitude and FWHM of the (√2×√2)R45° diffraction spot as a function of temperature. Reprinted figure with permission from [265]. Copyright 2013 by the American Physical Society.

### 3.3.5 Surface Defects and Impurities

### 3.3.5.1 Surface Hydroxyl Groups

The most common defects observed on a freshly prepared (√2×√2)R45°-Fe$_3$O$_4$(100) surface are shown in Figure 32. Bright protrusions on the Fe rows (red arrow) observed in empty states STM images are due to surface OH groups [98]. These are easily distinguished because they exhibit a characteristic diffusion at room temperature; hopping backward and forward between two Fe$_{oct}$ rows. This behaviour occurs because the H atom binds to one of the O atoms in the "not-blocked" site (i.e. where there is no second layer Fe$_{tet}$ neighbour) and forms a hydrogen bond to the other symmetrically equivalent O atom (Figure 32b). The OH itself is not directly observed in STM, but the density of states of the neighbouring Fe$_{oct}$ atoms is modified such that they appear brighter in empty-states images (discussed in the section on Hydrogen adsorption). When the H atom diffuses to the opposite O atom, the bright Fe$_{oct}$ contrast is transferred to the opposite row (Figure 32c). Typically two such featured are observed in neighbouring (√2×√2) unit cells following sample preparation because H$_2$O from the residual gas reacts with surface oxygen vacancies, filling the vacancy and depositing two H atoms onto the surface. While the DFT+U calculations underlying this assignment were performed for a bulk terminated structure [98], very little is different in the vicinity of the adsorbed H atom so the conclusions regarding the adsorption geometry and the origin of contrast change in STM are most likely valid.

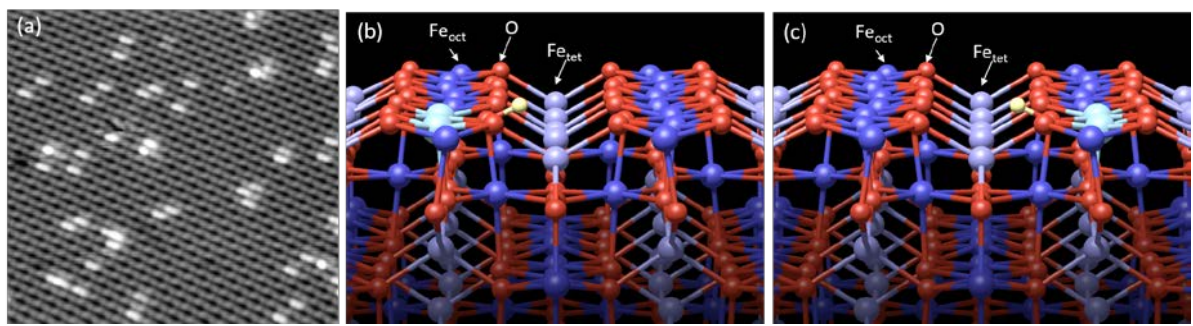

Figure 32: STM image of the clean Fe$_3$O$_4$(100) surface following exposure to 0.06 L water. Pairs of OH groups in neighbouring unit cells are formed through dissociative adsorption in V$_O$s. The H atom is not directly imaged, but the neighbouring Fe atoms become brighter in empty states images (cyan in the schematic). An adsorbed H atom binds to surface O in the "not-blocked" site, but can easily diffuse to the symmetrically equivalent O atom on the opposite row (compare b,c) resulting in a characteristic hopping between Fe$_{oct}$ rows observable in STM movies .

### 3.3.5.2 Other Defects

When a Fe$_3$O$_4$ single crystal is introduced into a vacuum chamber from the ambient environment XPS spectra show the surface to be hydroxylated, with significant carbon contamination. LEED shows a (1×1) pattern, but a (√2×√2)R45° pattern can be obtained already simply be heating above 500 K. To achieve a clean surface requires sputter/anneal cycles. With natural samples, elements commonly found in rocks (K, Na etc.) often segregate during the initial cycles making ordered surface phases [215; 216].

In Figure 33 two STM images of a defective $Fe_3O_4$(100) surfaces are shown. These images were selected both to show the common defects, but also to give an indication of the most common resolution obtained in STM images in our experiments. In a typical image, pairs of $Fe_{oct}$ atoms along the row are imaged as a single protrusion, which tends to accentuate the undulating appearance of the $Fe_{oct}$ rows. A nice way to assess the quality of the STM tip on $Fe_3O_4$(100) is to compare the resolution of the $Fe_{oct}$ rows on neighbouring terraces, since the rows rotate by 90°. In Figure 33b, the rows on the lower terrace appear as a continuous line, while the upper terrace is resolved into pairs of $Fe_{oct}$ atoms. Various defects are observed in the images, labelled APDB, OH, and 1-5. Since the OH groups and APDBs are discussed in detail in their own right in sections 3.2.5.1 and 3.2.5.4, here we focus on defects 1-4.

1. When a sample is new to the chamber an atomic-scale defect is observed in STM, occupying the "blocked" site in the SCV reconstruction. This defect is mobile on the surface at room temperature and is therefore most likely an adsorbate related to the C observed in XPS. Over the course of 10-15 sputter/anneal cycles the density of such defects reduces significantly.

2. Bright protrusions on opposite $Fe_{oct}$ rows straddling the blocked (or "wide") site of the SCV reconstruction appear when the sample surface is reduced. The defect resembles a pair of surface OH groups, but the protrusions do not diffuse at all at room temperature, in contrast to OH groups. These defects are linked to an unreconstructed unit cell, in which an excess Fe atom and the $Fe_{int}$ atom from the SCV reconstruction occupy bulk-like positions within the spinel structure. In Figure 33b, the density of these defects has been artificially increased by deposition of Fe onto the surface at room temperature. Interestingly, species from the residual gas adsorb more readily at these defects than the regular surface. Further details will be published in the near future.

3. $Fe_{tet}$ adatoms occur as bright protrusions at the not-blocked (or "narrow" [212]) phase of the SCV reconstruction when the sample is reduced. In Figure 33b, the density of $Fe_{tet}$ adatoms and unreconstructed unit cells is similar, consistent with the DFT+U prediction that the DBT surface is similar in energy to a SCV surface with $Fe_{tet}$ adatoms (Figure 28e).

4. Alternating bright/dark features frequently appear in STM images of the $Fe_3O_4$(100) surface, and can extend for four (or more) unit cells perpendicular to the $Fe_{oct}$ row direction. The density of this defect appears to be sample specific, so we speculate that it is related to contamination rather than preparation. It is known that the incorporation of foreign cations in subsurface layers can cause the surface to revert to an inverse spinel-like structure [260], but the metals studied to date (Fe, Zr, Ti, Mn, Co, Ni) all produce bright/bright defects. Incorporation in every 4[th] $Fe_{oct}$ site in the subsurface layer could produce the observed ordering.

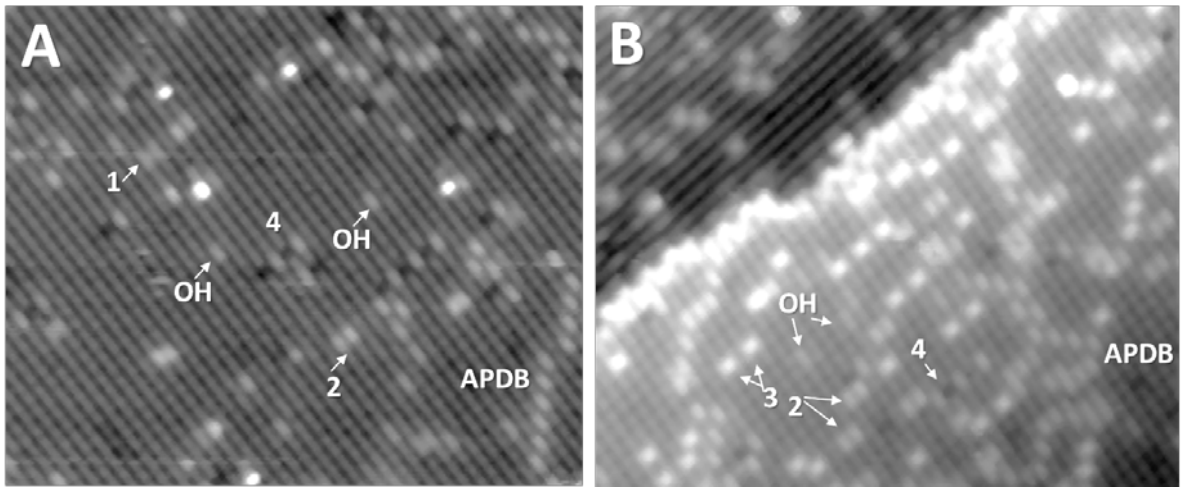

Figure 33: STM images (24x20 nm$^2$, $V_{sample}$ = 1.5 V, $I_{tunnel}$ = 0.3 nA) of a defective Fe$_3$O$_4$(100)-($\sqrt{2}\times\sqrt{2}$) surface obtained with an imperfect STM tip. Image A was acquired from a new natural Fe$_3$O$_4$(100) single crystal following ten sputter/anneal cycles. In image B, the fraction of unreconstructed unit cells (defect 2) and Fe adatoms (defect 3) has been artificially increased over the pristine surface by the deposition of 0.15 ML Fe at room temperature. For a description of the different defects see the main text.

### 3.3.5.3  Step edges

Step edges on Fe$_3$O$_4$(100) have received little attention, which is surprising given that steps are typically active sites for adsorption and dissociation of molecules and anchoring sites for metal clusters on other metal-oxide surfaces. Metal clusters, for example Ag [261], are indeed stable at steps on Fe$_3$O$_4$(100). The most common step height observed in STM is 2.1 Å (one example is shown in Figure 33B), and these steps separate terraces on which the Fe$_{oct}$ row direction rotates by 90° (as expected for Fe$_{oct}$-O layers in the spinel structure). The steps that form with the Fe$_{oct}$ rows parallel to the step edge on the upper terrace (α-type) are typically straight, whereas steps on which the Fe$_{oct}$ rows are perpendicular to the edge (β-type) are jagged. Steps with a height of 4.2 Å are also observed, and these also tend to be straight. Wang et al. [266] summarized the various possibilities for the structure of each step on the basis that any Fe atom missing more than half of its O ligands should not be stable at a step edge. They then compared the different possibilities in terms of the number of dangling bonds and the excess charge, and found that the two methods predict a different stability hierarchy, in contrast to similar work on Fe$_3$O$_4$(111) [267]. Noting that the surface Fe atoms are now known to possess charges different to the bulk, it perhaps makes more sense to consider the predictions based on coordinative unsaturation. With this approach, the α-type steps are more stable than the β-type, and the α step terminated with the O atoms (B-α*, see figure 34) was found to be most stable parallel to the Fe$_{oct}$ rows. Perpendicular to the Fe$_{oct}$ row, the B-β* and B-β are similar, and undercoordinated Fe$_{oct}$ are exposed at the step edge.

Given that the structural model and charges on the surface atoms have been revised since this work the issue of step edge termination requires further work.

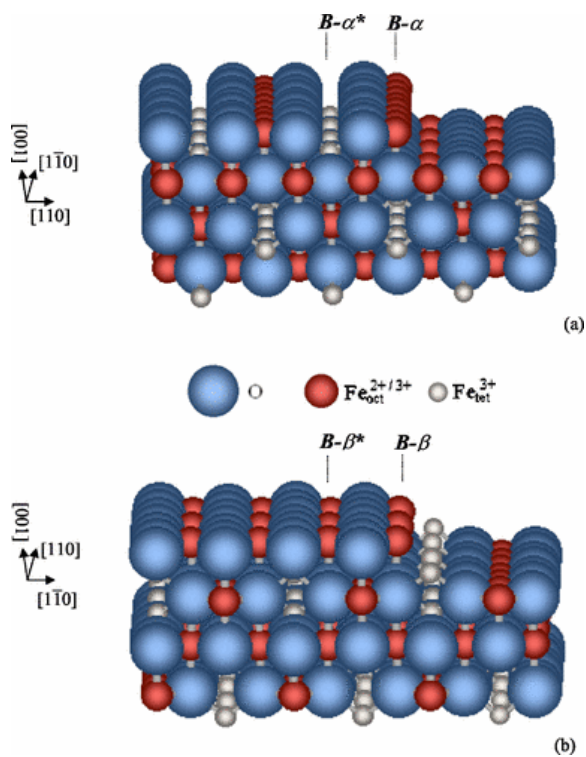

Figure 34: Potential step edge geometries based on a bulk termination at the $Fe_{oct}$-O plane. Based on the coordinative unsaturation, Wang et al. [266] concluded B-α* (i.e. removing the last rows of atoms from the right of the B-α* line) are the most stable steps parallel to the $Fe_{oct}$ row geometry. Steps terminated with the B-β* and B-β structures are similar for the direction perpendicular to the $Fe_{oct}$ row. Reprinted figure with permission from ref. [266]. Copyright 2006 by the American Physical Society.

### 3.3.5.4 Antiphase Domain Boundaries (APDBs)

Antiphase domain boundaries (APDB) in the (√2×√2)R45° reconstruction are a common defect observed in STM images of the sputter/annealed $Fe_3O_4$(100) surface [268]. APDBs occur because the (√2×√2)R45° reconstruction can nucleate with two distinct registries with respect to the underlying bulk. APDBs appear as bright line defects that zig-zag across the surface, typically beginning and ending at a step edge. On close inspection it is clear that the protrusion associated with the APDB is located on the $Fe_{oct}$ row, and that the "not-blocked" site of the reconstruction is always located to either side. Previous DFT+U calculations based on the DBT model suggested that the APDB represents a disruption in the bimodal subsurface charge order [268]. In the light of the SCV structure, it is clear now that the APDB represents a disruption in the subsurface cation vacancy distribution. The meeting of two "not-blocked" sites of the (√2×√2)R45° reconstruction is consistent with four $Fe_{oct}$ cations in the second layer row across the junction, which can alternatively be seen as one unit cell in which the subsurface $Fe_{oct}$-O layer is bulk-like (see Figure 35). The alternative, a meeting of "blocked" sites of the unit cell would result in removal of all $Fe_{oct}$ in the second layer and create twofold coordinated oxygen atoms, which is almost certainly unfavourable. The protrusion that marks the boundary may be linked to an intrinsic change in the electronic structure in the surface $Fe_{oct}$ atoms, and/or linked to a surface OH group if the APDB is reactive. Again, clusters and molecules are found to interact preferentially with APDBs. Given the discovery of the high temperature phase transition by Bartelt et al. [265] it now seems obvious that the APDBs form when the reconstruction re-nucleates on cooling through 723 K at the end of each annealing cycle.

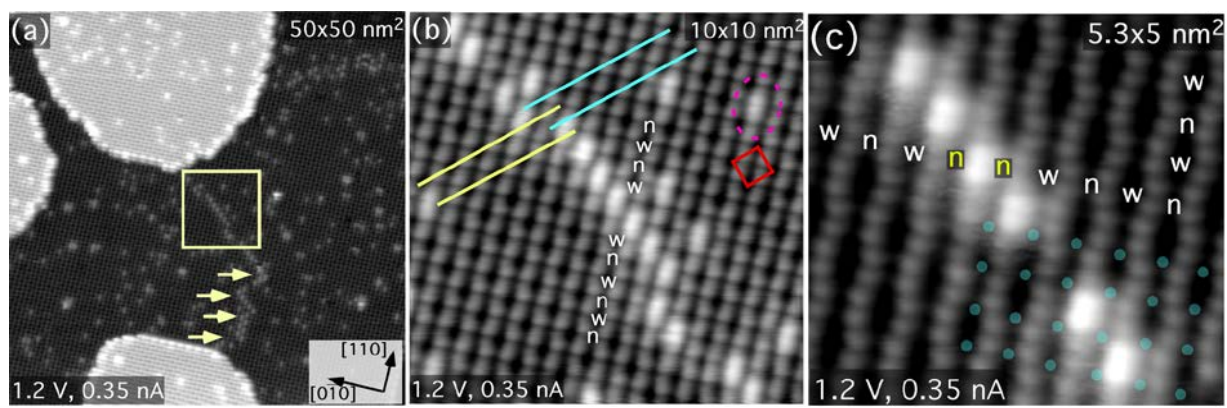

Figure 35: STM images showing APDBs in the (√2×√2)R45° reconstruction. These features appear as line defects with bright protrusions that zig-zag across a terrace. (b) Either side of the protrusions the surface reconstruction is ½ of a unit cell out of phase (cyan and yellow lines). (c) High-resolution images of APDBs reveal that the boundary always occurs at the junction between not-blocked "n" sites of the reconstruction. Note that the original model for APDBs was based on the outdated DBT model of the (√2×√2)R45° reconstruction. Adapted with permission from Ref. [268]. Copyrighted by the American Physical Society.

### 3.3.6 The "Fe dimer" Termination

Even in the earliest papers it was clear that more than one termination of $Fe_3O_4$(100) could be stabilized depending on the preparation conditions. Using STM, Wiesendanger and co-workers showed that an array of bright protrusions with (√2×√2)R45° periodicity emerges when a natural sample is sputtered with $Ar^+$ ions and subsequently annealed in UHV [203]. Considering the bulk structure, the simplest explanation of such protrusions is that the surface is terminated by one $Fe_{tet}$ atom per (√2×√2)R45° unit cell (Figure 36(a), frequently termed a ½ ML Fe(A) termination). Indeed, we have shown that an $Fe_{tet}$ adatom is a common defect on the $Fe_{oct}$-O terminated surface. If we assume bulk-like charge states for the atoms, a ½ ML $Fe_{tet}$ surface would indeed provide the +3 charge per unit cell required to compensate polarity at the surface [203; 222; 269]. However, on the basis of molecular dynamics simulations with empirical potentials [270] Rustad and co-workers found that the 1/2 ML $Fe_{tet}$ surface is unstable against one of the subsurface $Fe_{tet}$ atoms migrating to the surface to form an Fe dimer in the surface layer (Figure 36(b)). Moreover, XPS measurements of such a sputter/ UHV annealed surface find an enrichment in $Fe^{2+}$ relative to the $Fe_{oct}$-O termination [212], not $Fe^{3+}$. Subsequently, several groups have shown that the oval-like protrusions observed in early STM images [203; 271; 272] can be resolved into pairs of protrusions located between the $Fe_{oct}$ rows. The structure of the "Fe dimer" surface, as it has become known, remains somewhat controversial because assigning the position of the additional Fe atoms with respect to the underlying structure on the basis of STM is not straightforward; a quantitative structure determination is probably necessary for the issue to be ultimately resolved.

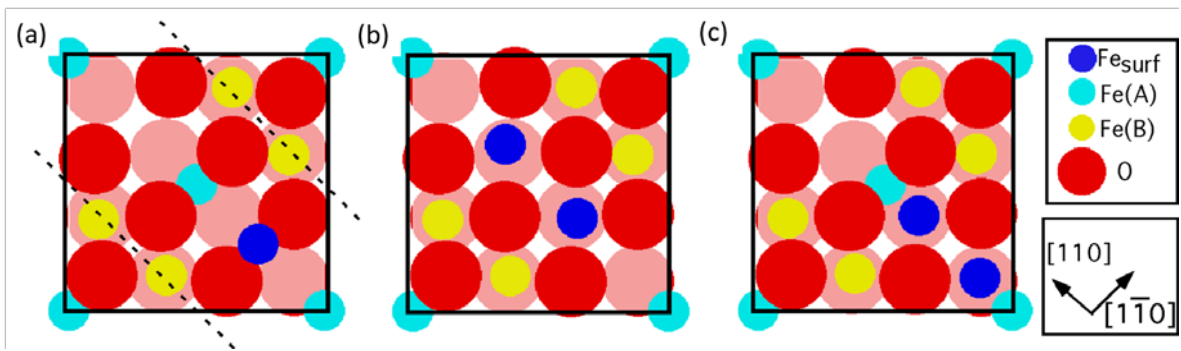

Figure 36: Iron rich surfaces in top view. Here, Fe(A) is $Fe_{tet}$, Fe(B) is $Fe_{oct}$ and $Fe_{surf}$ represents the additional surface Fe atoms. (a) The ½ ML Fe(A) termination. (b) The "Fe dimer" model proposed by Rustad [270]. Note, the Spiridis model has the same dimer position with respect to the unit cell, but with an additional $Fe_{tet}$ located immediately beneath the dimer (centre). (c) Model of the Fe dimer surface proposed by the present author's research group, with the Fe atoms located in fourfold hollow sites between subsurface $Fe_{tet}$ atoms (right). Adapted with permission from Ref. [212]. Copyrighted by the American Physical Society.

The Rustad model [270] of the Fe-dimer surface, shown in Figure 36 (b), is attractive from a polarity point of view because the outermost 3 layers maintain the same net charge as the 1/2 ML $Fe_{tet}$ termination. Spiridis et al. propose a different model on the basis of their studies of $Fe_3O_4$(100) thin films grown on an Fe buffer layer on MgO(100). In their model the Fe dimers reside on a bulk terminated surface, but with the dimer straddling one of the subsurface $Fe_{tet}$ atoms [210; 227]. In this case, the driving force for dimerization is proposed to be the relaxation of surface $Fe_{tet}$ atoms due to the distortions of the underlying ($\sqrt{2}\times\sqrt{2}$)R45° reconstruction. The authors claim that, on average, only 50 % of the unit cells are occupied over a large scale to ensure polarity compensation. Again, it must be stressed that this rests on the assumption that the surface atoms assume bulk-like charges, as Fe rich surfaces are strongly enriched in $Fe^{2+}$ [68]. Very recently, Davis et al. prepared $Fe_3O_4$(100) thin films using a similar Fe-buffer layer approach, although Pt(100) was utilized as the substrate in place of MgO(100). In this case the coverage of Fe dimers is significantly greater than 0.5 ML Fe, and the authors demonstrated that several different Fe-rich terminations could be formed depending on the thickness of the buffer layer. Since the Fe buffer layer most likely diffuses into the $Fe_3O_4$ film at the growth temperatures forming Fe interstitials, the Fe-rich termination is most likely the result of non-stoichiometry within the film. Similar terminations occur when the near-surface region is enriched by Fe following sputter/anneal cycles [212]. In general the evidence suggests the termination is a function of both the stoichiometry of the $Fe_3O_4$ and the external chemical potential (whether one anneals in $O_2$ or in UHV). Ceballos et al. [215; 216] observed the coexistence of Fe and $Fe_{oct}$-O terminated planes when a natural $Fe_3O_4$(100) single crystal was annealed at 990 K and exposed to 2000 Langmuir $H_2$, whereas the same procedure in oxygen results solely in a $Fe_{oct}$-O termination.

An alternative approach to study the structure of Fe-rich terminations, pursued by this author's group, is to deposit Fe directly onto the $Fe_{oct}$-O surface. For low Fe doses at room temperature, Fe is accommodated as isolated $Fe_{tet}$ adatoms, but also incorporates within the surface, locally lifting the ($\sqrt{2}\times\sqrt{2}$)R45° reconstruction (see Figure 33b, defect #2). With increasing Fe exposure, patches of unreconstructed (1×1) unit cells form. At higher coverages still, Fe dimers nucleate on the (1×1) patches (see Figure 37). Full details of this work will be published separately. In Figure 28e, we have updated the surface phase diagram from ref. [35] to include the Fe-rich surface structures calculated previously in ref. [68], and added two new surfaces; an Fe adatom and Fe dimer on the SCV structure.  Essentially, the adatom on SCV and the DBT structure have the same stoichiometry and almost identical surface energy, consistent with the coexistence of adatoms and incorporated atoms with Fe atom deposition at room temperature. The Fe dimer on the SCV structure is very unfavourable. One final note on terminology; the term "Fe dimer" suggests an Fe-Fe bond, and it is used here for compatibility with the prior literature. However, our DFT+U calculations reveal that no such bond exists, and the Fe atoms are rather fourfold coordinated to surface oxygen atoms, filling two of the four octahedral interstitial sites per unit cell. Filling the final two such sites produces a surface with FeO stoichiometry [68].

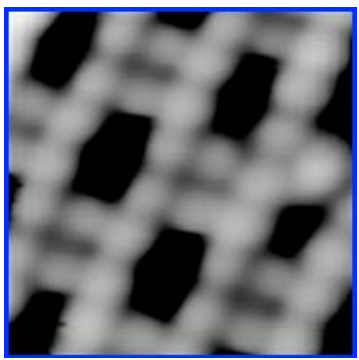

Fig 37: High resolution STM image of the Fe dimer surface. The rows of protrusions are the $Fe_{oct}$ atoms, and the Fe-dimers are located in between with a dimer-dimer distance of 10.8 Å. Adapted with permission from ref. [68]. Copyrighted by the American Physical Society.

### 3.3.7  Other reduced terminations

As the $Fe_3O_4$(100) selvedge becomes further reduced more Fe-rich terminations can be stabilized. The same phenomenon occurs with repeated sputter/anneal cycles without $O_2$ annealing on a single crystal, by deposition of Fe, or by growing the film on an Fe buffer layer. Davis et al. [211] recently published STM images of several such intermediate structures (see Figure 38). The structure of such surfaces is not yet known, but it seems likely that such structures will retain the close packed O lattice of the iron oxide, and incorporate Fe cations in interstitial sites. Such surfaces are reduced, midway between $Fe_3O_4$ and $Fe_{1-x}O$, and thus contain very high amounts of $Fe^{2+}$ in XPS. An $Fe_{1-x}O$-like termination can be formed in the ultimate limit [271; 272], although this surface appears highly defective.

Interestingly, a DBT surface with $V_O$s is found to be energetically favourable under reducing conditions in DFT calculations (Figure 28e), as suggested previously [258]. Despite much effort (extended high temperature annealing, e$^-$ bombardment, this author's group have been unable to create such a surface experimentally. In this author's opinion, the issue could be that the DBT calculations are performed on a symmetric slab which is non-stoichiometric (too much oxygen), and/or that the experiments are performed on sputter/annealed samples that are most likely Fe-rich in the selvedge.

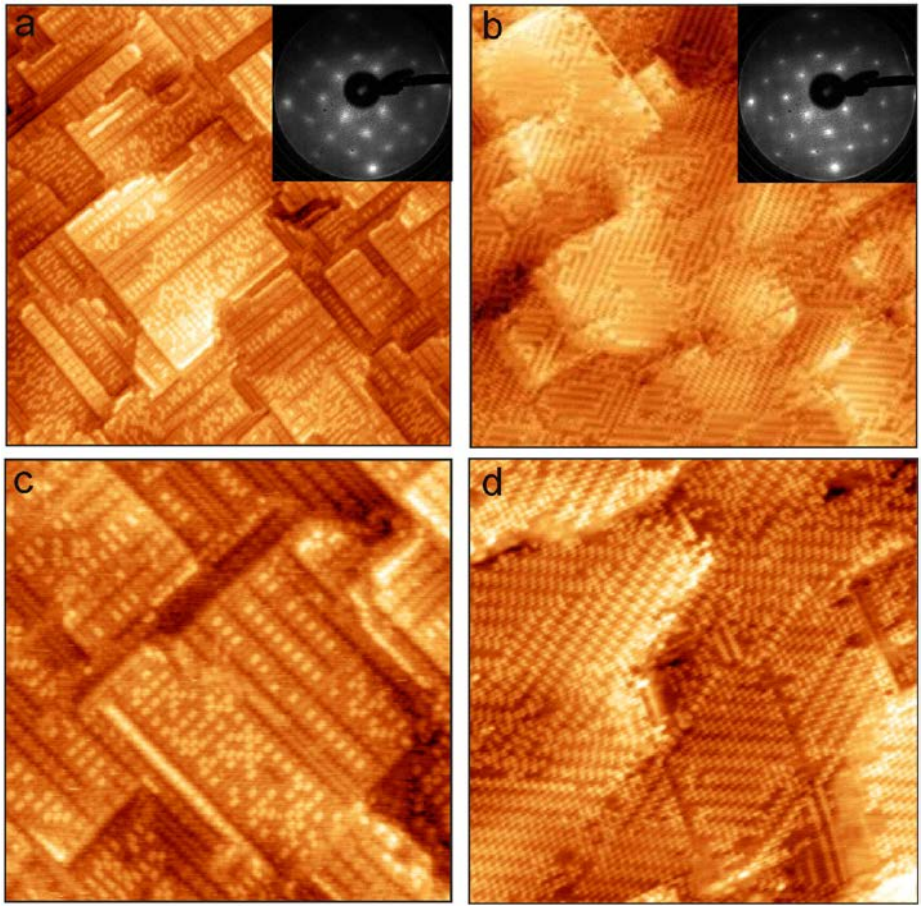

Figure 38: STM images of intermediate surface structures observed on the Fe$_3$O$_4$(001) film grown on 4 nm-thick Fe buffer layer UHV-annealed at 775 K for 40 min (a, c) and 70 min (b, d). Image sizes: 100 nm × 100 nm (a, b), and 50 nm × 50 nm (c, d); sample bias = − 2 V; current: 2.0 nA (a, c), and sample bias = − 1.45 V; current: 0.08 nA (b, d). LEED patters are recorded at 95 eV. Figure reprinted from ref. [211] with permission from Elsevier.

### 3.3.8. How to Prepare and Recognise the SCV Surface

In this author's experience, the SCV surface is routinely achieved on stoichiometric Fe$_3$O$_4$(001) single crystals by cycles of Ar$^+$ sputtering (1 keV, 1 µA, 20 minutes) and annealing (900 K, 20 minutes). Typically four annealing cycles are performed with UHV annealing, solely to remove surface contamination, and the fifth anneal cycle is performed in 1×10$^{-6}$ mbar O$_2$ for 20 minutes. As discussed in section 3.2.1 the oxygen anneal results in the growth of many layers of pristine Fe$_3$O$_4$(001), and thus redresses the reduction of the surface that occurs through the standard cycles. An oxygen anneal is not performed on every cycle, however, in order to limit the growth of α-Fe$_2$O$_3$ inclusions as much as possible over the long term. Crucially, as long as the last annealing cycle is performed with the O$_2$ anneal, the SCV structure is always obtained. Surveying the literature, no reports could be found of a surface prepared by oxygen annealing displaying anything other than the SCV termination.

In order to take advantage of the precisely known structure of the SCV reconstruction [35], it is important to recognise when the surface is well prepared. While this is straightforward with STM, synchrotron beamlines, for example, do not typically have this capability in-situ, and many interesting experiments can be envisaged that require this setting. In this section, reference XPS and LEED data are provided for the clean SCV and Fe-dimer terminated surfaces. An STM image of the specific Fe-dimer surface measured by LEED, obtained by deposition of 2 ML Fe on the SCV surface at room temperature, is shown in Figure 39.

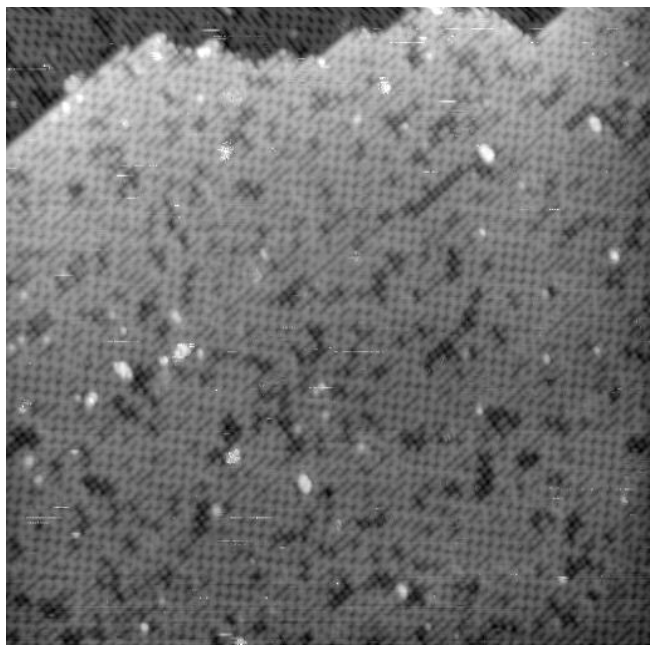

Figure 39: STM image (50×50 nm$^2$, 1.6 V, 0.3 nA) of the "Fe-dimer" surface prepared by deposition of 2 ML Fe on the SCV terminated Fe$_3$O$_4$(001) surface at room temperature. The internal structure of the dimer is not resolved at this resolution, and they appear as single protrusions spaced with (√2×√2)R45° periodicity.

Figure 40 shows the LEED *IV* curves acquired from the pristine SCV terminated surface at the University of Erlangen by the group of Lutz Hammer. The vacuum chamber includes in-situ STM, which allowed the condition of the surface to be checked both before and after the measurements. The surface was clean and SCV reconstructed, with no evidence of reduction in the form of Fe adatoms. Many more *IV* curves are included in the supplementary information available with ref. [35]. These data can be used as the "fingerprint" of the SCV reconstruction. A quick and rough guide to the quality of the SCV surface can be obtained by inspection of the LEED pattern at an electron energy of 90 eV (Figure 41). The clean and well-ordered surface exhibits half-order spots of almost equal intensity to the integer order spots (compare (2,0) and (3/2, 1/2) in Figure 41). When the surface is reduced, the "filling" of the cation vacancies weakens half order spots in the (√2×√2)R45° reconstruction significantly, to the point where they are not visible at all. With further reduction, Fe dimers populate the surface with (√2×√2)R45° periodicity, but even when the coverage is high the (2,0) and (3/2, 1/2) spots never recover the comparable intensity as observed for the clean surface at 90 eV. Figure 41 shows a direct comparison of the LEED images from the SCV and Fe dimer surfaces acquired using a commercial rear-view LEED with an electron energy of 90 eV at the TU Vienna.

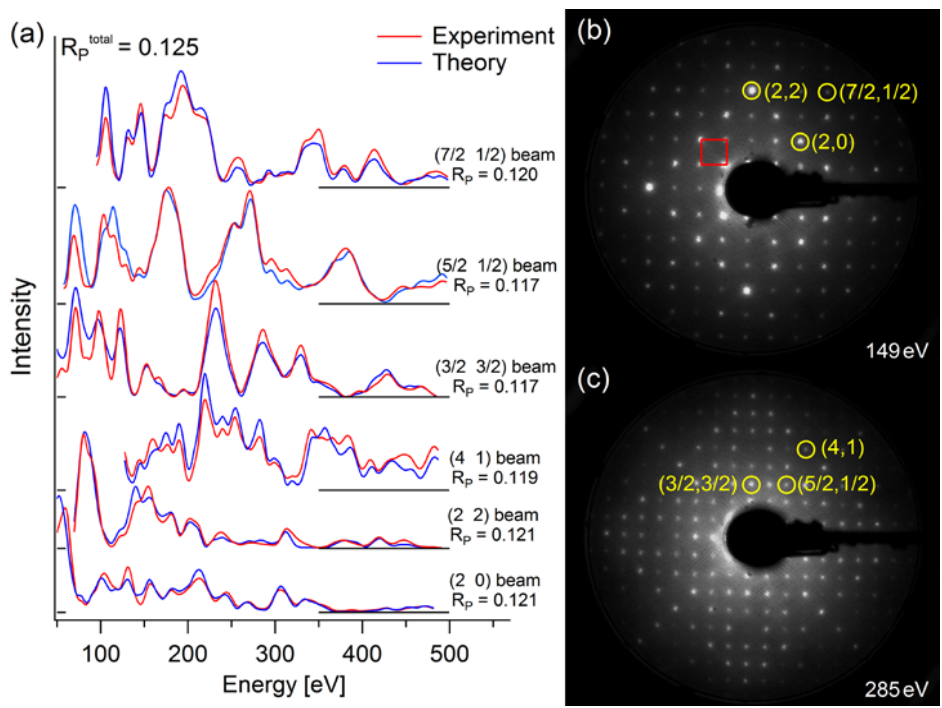

Figure 40: LEED *IV* curves and representative LEED patterns acquired for the SCV reconstructed Fe$_3$O$_4$(100)-($\sqrt{2}\times\sqrt{2}$)R45° surface. *IV* curves for other LEED spots are available in the supplementary information of ref. [35]. The ultimate $R_P$ of 0.125 indicates a high level of agreement between the experimental data and theoretical curves for the SCV reconstruction model. Redrawn after ref. [35]. Reprinted with permission from AAAS.

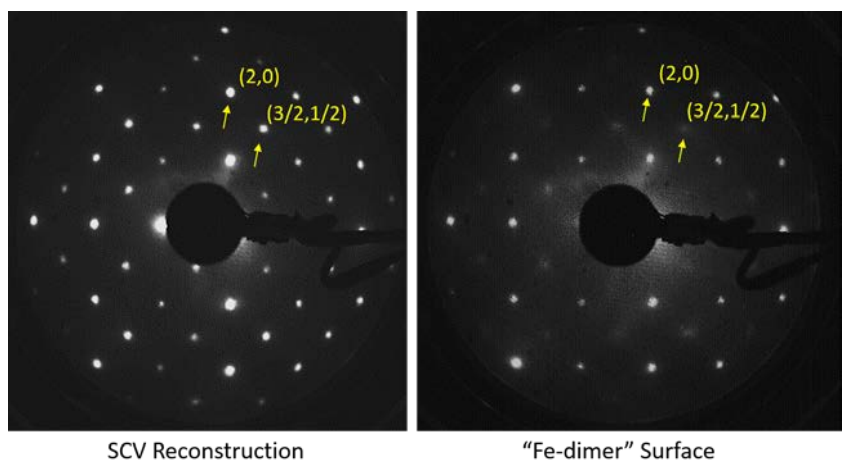

Figure 41: Direct comparison of LEED patterns acquired at an electron energy of 90 eV for the SCV reconstructed and "Fe-dimer" surface. The (2,0) and (3/2,1/2) have similar intensity at 90 eV for the SCV reconstruction, but the ($\sqrt{2}\times\sqrt{2}$)R45° spots are significantly weaker for the Fe-dimer surface.

The XPS data shown in Figure 42 were acquired at normal exit using non-monochromatized Al Kα radiation and a SPECS PHOIBOS 100 analyser operating with a pass energy of 20 eV. The Fe-dimer termination is known to be enriched in Fe$^{2+}$ compared to the SCV surface [68], and thus exhibits an enhanced Fe$^{2+}$ component in Fe 2p spectra. The easiest way to distinguish the surfaces is the satellite peak linked to Fe$^{2+}$ at 714 eV, which is absent for the SCV surface, and significantly increased for the Fe dimer surface. Higher resolution and angle-resolved Fe2*p* XPS spectra from the SCV reconstructed surface can be found in the supplementary information of ref. [35]. Note that the Fe 2*p* peak structure is particularly complex in iron oxides so determination of the $Fe^{2+}/Fe^{3+}$ ratio by peak fitting is not reliable. Indeed, the optimum methods for fitting such Fe XPS spectra are still under discussion [75; 273-275]. Note also a small shift in the O *1s* peak.

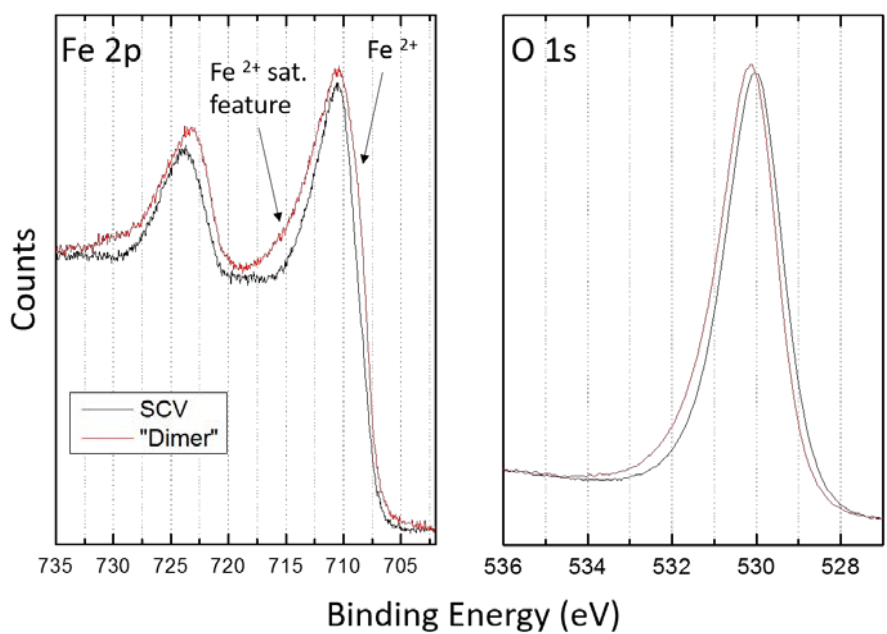

Figure 42: Normal exit XPS spectra acquired from the freshly prepared SCV and Fe-dimer terminated surfaces using Al Kα radiation and a SPECS PHOIBOS 100 analyser with a pass energy of 20 eV. The $Fe^{2+}$ satellite feature is evidence of surface reduction.

## 3.4 The $Fe_3O_4$(111) Surface

Surface science studies of the $Fe_3O_4$(111) surface have been performed on synthetic [276; 277] and natural [107; 278-280] $Fe_3O_4$ single crystals, and on α-$Fe_2O_3$(0001) single crystals reduced in UHV by sputter/anneal cycles [281-283] (see section 3.6.1 for details). Thin films have been grown on various substrates including Pt(111) [17; 284-288], Pt(100) [211], Au(100) and (111) [289], Cu(001) [290], sapphire (0001) [291], and TiN buffered Si [292], and, in general, the (111) is the most common growth direction. This occurs because a FeO(111) wetting layer often forms first at the interface when Fe is deposited in an $O_2$ atmosphere, and the formation of a close-packed O layer templates further growth [211]. The growth mode is described in great detail for the $Fe_3O_4$(111)/Pt(111) system in the previous review of Weiss and Ranke [17], but essentially, triangular $Fe_3O_4$(111) islands grow on the ultrathin FeO(111) wetting layer, eventually coalescing to form a closed surface. Most studies are performed on films at least 100 Å thick, and they will be discussed in the context of this review, but it is important to note that the film properties can be affected by strain or a high concentration of domain boundaries. The lack of a semi-infinite Fe reservoir could also lead to differences between single crystal and thin film results, particularly if a similar growth of new $Fe_3O_4$ layers occurs when the $Fe_3O_4$(111) surface is annealed in $O_2$ at high temperature, as has been observed for the (100) surface [213].

## 3.4.1 The "Regular" Termination

In the (111) direction the $Fe_3O_4$ structure is made up of 6 distinct planes of atoms, commonly denoted $Fe_{tet1}$, $O_1$, $Fe_{oct1}$, $O_2$, $Fe_{tet2}$, and $Fe_{oct2}$ in the literature (see Figure 42). The planes of fcc close packed $O^{2-}$ ions are negatively charged, while the Fe-containing layers are positively charged, which makes $Fe_3O_4$(111) a Tasker type-3 polar surface [293]. Consequently, as for $Fe_3O_4$(100), a simple bulk truncation is expected to be energetically unstable, and early studies focused on finding structural models consistent with polarity compensation and the minimisation of dangling bonds [278; 283; 285]. However, a consensus quickly emerged that the most commonly reported termination, frequently termed the "regular" termination in the literature, exhibits a so-called (2×2) periodicity in LEED (with respect to the O plane) and a hexagonal lattice of protrusions with a periodicity of 6 Å in STM images [107; 277]. These images are interpreted as a bulk termination at the $Fe_{tet1}$ plane, as shown in Figure 43, and thus it can also be viewed as a (1×1) periodicity.

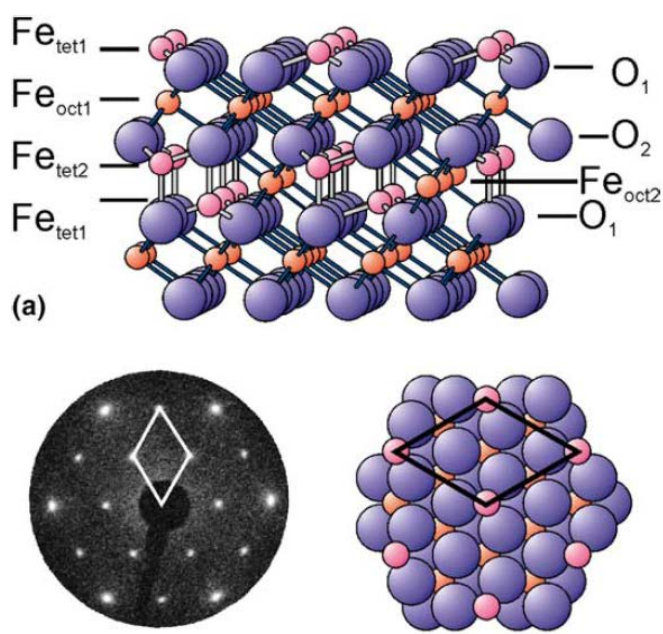

Figure 43: Side view of the $Fe_{tet1}$ termination of $Fe_3O_4$(111) (a), the so-called (2x2) LEED pattern observed for this surface (b) and a top view illustrating the (2x2) unit cell. Images reproduced from ref [284].

Table 5: Layer relaxations in the $Fe_{tet1}$ termination of $Fe_3O_4$(111) as determined by LEED *IV* [285] and PBE+*U* [294].

|  | Layer Distances (Å) | | Layer Relaxations (%) | |
| --- | --- | --- | --- | --- |
| **Layers** | **Bulk** | **Surface (LEED)** | **LEED** | **PBE+*U*** |
| $Fe_{tet1}$ - $O_b$ | 0.64 | 0.38 ± 0.05 | -41 ± 7 | -37 |
| $O_b$-$O_a$ (within $O_1$ layer) | 0.04 | 0.08 ± 0.09 | | |
| $O_a$ – $Fe_{oct1}$ | 1.18 | 0.87 ± 0.05 | -26 ± 4 | -22 |
| $Fe_{oct1}$-$O_b$ | 1.18 | 1.36 ± 0.05 | 15 ± 4 | 13 |
| $O_b$-$O_a$ (within $O_2$ layer) | 0.04 | 0.12 ± 0.09 | | |
| $O_a$ – $Fe_{tet2}$ | 0.64 | 0.57 ± 0.05 | -11 ± 7 | -16 |
| $Fe_{tet2}$ – $Fe_{oct2}$ | 0.6 | 0.6 | 0 | -3 |

LEED *IV* experiments conducted on thin film samples also favour the Fe$_{tet1}$ termination [285; 287; 288], but measure strong relaxations in the surface layers (see Table 5). The Pendry R-factor of 0.19 obtained by Shaikhutdinov et al. [287] suggests this structural model is likely reliable. Indeed, the surface was pre-characterized by STM and a hexagonal lattice of protrusions was imaged. Including Fe$_{tet1}$ vacancies (observed as missing protrusions in the STM images) in the model led to an improvement in the agreement over a stoichiometric surface. Note however, that the quantitative level of agreement is not perfect, which may be because a small fraction of the surface exhibited the Fe$_{oct2}$ termination, which is now known to be close in energy and can be difficult to distinguish by STM [107]. The most recent LEED study of the Fe$_3$O$_4$(111) surface was performed by Sala et al. [295], who investigated the structural evolution of thin films grown epitxially on Pt(111). The Fetet1 termination was concluded to form only after a final annealing step at 900 K in UHV, with FeO$_x$ agglomerates also present on the as-grown surface, and after cooling in O$_2$.

Theoretical calculations have been performed to study the termination of Fe$_3$O$_4$(111) by several groups [198; 296-298] [198; 299-302]. The most recent theoretical calculations, which compare results of DFT+U and HSE functionals, favour the Fe$_{tet1}$ termination for chemical potentials corresponding to UHV conditions (see Figure 44) [294], and predict layer relaxations in good agreement with those obtained in the original LEED *IV* investigation (see Table 5). The Fe$_{oct2}$ termination only becomes stable under very reducing conditions, nominally less than $10^{-20}$ mbar O$_2$ at a typical annealing temperature (assuming a stoichiometric sample). Consequently, the available evidence suggests that the "regular termination", insofar as it can be prepared in isolation, is most likely a strongly relaxed Fe$_{tet1}$ termination.

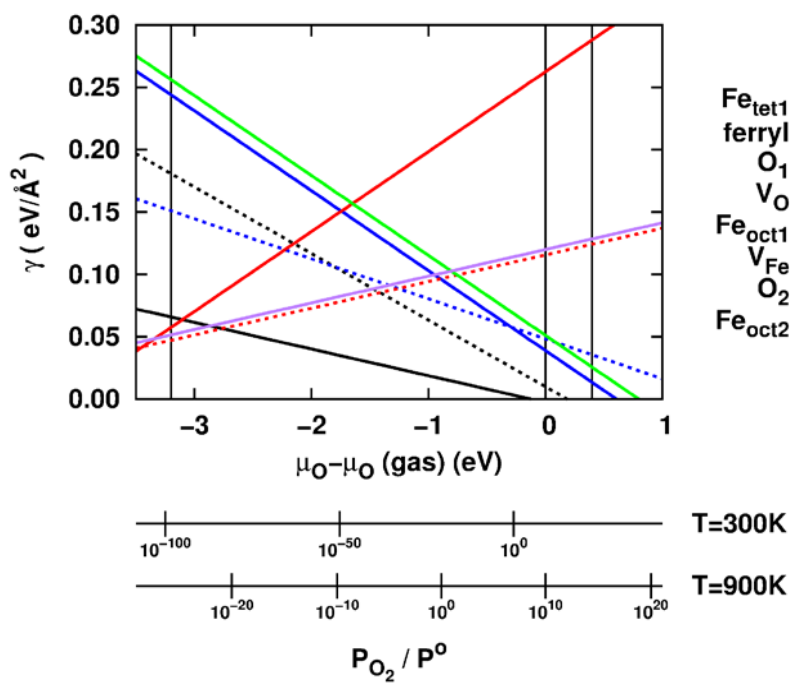

Figure 44. Theoretical (PBE+*U*, $U_{eff}$ = 4 eV) surface phase diagram for Fe$_3$O$_4$(111) reproduced from Ref. [294]. A strongly relaxed Fe$_{tet1}$ termination is found to be the most stable of the bulk-like terminations over a broad range of oxygen chemical potentials. The Fe$_{oct2}$ termination is favoured

over the Fe$_{oct1}$ termination under very reducing conditions, but including cation vacancy defects in Fe$_{oct1}$ termination (V$_{Fe}$) makes the two surfaces competitive.

### 3.4.2 Other Terminations

The major difficulty encountered by researchers working with Fe$_3$O$_4$(111) is that multiple terminations can coexist, and the surface obtained appears to be a strongly dependent on the preparation conditions, but also sample history [44; 107; 277; 287; 295]. STM images of single crystal surfaces sometimes exhibit a honeycomb-type lattice that coexists with the regular termination. Shimizu et al. [107] report that this situation arises when the sample surface is reduced by sputter/anneal cycles, and suggest it can be explained as the coexistence of Fe$_{tet1}$ and Fe$_{oct2}$ terminations. The honeycomb images arise when both Fe$_{oct}$ and Fe$_{tet}$ atoms in the Fe$_{oct2}$ termination are imaged as bright protrusions. It is important to note, however, that the appearance of the different regions depends strongly on the STM tip condition and bias voltage, and that the most common appearance of both the Fe$_{tet1}$ and Fe$_{oct2}$ surfaces is a hexagonal lattice of protrusions (Figure 45). Thus, the observation of a hexagonal lattice of protrusions in STM is a necessary, but insufficient proof that the Fe$_3$O$_4$(111) sample is uniformly terminated with the Fe$_{tet1}$ surface. The theoretical calculations presented in Figure 43 [294], like many others [198; 299-302], find that the Fe$_{oct2}$ termination is competitive with the Fe$_{tet1}$ termination at low oxygen chemical potentials.

In recent years there has been increasing evidence that the Fe$_{oct2}$ termination is obtained under conditions that would be expected to produce the Fe$_{tet1}$ termination. For example, TPD, IRAS and HREELS measurements on thin film Fe$_3$O$_4$(111) samples revealed three distinct adsorption sites for CO [284]. Water TPD acquired for the same surface exhibited more peaks than equivalent spectra published by Joseph et al.[303]. Both results suggest that a different surface was measured. In very recent work, Dementyev et al. [70] invoked the Fe$_{oct2}$ termination to explain their water adsorption data (these water/Fe$_3$O$_4$(111) studies are discussed in section 3.2.2). Finally, Kaya et al. [301] found that their Fe$_3$O$_4$(111) thin films were enriched in Fe$^{3+}$ at the surface, and involved octahedrally coordinated Fe atoms, also consistent with a Fe$_{oct2}$ termination.

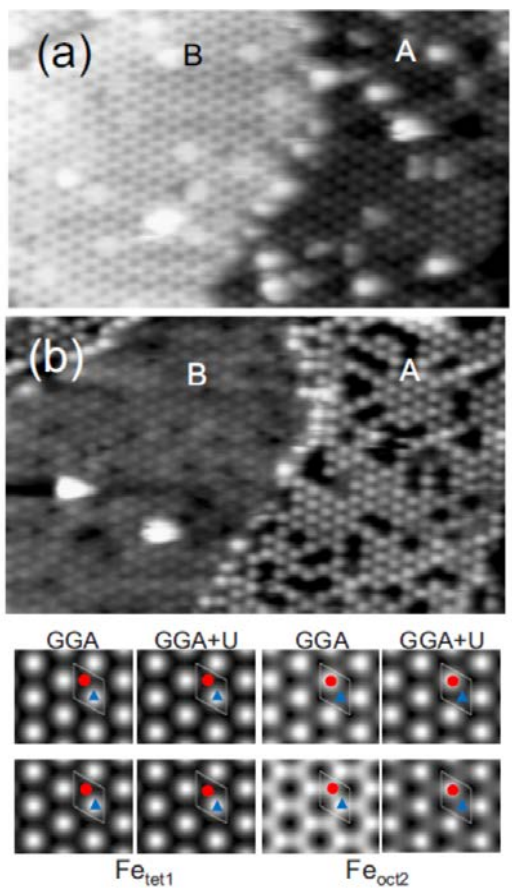

Figure 45: (a,b) STM images (9×15 nm$^2$) of the Fe$_3$O$_4$(111) surface acquired with (a) V$_{sample}$ = +2.0 V, I$_{tunnel}$ = 0.2 nA and (b) V$_{sample}$ = -2.0 V, I$_{tunnel}$ = 0.2 nA. At negative sample bias regions "A" and "B" appear as a hexagonal lattice of protrusions, whereas at positive sample bias region "B" exhibits a honeycomb structure. STM simulations suggest that region A is the Fe$_{tet1}$ termination, but region B can be assigned as the Fe$_{oct2}$ termination. The upper (lower) row of simulated images correspond to STM images with V$_{sample}$ = +2.0 V (-2.0 V). Optimum agreement between experiment and theory is found for STM simulations based on GGA rather than GGA+U results.

Although the evidence that the Fe$_{tet1}$ and Fe$_{oct2}$ terminations dominate at the Fe$_3$O$_4$(111) surface is growing, they are not the only structures under consideration. Cutting et al. [304] studied a Fe$_3$O$_4$(111) single-crystal surface prepared by sputter / anneal cycles (1023 K, 45 minutes, without O$_2$ annealing), which can be considered very reducing conditions. Indeed three coexisting terminations were observed on the surface, one of which (denoted A') was considered to be the Fe$_{tet1}$ termination. A second region (denoted A) had the same periodicity but was found on terraces 1.2 Å above those exhibiting the Fe$_{tet1}$ termination. The authors ascribed this to a surface in which each Fe$_{tet1}$ cation is capped by a terminal O atom, forming a ferryl group. This termination was considered in recent theoretical calculations, and found to be some 0.05 eV/Å$^2$ less stable than the Fe$_{tet1}$ termination in the relevant range of chemical potentials (see Figure 43) [294]. The third termination observed by Cutting et al. [304] appears similar to the honeycomb structure observed by Shimizu et al. [107]. However, since this surface was found to be completely inert in adsorption experiments, Cutting et al. [304] suggest their honeycomb surface results from imaging O atoms that cap both the Fe$_{tet1}$ and Fe$_{oct2}$ atoms of an Fe$_{oct2}$ termination. IRAS experiments are ideal to search for the signature of ferryl groups on such a surface (see Section 3.7.1.3 for a discussion of ferryl groups on the α-Fe$_2$O$_3$(0001) surface), and to date the experiments performed on $Fe_3O_4$(111) have not reported the relevant vibrational frequency.

### 3.4.3 The Bi-Phase Termination

When a $Fe_3O_4$(111) single crystal is subjected to multiple sputter/anneal cycles without an oxygen annealing step, i.e. very reducing conditions, several long-range superstructures occur [107; 276; 277; 280]. Similar effects have been observed for thin $Fe_3O_4$(111) films prepared at 870 K, rather than the more common temperature of 1000 K [287; 305]. The superstructure can coexist with the regular termination, and appears as triangularly shaped islands with a periodicity of approximately 50 ±5 Å. The distance between protrusions inside the islands differs, some remain identical to $Fe_3O_4$(111), while others exhibit an expanded lattice consistent with $Fe_{1-x}O$. The most common interpretation is a co-existence of $Fe_{1-x}O$ and $Fe_3O_4$(111) islands at the surface [44; 280]; a phenomenon known as bi-phase ordering, first observed at the α-$Fe_2O_3$(0001) surface [280]. The ordered mosaic of $Fe_{1-x}O$ and $Fe_3O_4$(111) islands is thought to be defined by strain (the O-O distance in $Fe_{1-x}O$ is slightly larger (3.04 Å) than in $Fe_3O_4$ (2.97 Å)). With further cycles the selvedge becomes more and more reduced until only $Fe_{1-x}O$ exists several layers deep into the crystal [277; 280]. Fe 2$p$ XPS spectra acquired on such a surface reveal a significant enhancement in the proportion of $Fe^{2+}$ relative to $Fe^{3+}$ (seen most clearly in the strength of the satellite features), consistent with the surface reduction and formation of $Fe_{1-x}O$ [277].

In summary, it is clearly not straightforward to ensure that the $Fe_3O_4$(111) surface is terminated by a single phase, and this must be borne in mind in the interpretation of data from area averaging techniques. Moreover, other than the $Fe_{tet1}$ termination, the structure of the other frequently observed terminations has not been determined using quantitative structural techniques, and further work is needed to establish the details of the $Fe_{oct2}$ termination. In this author's opinion, complex surface structures such as $Fe_3O_4$(111) will benefit greatly from study by the new generation of low-temperature STM/AFM machines, where frequency shift (force) and electronic structure can be simultaneously imaged. The additional channel of information will help remove ambiguity from the interpretation of STM images and help screen possible models to use as a basis for theoretical calculations and quantitative structural measurements. In general, the experience with $Fe_3O_4$(111) is in line with that of the (100) surface; i.e. Fe-rich terminations occur when a sample is reduced [68], which makes sense when we consider that iron oxides prefer Fe interstitials to $V_O$ formation in the bulk under such conditions. Surface reduction comes about through sputter/anneal cycles, but in both cases can be (mostly) redressed if the sample is annealed in $O_2$ at sufficiently high temperature. A surface with near mono-phase $Fe_{tet1}$ termination has been created and measured by LEED $IV$, but the $Fe_{oct2}$ termination is close in energy and coexists when the conditions become slightly reducing. Similar issues with the termination occur for thin films grown on metal, particularly if the oxidation temperature is too low. This suggests that the kinetics of oxidation are equally important in obtaining a stoichiometric sample in this case [287].

### 3.4.4 Step edges

Henrich and Shaikhutdinov [267] analysed the $Fe_3O_4$(111) surface to demonstrate how coordinative unsaturation and excess charge can be used to predict stable step edge geometries on metal oxides. STM images of $Fe_3O_4$(111) reveal triangular or hexagonal islands with step edges aligned with the $<\bar{1}10>$ type directions (see Figure 46). The steps are typically 4.8 Å high [287], which corresponds to the repeat distance in the (111) direction. The position of the most stable step edges are indicated using solid black lines in a top view schematic (Figure 46b). In the paper, the authors assume a $Fe_{oct2}$ termination, and form the step edge models by removing two layers of O atoms (blue) outside the enclosed area. Note that the longer steps (termed A*-1 in ref. [267], shown in Figure 46(c)) are not equivalent to the shorter ones (B-1, shown in Figure (d)). The longer steps terminate at surface $Fe_{tet}$ atoms (grey), while the shorter steps terminate at the $Fe_{oct}$ atoms (red).

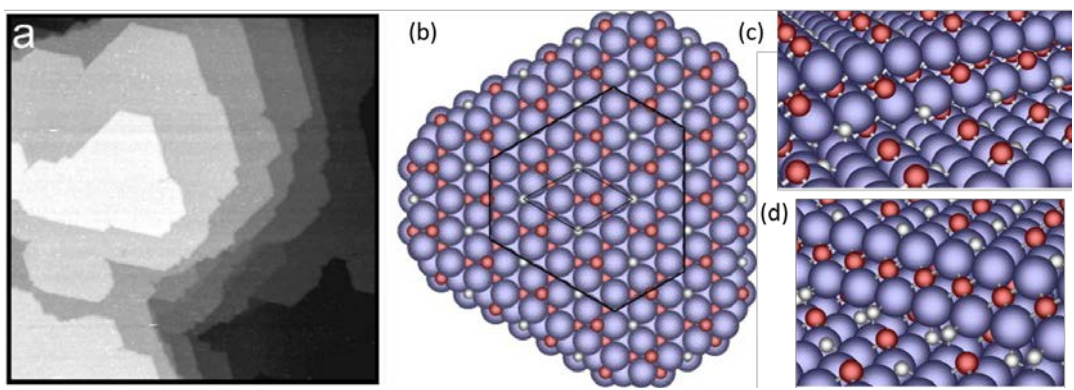

Figure 46: (a) STM image (3000×3000 Å$^2$, $V_{sample}$ = 1 V, $I_{tunnel}$ = 0.8 nA) of a $Fe_3O_4$(111) film grown on Pt(111). Step edges run parallel to the <$\bar{1}$10> directions. (b) Top view model of $Fe_3O_4$(111) in which oxygen is blue, $Fe_{tet}$ atoms are grey and $Fe_{oct}$ are red. The most stable step edges, as determined by coordinative unsaturation and excess charge arguments, are marked by the solid black lines. The surface unit cell is marked by a thinner black line. (c, d) Side views of the most stable A*-1 (c) and B-1 (d) steps. Adapted with permission from Ref. [267] and [287]. Copyrighted by the American Physical Society and Elsevier, respectively.

## 3.5 The $Fe_3O_4$(110) Surface

The $Fe_3O_4$(110) surface has been less studied than the (111) and (100) surfaces, both experimentally [306-308] and computationally [309-311]. In principle, there are two possibilities to truncate the bulk at the (110) plane (Figure 47), exposing either a layer of $Fe_{oct}$ ($Fe_B$), $Fe_{tet}$ ($Fe_A$), and O atoms, or a layer exposing only $Fe_{oct}$ and O (see Figure 46). The separation between dissimilar and similar layers is 1.6 Å and 3.2 Å, respectively.

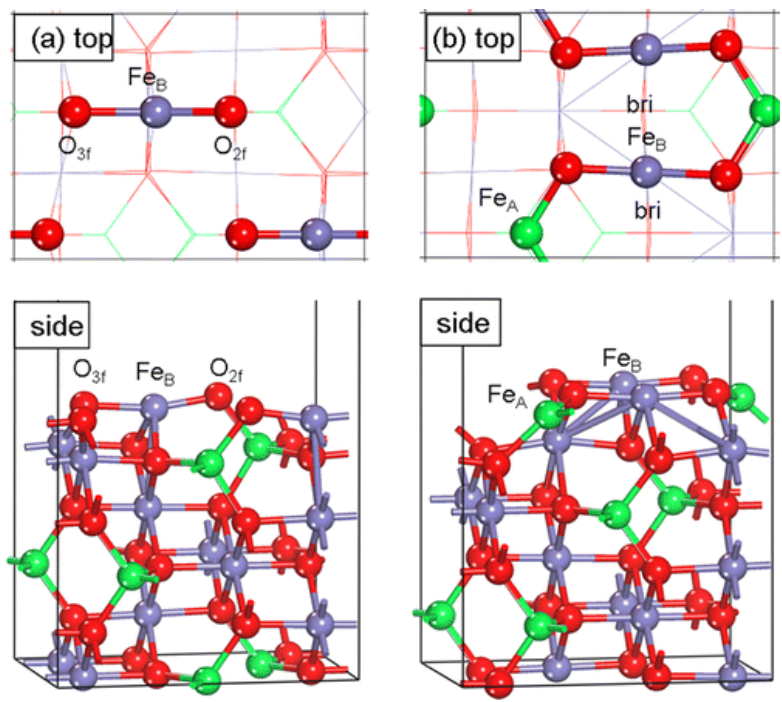

Figure 47: Top and side views of the Fe$_3$O$_4$(110) surfaces: (a) A termination; (b) B termination in a p(1 × 1) surface unit cell (O$^{2-}$ are red, Fe$_{tet}$ green, and Fe$_{oct}$ blue). Reprinted with permission from ref. [311]. Copyright 2013 American Chemical Society.

Jansen et al. studied a bulk single-crystal sample [307] and thin Fe$_3$O$_4$(110) films grown on MgO(110) [312] and found that annealing at 850 K was required to produce a clean surface after exposure to air. However, this led to Mg diffusion through the film. Nevertheless, on both thin film and single crystal surfaces, a one-dimensional reconstruction was observed in STM images (Figure 48), with rows of protrusions in the [-110] direction separated by some multiple of the lattice constant (8.4 Å). The most common separation was 25 Å, which is equivalent to three unit cells. Terraces were separated by steps with an apparent height of 3.2 Å or 6.4 Å, consistent with steps between similar layers of the bulk structure. Along the rows, a periodicity of 3 Å was observed, which the authors interpreted as the Fe$_{oct}$ atoms of the Fe$_{oct}$-Fe$_{tet}$-O termination. Occasionally, steps of 1.6 Å or 4.8 Å were observed, to terraces with a similar striped appearance, but with a different row separation of 17 Å (2×8.4 Å) or 34 Å (4×8.4 Å). On the basis of the STM images, a model was proposed in which additional rows with a Fe$_2$O$_3$-like stoichiometry form on the Fe$_{oct}$-Fe$_{tet}$-O termination.

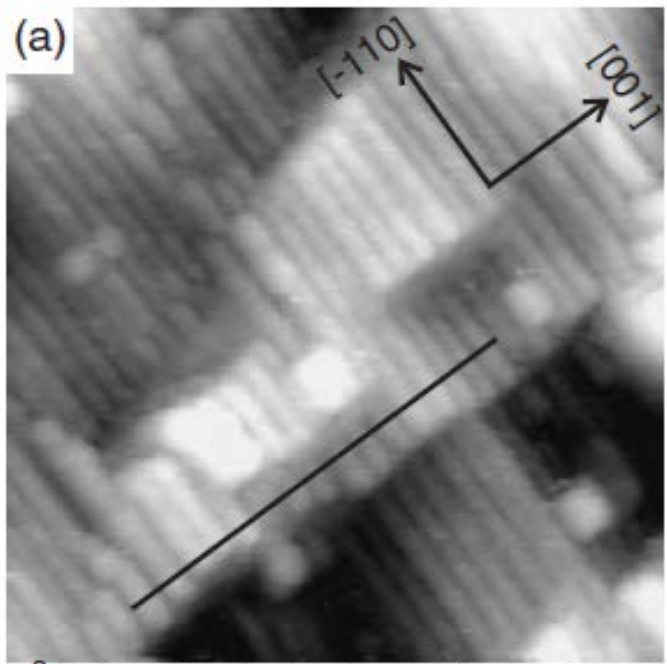

Figure 48: 700x700 Å² STM image showing the 1-dimensional reconstruction on the $Fe_3O_4$(110) surface. The rows are separated by approximately 25 Å. Reproduced from ref. [313].

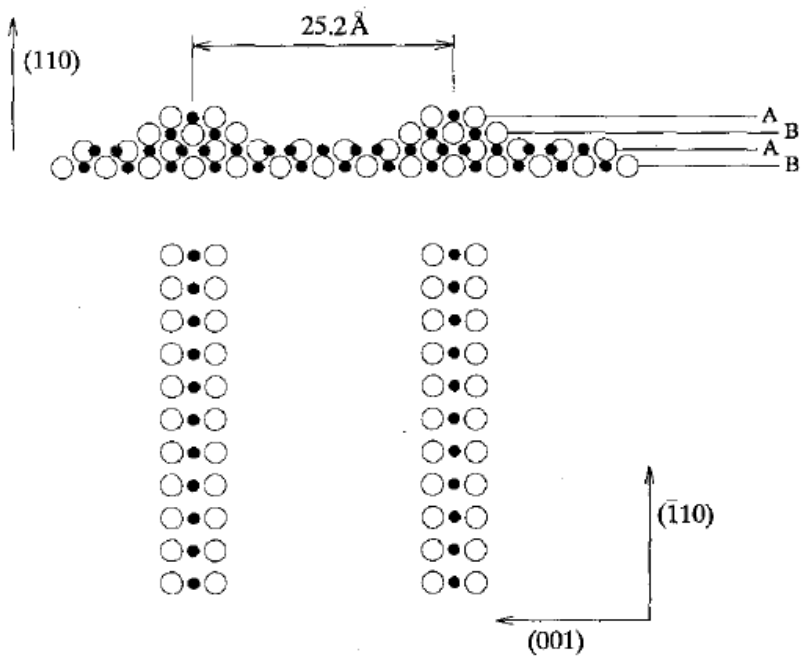

Figure 49. Side and top views of the structural model of $Fe_3O_4$(110) proposed by Jansen et al. [307] based on STM images. The rows observed in STM are thought to have oxidised stoichiometry with respect to the underlying bulk. Fe cations and O anions are filled and open circles, respectively. Reprinted from [307], Copyright 1995, with permission from Elsevier.

Later, Maris et al. [308; 313; 314] studied the surface of $Fe_3O_4$(110) films grown on MgO(110). They found that annealing to 1000 K in UHV was required to produce a well-ordered surface, but again, this procedure led to Mg interdiffusion, resulting in 0.06 ML Mg in AES spectra. In spite of this difficulty the authors observed a similar surface reconstruction to Jansen et al. in STM, with rows of protrusions in the [-110] direction separated by 25 Å. However, the appearance of the surface in atomically resolved images was different; each stripe along the [-110] direction was found to be 12.5 Å wide and exhibited an internal structure. The authors claim that the images show the $Fe_{oct}$ and O atoms as protrusions, with a shift by 1/4 of one unit mesh along the [-110] direction within the surface mesh. Based on the STM images, a structure based on an $Fe_{oct}$-$Fe_{tet}$-O layer was proposed. While the assignment of protrusions to oxygen is doubtful (the oxygen states are several eV away from the Fermi level and these atoms are not observed on the (100) and (111) surfaces), the influence of the Mg impurities on the surface is not clear and thus it is not possible to state that the imaged surface is representative of the "real" $Fe_3O_4$(110) surface. Mg has been shown to dramatically affect the structure of the $Fe_3O_4$(100) surface, for example [228]. Further experiments are necessary to determine the nature of the surface reconstruction.

In early 2016, the present authors group revisited the structure of the $Fe_3O_4$(1×3) surface with STM, LEED and RHEED experiments [315], and concluded that the one dimensional reconstruction reported by previous studies results from periodic nanofaceting of the surface to expose the lower energy (111) surface planes (see Figure 50). A line profile (Figure 51b) of the STM image in the [001] direction reveals that the ridge-trough height is 0.46 Å, greater than the distance between similar layers in the (110) direction, and thus inconsistent with the model in Figure 49. RHEED patterns acquired along the [$\bar{1}$10] azimuth at various electron energies between 15 and 35 keV show strong diffracted beams at a scattering angle of 70°, consistent with scattering from the (111) planes (inclined by 35.26° from (110), see Figure 50e). The termination of the (111) planes could not be determined by the experimental methods used, but given the discussion above it seems likely that either an $Fe_{tet1}$ or $Fe_{oct2}$ termination would be present.

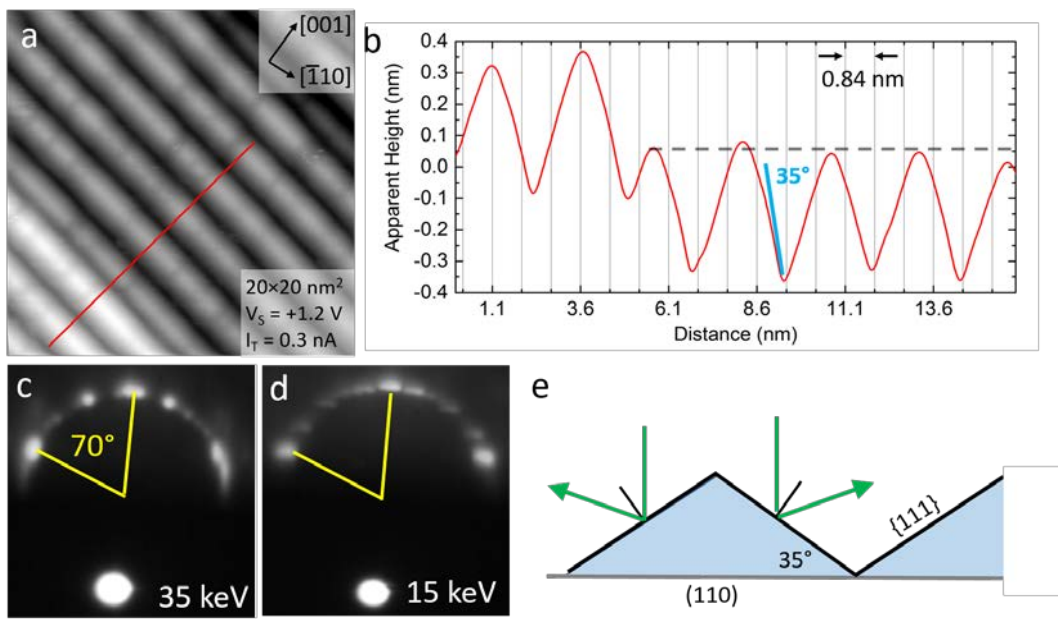

Figure 50: Experimental evidence that the $Fe_3O_4$(110) surface facets to expose (111) planes. (a) STM image of the $Fe_3O_4$(110)-(3×1) surface. (b) Line profile corresponding to the red line in (a). The depth of the trough is greater than the distance between similar layers in the (110) direction, and the maximum slope measured by STM is ≈35°. (c,d) RHEED patterns obtained from the $Fe_3O_4$(110)-(3x1) surface at 35 and 15 keV with the beam aligned with the [$\bar{1}$10] azimuth. The beam is strongly scattered through 70°, consistent with the existence of (111)-type nano-facets.

Density functional theory based calculations have been used to investigate the stability of the different terminations of $Fe_3O_4$(110) using the GGA [197; 309], LDA+U [310], and GGA+U [198] approach. In ref. [309], 6 alternative (110) surfaces were compared including the two bulk truncations, as well as 4 variations including oxygen or iron vacancies. The calculations predict an $Fe_{oct}$-O layer with 1/6 ML oxygen vacancies to be the lowest energy configuration across all oxygen chemical potentials (see Figure 51). Stoichiometric bulk truncations were found to be similarly stable at an oxygen chemical potential corresponding to UHV conditions. In general, the consensus appears to be that including a In the paper that utilizes GGA+U [198], only the bulk truncations were considered. The calculations suggest that both surfaces undergo significant relaxation with strong changes in the interlayer spacing. For the $Fe_{oct}$-$Fe_{tet}$-O termination, the first interlayer spacing contracts by -3.53%, while the second expands by 5.19%. Furthermore, the outer layer splits into 3 distinct sub-layers; the oxygen atoms are outermost, with the $Fe_{oct}$ atoms 0.113 Å below, followed by the $Fe_{tet}$ atoms a further 0.228 Å below. For an $Fe_{oct}$-O termination the first interlayer spacing contracts -6.48 % while the second expands by 4.04 Å. Again, the outermost layer splits into two distinct planes separated by 0.167 Å. In terms of electronic structure, the theoretical works predict a half-metallic surface for the $Fe_{oct}$-$Fe_{tet}$-O termination, while there is a disparity for the $Fe_{oct}$-O termination, which is half-metallic in GGA [309] but metallic in GGA+U [198].

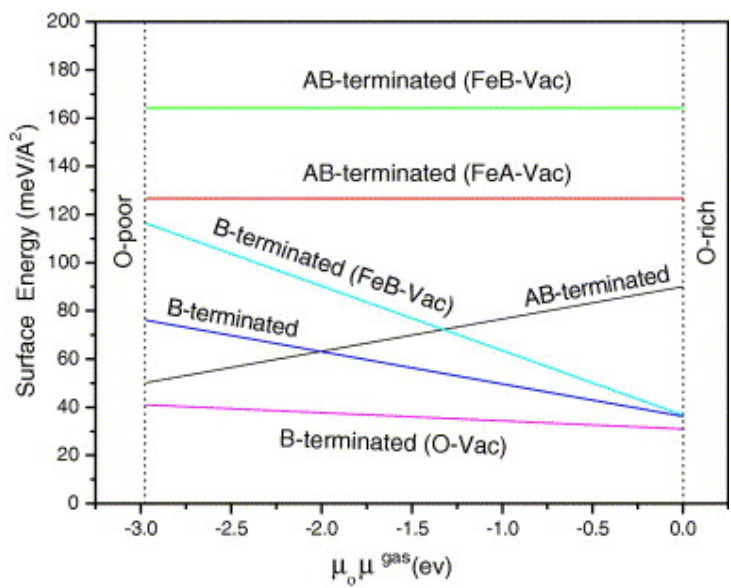

Figure 51: Theoretical surface phase diagram for $Fe_3O_4$(110) calculated by GGA calculations. Note that only surfaces with a (1×1) unit cell were considered. Reprinted from ref. [309], Copyright 2007, with permission from Elsevier.

In summary, the $Fe_3O_4$(110) surface appears to be unstable against faceting to (111), consistent with DFT+U calculations that predict (111) to have a lower surface energy [146; 198]. Theoretical calculations have been performed, but they ignore the experimental results and focus solely on comparing the stability of bulk truncations without taking the (1×3) reconstruction into account.

## 3.6   Maghemite Surfaces

To date, only one surface science study of a maghemite γ-$Fe_2O_3$ single crystal has been performed [316]. However, γ-$Fe_2O_3$ surfaces have been prepared on thin films, grown by plasma assisted MBE [224; 317; 318], directly and by oxidation of an existing $Fe_3O_4$(001) thin film [224], by $NO_2$ MBE [319], and by reactive sputtering [320; 321]. A general conclusion of this work to date is that molecular oxygen is not oxidising enough to prepare γ-$Fe_2O_3$ under UHV conditions [321].

The termination of the γ-$Fe_2O_3$(001) surface is controversial. Voogt et al. [255] suggest that the surface exhibits the same (√2×√2)R45° periodicity as $Fe_3O_4$(001), but Chambers and co-workers [224] demonstrated that oxidising the $Fe_3O_4$(001)-(√2×√2)R45° surface in an oxygen plasma at 520 K changes the surface to a (1×1) structure, which exhibits a Fe 2$p$ photoemission lineshape identical to an α-$Fe_2O_3$ reference sample. This suggests the film is fully oxidised. Interestingly, the (√2×√2)R45° periodicity was recovered by annealing at 523 K in UHV. On the basis of their measurements, Chambers and co-workers proposed a model for the γ-$Fe_2O_3$(001) surface in which the spinel structure is terminated at the $Fe_{tet}$ plane, and $Fe_{oct}$ vacancies occur in the first subsurface $Fe_{oct}$-O plane. Such a surface would exhibit a (1×1) LEED pattern if the vacancies are disordered, as in bulk γ-$Fe_2O_3$. As yet, this model has not been verified by other further investigations. Useful reference data for the core level shifts for the different iron oxide phases were published in ref. [322].

The structure of the γ-$Fe_2O_3$(111) surface was investigated recently by Bowker and co-workers [316] using a natural γ-$Fe_2O_3$(111) single crystal. Two different preparation recipes were used: (1) sputtering with $Ar^+$ followed by UHV annealing at 873 K for 20 minutes, and (2) annealing in 1×10$^{-6}$ mbar $O_2$ for 20 minutes at 873 K. The former treatment led to a (2×2) LEED pattern (NB. with respect to the oxygen lattice), while the more oxidising preparation led to a (√3×√3)R30° superstructure. STM images could only be acquired from the (2×2) surface, and appeared to resemble images of the $Fe_{tet1}$ and $Fe_{oct2}$ terminations of $Fe_3O_4$(111). The (√3×√3)R30° termination is proposed to be α-$Fe_2O_3$–like. Interestingly, the authors report that it is impossible to change between the two structures by annealing, and that sputtering was required to induce such a change. This might be linked to the different O stacking between the corundum and spinel structures.

## 3.7 Hematite α-$Fe_2O_3$ Surfaces

### 3.7.1 α-$Fe_2O_3$(0001)

#### 3.7.1.1 The Reduced $Fe_3O_4$(111)-like Termination

The current knowledge of the α-$Fe_2O_3$(0001) surface was recently discussed in the 2013 *Chemical Reviews* special issue on oxide surfaces in articles by Kuhlenbeck, Shaikhutdinov, and Freund [36] (focussing on thin-film studies of several metal oxides), as well as Woodruff [323] (comparing corundum and rocksalt oxide surfaces, with a focus on quantitative structural determinations). The former authors describe α-$Fe_2O_3$(0001) as "the most challenging and controversial among the iron oxides". The major problem facing single-crystal work is that in-situ cleaning by sputter/anneal cycles always produces reduced terminations, [324-327] even when annealing is performed in an $O_2$ atmosphere of 1×10$^{-6}$ mbar. This is because the corresponding chemical potential is close to the border with $Fe_3O_4$ (see Figure 2). Indeed, it is for this reason that the most commonly reported surface obtained for annealing temperatures up to 1050 K is a $Fe_3O_4$(111)-(2×2) surface [324; 326; 328] (see Figure 52).

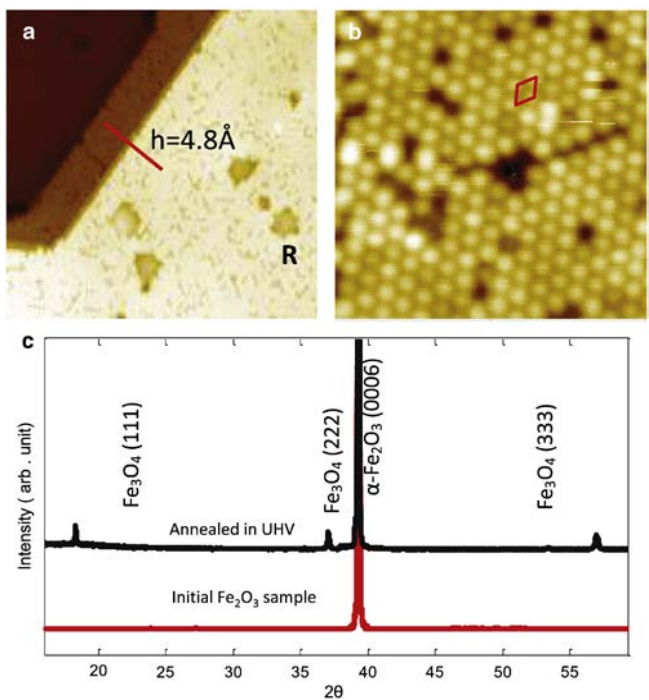

Figure 52: (a,b) STM images of the α-Fe$_2$O$_3$(0001) surface prepared by sputter/anneal cycles appear identical to Fe$_3$O$_4$(111). (c) After repeated cycles the Fe$_3$O$_4$ surface layer becomes sufficiently thick that it can be seen in XRD analysis of the sample. Reprinted from ref. [328]. Copyright 2007, with permission from Elsevier.

The Fe$_3$O$_4$(111) termination is also routinely observed on thin α-Fe$_2$O$_3$(0001) thin films grown on Pt(111). It makes sense, of course, that the α-Fe$_2$O$_3$(0001) surface might be reduced to Fe$_3$O$_4$(111) under UHV conditions, and the model seems to have been accepted. Indeed, the Osgood group [279; 281; 329-333] has frequently used this preparation procedure as a means to study the surface chemistry of the Fe$_3$O$_4$(111) surface. However, given the current controversy regarding the termination of Fe$_3$O$_4$(111) (see section 3.3.2), there must be similar doubt about whether this surface is terminated at the Fe$_{tet1}$ or Fe$_{oct2}$ layer. Indeed, STM images published by Condon et al. [326] reveal that two terminations can exist concurrently on this surface, just as on Fe$_3$O$_4$(111). Very recently, the Freund group has shown that an entire α-Fe$_2$O$_3$(0001) film on Ag(111) can be converted to Fe$_3$O$_4$(111) by UHV annealing [334] (see Figure 53), and measured the velocity of reaction fronts as part of a very thorough investigation of the transformation [335]. The structure of the fronts were found to depend on features such as step bunches on the underlying metal support.

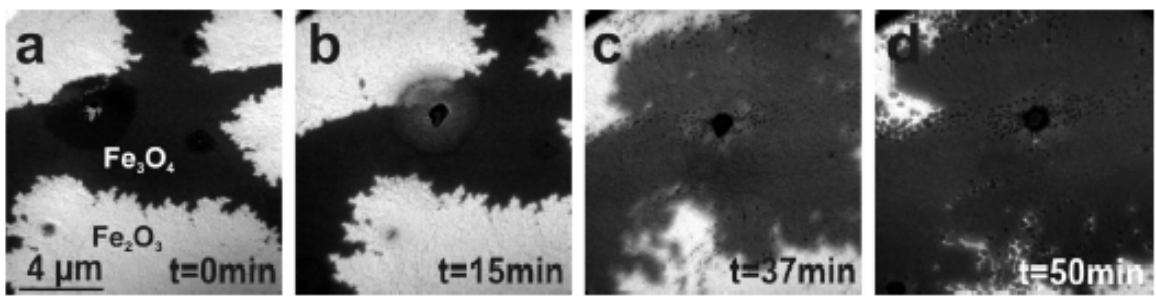

Figure 53: LEEM images showing the transformation of a mixed α-Fe$_2$O$_3$(0001)/Fe$_3$O$_4$(111) thin film on Ag(111) into Fe$_3$O$_4$(111) by annealing UHV at 823 K. Reprinted with permission from ref. [334]. Copyright 2014 American Chemical Society.

### 3.7.1.2 The Bi-phase Termination

The so-called bi-phase termination was first observed by Lad and Henrich in 1988 [324] on an α-Fe$_2$O$_3$(0001) single crystal prepared by sputtering and annealing at 1173 K in 1x10$^{-6}$ mbar O$_2$. A LEED pattern acquired from the surface exhibited six hexagonal spots around the integer order spots, and the authors proposed that such a LEED pattern emerged from multiple scattering across a Fe$_{1-x}$O(111)/α-Fe$_2$O$_3$(0001) or Fe$_3$O$_4$(111)/α-Fe$_2$O$_3$(0001) interface [324]. In the 1990's, Thornton's group [280; 326; 327; 336] studied both the Fe$_3$O$_4$(111) and α-Fe$_2$O$_3$(0001) surfaces by STM (amongst other techniques), and coined the term bi-phase ordering to describe the phenomenon of ordered islands of different phases coexisting on the surface. In the case of α-Fe$_2$O$_3$(0001), the surface prepared by Ar$^+$ sputtering followed by annealing at 1073 K in 7.75×10$^{-7}$ Torr O$_2$ was proposed to be based on Fe$_{1-x}$O and α-Fe$_2$O$_3$(0001) islands arranged in a 40 ± 5 Å superlattice, rotated 30° relative to the underlying α-Fe$_2$O$_3$ bulk. This interpretation was supported by the observation of domains containing protrusions with 5 Å and 3 Å periodicity, which were attributed to the α-Fe$_2$O$_3$ and Fe$_{1-x}$O, respectively. It should be noted that the bi-phase surface is not typically found to exist alone, but often coexists with the Fe$_3$O$_4$(111)-(2x2) termination, as shown in Figure 54. Interestingly, this STM image was obtained by oxidising the Fe$_3$O$_4$(111) surface shown in Figure 52 in 1x10$^{-6}$ mbar O$_2$ at 923 K, rather than reducing α-Fe$_2$O$_3$(0001), which highlights that reversibility of the reduction/oxidation processes and the clear link to the O$_2$ chemical potential during preparation.

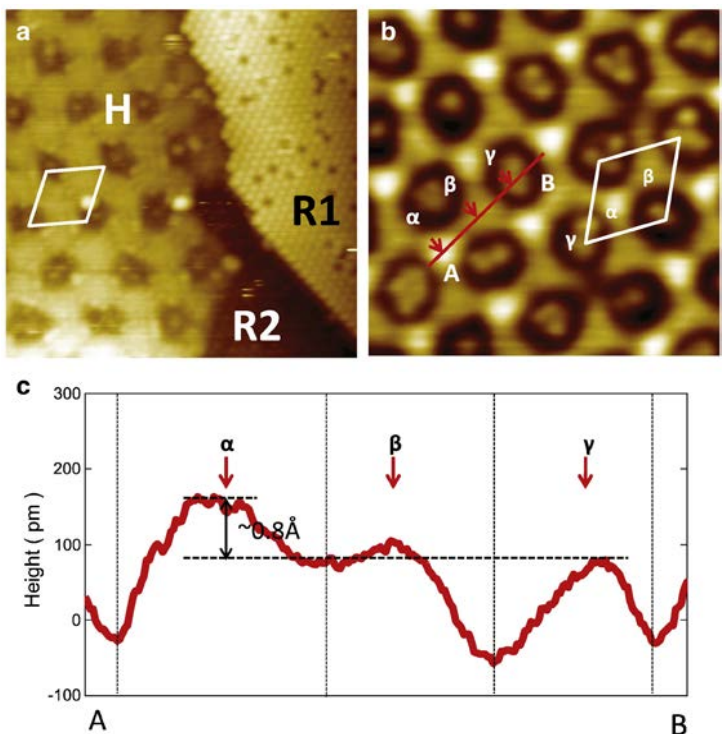

Figure 54: STM image of the "bi-phase" termination of α-Fe$_2$O$_3$(0001) labelled "H" in coexistence with the Fe$_3$O$_4$(111) termination ("R") following annealing of the surface shown in Figure 50 at 923 K in 1×10$^{-6}$ mbar O$_2$. The bi-phase is characterised by a long-range superstructure with a periodicity of 40 Å. α, β, and γ represent different parts of the unit cell. Reprinted from ref. [328]. Copyright 2007, with permission from Elsevier.

In 2009, Lanier et al. [337] challenged the $Fe_{1-x}O(111)/\alpha$-$Fe_2O_3(0001)$ interpretation of the bi-phase surface, and instead suggested that the reconstruction is related to a surface transformed to a $Fe_3O_4(111)$-like structure (Figure 55). Essentially, the authors discount the possibility of the $Fe_{1-x}O(111)/\alpha$-$Fe_2O_3(0001)$ bi-phase model [326] on the basis of TEM diffraction measurements, suggesting that a complex diffraction pattern would be observed for such a structure. Instead, they propose that the "floretting" observed in diffraction studies emerges from interfacial scattering from a subsurface $Fe_3O_4/\alpha$-$Fe_2O_3$ interface, much like the Moiré pattern often observed in thin film studies, including the FeO(111)/Pt(111) system [17]. However, It is not completely clear that the bi-phase termination was measured in this study, although the authors point out that the preparation conditions should produce the bi-phase surface, and that this is close to the boundary of $Fe_3O_4$ stability, not the $Fe_{1-x}O$ phase. Very recent work has shown that mild reduction of the bi-phase termination indeed leads to an increase in subsurface $Fe^{2+}$ [338], and it is well known that further reduction leads to the growth of Fe3O4(111) into the sample. In this author's opinion, it seems plausible that a thin $Fe_3O_4(111)$ structure could exist as a reduced skin on $\alpha$-$Fe_2O_3(0001)$ under reducing conditions, much like the SCV reconstruction forms as an oxidised $Fe_{11}O_{16}$ skin on $Fe_3O_4(100)$. The stabilization at two layers of $Fe_3O_4(111)$ under mildly reducing conditions might be linked to the change in stacking between the corundum (hcp, i.e. ABAB stacking) and spinel structures (fcc, ABCABC stacking). Ultimately though, the structure remains an open question, and the large size of the reconstruction makes it difficult to unambiguously prove either way with a technique such as LEED or SXRD, and the unit cell is large for DFT-based calculations, although there are examples of similarly complex long-range reconstructions being treated theoretically [339].

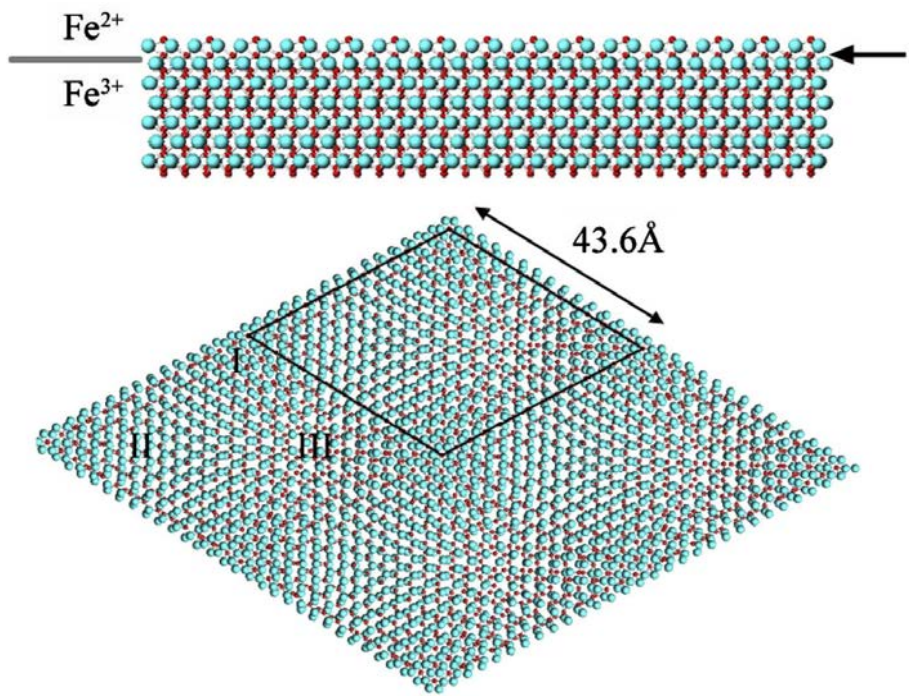

Figure 55: The model of the bi-phase reconstruction of $\alpha$-$Fe_2O_3(0001)$ proposed by Lanier et al. [337] is based on a thin layer of $Fe_3O_4(111)$-like structure (two O layers) formed over a $\alpha$-$Fe_2O_3(0001)$ bulk. Oxygen atoms are blue, and $Fe_{tet}^{3+}$ are red. Reprinted from ref. [337]. Copyright 2009, with permission from Elsevier.

### 3.7.1.3 Stoichiometric terminations of α-Fe$_2$O$_3$(0001)

Clearly, to stabilize a termination truly representative of α-Fe$_2$O$_3$(0001) requires higher oxygen chemical potential than is easily achieved in a UHV experiment. DFT-based calculations [165; 298; 340-343] suggest that an Fe termination containing ½ of the inter-plane Fe, sometimes termed a "half-metal" termination, should be stable at low oxygen pressures (note, formation of Fe$_3$O$_4$(111) surfaces not generally considered in the calculations). This surface is denoted Fe-O$_3$-Fe- by most authors, where the final "-" indicates the remainder underlying crystal, which has a -O$_3$-Fe-Fe-O$_3$-Fe-Fe- stacking. At high oxygen pressures however, the stable termination depends quite strongly on the details of the calculation. For example, the most recent calculations by Kiejna et al [344] find that a truncation at the oxygen plane (known as O$_3$-Fe-Fe-) and ferryl terminated surfaces (O=Fe-O$_3$-) can be stable at high O$_2$ pressures using GGA, but not with GGA+U (see Figure 56). This latter result differs somewhat from the earlier calculations of Bergermayer et al. [340], who found that a ferryl termination could be stabilized under a narrow window of conditions irrespective of the calculation details. Kiejna et al [344] suggest the discrepancy arises from the smaller basis set used in the prior work. The calculations universally find that large relaxations of the surface atoms occur (illustrated in Figure 57).

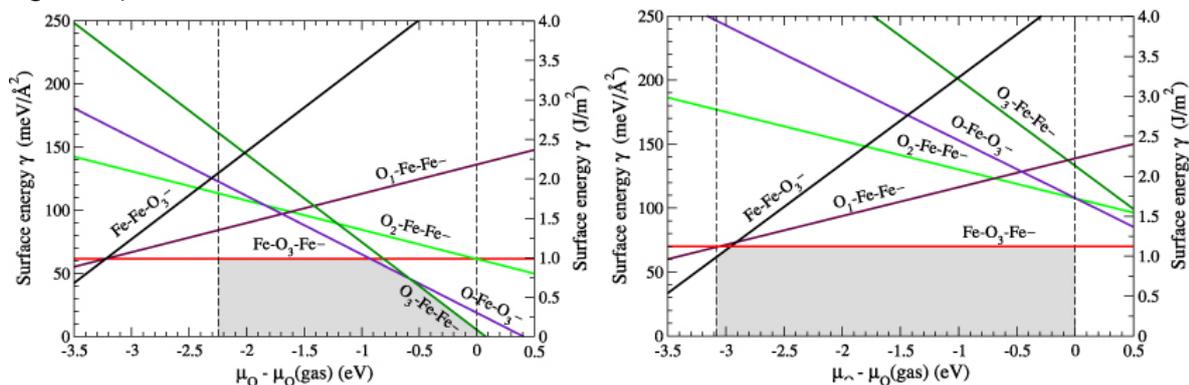

Figure 56: Surface phase diagram for Fe$_2$O$_3$(0001) calculated using GGA (a) and GGA+U (b). Note that the stability range varies significantly depending on the method, but the preference for a half metal (Fe-O$_3$-Fe-) termination is common to both. The oxygen (O$_3$-Fe-Fe-) and ferryl (O-Fe-O$_3$-) terminations are only favourable for GGA. Reprinted figure with permission from ref. [344]. Copyright 2012 IOP Science.

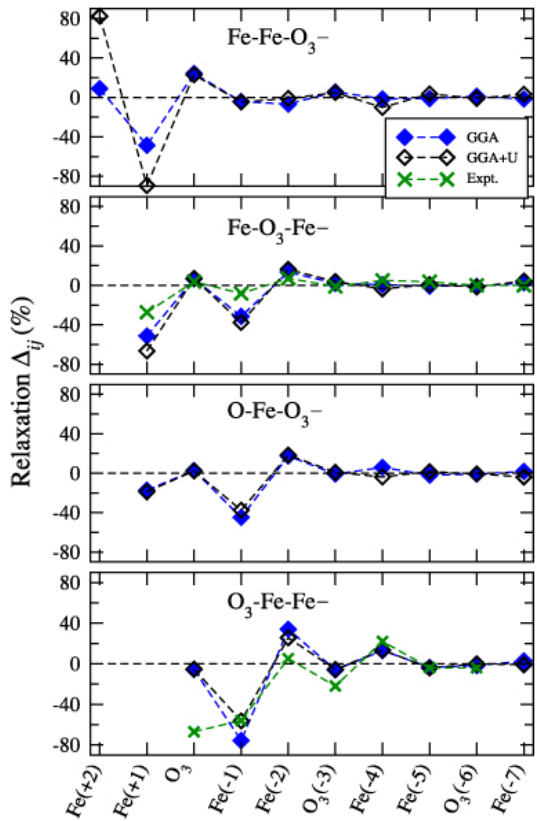

Figure 57: Layer relaxations for various α-Fe$_2$O$_3$(0001) terminations as determined by recent GGA and GGA+U calculations [298]. Experimentally derived comparison for the Fe-O$_3$-Fe- termination comes from the LEED IV study of Lübe and Moritz [345], while the comparison for O$_3$-Fe-Fe- comes from the work of Barbier et al [346]. Reprinted figure with permission from ref. [298] Copyright 2012 by the American Physical Society.

To date, there is experimental evidence for all the theoretically predicted terminations, but no clear recipe has been established to prepare a monophase termination of any. Shaikhutdinov and Weiss [347] grew α-Fe$_2$O$_3$(0001) thin films on Pt(111) under UHV-compatible conditions, and performed a final anneal at 1100 K in O$_2$ pressures between $10^{-6}$ and 1 mbar in a separate chamber to fully oxidise the film. Two distinct regimes were observed: at 1 mbar the surface exhibited a hexagonal lattice with 5 Å periodicity, with 1.8 Å between neighbouring terraces, while below $10^{-1}$ mbar this structure coexisted with a more highly corrugated hexagonal lattice of the same dimension. The authors proposed the former surface to be O$_3$-Fe-Fe terminated, with subsurface Fe cations imaged by STM. The structure that emerged at lower O$_2$ pressure was suggested to be Fe-O$_3$-O- terminated, with the surface Fe atoms imaged. The first LEED *IV* study [348] of this system appeared to confirm the high-pressure O$_3$-Fe-Fe termination, but the film prepared at $10^{-5}$ mbar was thought to be hydroxylated through reaction with water in the residual gas.

A more recent LEED *IV* investigation of this surface performed by Moritz and co-workers [345] found best agreement (Pendry R-factor 0.34 ) for a half-metal (Fe-O$_3$-Fe-) termination, but with only 50% occupation of the outermost Fe layer. Unfortunately no images of the corresponding surface were acquired to corroborate this aspect of the structure. In this study, a synthetic single crystal sample was annealed at 773 K in $10^{-8}$ mbar O$_2$ for several hours without sputtering (the authors recognised this would likely lead to reduction of the surface). The preparation conditions correspond to an oxygen chemical potential of -1.8 eV, which is in the range where both GGA and GGA+U predict the half-metal termination to be stable. Interestingly though, the relaxations determined from the structure optimization were significantly smaller than those predicted by theory (see Figure 57). The half-metal termination has also been reported based on an XPD study [349] conducted on an α-$Fe_2O_3$(0001) thin film grown on α-$Al_2O_3$(0001) (in very oxidising conditions), but the measured relaxations were again very different.

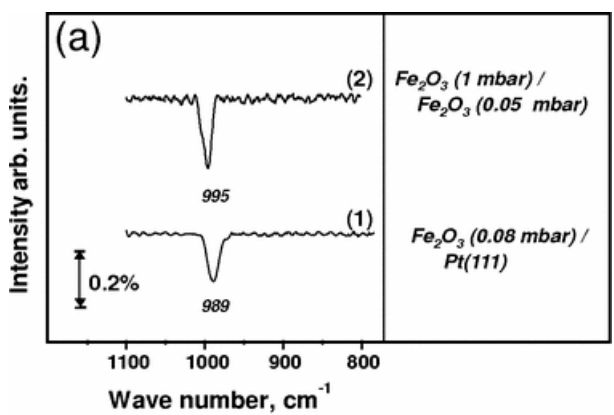

Figure 58: IRAS data reveal an absorption peak close to 1000 cm$^{-1}$ attributable to surface ferryl groups when an α-$Fe_2O_3$(0001) thin film is annealed in at 1050 K in $O_2$ pressures above 10$^{-3}$ mbar. Figure adapted with permission from Ref. [350]. Copyrighted by the American Physical Society.

Ferryl species, not considered in early experimental work, were detected [346] in IRAS experiments conducted on an α-$Fe_2O_3$(0001) thin film prepared at 10$^{-3}$ mbar $O_2$ (see Figure 58). Similar vanadyl species have been reported on $V_2O_3$(0001), and although controversial [351; 352], recent work from the Freund group has confirmed their presence [353-355]. SXRD experiments by Barbier et al. [346] find evidence that a natural α-$Fe_2O_3$(0001) single crystal exhibits the $O_3$-Fe-Fe- termination after heating in 10$^{-5}$ mbar $O_2$ for 1 hour, which remains when the sample is heated at 573 K at different $O_2$ pressures, but that a ferryl termination emerges when the temperature is increased to 873 K in UHV. This can be removed by increasing the $O_2$ pressure above 10$^{-6}$ mbar. At this temperature, the $O_3$-Fe-Fe survives all the way up to 1 mbar.

In concluding this section, this author wishes to echo the sentiment of Kuhlenbeck, Shaikhutdinov and Freund [36]; α-$Fe_2O_3$(0001) is certainly the most difficult and controversial of the iron oxide surfaces, partly because the preparation conditions required to create a stoichiometric surface are out of the comfort zone of surface science experiments. The available evidence suggests that half-metal (Fe-$O_3$-Fe), $O_3$-Fe-Fe and ferryl terminations can exist, but much work remains to be done to develop procedures to create monophase, well-characterised terminations in conditions compatible with UHV. Thin-film growth by MBE is complicated by the high $O_2$ pressures required to fully oxidise the film, and sputter/anneal cycles of single crystals are to be avoided as they lead to reduced surfaces. Annealing single crystals in the range of chemical potential linked to α-$Fe_2O_3$(0001) stability without sputtering appears to result in clean surfaces with a (1×1) LEED pattern [345], but as yet there is no atomic-scale imaging of this preparation technique to assess the defect density or whether multiple terminations coexist (the relatively poor LEED *IV* R-factor of 0.34 suggests this may be the case [345]). An alternative approach could be that of Bedzyk and co-workers [356] who report obtaining a sharp (1×1) LEED pattern when a natural single crystal was Ar$^+$ sputtered and then annealed at 723 K in a stream of atomic oxygen. The atomic species were produced by passing molecular $O_2$ at $1\times10^{-6}$ torr through a refractory capillary held at 1250 K. Pratt et al. [357] have also recently described a "safe, low-cost, and practical" low concentration ozone apparatus that is claimed to clean oxide surfaces of C without the need for high temperature annealing.

### 3.7.2 The α-Fe$_2$O$_3$(1$\bar{1}$02) "R-Cut" Surface

Given that the (1$\bar{1}$02) surface (also known as (012) or the "R-cut" surface) is prevalent on nano-hematite (see Figure 23) it is somewhat surprising that this surface has not received more attention from the surface science community. One possible reason is that many groups grow (insulating) α-Fe$_2$O$_3$ as a thin film on metal substrates, thereby taking advantage of the conductivity of the substrate. However, as discussed above and in Section 3.8, such growth often begins with formation of an FeO(111) bilayer (even on Ag(100) [358]), which then sets the (0001) basal plane parallel to the surface and templates further growth in the (0001) direction. The α-Fe$_2$O$_3$(1$\bar{1}$02) surface has been studied on several occasions on single crystals, [324; 359-366] most notably by Henderson at PNNL. A (1×1) surface is best prepared by annealing at 750 K in $\geqslant 5\times10^{-7}$ Torr $O_2$, followed by cooling in $O_2$ to below 500 K. Annealing at 950 K leads to reduction in vacuum, with $Fe^{2+}$ observed in XPS and a (2×1) reconstruction in LEED. In contrast to (0001) it is reported to be straightforward to regenerate the (1×1) surface by annealing in $O_2$. Models based on a bulk-terminated surface and a missing-row structure are shown in Figure 59. Gautier-Soyer et al. proposed that the reconstruction extends some 30 Å into the bulk [359], which would be inconsistent with the model structure, but this finding was later refuted on the basis of HREELS data [366]. Unfortunately, neither SPM nor a quantitative structural determination has been performed on either termination to date, and thus the finer details remain to be determined.

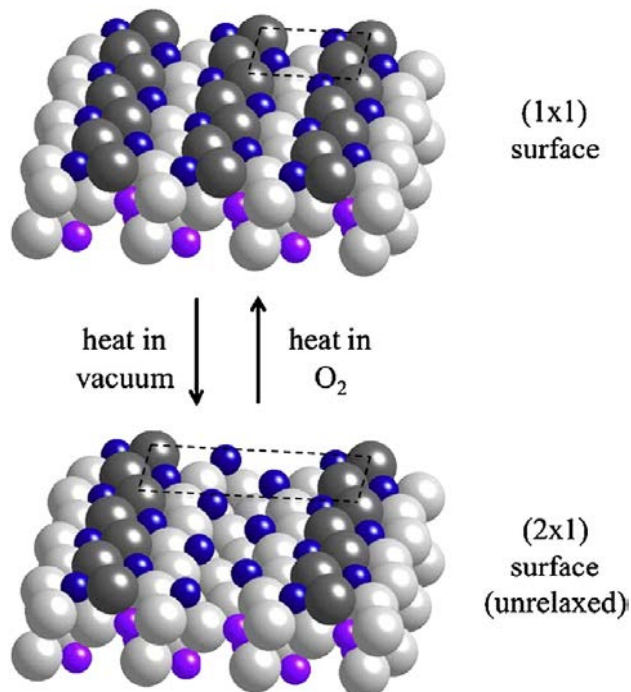

Figure 59: Proposed models of an ideal (1×1) (top) and the unrelaxed (2×1) α-Fe$_2$O$_3$(1$\bar{1}$02) surfaces. The (2×1) surface, obtained experimentally by annealing in UHV, is constructed by removing a row of bridging oxygen atoms from the (1×1) model. Large grey balls are oxygen atoms, smaller blue balls are surface Fe. Figure reprinted from ref. [362] with permission from Elsevier.

Interestingly, it has been reported that a fully hydroxylated stoichiometric surface can be prepared by wet chemical treatment followed by annealing in air at 700 K [364; 365]. The treatment consisted of polishing with a high pH (>10) colloidal silica solution, followed by washing with a pH 10 NaOH solution and multiple rinses with ultra-pure (>18 MΩ cm) water. The samples were then etched in 0.01 M $HNO_3$ for approximately 2 h followed again by multiple rinses with ultra-pure water. This produced a clean, but fully hydroxylated surface with terrace widths on the order of 200 nm and predominant step heights of approximately 3.6 Å. The RMS roughness of the surface as measured by AFM was reported to be less than 5 Å.

## 3.8 Wüstite ($Fe_{1-x}O$) Surfaces

When Fe is deposited on a single-crystal metal surface in a partial pressure of oxygen a monolayer film of FeO(111) wets the surface prior to the growth of other iron oxides. The properties of this classic ultrathin-film system have been extensively studied in the surface-science literature (e.g. [358; 367-372]). The surface is easily and reproducibly prepared in UHV, and the ultrathin film on Pt(111), in particular, has become a model system through which much exciting physics and chemistry was discovered [373] [374]. One of the most important lessons learned about this system is that the most exotic properties are linked to the close proximity of the metal support to the surface, and are not simply due to the FeO(111). Indeed, the O-terminated FeO(111) bi-layer film is particularly inert [303; 375-380] because undercoordinated cations/anion pairs are generally required for dissociative adsorption on metal oxide surfaces [381; 382]. However, the thin film can be reduced through exposure to H and alcohols via a Mars-van Krevelen mechanism at elevated temperature [383-385], with undercoordinated Fe dissolving into the Pt crystal.

Since this article is intent on describing the properties of iron-oxide surfaces, the interesting properties of the ultrathin system lie somewhat beyond the scope. Indeed, a proper treatment of this field would no doubt require a review article in its own right. However, it would be remiss not to at least mention a few highlights of the recent literature. Giordano et al. [372] have shown that the properties of the thin film are strongly influenced by the coupling to the metal substrate, and that properties such as the work function vary across the resulting Moiré structure. Au and Pd adatoms become charged by electrons that tunnel through the oxide [386; 387]. A fascinating recent development came with the discovery that under strongly oxidising conditions the FeO monolayer can be transformed to a O-Fe-O (i.e. $FeO_2$) trilayer structure [388]. This unusual iron oxide has no bulk counterpart and is stabilized by the interaction with the substrate. Of course, such an oxidised surface is easily reduced, and surface reacts readily with CO to form $CO_2$ [389] via a Mars-van Krevelen type mechanism [390]. NO can also be used to initially form the trilayer, but $O_2$ performs better in the CO oxidation reaction because it reacts with $V_O$s [391]. Until recently, it was thought that the growth of FeO(111) films terminated at two FeO bilayers on Pt(111) due to the intrinsic polarity of the structure [254]. Above this coverage, $Fe_3O_4$ islands were observed to form [17]. However, Spiridis et al. recently found that FeO films up to 17 bilayers thick could be stabilized if a thin film was built up layer by layer by deposited of Fe, with subsequent post anealing at 570 K in 10 L of $O_2$ [392]. However, the properties of these thicker films are also very different to those of bulk $Fe_{1-x}O$ [393].

There are no surface science studies on $Fe_{1-x}O$ single crystals (one imagines this material would be difficult to prepare in situ and would most likely transform to $Fe_3O_4$ on heating, see Figure 2). To get a better idea of the intrinsic properties of $Fe_{1-x}O$ surfaces one might think it easier to grow the a thin film of the non-polar $Fe_{1-x}O(100)$ orientation. However, even growth on Ag(100), which has a close lattice match to FeO(100), results in the growth of a hexagonal FeO(111) film, which clearly demonstrates that the polar orientation is strongly favoured in the few-monolayer regime. However, Lundgren and co-workers [358] found that grains of $Fe_{1-x}O(100)$ were formed via the oxidation of metallic Fe grains that formed during high temperature (623 K) Fe growth at $2\times10^{-7}$ mbar $O_2$ on Ag(100), while Abreu et al. [394] report that growth of FeO(100) can be achieved with a substrate temperature of approximately 773 K and an oxygen pressure between $5\times10^{-8}$ mbar and $1\times10^{-7}$ mbar. However, some $Fe_3O_4$ was always present within the film. A subsequent LEED *IV* investigation [395] ($R_P = 0.23$) of a film determined to be 22 ML thick and 90 % FeO(100) found the surface to be bulk terminated with a small rumple of 3.9 ±3.2 % of the equivalent bulk interlayer spacing.

## 4  Molecular Adsorption on Iron-Oxide Surfaces

### 4.1  Hydrogen

Hydrogen is an important adsorbate on iron oxide surfaces, partly because OH groups form through dissociation of water in a humid or aqueous atmosphere, but also because $H_2$ is often used to reduce iron oxides at high temperature [38; 396]. The reduction of iron oxides has been extensively studied outside the vacuum chamber, and the interested reader is referred to the reviews by Pinaeu et al. regarding the reduction of hematite [397], and magnetite [193]. Briefly, 3-stage ($3Fe_2O_3 \rightarrow 2Fe_3O_4 \rightarrow 6FeO \rightarrow 6Fe$) and 2-stage ($3Fe_2O_3 \rightarrow 2Fe_3O_4 \rightarrow 6Fe$) reactions are considered. Generally, reduction in $H_2$ occurs above 600 K (see Figure 60). A key issue is whether FeO is formed as an intermediate below 570 °C, although it should not be thermodynamically stable under such conditions. Pinaeu et al. [193] suggest FeO can form at temperatures as low as 30 °C under non-equilibrium conditions. In the absence of FeO, metallic Fe is thought to nucleate directly from $Fe_3O_4$. The reaction of $Fe_3O_4$ with $H_2$ is known to produce water ($Fe_3O_4 + 4H_2 \rightarrow 3Fe + 4H_2O$), so it seems likely that $H_2$ reacts with $O_{lattice}$ at the surface, and that excess Fe is accommodated in interstitial sites in the bulk up to some limit when metallic Fe nucleates. The details of the process are not well understood however, for example, it is not known whether $H_2$ dissociates to form surface OH groups prior to extraction of $O_{lattice}$. Theoretical calculations [398] for $H_2$ adsorption on $Fe_3O_4(111)$ find that homolytic dissociation to create two surface OH groups is preferred, whereas on the $Fe_{oct2}$ termination heterolytic dissociation is also possible with the formation of Fe-H bonds. The only instance of the high-temperature $H_2$ reduction being studied in the vacuum chamber was a study by Ceballos et al. [215; 216], who observed the coexistence of Fe dimer and $Fe_{oct}$-O terminations when they annealed a natural $Fe_3O_4(100)$ single crystal at 990 K in 2000 Langmuir $H_2$. As discussed in section 3.3.4, Fe-rich terminations are clear evidence of surface reduction.

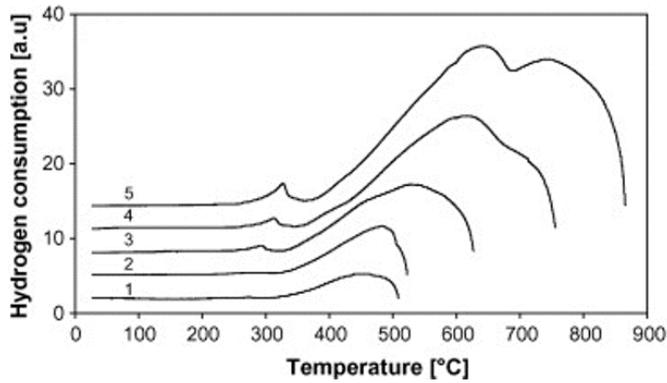

Figure 60. TPR$_{H2}$ curves for Fe$_3$O$_4$ in a 5% H$_2$–95% Ar gas flow with a heating rate of (1) 0.58; (2) 1.07; (3) 2.57; (4) 5.52; (5) 10.7 °C/min. The small peak at 300 K is attributed to hematite impurity. Rapid reduction of Fe$_3$O$_4$ begins above 573 K (300 °C). Based on an Arrhenius plot of these data, an activation energy for H$_2$ reduction of Fe$_3$O$_4$ was estimated at 55 kJ/mol in ref. [38]. Reprinted from ref. [38], Copyright 2007, with permission from Elsevier.

### 4.1.1 H/Fe$_3$O$_4$(100)

Under UHV conditions molecular H$_2$ does not adsorb on any magnetite surface at room temperature. Although not strictly a molecule, atomic hydrogen does adsorb and results in the formation of surface hydroxyl groups. This is typically achieved by backfilling the vacuum chamber with molecular H$_2$, and then cracking some fraction of the molecules with a hot tungsten filament or commercial gas cracker such that radicals are incident on the sample surface. As discussed in Section 3.3.3, surface hydroxyl groups are visible in STM images of the Fe$_3$O$_4$(100) surface because they modify the DOS of nearby Fe$_{oct}$ atoms, making them appear brighter in empty states images (see Figure 32) [98; 399; 400]. DFT+U calculations [66; 91; 98] show that charge donated to the system by the H adsorption is transferred into t$_{2g}$ orbitals of the Fe$_{oct}$ atoms, which become more Fe$^{2+}$-like, and the surface becomes progressively reduced. STS spectra acquired over OH-related protrusions at the Fe$_{oct}$ rows reveal a state 0.2 eV below E$_F$ consistent with this picture (see Figure 61) [399; 400].

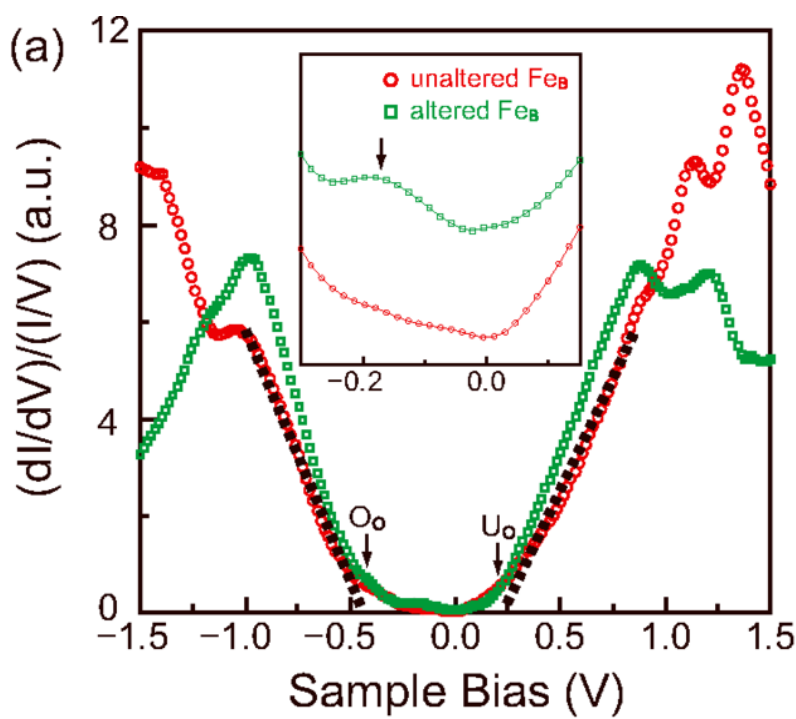

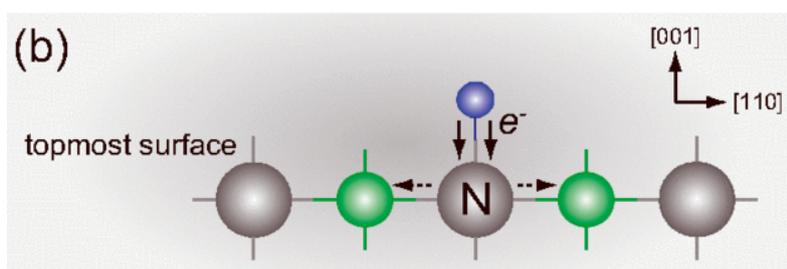

Figure 61: (a) STS spectra acquired over a regular $Fe_{oct}$ atom (red) and an OH-altered $Fe_{oct}$ atom (green). The inset reveals the emergence of a small state 0.2 eV below the band gap. (b) Schematic model of an H-adsorbed $Fe_3O_4(001)$ surface. Surface $Fe_{oct}$ atoms are green, oxygen atoms grey, and the H atom blue. The arrows indicate the directions of electron flow. Reprinted figure with permission from ref. [399]. Copyright 2015 by the American Physical Society.

At low coverages OH groups adsorb preferentially at the not-blocked sites of the SCV reconstruction, i.e. the O atoms without a subsurface $Fe_{tet}$ neighbour (Figure 62c). DFT+U calculations based on the DBT surface model suggest that an isolated OH group adsorbs with a binding energy of -0.85 eV, and forms an angle 18° from the surface plane. The H atom forms a hydrogen bond with the symmetrically equivalent $O_{lattice}$ atom opposite [98], which facilitates diffusion between the two sites. This diffusion is seen in STM movies as the bright features associated with the OH group switching back and forth between the $Fe_{oct}$ rows across the not-blocked site [98]. Since the vicinity of the adsorption site does not differ much between the DBT and SCV structures, no major changes are expected if the calculations are performed with the SCV model. This is not the case as the coverage increases though, because the SCV model only has two surface O atoms per unit cell without a subsurface $Fe_{tet}$ neighbour, while the DBT model has four. DFT+U calculations [66] suggest that OH formation at $O_{lattice}$ atoms with a surface $Fe_{tet}$ neighbour is not preferred. In experiment, increasing

OH coverage leads to the Fe$_{oct}$ rows appearing straighter and brighter in empty states STM images (Figure 62 e).

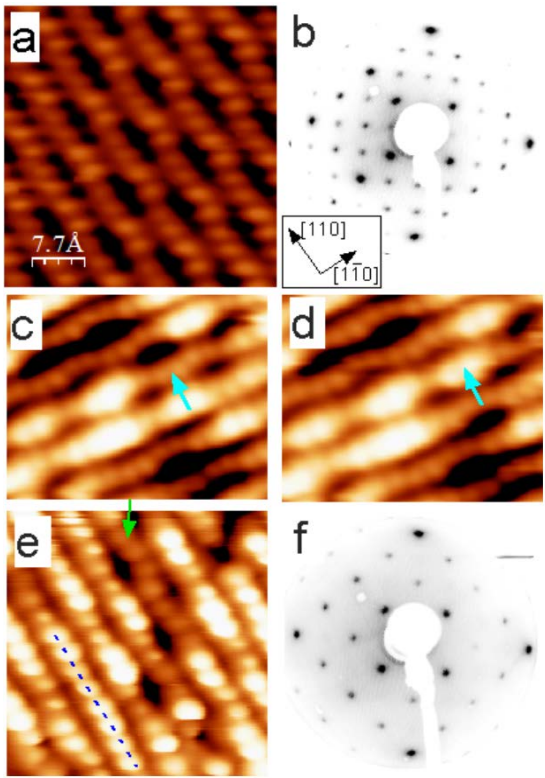

Figure 62: Hydroxylation of the Fe$_3$O$_4$(100) surface by atomic H. (a) STM image (V$_{sample}$=+1.7 V, I$_{tunnel}$=0.14 nA) and (b) LEED pattern (E$_{el}$=90 eV) from the clean Fe$_3$O$_4$(100) surface. (c) and (d) Consecutive STM images (4 nm×3.5 nm$^2$, 1.4 V, 0.14 nA) at very low atomic H coverage. The arrows mark the motion of an OH-induced bright double protrusion to a neighbouring Fe$_{oct}$ row. (e) STM image at a coverage close to 2 OH per unit cell (4×4 nm$^2$, 0.76 V, 0.21 nA). One row of Fe$_{oct}$ atoms is bright and straight (dashed blue line), an uncovered section, marked by the green arrow, displays the undulations characteristic of the clean SCV surface. (f) LEED pattern from the H saturated surface displaying (1×1) symmetry. Reprinted figure with permission from ref. [98]. Copyright 2010 by the American Physical Society.

At saturation coverage (Figure 62 f), the (√2×√2)R45° reconstruction is lifted in LEED [97; 98] and a significant increase in the density of states near E$_F$ is observed in UPS [98]. XPS measurements reveal that this surface is strongly enriched in Fe$^{2+}$ [98]. Interestingly, metastable helium beam scattering performed on a Fe$_3$O$_4$(100) thin film suggests that the surface spin-polarization increases from < 5 % to > 50 % [97]. The phenomenon has been studied theoretically [66; 91; 98], and DFT+U calculations based on the DBT structure suggest that Fe$^{2+}$-like states just below E$_F$, similar to those found in bulk Fe$_3$O$_4$ are formed. Note, the STS spectra shown in Figure 61 showed the emergence of such a state with a single OH group.

Lifting the (√2×√2)R45° reconstruction by H adsorption is a very interesting unresolved issue linked to the discovery of the SCV structure. Either the adsorption of the surface OH groups induces a rearrangement in the subsurface cations, or H is somehow incorporated in the subsurface lattice and directly affects the structure. Further calculations and experiments are clearly required to address this, and it is particularly important because similar reconstruction lifting has been observed following deprotonation of HCOOH [401] and H$_2$O [71] at this surface.

Heating a hydroxylated Fe$_3$O$_4$ surface above 520 K results in desorption of water (not H$_2$), and the observation of single-monolayer deep holes in the oxide support [257; 402]. This occurs when two OH groups react with the surface [257], removing an O$_{lattice}$ atom and creating a V$_O$, a well-known process on oxide surfaces such as TiO$_2$(110) [34]. However, since V$_O$s do not appear to be stable on Fe$_3$O$_4$(100), excess Fe diffuses inside the sample crystal when the surface is reduced [402].

### 4.1.2 H/Fe$_3$O$_4$(111)

On the Fe$_3$O$_4$(111) surface, Huang and Ranke [384] observed irregular bursts of H$_2$ and H$_2$O while depositing atomic H at 335 K (the origin of the bursts is not clear), but still observed a small OH peak in XPS and an increase in Fe$^{2+}$ relative to the clean surface. Two peaks for H$_2$O were observed in TPD, at 440 and 560 K, suggesting that OH groups combine on the surface and extract an O$_{lattice}$ atom to create water. Since no change in XPS spectra was detected, the authors proposed that Fe atoms diffused into the bulk, as observed when the Fe$_3$O$_4$(100) is reduced in this way [35]. At present, there is no STM study of atomic H adsorption on Fe$_3$O$_4$(111). It could be possible to image OH if additional Fe$^{2+}$ is formed in the surface, as the experiments of Huang and Ranke [384] suggest, as this is how it is imaged on the (100) surface. Shaikhutdinov et al. [305] have published STM images showing several (as yet unidentified) mobile features related to adsorption from the residual gas, one of which could well be surface a surface OH group.

Pratt et al. [92] utilized their metastable helium beam scattering technique to Fe$_3$O$_4$(111), and found that the surface exhibits a positive spin polarization at the Fermi level. This somewhat surprising result was also found in DFT+$U$ calculations [92; 294], and attributed to filled majority spin states on the surface Fe$_{tet1}$ atoms. In contrast to the (100) surface however, hydroxylation had little impact on the magnitude of the spin asymmetry.

### 4.1.3 H/Fe$_3$O$_4$(110)

There has been one theoretical study of H$_2$ adsorption on Fe$_3$O$_4$(110) [403], albeit on bulk truncated surfaces inconsistent with experimental observation. Using GGA+U, Yu et al. determined that H adsorbs preferentially at O atoms to form hydroxyl groups. The adsorption was found to be stronger on the Fe$_{oct}$-O termination, which exposes twofold coordinated oxygen, in comparison to the Fe$_{oct}$-Fe$_{tet}$-O termination, which exposes threefold coordinated oxygen atoms.

### 4.1.4 H/α-Fe$_2$O$_3$(0001)

Huang et al. [404] deposited atomic H on a bi-phase terminated α-Fe$_2$O$_3$(0001) thin film and then studied the desorption of water and hydrogen as a function of temperature. O1s XPS spectra are consistent with hydroxylation of the surface at room temperature, with reduction of Fe to the 2+ oxidation state (Figure 63). Saturation coverage (1 H per unit cell) is obtained already at 1 L. Heating the 1 and 10 L atomic-H surfaces led to H$_2$O desorption, consistent with loss of oxygen from the surface, and in-line with observations for Fe$_3$O$_4$ surfaces discussed above. Further exposure led to the reduction of the film, and after 1000 L exposure at room temperature, the entire 3-4 nm film was thought transformed to Fe$_3$O$_4$. For the higher exposures H$_2$ desorption was also observed, and the sample remained in its reduced state following the TPD acquisition.

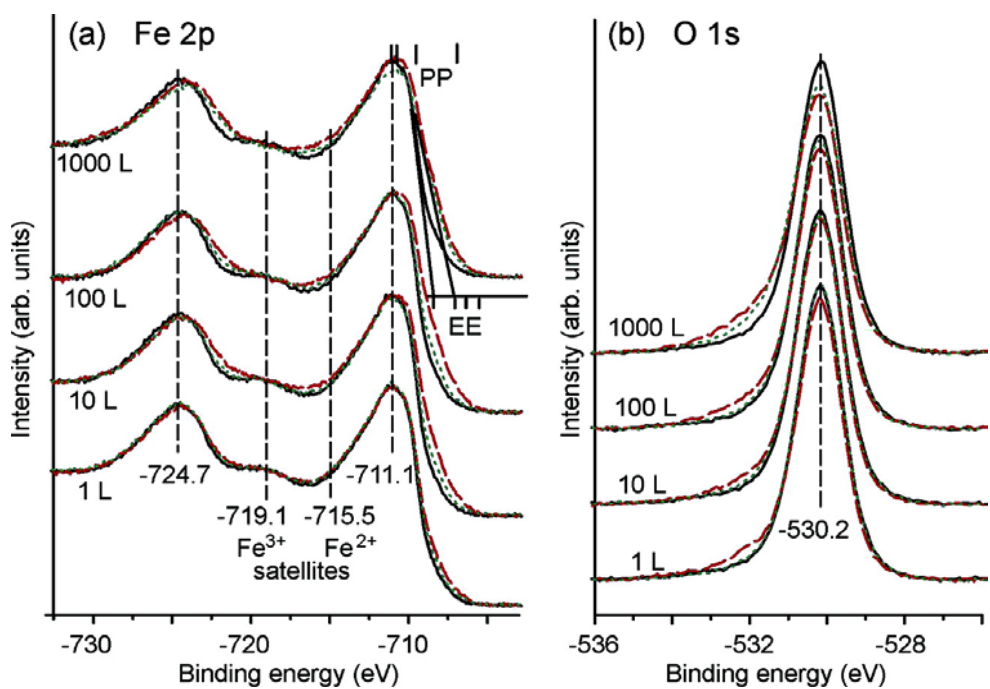

Figure 63: XPS spectra acquired for a bi-phase α-$Fe_2O_3$(0001) thin film (black), and following exposure to atomic H (red), and following TPD (green dashed). Reprinted with permission from ref. [404]. Copyright 2007 American Chemical Society.

## 4.2  Water ($H_2O$)

### 4.2.1  $H_2O$/$Fe_3O_4$(100)

The adsorption of water on $Fe_3O_4$(100) has been studied on several occasions, but a unified picture is yet to emerge. This is partly because older experimental data must be re-evaluated in the light of the SCV structure, but since the surface layer is quite similar to the previous DBT model, basic adsorption trends based on prior DFT calculations are probably valid. As discussed above, one of the most interesting questions to arise with the SCV structural model is how the (√2×√2)R45° reconstruction is lifted, given that a rearrangement of the subsurface cations is necessary to recover a (1×1) periodicity in LEED. That this phenomenon is also observed with exposure to atomic H [98] and following dissociative adsorption of formic acid [401], suggests that adsorbed H atoms are ultimately responsible.

Published TPD measurements of the $D_2O$/$Fe_3O_4$(100) system (in fact, the only TPD data from the $Fe_3O_4$(001) surface) reveal a broad multi-component desorption peak extending from 200 K up to 350 K, with a small peak at 520 K [405] linked to recombinative desorption [257]. Although the surface in question was not characterised by other techniques, the preparation conditions used (annealing a thin film sample in 1x10$^{-7}$ mbar $O_2$ at 573 K) probably resulted in the $Fe_{oct}$-O terminated surface. HREELS data show that a surface exposed to water at 150 K contains a mixture of OH groups and molecular water [406]. On heating, the signal due to molecular water was lost at approximately 240 K, while the peaks attributed to OH groups disappeared at approximately 400 K.

The adsorption of water close to room temperature can be separated into 3 regimes. At the lowest exposures, experiment and theory agree that water will dissociate at surface $V_{OS}$, should they be present [65; 257], repairing the vacancy and adsorbing two surface hydrogen atoms to lattice oxygen atoms; a well-known process on metal oxide surfaces [407; 408]. STM images of the as-prepared $Fe_{oct}$-O surface always exhibit pairs of $O_{surface}$H groups in neighbouring unit cells that result from the reaction of residual-gas water molecules with $V_O$s formed during sputter/anneal cycles [257]. The density of such sites is very low however [257], so the fact that the $V_O$s are rapidly saturated in $10^{-11}$ suggests a mobile precursor state of water on the surface.

The mechanism underlying further dissociation of water on the non-defective surface is not yet clear. DFT+U calculations based on the DBT structure suggest water dissociation (-0.76 eV) is favourable over molecular adsorption (-0.39 eV) [65], yielding a $O_{water}$H species bound to the $Fe_{oct}$ row and a hydrogen atom bound to the metal-oxide lattice (i.e. a $O_{surface}$H species). However, STM and LEIS experiments conducted at room temperature only observe the $O_{surface}$H species [257]. This strange result suggests that the $O_{water}$H disappears from the surface following room temperature adsorption. It could be that water-water interactions result in the desorption of $O_2$, as suggested in ref. [257], or that water oxidises unknown species on the surface or present in the residual gas.

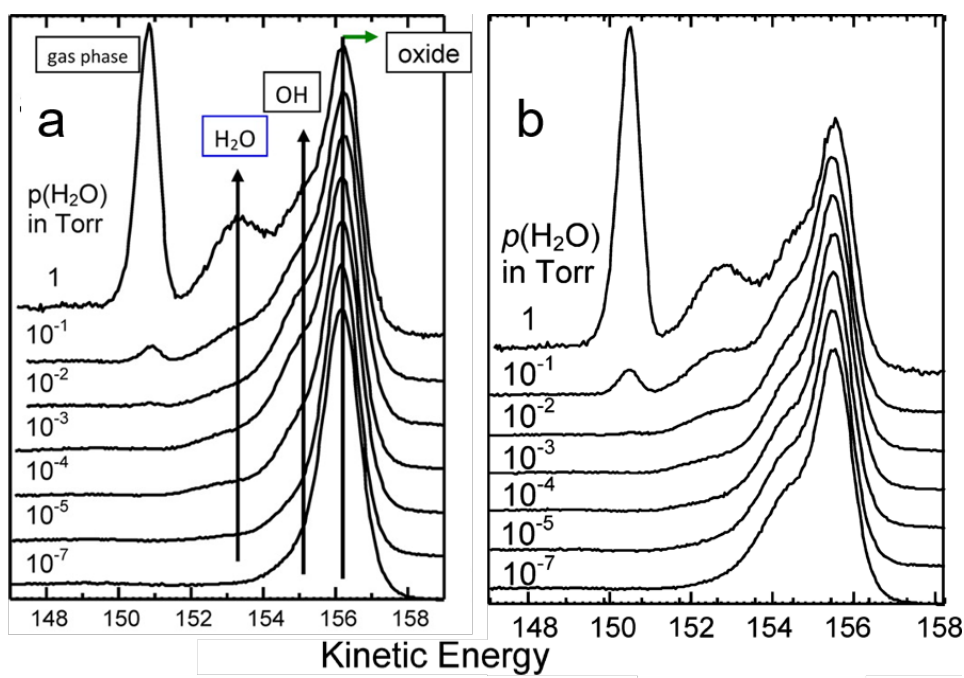

Figure 64. (a) Height normalized O1s XPS spectra acquired during exposure of an $Fe_3O_4$(100) single crystal to increasing pressures of water at 273 K. Hydroxylation of the surface is concomitant with the adsorption of molecular water. (b) Height normalized O1s XPS spectra acquired from a $Fe_3O_4$(100) single crystal during stepwise evacuation of high pressure water. Note that below $10^{-5}$ mbar no molecular water remains on the sample at 273 K. Reprinted with permission from ref [67]. Copyright 2013 American Chemical Society.

At high pressure (above $10^{-5}$ mbar) it was originally proposed [409] that the surface undergoes extensive hydroxylation, extending some 8 Å into the sample. However, this result has been questioned by a similar study by some of the same authors that utilized an in-situ instrument capable of measuring at ambient pressures [67]. In this study, the authors make a point that the C1s spectrum was carefully checked during data acquisition for evidence of contamination. O1s XPS data (Figure 64a), acquired at 273 K, reveal that significant hydroxylation coincides with the adsorption of molecular water above $10^{-5}$ mbar. When the system is evacuated (Figure 64b), the molecular water desorbs leaving only the hydroxyl groups below $10^{-5}$ mbar. The hydroxylation appears restricted to the surface. The interpretation of these data is that water dissociation proceeds via a cooperative process involving multiple water molecules, and that a stable mixed-mode adsorption results from hydrogen bonding between the adsorbed species. This is consistent with DFT+U calculations [65; 67], which find a mixed-mode adsorption to be energetically favourable at high coverage (see Figure 65).

A LEED *IV* experiment (Figure 66) has clearly shown that the (√2×√2)R45° reconstruction is lifted on exposure to water [71]. However, the LEED R-factor achieved ($R_P$ = 0.27) was based on a bulk terminated surface model with molecular and dissociated water. It is possible that the agreement can be improved upon once the mechanism of reconstruction lifting is known. Interestingly, the data were collected following evacuation of the water vapour from the vacuum system, conditions at which the later XPS study suggests the surface should have no molecular water adsorbed. At present there are no reported studies of the interaction of water with the Fe-terminated surfaces of $Fe_3O_4$(001).

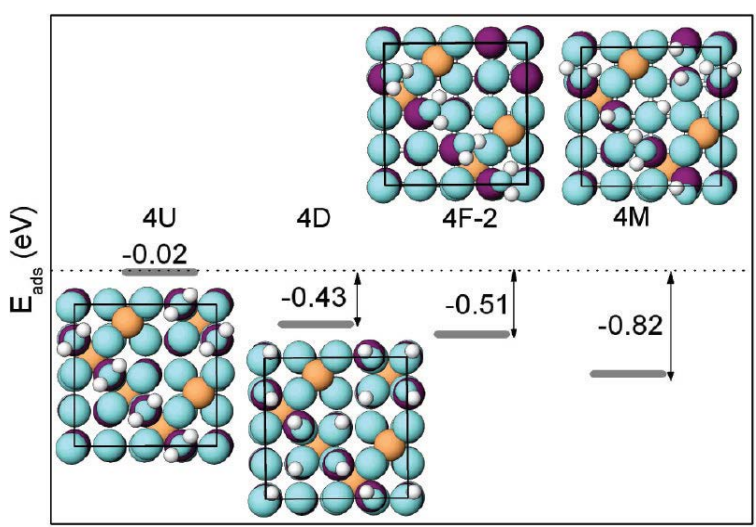

Figure 65: Calculated (DFT+U) adsorption geometries for 4 water molecules per $Fe_3O_4$(100)-(√2×√2)R45° unit cell ($Fe_{oct}$ atoms are purple, $Fe_{tet}$ are yellow, oxygen from the lattice and water are cyan, and hydrogen atoms are white). The geometries from left to right are water adsorbed oxygen down (4U), all water molecules dissociated with OH groups capping surface $Fe_{oct}$ (4D), four water molecules adsorbed flat with H bonding to surface O atoms, and mixed mode where every other water molecule is dissociated along the $Fe_{oct}$ row (4M). Reprinted with permission from ref [65]. Copyright 2010 American Chemical Society.

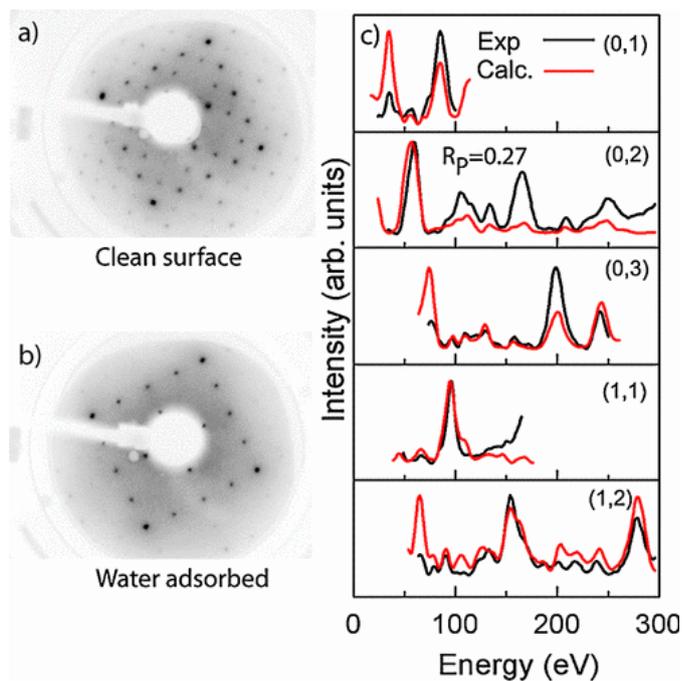

Figure 66: Exposure of the clean $Fe_3O_4(100)$ surface to $2\times10^{-6}$ mbar water at 273 K for 2 min is reported to lift the $(\sqrt{2}\times\sqrt{2})R45°$ reconstruction (a), resulting in a (1×1)-LEED pattern with an enhanced background (b). (c) A LEED *IV* analysis of this surface favoured a model based on the DBT structure with both molecular and dissociated water adsorbed. Figure reproduced from ref. [71].

### 4.2.2 $H_2O/Fe_3O_4(111)$

Ranke and co-workers studied the adsorption of water on thin $Fe_3O_4(111)$ films grown on Pt(111) using a variety of techniques [17; 44; 296; 303; 375; 410] and, given this is the same group that performed the LEED analyses of $Fe_3O_4(111)$ [283], it seems plausible that their sample preparation procedure resulted in a surface that was predominantly $Fe_{tet1}$ terminated. Indeed, the analysis of their data assumes this throughout. TPD curves obtained from the $Fe_3O_4(111)$ surface exhibit two peaks similar to those obtained from an O-terminated FeO(111) ultrathin film, water physisorbed on the FeO(111) surface and multilayer water. In addition, a peak appears at 282 K and shifts to 265 K with increasing coverage (peak γ, see Figure 67). This second order peak is attributed to recombinative desorption of water dissociated on the undercoordinated $Fe_{tet1}$ cations, which exist above a FeO(111)-like close-packed O-layer in the $Fe_{tet1}$ model. The estimated coverage of this species from TPD, XPS and UPS spectra corresponds to the density of $Fe_{tet1}$ atoms on the surface. Some subsequent theoretical calculations [282; 296] find that $O_{water}H$ binds to the $Fe_{tet1}$ atom, while the H atom binds to an undercoordinated surface O atom with an adsorption energy of 95 Kcal/mol, but the most recent calculations disagree with this assertion, finding only molecular water to be stable on the $Fe_{tet1}$ surface [70]. Further work by Ranke and co-workers using IRAS [375] suggests that dissociative adsorption occurs already at 110 K, but a later STM study by Rim et al. [282] claims that dissociation does not occur until 235 K, quite close to the desorption temperature. The second desorption peak from the first monolayer shifts from 210 to 190 K with increasing coverage. Due to the very low pre-exponential factor derived for this peak, Ranke and co-workers proposed that it corresponds to water desorbing from a $OH-H_2O$ network (see Figure 67) [303]. Such a mixed-mode adsorption is supported by theoretical calculations [296], and is similar to that found on $Fe_3O_4(100)$ [67].

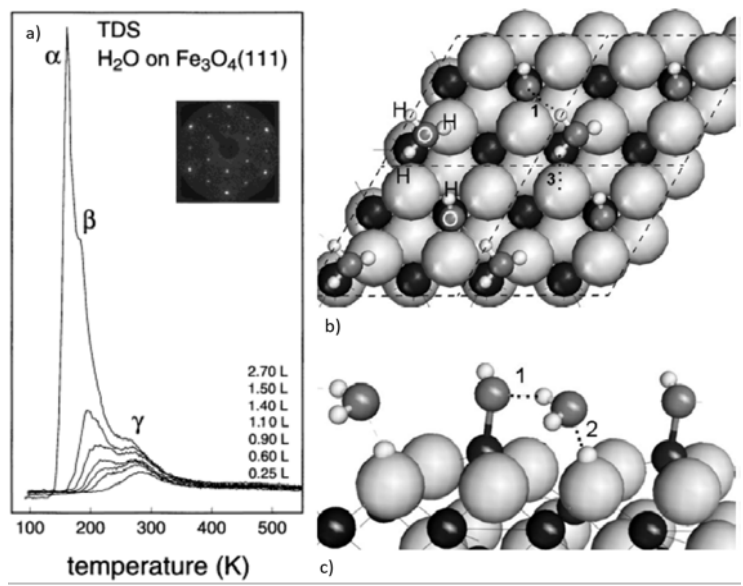

Figure 67: (a) Water TPD spectra acquired from $Fe_3O_4(111)$ thin film on Pt(111) [303]. The α and β peaks are common and linked to multilayer ice and physisorbed water, respectively. The γ is attributed to recombinative water desorption following the dissociative adsorption at the $Fe_{tet1}$ cations. (b, c) Top and side views of the mixed mode OH-$H_2O$ adsorption on $Fe_3O_4(111)$. Note oxygen atoms are gray, Fe atoms are black, and hydrogen atoms smallest and light gray. Reprinted figure with permission from ref. [296]. Copyright 2008 by the American Physical Society.

At first glance, the work of the Ranke group appears comprehensive, and presents a consistent picture of water adsorption on the $Fe_{tet1}$ termination of $Fe_3O_4(111)$. However, the conclusions are not consistent with the work of Cutting et al. [411], who studied water adsorption on a $Fe_3O_4(111)$ single crystal (exposing three different terminations). They suggest that water dissociation is coverage dependent because OH groups were only observed in XPS/UPS when the sample was cooled (from room temperature to 170 K) in $10^{-6}$ mbar $H_2O$, and not in $10^{-7}$ mbar [411]. When the sample was exposed to water at low temperature and heated (i.e. a similar process to a TPD experiment) the surface remained hydroxylated at 400 K, and STM images suggest this hydroxylation was restricted to the $Fe_{tet1}$ termination. Cutting et al. [411] suggest that these results show that Joseph et al. [303] were not measuring the $Fe_{tet1}$ termination. Given that Cutting et al. prepared their single crystal sample by sputter/anneal cycles without annealing in oxygen, it could also be that their surface exhibited reduced terminations. Adib et al. [331] studied water adsorption on the reduced selvedge of an α-$Fe_2O_3(0001)$ single crystal and observed five peaks in TPD, three similar to Joseph et al [303], with the addition of peaks at 300 and 340 K. In this case it seems certain that a second termination was present, on which dissociated water is more strongly bound.

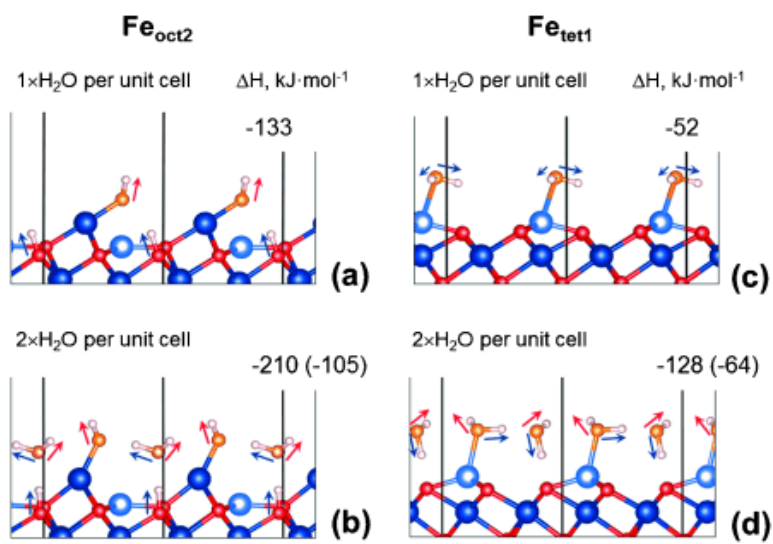

Figure 68: PBE+U calculated minimum-energy configurations for water on the Fe$_{oct2}$ (a, b) and Fe$_{tet1}$ termination of Fe$_3$O$_4$(111). Figure reproduced from ref. [70] with permission.

Very recently, Dementyev et al. [70] revisited the issue of water on Fe$_3$O$_4$(111) thin films using microcalorimetry, IRAS, and DFT+U calculations. The microcalorimetry data are particularly novel, giving a direct measurement of the adsorption energy of water on this surface, ≈ 101 KJ/mol at the lowest coverage. This adsorption energy is high, and indicative of dissociative adsorption, but decreases rapidly with coverage. The authors link this to either intermolecular repulsion, or competition for electrons for bonding. IRAS data for saturation water exposure show two bands attributable to OH groups, as observed previously [375], but when the surface is prepared with $^{18}$O, no corresponding isotopic shift was observed in the IRAS peaks, which suggests that no O$_{lattice}$-H bond is observed at this water coverage.

The new DFT+U calculations, discussed in further detail in ref. [412], predicts different behaviour to the prior results of Grillo et al. [296], despite that seemingly similar computational approaches were employed (PBE+U). The new results suggest that single water molecules do not dissociate on the Fe$_{tet1}$ termination after all, but dissociative adsorption is strongly exothermic (133 kJ/mol) on the Fe$_{oct2}$ termination (Figure 68). This is then taken as evidence that the surface measured experimentally using microcalorimetry was the Fe$_{oct2}$ termination. This conclusion is consistent with previous CO adsorption experiments from the same group conducted on similarly prepared films [284]. Interestingly, the calculated IRAS band frequency for an H-O$_{lattice}$ bond does not match the experimental observation, but a good match is obtained for a water dimer species in which one molecule is dissociated and one not (see Figure 68b), i.e. a mixed-mode adsorption. The adsorption energy calculated for the mixed-mode surface (-105 kJ/mol) is very similar to that measured by calorimetry at the lowest coverage, suggesting that water immediately forms the dimer configuration at low coverage (if kinetics allow) and that isolated molecular dissociation at Fe cations does not occur.

In general, the story of water adsorption on $Fe_3O_4$(111) illustrates the problem presented when the surface termination is in question. The quantitative agreement achieved between experiment and theory achieved by Dementyev et al. [70] is convincing, and adds weight to their claim that their surface was $Fe_{oct2}$ terminated, but it is not clear whether this new work supersedes the model developed by Ranke and co-workers, or if the differences arise because different terminations were measured. Irrespective though, the final adsorption models determined by theory are very similar (compare Figures 67 and 68), and thus it seems that water is adsorbed in a mixed-mode manner, as found on the $Fe_3O_4$(100) surface.

### 4.2.3 $H_2O$/$Fe_3O_4$(110)

A solitary theoretical study has been performed of water adsorption on $Fe_3O_4$(110) [311], by the same group that performed the only study of $H_2$ adsorption on this surface [403]. Again, the DFT+U calculations were performed for the $Fe_{oct}$-O and $Fe_{oct}$-$Fe_{tet}$-O bulk terminations (denoted A and B termination in ref. [403], respectively). $H_2O$ was found adsorb molecularly on the A termination at low coverage, and in a mixed mode at higher coverage, while dissociation to surface hydroxyls occurred on the B termination. Note that these terminations do not appear to be representative of the (1x3) reconstruction obtained in experiment.

### 4.2.4 $H_2O$/ α-$Fe_2O_3$(0001)

The interaction of water with α-$Fe_2O_3$(0001) surfaces has been studied both experimentally [413-416], and theoretically [416-420]. Kurtz and Heinrich [414] determined that water adsorbed dissociatively on an α-$Fe_2O_3$(0001) surface prepared by sputtering and 1100 K UHV annealing for pressures over $10^{-5}$ torr, while the XPS study of Liu et al. [349] suggested a threshold for hydroxylation at $10^{-4}$ torr (both studies conducted at room temperature). In a story similar to that described above for $Fe_3O_4$(100), the results have been superseded [413] by a new in-situ ambient-pressure XPS investigation, performed by some of the original authors. The observed threshold to hydroxylation is proposed to have emerged from sample contamination when the sample was exposed to high water pressures in a "transfer chamber". The recent study [413] was performed on a natural single crystal prepared by annealing at 723-773 K in $1x10^{-5}$ mbar $O_2$ (i.e. without sputtering). Given the controversy over the termination of this surface one cannot be absolutely sure, but the authors suggest the surface to be terminated by the Fe-$O_3$-Fe- termination, possibly with some coverage of ferryl groups. Note that the ferryl termination is essentially the Fe-$O_3$-Fe- with an additional O atom capping the Fe cation. With room temperature water exposure the surface becomes hydroxylated at pressures up to $10^{-6}$ mbar (see Figure 69), above which peaks due to molecular $H_2O$ and OH groups grow together. Further, the authors determine that OH and water exist only at the surface, and that no hydroxide phase is formed. A CTR study [416] was performed on a natural single crystal prepared by annealing in $1x10^{-5}$ mbar $O_2$. On exposure to a highly water-saturated atmosphere, evidence was found for a hydroxylated oxygen termination $(HO)_3$–Fe–Fe–R, and a multiple coordination $(HO)_3$–Fe–$H_3O_3$– termination (the Fe-$(OH)_3$ species is thought to occur in aqueous solution [421]).

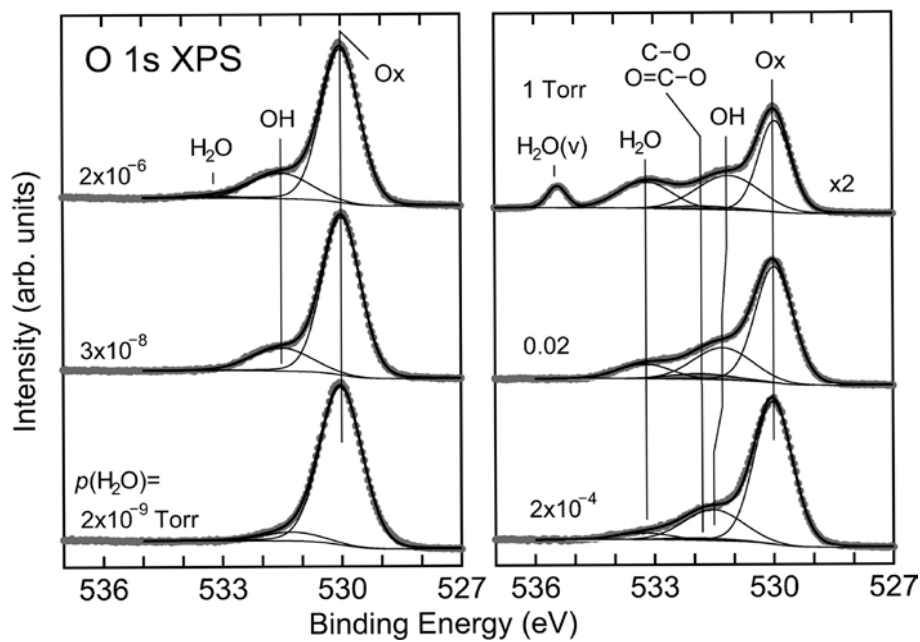

Figure 69: Near-ambient-pressure O1s XPS spectra acquired for a α-Fe₂O₃(0001) sample in different H₂O pressures. Reprinted with permission from ref. [413]. Copyright 2010 American Chemical Society.

Interestingly, α-Fe₂O₃(0001) single crystals have also been cleaved, and prepared by a wet-chemical method (etching in Nitric acid, washing in doubly deionized water). The samples were imaged in air [422], as well as in liquid water [423], and hexagonal lattices with 3 Å and 5 Å periodicity were observed. Although it is tempting to think these may be the O and Fe sublattices of a bulk-terminated surface, such a conclusion is probably not justified on the basis of the STM images presented, particularly in this environment. However, it is fascinating that a surface prepared in this way appeared flat and ordered. Crucially, no evidence was found for the formation of hydroxide phases. Generally speaking, there is no evidence that α-Fe₂O₃ surfaces transform to goethite, or any hydroxide, when exposed to air or aqueous solution [424].

Theoretical calculations of water on α-Fe₂O₃(0001) find water dissociation to be favoured over molecular adsorption on the Fe-O₃-Fe- termination [416-420]. Recent PBE+U calculations [418], which considered a variety of different terminations, suggest an adsorption energy of -0.75 eV for molecular water, and -1.0 eV for the dissociated state with $O_{water}H$ bound to the surface Fe atom, and formation of an $O_{surface}H$ group (Figure 70). Dissociation is also found to be facile on a ferryl-terminated surface (Figure 71), with a water molecule bound to the Fe atom of a ferryl group transferring a H atom to the ferryl oxygen. The final state is thus two OH groups attached to the surface Fe atom. The binding energy is slightly weaker than on the Fe-O₃-Fe termination though, -0.33 eV in the molecular state and -0.53 eV in the dissociated geometry. Further calculations [425] suggest that defects such as $V_{OS}$, $V_{Fe}$ on the Fe-O₃-Fe and Fe adatoms on the O₃-Fe-Fe termination are highly reactive toward water dissociation, so these sites would be expected to saturate first on water exposure.

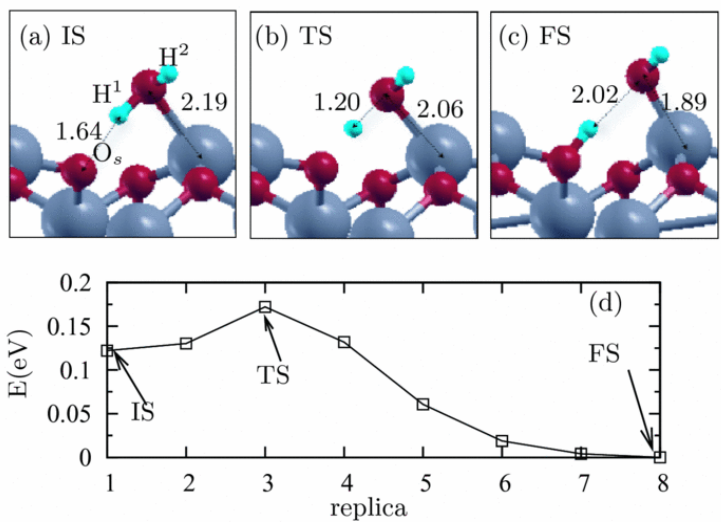

Figure 70: PBE+U calculations showing the optimum adsorption geometry for molecular water, the transition state, and dissociated water on the half metal (Fe-$O_3$-Fe-) termination of α-$Fe_2O_3$(0001). Reprinted with permission from ref. [418]. Copyright 2013, AIP Publishing LLC.

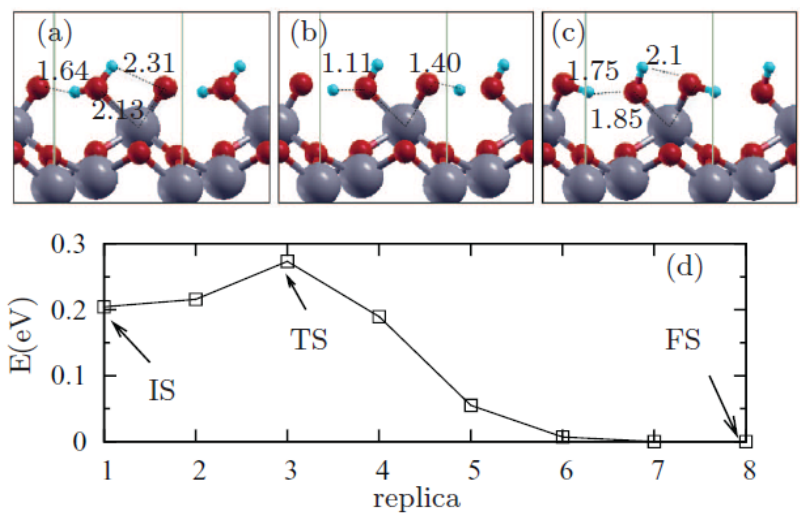

Figure 71: PBE+U calculations showing the optimum adsorption geometry for molecular water, the transition state, and dissociated water on the ferryl (O=Fe-$O_3$-) termination of α-$Fe_2O_3$(0001). Reprinted with permission from ref. [418]. Copyright 2013, AIP Publishing LLC.

### 4.2.5 $H_2O$/ α-$Fe_2O_3$($1\bar{1}02$)

Water adsorption has been studied on the α-$Fe_2O_3$($1\bar{1}02$) surface with a stoichiometric (1×1) termination and a reduced (2×1) termination using a combination of TPD, SIMS, LEED and HREELS [360]. Water was found to dissociate on both surfaces resulting in terminal and bridging OH groups. Interestingly, the shape of TPD peaks linked to recombinative desorption suggested a significant structure sensitivity in the process, with first-order behaviour observed at 350 K on the (1×1), and pseudo-zeroth-order kinetics at 400 K on the (2×1) surface. The authors suggest that the former behaviour is linked to hydrogen bonding between the dissociated OH groups, while the zero-order kinetics result from the formation of one dimensional arrays of hydroxyl groups in the missing row structure proposed for the (2×1) termination. Given the extensive spectroscopic information provided by Henderson on this surface [360-362; 366; 426], with and without adsorbates, it would appear to be an excellent choice for future studies with scanning probe spectroscopy, particularly as the (1×1) and (2×1) surfaces can be interchanged easily under UHV conditions [362].

## 4.3 Carbon Monoxide (CO)

Prior to studies on low-index surfaces, there were several studies of CO on $Fe_3O_4$ powders that utilized surface science techniques [427-429]. Often motivated by the use of $Fe_3O_4$ as a water-gas shift catalyst, these studies suggested that CO binds most strongly to the metal cations. This is similar to the behaviour on other metal-oxide surfaces [34].

The only studies of CO at the $Fe_3O_4(100)$ surface suggest that CO does not adsorb strongly at room temperature. STM images acquired at room temperature reveal CO adsorbs on Pd and Pt adatoms, resulting in their agglomeration into nanoparticles, but no adsorbate-related features were observed on the substrate [262; 402]. When a $Pt_{1-6}/Fe_3O_4(001)$ system was exposed to CO at 250 °C, however, large holes appeared on the surface because CO extracted $O_{lattice}$ atoms at the cluster perimeter, forming $CO_2$, and the undercoordinated Fe atoms diffused into the bulk [402]. Similar metal-catalysed reduction of the oxide in CO was observed on a $Pd/Fe_3O_4(111)$ model catalyst [430]. Both CO and $H_2$ are well known to reduce iron oxides at elevated temperatures [38], and here the presence of the metal appears to catalyse the process. The only studies performed below room temperature have focussed on the $Fe_3O_4(111)$ surface.

As discussed in the context of the termination of $Fe_3O_4(111)$, Lemire et al. [284] found three distinct adsorption states for CO on 5-7 nm thick films grown on Pt(111) with desorption peaks at (α) 110, (β) 180 and (γ) 210 K and stretching frequencies 2115–2140, 2080 and 2207 cm$^{-1}$, respectively (Figure 72). The α peak was interpreted in terms of loosely bound CO, while the β and γ peaks were attributed to chemisorbed CO at $Fe^{2+}$ and $Fe^{3+}$ cations, respectively. Given both cations appear to be present, the film was concluded to be $Fe_{oct2}$ terminated. STM images appear similar to the "regular termination", but were acquired from a sample grown and measured in another chamber. Indeed, the $Fe_{oct2}$ termination strongly resembles the $Fe_{tet1}$ termination under some imaging conditions [107], and a $H_2O$ TPD acquired from the surface has more structure than observed by Joseph et al. [303]. Consequently, it is possible that both $Fe_{tet1}$ and $Fe_{oct2}$ terminations may have existed on the surface. A recent LEEM study [295] has shown that thin-film samples can expose FeO areas on preparation, and that post-annealing in UHV is required to produce a uniform $Fe_{tet1}$ surface.

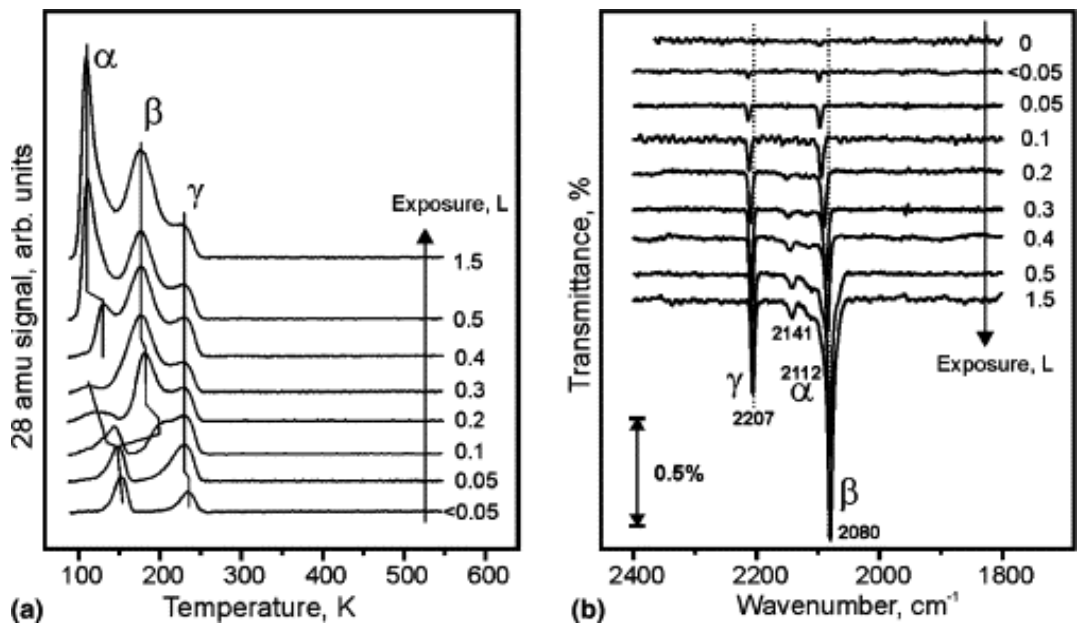

Figure 72: TPD and IRAS spectra acquired from the $Fe_3O_4(111)$ surface exposed to CO at 90 K. The α peak results from loosely bound CO. The β and γ peaks are attributed to chemisorbed CO at $Fe^{2+}$ and $Fe^{3+}$ cations, respectively. The existence of both peaks led the authors of ref. [284] to conclude that the surface was $Fe_{oct2}$ terminated. Reprinted from ref. [284], Copyright 2004, with permission from Elsevier.

DFT calculations (GGA-PBE) find that CO binds on metal cations on both the $Fe_{oct2}$ termination and the $Fe_{tet1}$ termination [297]. On the $Fe_{oct2}$ termination binding is stronger with the octahedral cations (-1.94 eV) than the tetrahedral cations (-1.09 eV), but a bridge configuration between $Fe_{oct}$ and $Fe_{tet}$ is favoured overall. On the $Fe_{tet1}$ terminated surface, CO is only stable on the $Fe_{tet}$ site (-0.8 eV).

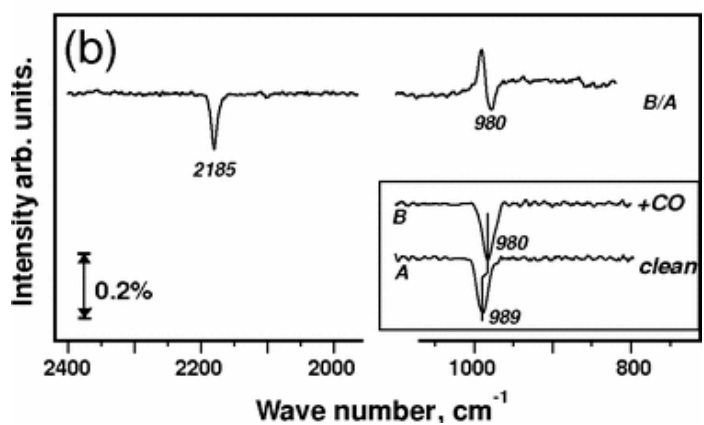

Figure 73: IRAS spectra for CO adsorbed on α-$Fe_2O_3(0001)$ at 90 K. Reprinted figure with permission from ref. [350]. Copyright 2005 by the American Physical Society.

Lemire et al. [350] also studied CO adsorption on 5-7 nm α-$Fe_2O_3(0001)$ thin film samples on Pt(111) that were prepared at 1050 K in $10^{-3}$-1 mbar $O_2$. These films exhibited a clear absorption peak for the ferryl vibration in IRA spectra (Figure 73). CO was reported to desorb at 140 K in TPD, although no spectra were published. Note this is lower than the CO desorption temperature (180 K) observed by the same group on $Fe_3O_4(111)$. IRAS data following exposure to CO at 90 K show a peak at 2185 cm$^{-1}$, characteristic of CO bound to $Fe^{3+}$, while the Fe=O peak shifts to lower wavenumbers. The authors propose that the CO molecules interact with the ferryl groups at the edge of domains of differing termination.

CO adsorption on the oxygen-terminated O$_3$-Fe-Fe surface of α-Fe$_2$O$_3$(0001) has been studied theoretically using DFT-based and molecular dynamics calculations [431]. Interestingly, the authors suggest that such a surface reacts with CO to form CO$_2$ at room temperature at low coverage (i.e. low pressures). If this is true, background CO in the residual gas might play a role in the preparation of α-Fe$_2$O$_3$(0001) surfaces.

## 4.4 Oxygen (O$_2$)

The oxidation of Fe$_3$O$_4$ at higher temperature has been long studied to establish the pathway to Fe$_2$O$_3$. Early work showed that submicron Fe$_3$O$_4$ particles oxidise to γ-Fe$_2$O$_3$ via a continuous change in composition through the diffusion of cations to the surface, and subsequent oxidation [41-43; 432]. This makes sense because γ-Fe$_2$O$_3$ and Fe$_3$O$_4$ share a common oxygen lattice, but differ through the Fe$_{oct}$ occupation. There is an ongoing debate on whether α-Fe$_2$O$_3$ formation requires a pre-existing α-Fe$_2$O$_3$ seed [37; 39; 40] or whether the corundum phase nucleates through strain [41; 432], this debate remains unresolved.

The first studies taking the surface orientation into account were performed by the Shvets group [433]. Using XRD, they determined that annealing Fe$_3$O$_4$ samples in air at 300 °C resulted in a capping layer of α- Fe$_2$O$_3$, and interestingly, that the (100) surface oxidised much faster than (111) and (110). Figuera and co-workers [213; 214] showed that annealing the (100) surface of an Fe$_3$O$_4$(100) single crystal in 10$^{-6}$ mbar O$_2$ at 650 °C resulted in the growth of hundreds of layers of new Fe$_3$O$_4$(100) surface. O$_2$ dissociation was determined to occur at regular terrace sites on the surface, and the Fe was provided by the growth of α-Fe$_2$O$_3$ inclusions [214]. Since the inclusions were observed to grow along the (111) directions, the authors propose the Fe$_3$O$_4$(111) surface would become quickly capped by an α- Fe$_2$O$_3$ layer, through which Fe cations cannot flow as freely as in Fe$_3$O$_4$. The (100) surface, in contrast, would remain largely Fe$_3$O$_4$ at the surface until oxidation extended a long way into the crystal [214].

Various authors have studied the oxidation of Fe$_3$O$_4$ thin films [224; 249; 334]. Chambers and co-workers [224] found that a Fe$_3$O$_4$(100) thin film could be converted to γ-Fe$_2$O$_3$(001) by exposure to O-plasma at room temperature, and then converted back by annealing in UHV at 250 °C, while Monti et al. [249] found that an Fe$_3$O$_4$(111) thin film could be converted to γ-Fe$_2$O$_3$(111) in NO$_2$ at 600 °C. A very recent paper by the Freund group [334] showed the oxidation behaviour of iron oxide thin films depends on the underlying metal substrate. They suggest that the oxidation to α-Fe$_2$O$_3$(0001) proceeds through cation diffusion through the film and oxidation at the surface, with the crucial difference that Pt(111) acts as a sink for Fe (potentially akin to the bulk of an Fe$_3$O$_4$ single crystal), whereas Ag(111) does not [334]. The former behaviour leads to the counter-intuitive observation that UHV annealing of a mixed Fe$_3$O$_4$/α-Fe$_2$O$_3$ film leads to α-Fe$_2$O$_3$(0001) on Pt(111).

The remaining studies involving O$_2$ and Fe$_3$O$_4$ are motivated by understanding heterogeneous catalysis, and Fe$_3$O$_4$ is the support for metallic clusters. Bliem et al. [402] showed that Pt clusters supported by Fe$_3$O$_4$(100) dissociate O$_2$, creating O atoms that spill over onto the support. There they react with Fe from the bulk creating new Fe$_3$O$_4$(100) islands. Similar metal-assisted oxidation of the substrate has been reported for Pd nanoparticles on Fe$_3$O$_4$(111) [430; 434].

It is perhaps surprising that $O_2$ chemisorption has not been more extensively studied on iron-oxide surfaces given their important role as a support for CO oxidation catalysts. Henderson investigated the interaction of $O_2$ with the (1×1) and (2×1) terminated α-$Fe_2O_3$(012) surface, finding no evidence of adsorption on the (1×1) surface, which contains only $Fe^{3+}$ cations. In contrast, $O_2$ was chemisorbed and dissociated the reduced (2×1) surface, which was linked to the $Fe^{2+}$ cations. It was suggested that these results reveal the importance of charge transfer from reduced cation sites in $O_2$ chemisorption. The only adsorption study mentioned in the $Fe_3O_4$ literature is TPD performed by the Freund group in 2004 [430]. There it was reported that molecular $O_2$ desorbs from the $Fe_3O_4$(111) surface in a single peak at 250 K (although the data were not shown). To put this into context, $O_2$ physisorbs on a stoichiometric $TiO_2$(110) surface, desorbing at 30 K [435]. However, charge transfer from defects (surface $V_O$'s, OH groups, and Ti interstitials) result in superoxo species, which desorb above 400 K. Dissociative adsorption can occur at $V_O$s, with an O atom filling the vacancy and the adsorption of an O adatom on the Ti rows [436]. Understanding the interaction of molecular $O_2$ with metal oxides remains one of the big challenges in the field in general.

## 4.5  Water – Gas Shift Reaction

The water-gas shift reaction ($H_2O + CO \rightarrow H_2 + CO_2$) is performed industrially over an iron-oxide based catalyst reported to contain 74.2% $Fe_2O_3$, 10 % $Cr_2O_3$, 0.2 % MgO. However, it is well established that under WGS reaction conditions $Fe_3O_4$ forms at the surface and is the active phase, and that the primary role of $Cr_2O_3$ is to stabilize the catalyst, and reduce thermal sintering. The topic has been reviewed many times (e.g. [437-439]), and the details are well beyond the scope of this review. Briefly though, the mechanism of the reaction at low temperatures (below 320°C) is still under debate, with some authors favouring an associative mechanism in which CO and $H_2O$ adsorb on the surface forming an intermediate (possibly formate), which then decomposes to desorb $CO_2$ and $H_2$, while other authors favour a redox-type reaction [439; 440]. Here, CO reacts with the oxide, extracting $O_{lattice}$ to form $CO_2$, and $H_2O$ oxidises the surface, regenerating the catalyst and forming $H_2$ in the process. At high temperature, where $Fe_3O_4$ is extensively utilized, most data appears to favour the redox mechanism. Interestingly, Rhodes et al. [441] showed that WGS activity is enhanced by the addition of various promotor metals, and concluded that strengthening CO adsorption was key to improving catalyst activity.

In late 2015 a very interesting computational study of WGS on the $Fe_{oct2}$ terminated $Fe_3O_4$(111) surface was published by Huang et al [442]. They identified that redox mechanism is the energetically most energetically favourable pathway, with COO* desorption was found to be the rate-limiting step (barrier of 1.04 eV). OH dissociation was determined to present the second-highest activation barrier of 0.81 eV.

Despite the huge interest in the WGS over many years, other than the above mentioned computational study [442], there has been little attempt to uncover the structure sensitivity of the reaction on well-defined $Fe_3O_4$ surfaces using the surface-science approach. In recent years however, motivation to understand the mechanisms of the WGS is quickly regaining momentum because of the vastly increased demand for $H_2$, and because cost-efficient water-gas shift catalysts without (toxic) chromium are required on environmental grounds. This has led to repeated attempts to study how and why the addition of different metals to $Fe_3O_4$ can promote the reaction [443-447]. Understanding WGS on any particular $Fe_3O_4$ surface at the atomic scale requires a detailed knowledge of CO and $H_2O$ adsorption, and how the surface responds to these reactants at the reaction temperature. These basic data simply do not exist at present, but what evidence there is sheds doubt on whether a traditional redox process would occur on the $Fe_3O_4$(100) surface. Bliem et al. [402] reported no reduction of the surface by CO in the absence of Pt clusters at 573 K, and even should it occur at higher temperatures, the data suggest oxygen vacancies are not stable species, and that excess Fe diffuses inside the crystal instead. Moreover, even if water did dissociate and fill vacancies before Fe was able to diffuse into the bulk, adsorbed hydrogen atoms are known to desorb as water, not $H_2$.

## 4.6 Fischer-Tropsch Synthesis

Fischer-Tropsch synthesis converts syngas ($H_2$ and CO) into hydrocarbon fuels, for example via the reaction: $(2n + 1) H_2 + nCO \rightarrow C_nH_{(2n+2)} + nH_2O$, where n is typically 10-20. The seemingly simple reaction scheme proceeds via several steps including dissociative adsorption of CO and $H_2$, water formation and desorption and the formation of surface $CH_2$. The typical catalysts are based on either cobalt or iron, with the iron-based variety produced through the reduction of $Fe_2O_3$ particles containing several promoters (K, Cu, $Al_2O_3$). In the strongly reducing environment of CO and $H_2$ at temperatures in excess of 600 K, various phases are formed at the surface during the reaction including $Fe_3O_4$ [448], metallic Fe [449], $\theta$-$Fe_3C$, $\varepsilon'$$Fe_{2.2}C$, $\chi$-$Fe_5C_2$ and $Fe_7C_3$. The relative abundance varies with the conditions, and each phase has been proposed to be the active component during the course of 90 years of studies. TEM [450], XPS [451; 452], and XAS [453] studies suggest that $Fe_2O_3$ and $Fe_3O_4$ are not active, but the transformation of $Fe_3O_4$ into carbide leads directly to the catalytic activity, with K speeding up the carburization process. Schroff et al. [450] found that $Fe_3O_4$ particles broke up during carburization into smaller carbide particles and suggested that the active phase might be a surface carbide residing on a $Fe_3O_4$ core, as also proposed by Huang et al. [454].

Despite the industrial importance of Fischer-Tropsch synthesis there is relatively little in the way of model studies of the relevant steps. The carburization and reoxidation of $Fe_3O_4$ surfaces would be a particularly interesting avenue for further study. A recent DFT+U investigation [455] suggests that carburization is thermodynamically favoured with several different sources of C, with the following preference (C>CH>$CH_2$>$CH_3$>CO). Furthermore, their work suggests that mixed phases might be particularly active if H atoms can spill over from the oxide to hydrogenate CO adsorbed on the carbide. Similar to the water-gas shift reaction discussed above, there is significant scope to study Fischer-Tropsch synthesis through the carburization of iron oxide surfaces, with the addition of promoter metals such as K and Cu providing an additional layer of complexity to be explored.

In recent years surface science has moved toward the study of model catalyst surfaces under reaction conditions with the development of new instrumentation such as ambient-pressure XPS [456] and STM that can function under reaction conditions [457]. A review of efforts in this area was published recently by Tao and Salmeron [458]. A particularly fascinating insight into what is possible for Fischer-Tropsch was recently demonstrated by Ehrensperger and Wintterlin [459] who imaged a cobalt single crystal under realistic reaction conditions. Surprisingly, the experimental data clearly show that the surface structure remained unchanged from that observed in UHV, shedding doubt on the formation of intermediate phases and the proposed reaction mechanism on this surface. Experiments such as this suggest that there need not always be a "pressure gap" that must be overcome for surface science studies to be directly relevant to real-world conditions.



## 4.7 Formic Acid (HCOOH)

Formic acid is an important adsorbate to study on Fe$_3$O$_4$ surfaces because it is representative of organic acids in general, and because it is proposed as an intermediate in the low-temperature water-gas shift reaction.

Formic acid was found to undergo dissociative adsorption on the Fe$_3$O$_4$(100) surface at room temperature, as on many metal-oxide surfaces. The close proximity of undercoordinated acidic and basic sites facilitates deprotonation of the molecule, resulting in adsorbed formate (HCOO) and surface OH groups [401]. A complete monolayer of adsorbed formate was observed in STM and XPS, but the surface hydroxyl groups were not detected by XPS, STM or IRAS. However, in each case there are experimental reasons underlying the lack of sensitivity to OH, and the co-adsorption of OH groups on the surface was assumed. This study is noteworthy because IRAS was used to show that the formate is bound in a bidentate configuration; it is the first example of IRAS being performed on a Fe$_3$O$_4$ single crystal, and one of a limited number of IRAS studies performed on metal oxide single crystals to date. Interestingly, the absorption peaks exhibit a Fano-like line shape (Figure 74) because the real and complex parts of the dielectric function of Fe$_3$O$_4$ are comparable in the IR regime (Re($\varepsilon$) = 13, Im($\varepsilon$) = 10) [52], and the sample is both a conductor and a dielectric. Similar peak shapes were observed on metallic samples with poor conductivity, but the phenomenon was not previously observed on a metal oxide. Importantly, the signal intensity is found to be comparable to that of Cu [401], so it should be possible to detect small coverages of adsorbates on Fe$_3$O$_4$ single crystal surfaces, which is promising and important for future studies of catalytic activity.

A second issue to arise from this work is the lifting of the ($\sqrt{2}\times\sqrt{2}$)R45° reconstruction observed following saturation exposure to HCOOH. Since similar behaviour was observed following exposure to atomic H and water, it seems probable that the rearrangement of the subsurface cations is linked to hydroxylation of the surface. However, the mechanism and the structure of the (1×1) surface are presently unknown.

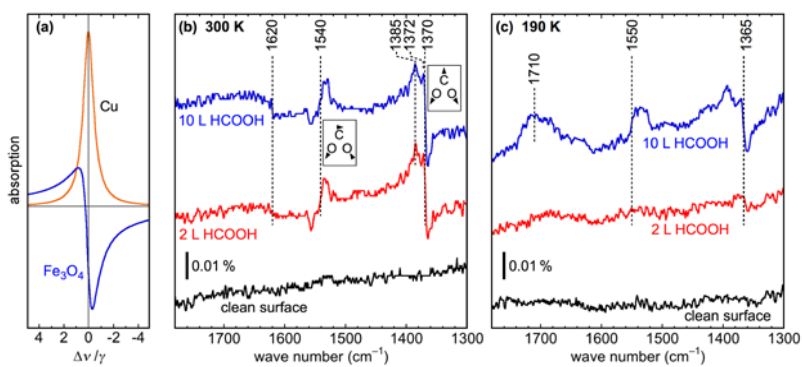

Figure 74: (a) Calculation of IRAS spectra for a simple oscillator on Fe$_3$O$_4$ and Cu surfaces. On Fe$_3$O$_4$, a Fano-like peak shape is observed. (b) Experimental IRAS spectra obtained at room temperature from the clean Fe$_3$O$_4$(001)-($\sqrt{2}\times\sqrt{2}$)R45° surface and following exposure to 2 L and 10 L HCOOH. The asymmetric (1540 cm$^{-1}$) and symmetric (1370 cm$^{-1}$) vibration modes are sketched in small boxes. Reprinted with permission from ref. [401]. Copyright 2015 American Chemical Society.

HCOOH adsorption has also been studied on a Fe$_3$O$_4$(111) single crystal prepared by sputter/anneal cycles in UHV. As discussed above, these conditions are reducing, and lead to multiple surface



structures on $Fe_3O_4(111)$. Three different, coexisting terminations were observed in STM images [304], and each interacted differently with HCOOH. However, it is important to note that the interpretation of the STM images relies heavily on the surface model assumed. The surface corresponding to the $Fe_{tet1}$ termination (labelled A') was found to adsorb formic acid resulting in protrusions associated with the $Fe_{tet1}$ atoms. A bidendate non-bridging configuration was preferred, since the $Fe_{tet1}$ atoms are too far apart (5.92 Å) to facilitate a bridging configuration, while a monodentate form was dismissed on the basis that the protrusion was symmetric. The latter conclusion is perhaps a little strong based on STM images alone, and IRAS measurements would be best suited to determine the configuration. HCOOH was also found to adsorb on the A-type (ferryl terminated) surface but since it was clearly mobile and easily swept away by the STM tip, a molecular adsorption was concluded. Stronger adsorption was found at a few point defects, and assumed to be dissociative in this case.

## 4.8   Methanol ($CH_3OH$)

The adsorption of methanol on iron oxides is of interest due to the possible application in $H_2$ production. Li et al. [279] investigated $CD_3OD$ adsorption on an $Fe_3O_4(111)$ single crystal prepared by sputter/anneal cycles using TPD, STM, and DFT+U calculations. The sample primarily exhibited the "regular termination" (assigned to $Fe_{tet1}$), but there were also areas of bi-phase reconstruction. A desorption peak at 360 K that shifted to 330 K with increasing coverage was attributed to recombinative desorption of methanol deprotonated at $Fe_{tet1}$ sites, while a further peak of similar size at 290 K was assigned to molecular methanol. The authors proposed that each $Fe_{tet1}$ cation binds both a methoxy and methanol group at 1 ML coverage. Molecular methanol bound at $Fe_{tet1}$ sites was calculated to have an adsorption energy of 0.84 eV, but dissociation to $Fe_{tet1}$-bound methoxy and $O_{surface}D$ was favoured with an adsorption energy of 1.38 eV. A small peak detected at 630 K was linked to $CD_2O$, supposed to be formed through the reaction: $CD_3OD + O_{surface} \rightarrow CD_2O + D_2O$. $D_2O$ desorption was observed at 280 K, and linked to a reaction of methanol derived D atoms and "weakly bound" surface oxygen. Unfortunately, the location of the weakly bound O could not be determined. A final peak at 255 K was attributed to desorption from the bi-phase areas of the sample. These data can be reconciled with the only previous report of methanol adsorption on $Fe_3O_4(111)$; Bäumer et al. [460] observed two methanol related species on thin $Fe_3O_4(111)$ films at low temperature, one of which desorbed below room temperature and the other due to decomposition between 300 and 600 K. Further details are contained within the PhD thesis of Brandt [461].

Henderson [361] has studied methanol adsorption on the (1×1) and (2×1) terminations of α-$Fe_2O_3(1\bar{1}02)$. Methanol was found to deprotonate on both surfaces forming adsorbed methoxy and OH groups. Most of the dissociated methanol molecules recombine and desorb as methanol in TPD at 365 and 415 K for the (1×1) and (2×1) surfaces, respectively, but some $H_2CO$ and methanol is also observed at 550 K. During this reaction, which occurs preferentially on the reduced (2×1) surface, some surface $Fe^{3+}$ is reduced to $Fe^{2+}$. The author suggests that the results show that exchange between organic ligands and OH groups may help stabilize organics on hematite mineral surfaces.

## 4.9   Carbon Tetrachloride ($CCl_4$)

The Osgood group published a series of papers [281; 329-332] in the early 2000's looking into the interaction of $CCl_4$ with iron-oxide surfaces, motivated primarily by the use of $Fe_3O_4$ in groundwater remediation. The experiments were performed on the reduced selvedge of α-$Fe_2O_3$ single crystals,



which exhibited both the so-called (2×2) termination of $Fe_3O_4$(111) (approx. 75 % of the surface) and bi-phase domains (which they believed to consist of hematite and FeO) [332]. XPS results show that $CCl_4$ dissociation occurs on adsorption even at 100 K [281], resulting in $CCl_2$ and Cl atoms on the surface. The reaction occurs at regular lattice sites on the $Fe_3O_4$(111) surface, but does not on the bi-phase areas, which were assumed to be O-terminated [329]. TPD data reveal four desorbing products; $CCl_4$, $C_2Cl_4$, $OCCl_2$ and $FeCl_2$, and two competing reaction pathways were determined.

$$CCl_4(g) \xrightarrow{100\ K} CCl_2(ad) + 2Cl(ad) \xrightarrow{\Delta} CCl_4(g) \quad (3)$$

$$CCl_2(ad) \xrightarrow{\Delta} {}^1\!/_2\, C_2Cl_4(g) \quad (4)$$

$$CCl_2(ad) + O_{lattice} \xrightarrow{\Delta} OCCl_2(g) \quad (5)$$

When the $CCl_4$ was dosed at low temperature (100 K), most desorption was in the form of $C_2Cl_4$ and $CCl_4$ (eq. 3 and 4) but when $CCl_4$ was dosed at room temperature, much $CCl_2$ reacted with the surface and extracted lattice $O_{lattice}$ atoms to desorb phosgene ($OCCl_2$, eq. 5). In the former case Cl atoms are observed atop Fe atoms in room temperature STM images, while in the latter Cl primarily occupies the oxygen vacancy sites created as part of the reaction. The barrier for phosgene formation was determined to be 0.16 eV [332]. A surface pre-treated with $D_2O$ has very limited reactivity to $CCl_4$, suggesting OH groups block the active $Fe_{tet1}$ sites [331].

## 4.10   Methyl Iodide ($CH_3I$)

In addition to $CCl_4$ (described above), the Osgood group also studied the adsorption of $CH_3I$ on the reduced selvedge of their α-$Fe_2O_3$(0001) single crystal (i.e. predominantly a $Fe_3O_4$(111) "regular" termination) [333]. TPD spectra obtained following adsorption at 100 K reveal 4 desorption peaks, two of which are thought to arise from the first monolayer. At the lowest coverages a peak is observed at 290 K that extends all the way up to 500 K. Desorption peaks were also observed at 240 and 200 K. Since these peaks grew simultaneously they are most likely from different sites. A small peak due to $I^+$ ions was observed at 865 K, possibly due to dissociative adsorption at defect sites. Irradiation with UV light desorbed $CH_3$ fragments, leaving behind I which desorbed at 865 K.

## 4.11   Dehydrogenation of Ethylbenzene to Styrene

The dehydrogenation of ethylbenzene to styrene is catalysed by both pure and K-promoted iron-oxide catalysts, and the reaction has been intensely studied via the surface-science approach since the group of Ertl [462; 463] proposed that the active phase of the catalyst is a $KFeO_2$ surface phase. Since the issue was the focus of a major review by Weiss and Ranke in 2002 [17], only the basics will be discussed here. For further details the reader is referred to the previous review, and the many references contained therein [17].

The adsorption of ethylbenzene and styrene on FeO(111), $Fe_3O_4$(111) and α-$Fe_2O_3$(0001) oriented thin films was studied using a variety of different techniques including TPD, UPS, and NEXAFS. Essentially, ethylbenzene is chemisorbed in a barrier-less process in the first monolayer on $Fe_3O_4$(111) and α-$Fe_2O_3$(0001), but not on FeO(111). This suggests that chemisorption is related to the presence of undercoordinated Fe atoms at the surface (a similar argument was made about the dissociative adsorption of water, above), and the authors suggest the π-electron system acts as a Lewis base in the interaction with the acidic Fe atoms (see Figure 75). The adsorption behaviour of styrene is similar to ethylbenzene, but higher desorption temperatures indicate that the molecule is slightly more strongly bound. Both adsorbates are more strongly bound on $Fe_3O_4(111)$ than on $\alpha$-$Fe_2O_3(0001)$.

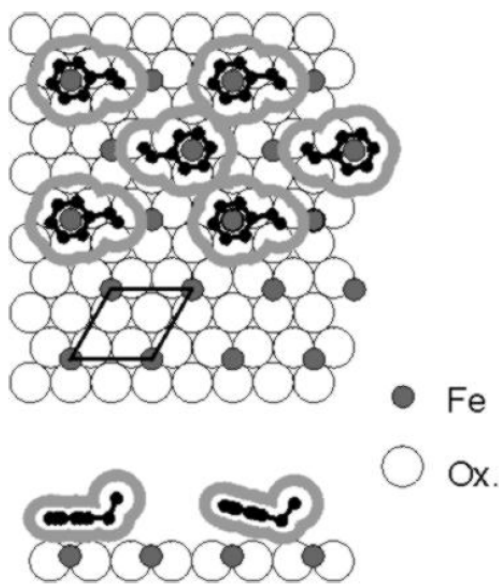

Figure 75: Model for ethylbenzene adsorption on the $Fe_{tet1}$ termination of $Fe_3O_4(111)$. The phenyl ring of the molecule is situated directly above the $Fe_{tet}$ cation. Figure reprinted from ref. [17] with permission from Elsevier.

Batch reactor experiments utilizing the thin film samples found that no styrene was produced from ethylbenzene at low pressures, but that significantly more carbonaceous deposits were left on $\alpha$-$Fe_2O_3(0001)$ than $Fe_3O_4(111)$. At higher pressures, no styrene production was observed from $Fe_3O_4(111)$. The authors attribute this to quick saturation of the active sites by a monolayer of styrene, which prevents further ethylbenzene adsorption. Significant catalytic activity was only observed from a defective $\alpha$-$Fe_2O_3(0001)$ surface. The nature of the defects was not known, but it is likely that the defects expose undercoordinated O atoms that are able to efficiently deprotonate the ethylbenzene. Based on these results, a reaction scheme was proposed (Figure 76) where ethylbenzene is adsorbed at regular cation sites (step 1), and dehydrogenation occurs most easily if a basic defect is located nearby, for example, at a step edge (step 2). The deprotonated molecule desorbs as styrene in step 3, before the two adsorbed H atoms react with the surface to desorb as water in step 4.

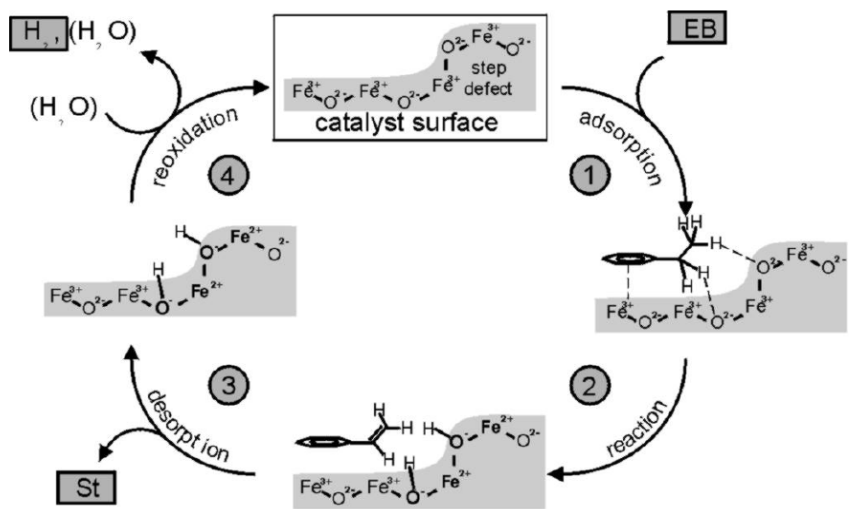

Figure 76: Reaction scheme proposed by Weiss and Ranke for the dehydrogenation of ethylbenzene (EB) to styrene over a defective α-$Fe_2O_3$(0001) surface. Figure reprinted from ref. [17] with permission from Elsevier.

## 4.12 Sulphur Dioxide $SO_2$

The interaction of $SO_2$ with the $Fe_3O_4$(100) surface was studied by Stolz et al. [464] with a view to understanding how S enhances rates of corrosion. Interestingly, a natural $Fe_3O_4$(100) sample, cleaned by sputter/anneal cycles, exposed to 50 L $H_2O$ and then 50 L $SO_2$ exhibited arrays of protrusions that were ordered with the same (√2×√2)R45° periodicity as the clean surface reconstruction (Figure 77). Complementary XPS data reveal that both $SO_3$ and $SO_4$ are present on the surface, so it seems likely that S binds to two O atoms from the support to make sulphate, perhaps across the "not blocked" sites of the SCV reconstruction where many other species also bind. Under these conditions (i.e. room temperature, low pressure) the 50 L exposure to water likely had little effect on the surface.

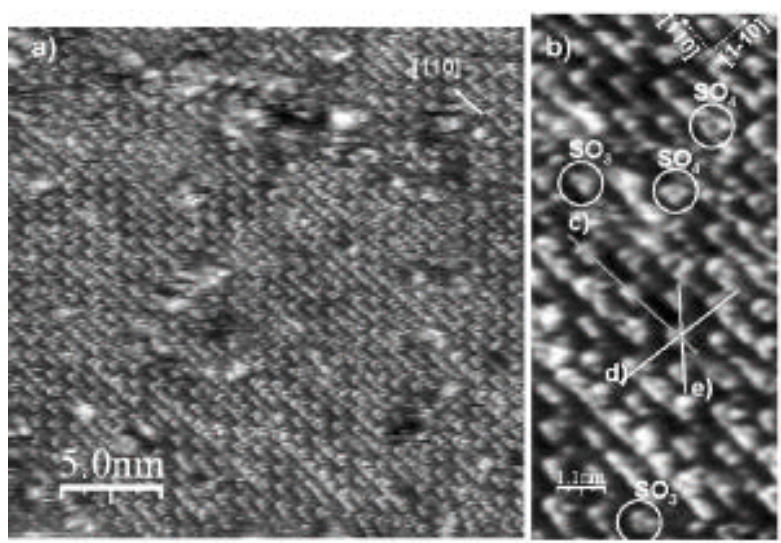

Figure 77: STM images ($V_{sample}$ = 1 V, $I_{tunnel}$ = 0.3 nA) of a surface following exposure to 50 L $H_2O$ and then 50 L $SO_2$ appear to show an overlayer ordered with the underlying (√2×√2)R45° reconstruction.. Reprinted with permission from ref. [464]. Copyright 2007, AIP Publishing LLC.

## 4.13 Hydrogen Sulphide $H_2S$

The interaction of hydrogen sulphide with iron oxides is of interest because $H_2S$ can be thermally decomposed to create $H_2O$ and S [465; 466] ($3H_2S+\alpha\text{-}Fe_2O_3 \rightarrow Fe_2S_3+3H_2O$, followed by $Fe_2S_3 \rightarrow FeS+FeS_2$), and because the iron oxide surface can be recovered by heating in air or oxygen ($4FeS+3O_2 \rightarrow 2Fe_2O_3+4S$ and $4FeS_2+3O_2 \rightarrow 2Fe_2O_3+8S$). Kim et al. [467] studied this process on a natural $\alpha\text{-}Fe_2O_3$(0001) single crystal prepared by cycles of $Ar^+$ sputtering and annealing at 723 K in the presence of atomic O, which produced a (1×1) LEED pattern with no trace of contamination in XPS. Following exposure to $H_2S$ at room temperature the surface began to exhibit S and Fe was partially reduced to $Fe^{2+}$. At 723 K this process was accelerated and, after a 20 min annealing at 723 K with a partial pressure of $1 \times 10^{-7}$ torr, new spots appeared in LEED. A model based on a surface pyrite $FeS_2$ phase was proposed, and subsequent reoxidation of the surface indeed led to the recovery of the $\alpha\text{-}Fe_2O_3$(0001) surface through the formation of $SO_4^{2-}$ (Figure 78)

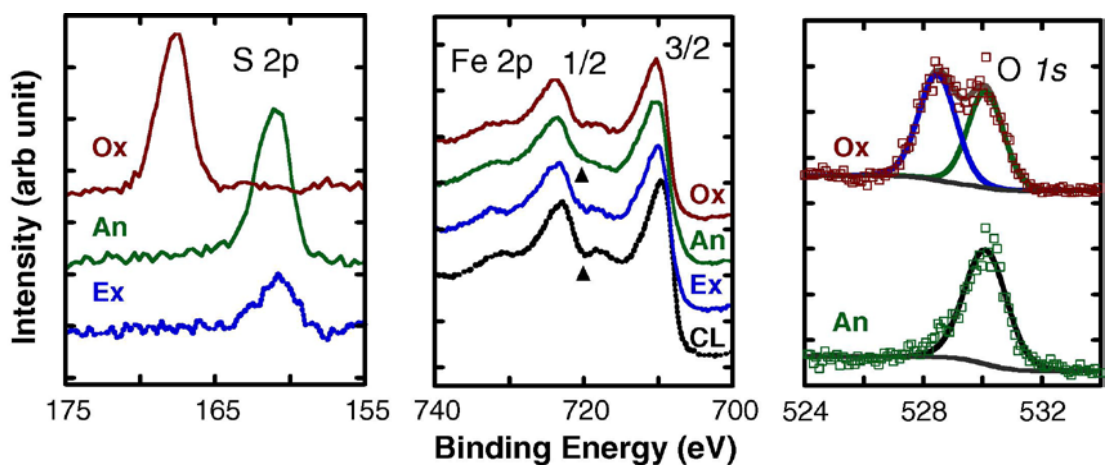

Figure 78: S 2$p$, Fe 2$p$ and O 1$s$ XPS spectra from $H_2S$ treated $\alpha\text{-}Fe_2O_3$(0001) surfaces. The surface treatments are cleaned (CL), RT exposed (Ex), annealed (An) at 430 °C, and oxidized (Ox) by exposure to atomic oxygen. Arrows are drawn to mark changes in the Fe 2$p$ spectra. Figure reprinted from ref. [467] with permission from Elsevier.

## 4.14 Benzene and Alq$_3$

The adsorption of benzene and Alq$_3$ ($C_{27}H_{18}AlN_3O_3$) have been studied by Pratt and co-workers as part of a series of papers investigating how adsorbates modify the spin polarization of the $Fe_3O_4$(100) surface [90-94; 99]. Such molecules are promising for spintronics because the low Z leads to very weak spin-orbit interaction, and long spin coherence times [468]. If the adsorption increases the $E_F$ spin polarization of the $Fe_3O_4$(001) surface, as has been observed for other adsorbates (e.g. atomic H [90]) the system might be useful for the injection of spin-polarized current into the organic molecule. The simplest π-conjugated molecule, benzene ($C_6H_6$), was deposited on 25 nm thick films of $Fe_3O_4$(100) grown on MgO(100), and an enhanced spin polarization at $E_F$ was indeed observed using metastable helium beam scattering [90]. The mechanism of the enhancement is not determined, however.

Later, the same group studied the adsorption of Alq$_3$ [99], one of the most popular organics in the spintronics field due its high electron mobility, on the Fe$_3$O$_4$(100) surface. An important factor governing efficient carrier injection is energy level alignment, and Alq$_3$ is known to possess a large dipole moment, which results in an interfacial dipole. The authors find that the secondary electron cut-off shifts quickly to lower binding energies on Alq$_3$ adsorption (see Figure 79), reducing the work function by 1.2 eV compared to the clean surface value (5.2 eV [87]). The HOMO level shifts to 1.8 eV below the Fermi level of the substrate representing the barrier for hole injection at this interface.

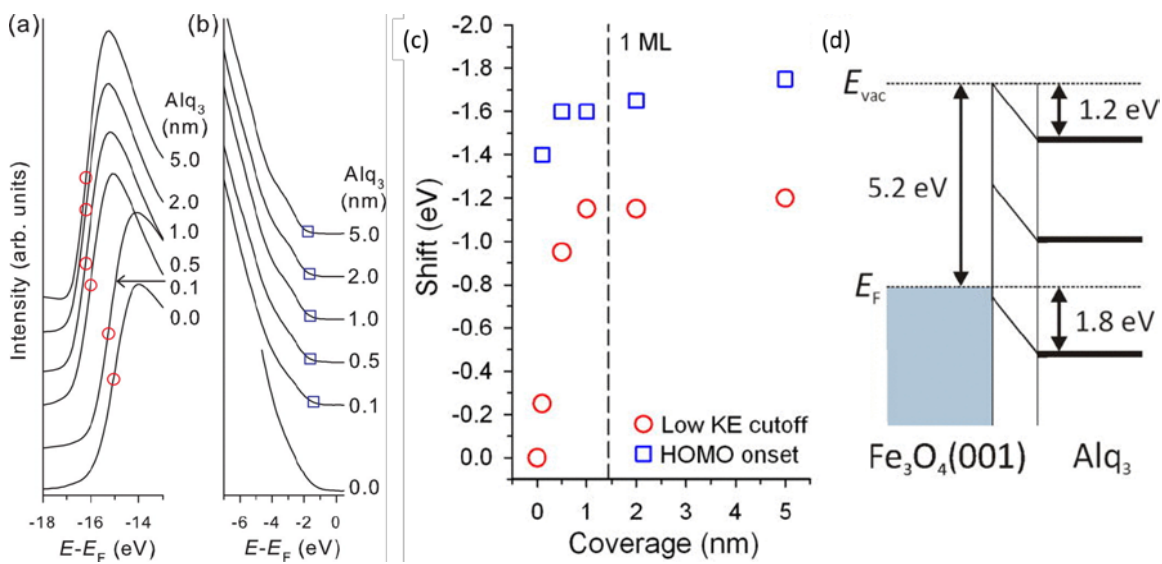

Figure 79: MDS spectra focusing on the low- and high-energy regions allowing determination of (a) the secondary electron cut-offs and (b) the HOMO onset position for each coverage of Alq$_3$, marked by red circles and blue squares, respectively. (c) Shifts of the secondary electron cut-off and the HOMO onset position, extracted from a and b. (d) The corresponding energy-level diagram for the Alq$_3$/Fe$_3$O$_4$(001) system. Reprinted with permission from ref. [99]. Copyright 2012, AIP Publishing LLC.

## 5 Metals

The adsorption of metals on Fe$_3$O$_4$ has been intensely studied due to the use of iron oxides as a support material in catalysis. Consequently, most work has been devoted to Pt-group metals. In these systems, the formation of 3D clusters is preferred on thermodynamic grounds (as with most metal-on-oxide systems [469; 470]) because the surface energy of Fe$_3$O$_4$ is lower than that of the metal. The Freund group has performed extensive investigations of the formation and catalytic activity of Pd and Pt clusters on Fe$_3$O$_4$(111) (typically the Fe$_3$O$_4$ is a thin film grown on Pt(111) e.g. [430; 434; 460; 471-479]), and have convincingly shown that the support plays a role both in the reaction mechanism, and also in important processes such as SMSI [474; 478; 479]. In general there has been much debate about the nano-Au catalysis. Hutchings and co-workers proposed that bilayer gold clusters become charged when supported on iron oxide (powder), and that cationic Au species catalyse CO oxidation [480].

In the last few years, the field of single-atom catalysis has emerged [481; 482], and iron oxides have been frequently utilized as the support to stabilize single-atom active sites (together with other oxides such as alumina and ceria). Activity for various reactions has so far been demonstrated for metals including Ir [483], Au [484], Pt [485-488], and this has motivated theoretical studies aimed at understanding how it is that a single atom can function as a catalyst [489]. Unfortunately, the complexity of the "FeO$_x$" support, usually a reduced α-Fe$_2$O$_3$, means that the termination of the oxide is not known. Consequently, the local environment of the active site, crucial to understand the stability, reaction mechanism, and deactivation processes, is not known either. As will be shown below, the Fe$_3$O$_4$(100) and (111) surfaces can stabilize metal adatoms at reaction temperatures, and thus are potentially ideal model systems to study single-atom catalysis. On (100), the SCV reconstruction provides one twofold coordinated adsorption site per unit cell [35], while V$_O$s [490], or maybe cation vacancies [491], are proposed to stabilize Au adatoms on Fe$_3$O$_4$(111).

In addition to serving as a catalyst support, iron oxides are also the active phase for several important reactions including WGS. Promoters are always present, Cr in the case of WGS, but usually alkali or transition metals. An important consideration with such metals is that they form stable bulk phases with iron oxides based on solid solution [492]. Therefore, when a non-noble metal is deposited onto an iron oxide it reacts with the surface, becoming oxidised. On heating, the metal then diffuses into the bulk. Such behaviour was recently demonstrated for Ni, Co, Mn, and Ti on Fe$_3$O$_4$(100) [260]. The initial incorporation occurred already at room temperature due to the presence of the cation vacancies in the SCV reconstruction, and bulk diffusion into the bulk coincided with the onset of bulk diffusion at around 550 K. Atomistic simulations [493] suggest that Al$^{3+}$ Cr$^{3+}$, Eu$^{3+}$, Gd$^{3+}$, Ho$^{3+}$, La$^{3+}$, Lu$^{3+}$, Nd$^{3+}$, Tb$^{3+}$, and Y$^{3+}$ dissolved in the bulk segregate to the surfaces of α-Fe$_2$O$_3$ on annealing.

## 5.1 Non-Ferrite-Forming Metals (Au, Ag, Pd, Pt)

### 5.1.1 Gold

Au adsorption has been studied on several different iron oxide surfaces including Fe$_3$O$_4$(100) [35; 259; 264; 494-497], Fe$_3$O$_4$(111) [490; 498; 499], while Au/Fe$_3$O$_4$ composite nanoparticles are frequently studied [500; 501].

When Au is evaporated onto the freshly prepared Fe$_3$O$_4$(100)-(√2×√2)R45° surface at room temperature it adsorbs exclusively as Au$_1$ adatoms up to a coverage of 0.13 ML (here, ML is defined as 1 atom (√2×√2)R45° unit cell, see Figure 80). The adatoms appear between the Fe$_{oct}$ rows in STM images and are immobile at room temperature [259]. DFT+U calculations suggest the Au is twofold coordinated to the surface oxygen atoms that do not have a subsurface Fe$_{tet}$ neighbour within the SCV reconstruction (referred to here as the "not-blocked" site, but termed the "narrow site" in ref. [259], see section 3.3.3 for discussion). Thus their nearest-neighbour distance corresponds to the periodicity of the reconstruction (8.4 Å) [35]. Furthermore, DFT+U calculations suggest the Au adatoms protrude ≈0.5 Å above the surface, are positively charged Au$^{1+}$ [35] (consistent with XPS data [259; 495]) and have a binding energy of 2 eV. As yet, there is no quantitative experimental measurement of the adatom geometry. Interestingly, Au adatoms remain stable against agglomeration into clusters up to ≈700 K, which correlates with the lifting of the (√2×√2)R45° reconstruction observed by Bartelt et al. [265]. As such the Au/Fe$_3$O$_4$(100) is a promising model system to study the fundamentals of single-atom catalysis.

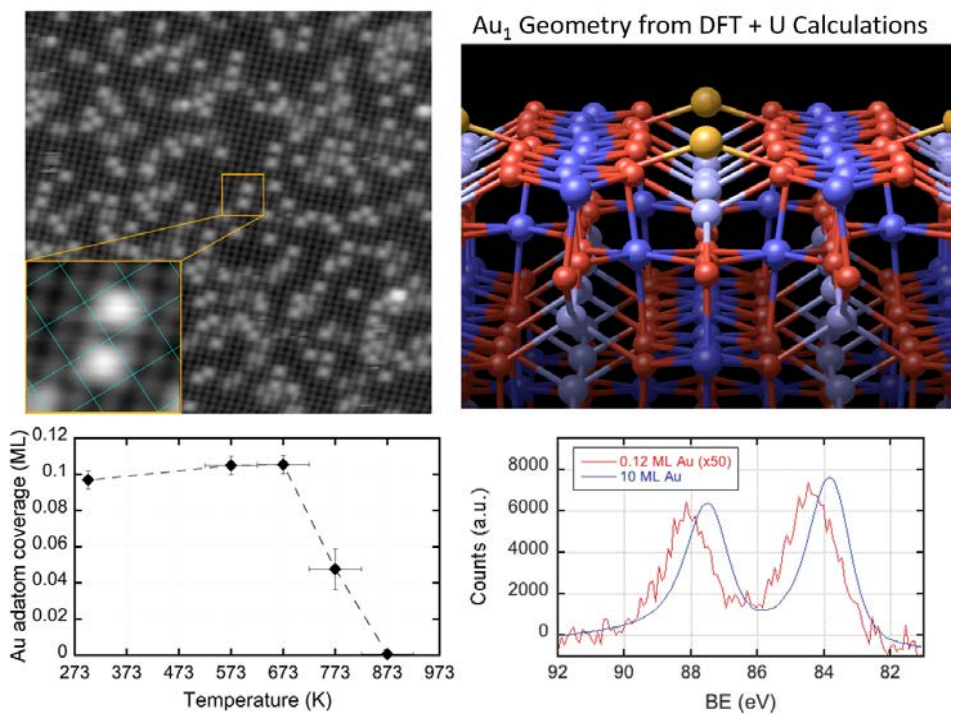

Figure 80: STM images reveal that Au deposited on the Fe$_3$O$_4$(100) surface adsorbs as isolated adatoms (top left), and this configuration is thermally stable up to 700 K (bottom left). DFT calculations find that Au binds with an energy of 2 eV on the subsurface cation vacancy (SCV) termination of Fe$_3$O$_4$(100) (top right) and has an oxidation state of 1+. A positive binding-energy shift is observed in Au 4f XPS (bottom right). Figures adapted from refs. [259] and [35] with permission.

For coverages above 0.13 ML Au, clusters nucleate and coexist with adatoms [259]. Simple Monte Carlo simulations based on a hit-and-stick model suggest that a significant number of Au atoms impinging on the surface begin to land in a unit cell already occupied by a Au atom at a coverage of ≈0.13 ML [502], which likely results in Au dimer formation. DFT+U calculations suggest that an Au dimer is not strongly bound to the Fe$_3$O$_4$(100) surface [502] and therefore will probably diffuse. Tracks in the vicinity of small clusters observed at 0.4 ML suggest that growth proceeds through cluster diffusion and the capture of otherwise stable Au adatoms [259]. Jordan et al. suggested that the smallest Au clusters were 2D, and it is possible that Au$_1$ adatoms were imaged in their STM study [494]. Spiridis and co-workers [495], using a combination of STM, XPS and CEMS experiments, concluded that clusters could be both 2D and 3D, and both are positively charged on Fe$_3$O$_4$(100). Yu et al. [497] on the other hand, suggest strong interaction with surface Fe$_{oct}$ cations and a negative charge for small Au clusters.

The growth of larger Au nanoparticles have been studied [496; 503]. Matek and Snoeck [496], using TEM, found two growth modes; between 200 and 400 °C nanoparticles grow via a Volmer-Weber mode resulting in particles exhibit a (111) texture. At a growth temperature of 750 °C Au grows 3D islands epitaxially on Fe$_3$O$_4$(100). Interestingly, Au cluster diffusion and coalescence was found to be the dominant growth mechanism at 200 °C. More recently, Munoz-Noval and co-workers [503] grew Fe$_3$O$_4$(100) thin films on STO using PLD, and found that Au(100) growth is incommensurate with the substrate and the Au(100) layer fully relaxed.

Spiridis and co-workers [264] have also deposited Au on the Fe-dimer termination of Fe$_3$O$_4$(001) using STM at room temperature, and observed protrusions related to Au$_1$ adatoms adsorbed alongside the Fe-dimers at low coverage (see Figure 81). This interesting result suggests that the strong binding of Au is not restricted to the twofold coordination shown in Figure 80, since these sites are blocked by the Fe atoms of the "dimer". At higher coverages Au clusters are formed, which eventually coalesce into a closed Au(111) film.

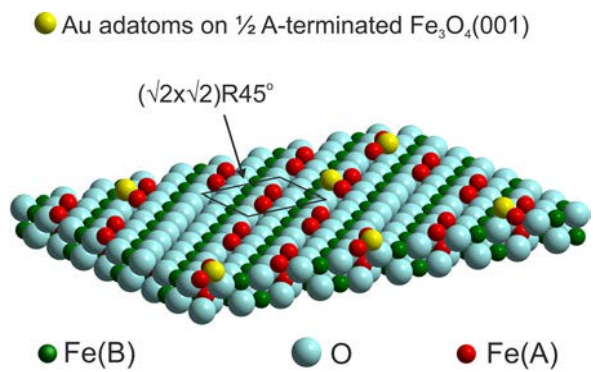

Figure 81: Schematic representation of $Au_1$ adatoms adsorbed on the Fe-dimer termination of $Fe_3O_4$(100), as proposed by Spiridis and co-workers on the basis of STM measurements. The Fe-dimer atoms are red, $Fe_{oct}$ are green, and $O_{lattice}$ are light blue. Reprinted with permission from ref. [264]. Copyright 2014 American Chemical Society.

Rim et al. [490] studied the adsorption of Au on the $Fe_3O_4$(111) surface using a reduced selvedge of a α-$Fe_2O_3$(0001) single crystal. Initially more than 1 ML of Au was deposited, resulting in a high density of small 3D Au nanoparticles. After annealing at 500 °C in UHV for 15 minutes many of the nanoparticles agglomerated to form large faceted nanoparticles, as observed previously on $Fe_3O_4$(111) thin films [498; 499], but $Au_1$ adatoms were also observed. On the basis of STM and STS, the authors propose the adatoms bind atop undercoordinated O atoms, and are positively charged [490]. DFT+U calculations [298; 491] however, find the most stable geometry to be atop a surface $Fe_{tet1}$ atom (1.66-1.98 eV binding energy, depending on method), although the Au adatom was found to bind very strongly (-3.45 eV) in a threefold hollow sites with a Bader charge of +0.83 electrons if a surface with $Fe_{tet1}$ vacancies was considered. Therefore, it is possible that the strong adatom binding observed by Rim et al. [490] may be linked to $Fe_{tet1}$ vacancies, particularly as their clean surface appeared to exhibit many such defects.

Most of the above-mentioned studies were performed with a view to understanding the performance of CO oxidation or water-gas shift on Au/$Fe_3O_4$. Novotny et al. [259] found no evidence of CO adsorption at room temperature on $Au_1$/$Fe_3O_4$(100) under UHV conditions, but Rim et al. [490] did when a $Au_1$/$Fe_3O_4$(111) sample was cooled to 260 K. Herzing and co-workers suggested that bilayer Au clusters supported on $FeO_x$ nanoparticles are optimal for CO oxidation [480] (not single atoms), and the DFT+U calculations of Li et al. [489] suggest that $Au_1$ atoms would quickly be poisoned by strong CO adsorption. Shaikhutdinov et al. [498] found that CO adsorbs more strongly on smaller Au clusters, and observed a desorption peak close to room temperature from the smallest particles, which should be ideal for CO oxidation. Finally, small islands of $Fe_3O_4$ deposited on Au(111) have been shown to be particularly active for CO oxidation [504].

There has been interest in the adsorption of Au on $Fe_2O_3$ ever since Haruta demonstrated that $Fe_2O_3$-supported Au nanoparticles are catalytically active for CO oxidation [505]. There are no experimental studies of $Au_1$ adatoms on α-$Fe_2O_3$(0001), but theoretical calculations [298; 506] predict the most stable geometry to be atop a surface Fe atom of the half metal termination, negatively charged, with approximately 0.2 e$^-$ transferred from the substrate. The strongest binding energy (-1.02 eV) is found at a Fe vacancy site, where the Au adatoms are threefold coordinated to oxygen, and this site is the most thermodynamically favourable adsorption site for CO [506]. As on $Fe_3O_4$ surfaces [35; 259; 490], binding to oxygen results in a positively charged $Au^{1+}$ ion. The most stable adsorption position on the $O_3$-Fe-Fe surface is also at a threefold hollow site (the binding energy varies from -2.11 eV using GGA to -3.59 eV using GGA+U [298]). Hoh et al. [507] found that the oxygen-vacancy formation energy decreased from 3.04 eV to just 0.86 eV underneath $Au_{10}$ clusters, suggesting that the Mars-van Krevelen mechanism might be facile for oxidation reactions in this system. Wong et al. [508] calculated the structure of Au, Pt, Pd and Ru bilayers on α-$Fe_2O_3$(0001) and found that Au was most stable as it caused least structural deformation of the support.

### 5.1.2 Silver

Ag/$Fe_3O_4$ nanoparticles are of interest to create catalysts for epoxidation reactions (e.g. [509-511]), and for use as a bactericide in medical applications (e.g. [512-516]). Interestingly, it has been shown [512] that calcination of Ag/$Fe_3O_4$ nanocatalysts at 400 °C results can lead to the dispersion of Ag atoms over the nanoparticle surface. The basis of such dispersion is not well known, and remains a potential topic of study for surface science.

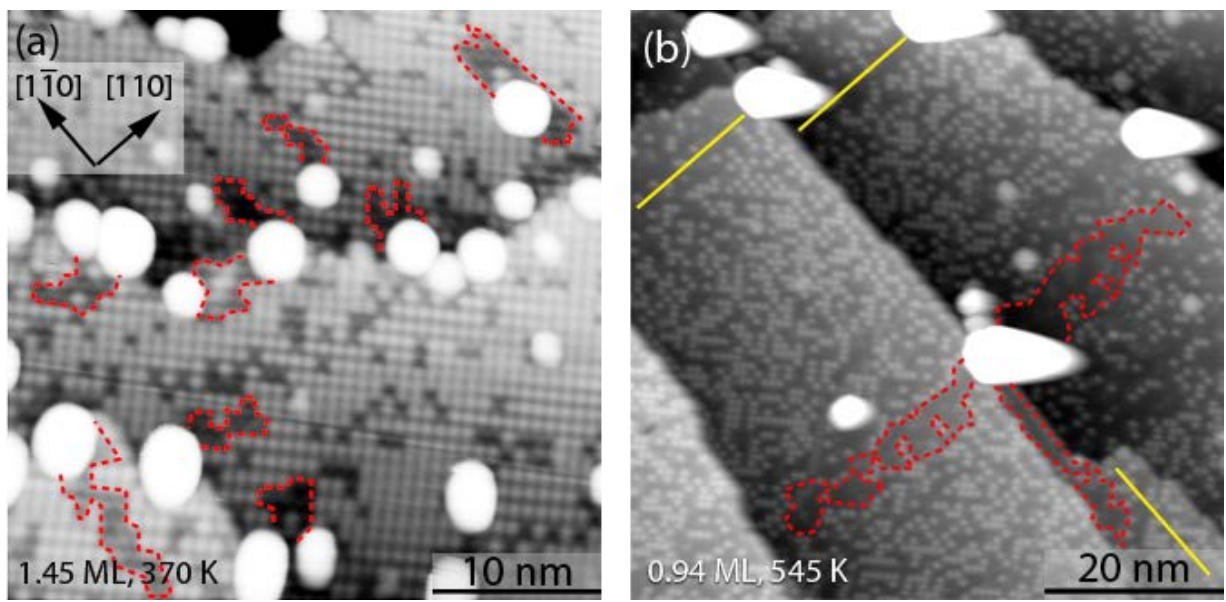

Figure 82: STM images of the Ag/$Fe_3O_4$(100) surface following post annealing in UHV. The Ag coverage and annealing temperature are indicated in the figure. Clusters nucleate spontaneously at room temperature for Ag coverages in excess of 0.5 ML, and then grow at the expense of the adatom phase when the system is heated. Trails (red dashed lines) of bare $Fe_3O_4$(100) surface appear because both the adatom and cluster diffusion is restricted to the local $Fe_{oct}$ row direction on each terrace (yellow lines). Following annealing to 545 K, the clusters are mostly located at step edges or defects such as APDBs. Note that the tails to the right of the larger clusters are likely an artefact of the measurement. Figure reproduced from ref. [261].

Ag adsorbs as $Ag_1$ adatoms on the $Fe_3O_4$(100)-(√2×√2)R45° SCV surface following deposition at room temperature, occupying the same "not-blocked" adsorption site as Au (Figure 80) [261]. However, in contrast to Au, the Ag adatoms are mobile at room temperature, and diffuse between neighbouring not-blocked sites along the Fe$_{oct}$ rows. The coverage of Ag adatoms that can be achieved prior to cluster formation is significantly higher (0.5 ML) than for Au, which is linked to the ability of the system to surpass the critical cluster size. In both cases the dimer is unstable with respect to two adsorbed adatoms, but the Au dimer is not strongly bound to the surface and can diffuse to meet other Au adatoms and grow. The Ag dimer, on the other hand, remains bound to the "not-blocked" adsorption site until it decays, and thus a critical cluster size is not reached. The Ag$_1$ adatom phase spreads through a dimer-decay mechanism, with excess Ag passed around until dimer decay occurs alongside an empty site in the (√2×√2)R45° reconstruction. In the absence of clusters, the adatom phase is stable up to 700 K (linked to the lifting of the SCV reconstruction [265]). Cluster nucleation occurs when the density of Ag is sufficiently high that multiple Ag atoms can meet on the surface, and a critical cluster size is exceeded. The corresponding coverage is approximately 0.5 ML. Once formed, Ag clusters grow at the expense of the adatom phase. With mild annealing, the clusters diffuse to the step edges, and continue to grow through the capture of Ag$_1$ adatoms. Interestingly, the denuded zones associated with the growth mechanism extend along the terrace in the local Fe$_{oct}$ row direction (see Figure 82). Gatel and Snoeck [496] performed AFM and TEM investigations of 3 and 7 nm Ag deposited at 200, 400 and 750 °C on Fe$_3$O$_4$(100). Large octahedral islands were observed at 200 and 400 °C, and the authors concluded a Volmer-Weber growth mode.

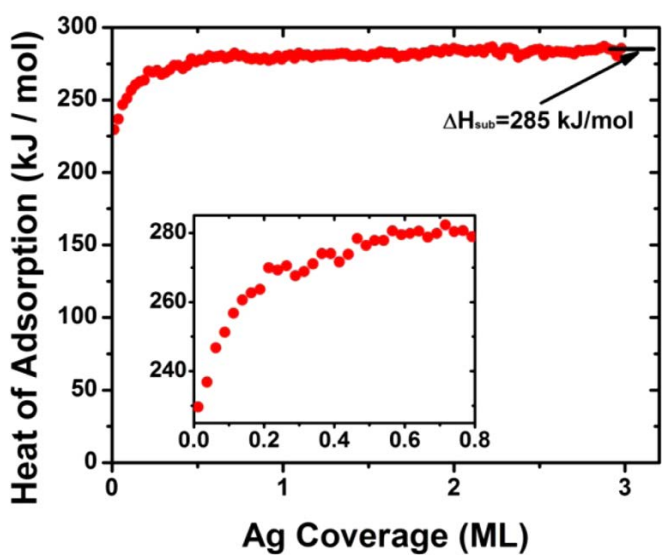

Figure 83. Heat of adsorption versus coverage for Ag on Fe$_3$O$_4$(111) at 300 K measured by microcalorimetry. Reprinted with permission from ref. [517]. Copyright 2013 American Chemical Society.

On the Fe$_3$O$_4$(111) surface, Sharp et al. [517] studied Ag adsorption on 6 nm films at room temperature using adsorption microcalorimetry (see Figure 83), AES and LEIS. The measurements reveal that Ag grows as clusters on Fe$_3$O$_4$(111), with a heat of adsorption that increases from 2.3 eV initially up to 3.0 eV at 2 ML coverage, which is close to the bulk heat of sublimation. No other surface science studies of Ag adsorption on Fe$_3$O$_4$(111) or any α-Fe$_2$O$_3$ surface could be found.

### 5.1.3 Palladium

Pd also adsorbs as a Pd$_1$ adatom on Fe$_3$O$_4$(100) at the "not-blocked" site of the SCV reconstruction (see section 3.3.3). The published binding energy of 2.25 eV and charge state of +0.6 e$^-$ were calculated using the DBT structure [262], but these values do not change significantly for the SCV structure as the local environment around the adsorption site is similar. High coverages of Pd adatoms can be achieved, which suggests the Pd-Pd dimer is unstable with respect to two adatoms (as discussed for Ag, above). However, in contrast to Au and Ag, the stability of the $Pd_1/Fe_3O_4(001)$ system is highly sensitive to CO [262]. Indeed, even in a pressure of $5\times10^{-10}$ mbar CO, Pd adatoms sinter completely to form Pd nanoparticles in under an hour (see Figure 84). Essentially, CO adsorption weakens the bond between $Pd_1$ and the surface, and the $Pd_1$-CO moiety diffuses rapidly at room temperature, but is trapped temporarily at stationary $Pd_1$ adatoms. This species appears as a "scratchy" feature in Figure 81 B because it is loosely bound and moves underneath the STM tip during scanning. After a short time the Pd-Pd bond breaks and the Pd-CO species diffuses further. This suggests that the Pd-Pd bond is not strong enough to induce cluster nucleation. STM movies of the sintering process (see Figure 84) reveal that cluster nucleation is homogeneous, and occurs when multiple carbonyl species exist close together on the surface. Such nucleation is unusual on metal oxide surfaces, where heterogeneous nucleation at defect sites is more common. Once a cluster is nucleated, further growth occurs through a diffusion and capture mechanism.

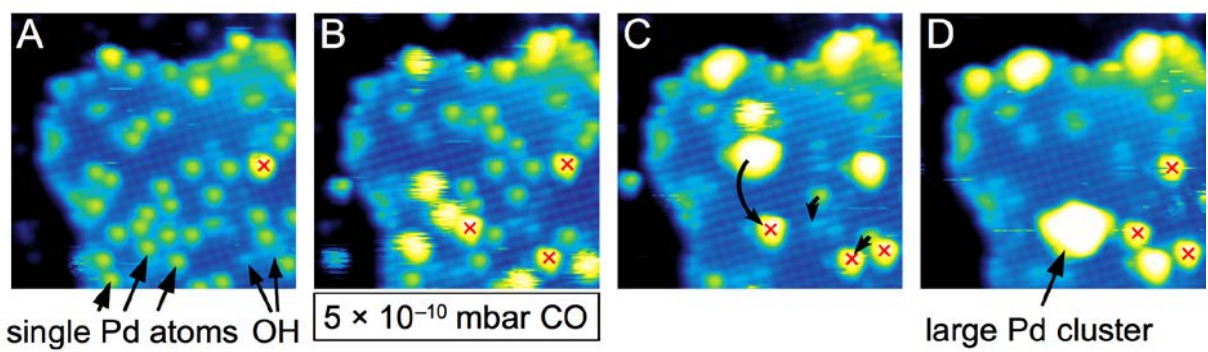

Figure 84. Selected frames from an STM movie showing the CO-induced sintering of $Pd_1$ adatoms on the $Fe_3O_4(100)$ surface. (A) $Pd_1$ adatoms are stable on the surface in the absence of CO. The feature marked by an x is a $Pd_1$ adatom bound to a surface OH group. (B) "Scratchy" features observed in a CO environment are $Pd_1$-CO species bound temporarily at stable $Pd_1$ adatoms. (C) Pd clusters diffuse across the surface collecting other $Pd_1$ species. (D) Since the agglomeration was followed in an atom-by-atom fashion, the "large Pd cluster" is known to include 19 Pd atoms. Image reproduced with permission from ref [262].

On the $Fe_3O_4(111)$ surface, theoretical calculations [298] predict that $Pd_1$ adatoms bind preferentially to oxygen at the 3-fold hollow site on the $Fe_{tet1}$ surface with a binding energy of 1.76 eV (DFT) or 2.2 eV (DFT+U). This is very similar to the binding energy of $Pd_1$ adatoms on $Fe_3O_4(100)$, which have been observed experimentally [262]. However, STM investigations suggest the formation of small clusters at 300 K following deposition of Pd at 100 K [430]. Freund's group performed several studies of CO oxidation over the $Pd/Fe_3O_4(111)$ system [430; 471-473], with Pd particles ranging from 2 to 100 nm in size. The particles grow in (111) registry with the support and are terminated by (111) facets. Annealing in $10^{-6}$ mbar $O_2$ at 500 K results in some sintering, which the authors propose to be mediated by a mobile Pd-oxide species [472]. The smallest particles are oxidised completely to palladium oxide, while PdO is formed at the particle/support interface of particles larger than 3 nm [473]. This PdO provides an oxygen reservoir that can be utilized to oxidise CO molecules at 500 K (see Figure 85).

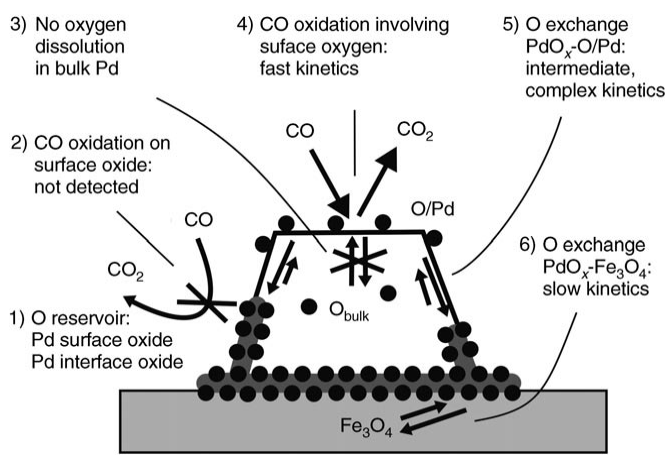

Figure 85. The mechanism of CO oxidation on a Pd/Fe$_3$O$_4$(111) catalyst. A PdO layer is formed at the interface between Pd and Fe$_3$O$_4$, which provides an oxygen reservoir that is utilized during the CO oxidation reaction at 500 K. Figure reproduced from ref [471] with permission from John Wiley and Sons.

Pd adsorption has been studied on α-Fe$_2$O$_3$(0001) from a theoretical viewpoint by Kiejna and Pabisiak [298]. They report that the most stable geometry is above a threefold hollow site (coordinated to O$_{lattice}$) on the Fe-O$_3$-Fe- termination, with the Pd atom positively charged (donating 0.22 e$^-$ to the surface) and a binding energy calculated to be 1.82 eV using GGA and 1.53 eV using GGA+U. Interestingly, the adsorption causes the surface layers to relax back towards bulk terminated coordinates. On the O$_3$-Fe-Fe- surface, Pd is found to bind strongly in a bulk continuation threefold hollow site with a binding energy of 2.85 eV (GGA) or 5.21 eV (GGA+U). No experimental investigations of Pd on α-Fe$_2$O$_3$ surfaces could be found.

### 5.1.4 Platinum

At the Fe$_3$O$_4$(100) surface, submonolayer coverages of Platinum adsorb as Pt$_1$ adatoms twofold coordinated to oxygen on the "not-blocked" site of the SCV reconstruction [402; 479]. In addition to single crystal work performed by this author's group [402], Zhang et al. [479] recently observed the same phenomenon on a Fe$_3$O$_4$(100) thin film grown on Pt(100) (see Figure 86). In the latter investigation, adatom coverages approaching 1 ML were observed. Similar to Pd [262], the Pt$_1$ adatom species is highly sensitive to sintering by CO [402], but in contrast to Pd the Pt$_1$-CO remains bound at the not-blocked site, and diffuses slowly between neighbouring sites. In a CO pressure of $2\times10^{-10}$ mbar, CO-induced sintering can be followed atom-by-atom and clusters in the size range of 2-6 atoms were identified. Therefore, this approach is an alternative to size-selected cluster deposition for STM studies of the size effect in catalysis. XPS data suggest that CO remains bound to the clusters following sintering at room temperature, but desorbs by 550 K [402]. Substantial sintering only occurs at 700 K [479], as observed for Ag and Au [259], which is probably linked to the lifting of the SCV reconstruction [265]. Interestingly, this temperature also coincides with the encapsulation of the Pt nanoparticles by iron oxide [479], as has been observed on Fe$_3$O$_4$(111). After annealing to 1000 K, cuboid nanoparticles are formed [479], consistent with the epitaxial (100) growth reported by Gatel and Snoeck [496].

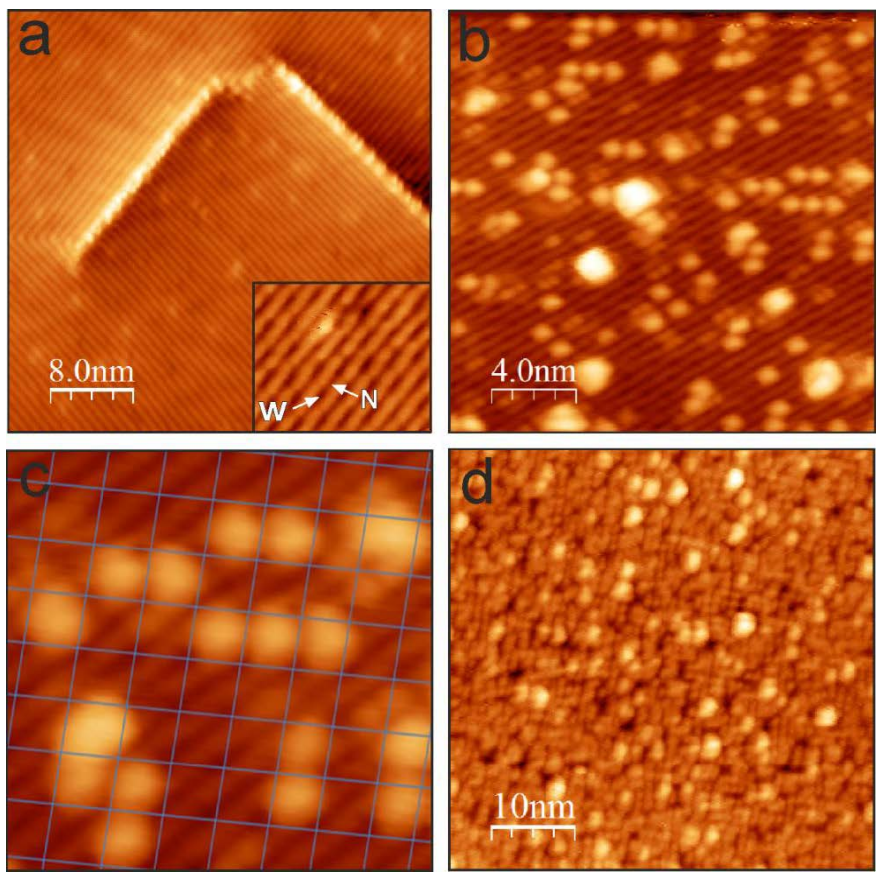

Figure 86: (a) STM image ($V_{sample}$ = 1 V, $I_{tunnel}$ = 0.5nA) of a clean $Fe_3O_4$(001) thin film grown on Pt(100). The surface exhibits the undulating rows of $Fe_{oct}$ cations associated with the SCV reconstruction. Note the imaged area exhibits a screw dislocation, around which the $Fe_{oct}$ row direction rotates. "Wide" (w) and "narrow" (N) adsorption sites (referred to as "blocked" and "not-blocked" sites in this review, respectively) within the ($\sqrt{2}\times\sqrt{2}$)R45°-$Fe_3O_4$(001) unit cell are indicated in the inset. (b,c) STM images ($V_{sample}$ = -1 V, $I_{tunnel}$ = 0.7 nA) acquired after deposition of 0.15 ML Pt at 300 K; $Pt_1$ adatoms adsorb in the "not-blocked sites", with some small clusters likely induced by CO in the background gas. The nodes of the grid in panel (c) are centred at the blocked sites. (d). STM images ($V_{sample}$ = -1 V, $I_{tunnel}$ = 1 nA) following deposition of 1 ML Pt. The surface is almost completely covered by $Pt_1$ adatoms, with some small clusters. Figure reproduced from ref. [479] with permission from John Wiley and Sons.

In a recently published paper [402], the present author's group found that the reduction and oxidation of the $Fe_3O_4$(100) surface could be catalysed by the presence of Pt clusters. Specifically, when a $Pt_{1-6}$/$Fe_3O_4$(100) model catalyst is exposed to CO at 550 K, $O_{lattice}$ atoms are extracted from the cluster perimeter to form $CO_2$. Since undercoordinated Fe diffuses into the bulk already at 550 K, the reduction process leads to monolayer holes in the support (Figure 87). The process appears to be autocatalytic because only some clusters are associated with holes, which suggests that removal of the first $O_{lattice}$ atom from the pristine surface is most difficult energetically. Similar holes occur when the sample is annealed in $H_2$ because the Pt clusters dissociate the molecules, and atomic H species spill over onto the metal oxide, where they react with the surface to desorb water. When the sample is annealed in $O_2$, $Fe_3O_4$(100) islands grow around the clusters because $O_2$ dissociates on the metal and atomic O spills over onto the oxide, where it reacts with Fe cations supplied from the bulk. These data represent further evidence that the $Fe_3O_4$ bulk acts as a sink for Fe atoms, which diffuse back and forth from the surface depending on the oxygen chemical potential. Since reduction and oxidation of the oxide occur at higher temperatures in the absence of Pt, these results were interpreted as the Pt clusters catalysing the rate-limiting step, i.e. CO adsorption, $O_2$ dissociation and $H_2$ dissociation.

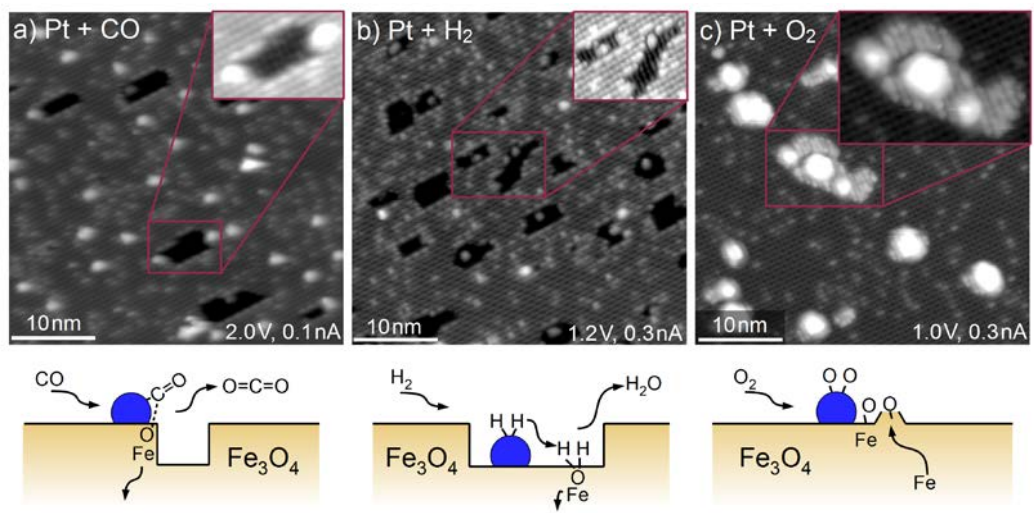

Figure 87: STM data reveal that reduction and oxidation of the support in CO (a), $H_2$ (b) and $O_2$ (c) is catalysed by Pt clusters in a $Pt_{2-6}/Fe_3O_4(100)$ model catalyst. Extraction of $O_{lattice}$ atoms to make $CO_2$ and $H_2O$ results in monolayer holes in the $Fe_3O_4(100)$ surface around the Pt clusters. Undercoordinated Fe diffuses into the $Fe_3O_4$ bulk. Dissociation of $O_2$ on the clusters results in O atoms that spill over onto the metal oxide and react with Fe from the bulk to create new $Fe_3O_4(100)$ islands. Figure reproduced from ref. [402] with permission from John Wiley and Sons.

On $Fe_3O_4(111)$ thin films, Qin et al. [476] found that Pt formed small, single-monolayer high islands at low coverage, as judged from STM images. Heating to 600 K led to some sintering, but the clusters remained one monolayer in height. When higher Pt coverages were deposited the density of clusters increases dramatically, but again they remain one monolayer height. Heating 1 ML Pt to 600 K resulted in bilayer islands with an irregular shape. Generally, these results are consistent with a high adhesion energy of Pt on $Fe_3O_4(111)$ (4 $Jm^{-2}$) [476]. Interestingly, TPD results show that the mono- and bilayer islands may be inert to CO adsorption above 100 K, which the authors tentatively attribute to the expansion of the Pt(111) lattice (i.e. larger lattice constant) during epitaxial growth on $Fe_3O_4(111)$.

Encapsulation of active material by metal oxide, sometimes termed SMSI, is one of the major causes of catalyst deactivation. Zhang et al. [479] recently utilized CO TPD to show that Pt particles on $Fe_3O_4(100)$ become encapsulated by an FeO(111) thin film at approximately 850 K, a phenomenon previously studied in detail for the $Fe_3O_4(111)$ surface (see Figure 88) [474]. Interestingly, the authors have subsequently shown that such an FeO overlayer on Pt can enhance CO oxidation activity [475; 478]. Sun et al. [477] found that CO can dissociate on low-coordination sites on supported Pt clusters resulting in a build-up of carbon, and found that Fe from the support diffuses onto (or into) the Pt clusters above 600 K.

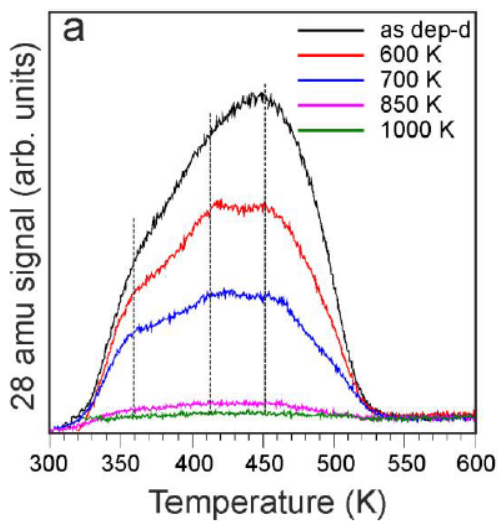

Figure 88: TPD spectra of 20 L CO adsorbed at 300 K on 1.5 ML Pt/Fe$_3$O$_4$(001) as deposited at 300 K and then UHV annealed to the indicated temperature. The heating rate is 2 K/s. The data show some reduction in CO adsorption capacity up to 700 K linked to sintering, before a dramatic reduction at 850 K linked to the encapsulation of the clusters in a thin film of FeO. Figure reproduced from ref. [479] with permission from John Wiley and Sons.

Platinum is one of several metals investigated to catalyse PEC water splitting at α-Fe$_2$O$_3$ surfaces (e.g. [518-520]). Very recently, Neufeld and Toroker [521; 522] studied the Pt(111)/α-Fe$_2$O$_3$(0001) system using a DFT+U approach and determined that the metal adheres with an energy of $\gamma_{adh} \approx 2.83 \pm 0.1$ J m$^{-2}$. Interestingly, the presence of the metal modifies the electronic structure of the oxide (see Figure 89), reducing the carrier effective mass from 3.98$m_e$ to 0.94$m_e$ (through a strong hybridization between Fe and Pt states). This could improve the charge transport within a PEC cell. However, the adsorption of Pt also creates a continuum of states in the band gap, which would likely increase the rate of interfacial recombination. The authors suggest the changes in electronic structure penetrate several nm from the interface.

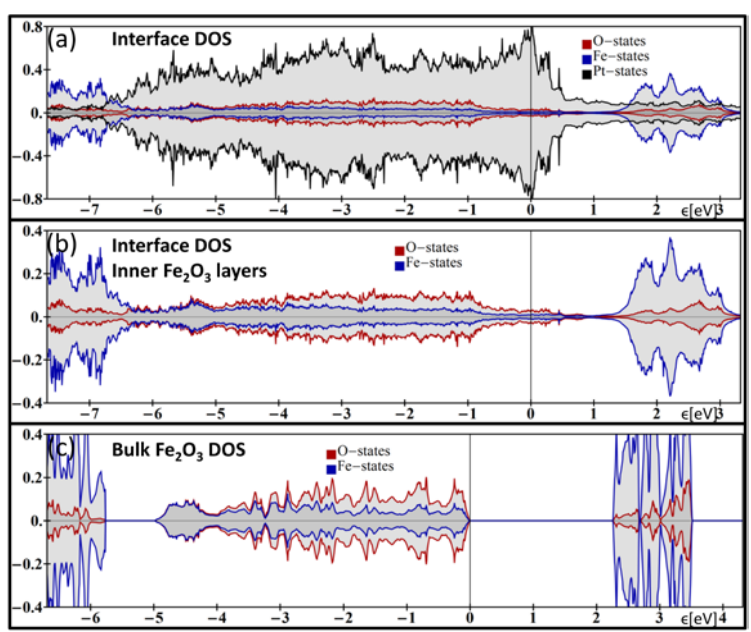

Figure 89: A significant change in the DOS is observed near the Pt/α-Fe$_2$O$_3$(0001) interface: (a) the entire interface, (b) the Fe$_2$O$_3$ layers closest to the Pt/oxide interface, and (c) bulk Fe$_2$O$_3$. The DOS is normalized to a maximum total density of one. The vertical line at zero is the last occupied orbital energy. Figure from ref. [521]. Reproduced by permission of the PCCP Owner Societies

## 5.2 Ferrite-Forming Metals (Co, Ni, Cu, Mn, Ti, Zr, Va, W, Cr)

As mentioned above, the majority of transition metals form well-known stable metal ferrite (M$_{1-x}$Fe$_{2+x}$O$_4$, M = metal) phases [523], and thus the thermodynamic preference can be for incorporation within the oxide rather than the formation of 3D nanoparticles. This trend was observed recently by Bliem et al. [260], who studied the adsorption of the transition metals Mn, Co, Ni, Zr and Ti on the SCV termination of Fe$_3$O$_4$(100) using STM, XPS, UPS, and DFT+U calculations. In that study, the temperature at which incorporation occurred differed depending on the metal, and it was suggested that the tendency for incorporation scales with the affinity of the metal towards oxygen. This can trend can be seen in the heat of formation of the most stable oxide (Ni < Co < Mn < Zr ≈ Ti) (see Table 6). Note that Au, Ag, Pt, Pt, at the top of this list, do not incorporate, and do not form solid solution compounds with Fe$_3$O$_4$. Ti and Zr, which form particularly stable oxide phases, were found to incorporate already at room temperature. Interestingly, the binding energy of the adatom in the "not-blocked" site of the Fe$_3$O$_4$(100) surface also appears to scale with this parameter, which can be seen as a measure of the desire of the metal to form bonds with oxygen. Consequently, it would be interesting to study the adsorption of Rh, Ru, and Cu on the Fe$_3$O$_4$(100) surface, which should bind very strongly to the oxide and thus may survive as adatoms on the surface up to high temperatures without sintering or incorporation.

Table 6: Standard heats of formation of the most stable metal oxide. Table adapted from ref. [469]

| HEAT OF FORMATION OF OXIDE ΔH$_F^0$ (KJ/MOL) | METAL |
| --- | --- |
| > 0 | Au |
| 0 – 50 | Ag, Pt |
| 50 – 100 | Pd |
| 100 – 150 | Rh |
| 150 – 200 | Ru, Cu |
| 200 – 250 | Re, Co, Ni, Pb |
| 250 – 300 | Fe, Mo, Sn, Ge, W |
| 300 – 350 | Rb, Cs, Zn |
| 350 – 400 | K, Cr, Nb, Mn |
| 400 – 450 | Na, V |
| 450 – 500 | Si |
| 500 – 550 | Ti, U, Ba, Zr |
| 550 – 600 | Al, Sr, La, Hf, Ce |
| 600 – 650 | Sm, Mg, Th, Ca, Sc, Y |

In what follows the details of the adsorption of different metals on iron oxide surfaces are described. In addition, the properties of the relevant bulk ferrite compounds will be described where possible, which provide context and are interesting in their own right. For example, ferrite nanoparticles are found to be more reactive than Fe$_3$O$_4$ [524], and thin films of Ni, Co, and Mn ferrite are studied as spin filters for spintronics devices [525]. The near-interface chemistry and structure are found to be crucial to device performance.

## 5.2.1 Cobalt

Co was found to adsorb as $Co_1$ adatoms at room temperature occupying twofold coordinated sites in the SCV reconstruction [260] (see Figure 90, 91), but approximately half of the deposited Co formed new species that appeared as bright protrusions on the $Fe_{oct}$ rows. The transition from adatom to "on-the-row" species was observed in room temperature STM movies (see Figure 91). DFT+U calculations and spectroscopic evidence suggest that Co is incorporated in subsurface octahedral sites, consistent with the site preference in the solid solution cobalt ferrite spinel $Co_{1-x}Fe_{2+x}O_4$. Note that the incorporation of Co causes the $Fe_{int}$ atom to diffuse to a $Fe_{oct}$ site resulting in the lifting of the SCV reconstruction locally (see dashed arrows in Figure 90. Gargallo-Caballero et al. [526] observed that deposition of 1 ML Co leads to the lifting of a (1×1) LEED pattern, consistent with this idea. Gargallo-Caballero et al also studied much higher Co coverages (between 1 and 25 ML) using LEED, PEEM, XMCD and XAS and DFT+U calculations, and determined that Co first occupies the subsurface vacancies for coverages up to one $Co^{2+}$ atom per unit cell, after which the $Co_1$ adatom sites are occupied by $Co^{2+}$ species. At higher coverages still, metallic Co accumulates on the surface. The distribution of the magnetic domains was found to be unchanged from the clean $Fe_3O_4$(100) surface at all coverages, but the coupling between Co and the substrate changes as the sample is annealed. This behaviour is linked to changes in the relative distribution of tetrahedral sites (spin up) and octahedral sites (spin down) as the metallic Co is incorporated in the spinel lattice. Annealing the system above 663 K leads to the diffusion of Co into the bulk oxide, with no trace remaining in XPS (see Figure 92). Consistent with this, DFT+U calculations find that $Co_{oct}^{2+}$ cations are preferably incorporated in bulk-like (S—6) layers of a $Fe_3O_4$ slab than in the surface (S) or subsurface layers (S—2) (see Table 7)

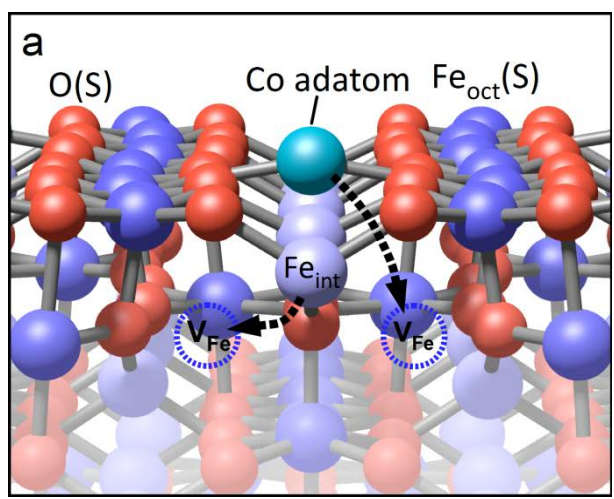

Figure 90: DFT+U structural model of Co adsorbed at the (√2×√2)R45°-reconstructed $Fe_3O_4$(001) surface. The Co adatom occupies an oxygen-bridging site between rows of octahedral Fe, the so-called "not-blocked" site. The dashed black arrows indicate one possible pathway of Co incorporation: One $V_{Fe}$ is filled by $Fe_{int}$, one by the Co adatom, restoring the number of cations in a bulk-truncated surface. Figure adapted from ref. [526].

Table 7: DFT+U calculations show that Co binds strongly at the $Fe_3O_4$(100) surface, and the binding energy in the adatom site is slightly smaller than that of incorporation in the second octahedral layer

(labelled (S—2) of a spinel lattice structure. Incorporation in bulk octahedral sites (here represented as the middle of the slab) is the most energetically favourable configuration [526].

| Configuration | Binding Energy (eV) | Oxidation State |
|---|---|---|
| $Co_1$ Adatom | 5.46 | 2+ |
| $Co_{oct}$ (S) | 5.41 | 2+ |
| $Co_{tet}$ (S—1) | 5.27 | 2+ |
| $Co_{oct}$ (S—2) | 5.59 | 2+ |
| $Co_{oct}$ (S—6) | 5.75 | 2+ |

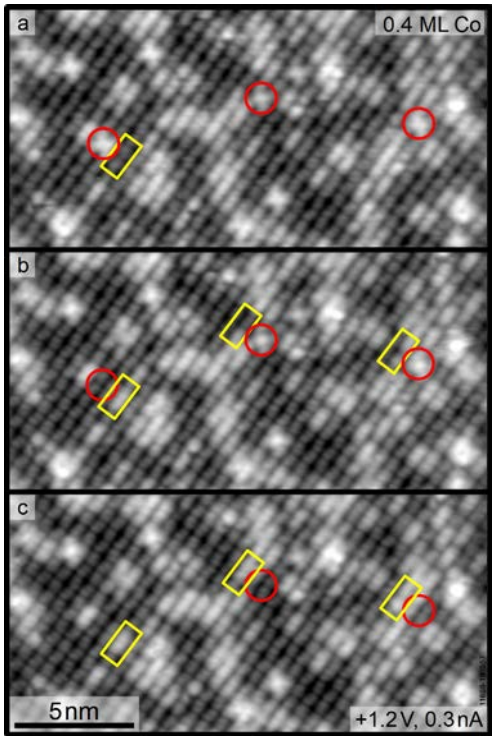

Figure 91: Sequential STM images acquired following the deposition of a submonolayer coverage of Co on $Fe_3O_4$(100) reveal that $Co_1$ adatoms (red circles) transition into bright "on the row" protrusions at room temperature. A combination of spectroscopies suggest that these protrusions are related to incorporation of Co in the subsurface $Fe_{oct}$-O layers. Figure adapted from ref. [260].

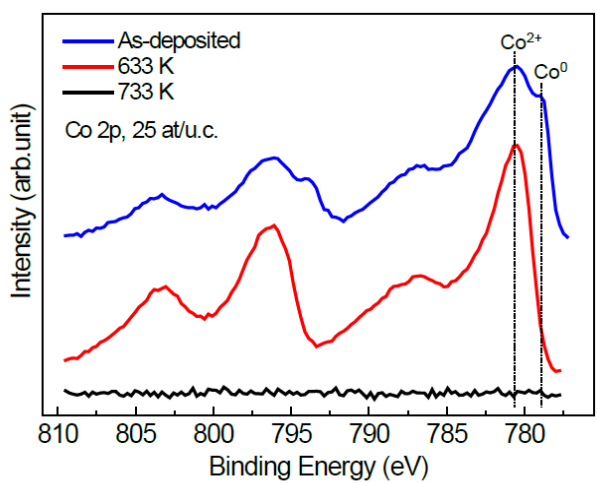

Figure 92: Co 2$p$ XPS spectra for 25 Co atoms per Fe$_3$O$_4$(100)-($\sqrt{2}\times\sqrt{2}$)R45° unit cell annealed to different temperatures. Figure adapted from ref. [526].

The electronic structure of Co-doped Fe$_3$O$_4$(100) thin films was studied by Ran et al. [527] up to a doping concentration of 33 %. A decrease in the intensity of the Fe$^{2+}$-related peak near E$_F$ was observed, consistent with Co$^{2+}$ replacing Fe$^{2+}$ in the octahedral sites. Bahlawane et al. [528] studied Co$_{3-x}$Fe$_x$O$_4$ films with x between 0 and 1.56 and found the spinel structure maintained up to 0.56, above which there was significant degree of inversion defects (i.e. cobalt on octahedral sites and iron on tetrahedral sites). Interestingly, they find that the Co concentration directly affects the ability of the sample to perform CO oxidation reaction via a Mars-van Krevelen process (see Figure 93), concluding that O atoms coordinated to Co are active sites for the reaction. Co-doped Fe$_3$O$_4$ has been found to exhibit higher activity toward H$_2$O$_2$ reduction [529; 530]. Experimental studies on CoFe$_2$O$_4$ nanoparticles and thin films suggest Co doping affects the magnetic properties [525; 531], while compressive (tensile) strain causes more (less) cation disorder [532].

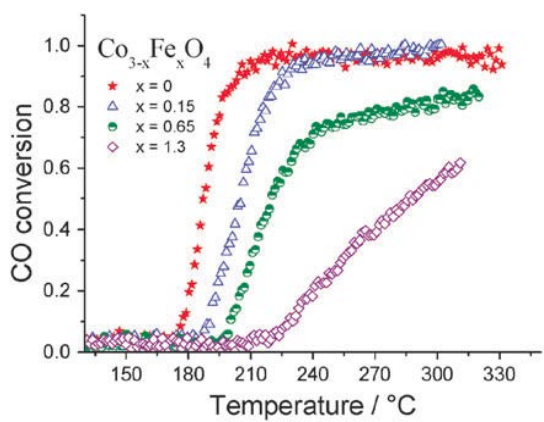

Figure 93: CO conversion as a function of temperature for polycrystalline Co$_{3-x}$Fe$_x$O$_4$ films grown on a glass substrate. Reproduced from ref. [528] with permission from the PCCP Owner Societies.

### 5.2.2 Nickel

Nickel adsorbs predominantly as Ni$_1$ adatoms on Fe$_3$O$_4$(100) at room temperature, but incorporates within the surface after annealing to 448 K [260]. DFT+U calculations find that the adatom binds to the surface with a binding energy of 3.21 eV, and that a further 0.18 eV is gained by incorporation in a subsurface Fe$_{oct}$ site within a bulk-terminated spinel-like structure. Again, this model suggests that the Ni fills one vacancy of the SCV structure, and that Fe$_{int}$ occupies the second (as depicted in Figure 90 for Co). The occupation of the octahedral sites is consistent with the site preference in NiFe$_2$O$_4$. Spectroscopic evidence and DFT+U calculations suggest the Ni is positively charged (1 electron is transferred to the surface) as an adatom, and is Ni$^{2+}$ in the incorporated geometry. The appearance of incorporated Ni was calculated using the Tersoff-Harman approach in ref. [260], and found to appear as an increase in the apparent height of the Fe$_{oct}$ row immediately above the location of the subsurface Ni (see Figure 94). For coverages in excess of 1 ML, Ni incorporation leads to the lifting of the ($\sqrt{2}\times\sqrt{2}$)R45° reconstruction, as discussed above for Co, which is further evidence that the Fe$_{int}$ of the SCV reconstruction fills the second vacancy. Heating above 448 K leads to the diffusion of Ni into the bulk of the sample, and the restoration of the SCV termination.

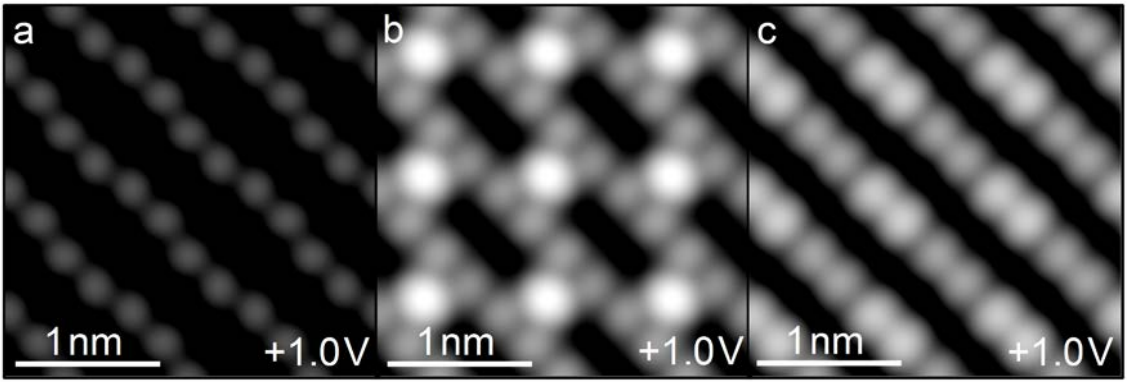

Figure 94: Simulated STM images for the clean $Fe_3O_4$(100) surface with SCV reconstruction (left), an adsorbed Ni adatom on the SCV reconstruction (centre), and following incorporation of Ni in the subsurface (S—2) $Fe_{oct}$-O layer of a bulk terminated spinel structure (right). Figure reproduced from ref. [260].

There are no further reports of Ni adsorption on $Fe_3O_4$ surfaces, but given the observed incorporation of Ni, it is interesting to consider theoretical works that have been carried out on $NiFe_2O_4$ surfaces to better understand how surface Ni atoms affect the reactivity. Selloni and co-workers [533] performed PBE+U calculations, and found the $NiFe_2O_4$ bulk to be a ferromagnetic inverse spinel with a band gap of 1.6 eV, in good agreement with experimental values. A surface phase diagram was calculated for $NiFe_2O_4$(100) in the presence of $O_2$ and $H_2O$ (see Figure 95), and the $NiFe_{oct}$-O plane was preferred. Interestingly, the formation of oxygen vacancies was determined to be much easier on the $NiFe_2O_4$(100) surface than in the bulk (0.34 eV vs. 2.8 eV), and easier than on a similarly terminated $Fe_3O_4$(100) surface. The authors suggest that O with two Ni neighbours is less strongly bound. In a humid environment, two water molecules dissociate in $V_O$s and a hydroxylated surface is favoured (labelled $P+2V_O+2H_2O$ in Figure 95). At elevated temperatures the water recombines leaving the surface with two $V_O$s ($P+2V_O$). This information suggests that a Ni-doped $Fe_3O_4$ surface may be more active for Mars-van Krevelen type reactions.

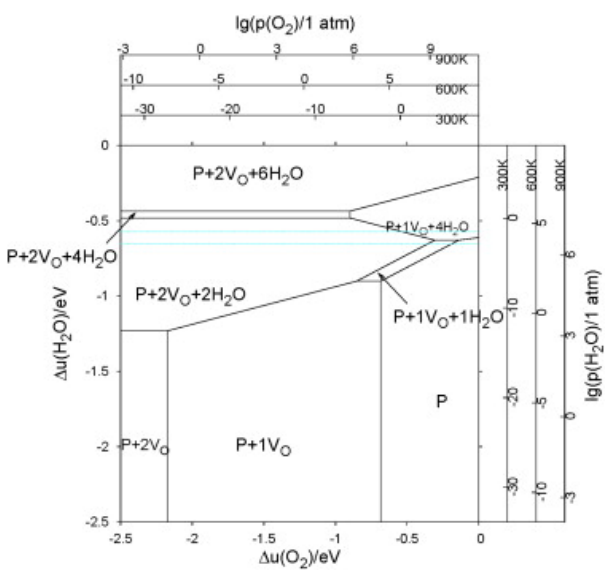

Figure 95. Surface phase diagram for NiFe$_2$O$_4$(001) in the presence of O$_2$ and H$_2$O calculated using PBE+*U*. Surfaces with V$_O$s are favoured, and these react with water to create adsorbed hydroxyls in a humid atmosphere. Reprinted from ref. [533], with permission from Elsevier.

Kumar et al. [534] performed DFT+U calculations of water adsorption on NiFe$_2$O$_4$(111), and found water dissociation to be more favourable than on the structurally equivalent terminations of Fe$_3$O$_4$(111). For example, 1.11 eV was calculated for a single molecule on NiFe$_2$O$_4$(111), which can be compared to the 0.95 eV on the Fe$_{tet1}$ termination [296] . An even bigger difference was found on the Fe$_{oct2}$ termination (terminated by Fe, although Ni could also be possible here as it occupies octahedral sites), 2.30 eV vs 1.33 eV [535]. A particularly interesting aspect of this paper is the discussion of transition-metal surface resonances (TSMR) [536] as a descriptor for the reactivity of a surface. The TMSR are obtained by plotting the difference in the PDOS for surface and bulk cations, and in Figure 96, the DFT+U calculated adsorption energy is plotted against the centre of mass of this difference plot. Clearly, as the TMSR shifts closer to the Fermi level the adsorption energy becomes significantly higher.

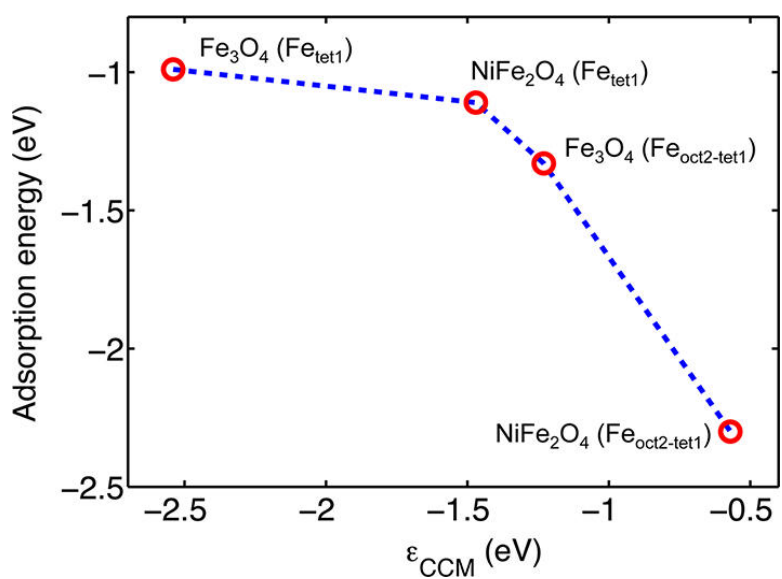

Figure 96. Adsorption energy plotted against the centre of mass of the transition-metal surface resonance. The closer the surface resonance to the Fermi level, the higher the adsorption energy. Reprinted with permission from ref. [534]. Copyright 2013 American Chemical Society.

### 5.2.3 Copper

Most studies of Cu adsorption on iron oxides are motivated to understand the promotional effect Cu on Fe$_3$O$_4$-based WGS catalysts [260; 302; 537-540]. Cu marks a significant departure from the other coinage metals (Au, Ag) because Cu forms a solid solution spinel with Fe$_3$O$_4$. The key question is if a mixed oxide phase (i.e. containing Cu$^{1+}$ or Cu$^{2+}$) is the active phase, or if metallic Cu catalyses the reaction. Rodriguez and co-workers [541] performed an in-situ study of the composition of CuO/CuFe$_2$O$_4$ catalysts and determined that Cu occupies octahedral sites in the spinel lattice only above 473 K. Surprisingly, they determined that Cu begins to leave the oxide in the presence of CO above 523 K to form metallic Cu and Fe$_3$O$_4$. Between 623 K and 723 K, a massive reduction of the oxide occurs. Since this structural change coincided with H$_2$ production in CO/H$_2$O mixtures, the authors conclude that $Cu^0$ (supported on $Fe_3O_4$) is the primary catalyst for WGS. Subsequent studies have come to similar conclusions [540].

Cu adsorption of $Fe_3O_4$ has been studied using theoretical calculations. Xue et al. [302] investigated the stability of adatoms and small clusters on $Fe_3O_4$(111) (tet1 termination). The most favourable adatom geometry was determined to be at the threefold hollow site binding to surface oxygen, while larger clusters were determined to lie flat and also interact with surface oxygen atoms. CO and $H_2O$ were found to interact more strongly with the Cu clusters than the $Fe_3O_4$ support. Interestingly, the interaction of CO and $H_2O$ with the surface $Fe_{tet1}$ atoms found to be very different for the clean surface 1.26 eV and 0.46 eV respectively, but in the presence of Cu clusters the binding energies come much closer (0.86 eV vs. 0.68 eV for $H_2O$ and CO respectively). The same group [542] also studied the effect of doping Cu atoms into $Fe_{tet1}$ sites of the $Fe_3O_4$(111) termination. In the presence of the Cu dopant the CO binds more strongly. The authors propose that the promotional effect of Cu could be linked to providing sites to adsorb CO, thereby preventing poisoning of the surface by adsorbed water (see Figure 97).

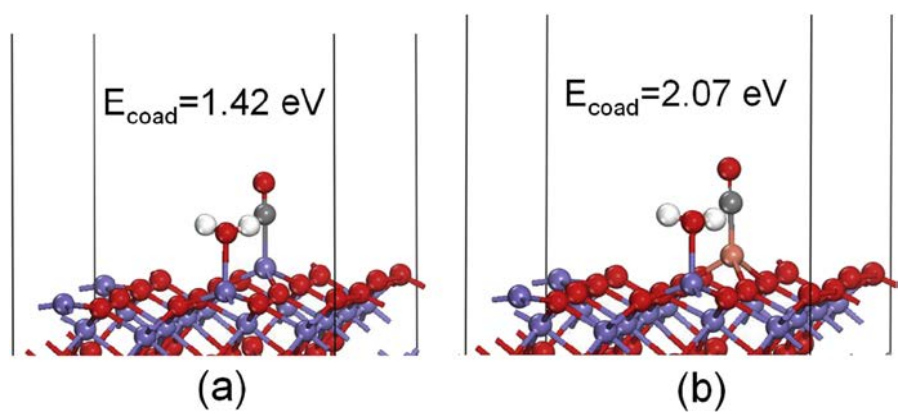

Figure 97: Co-adsorption configuration of $H_2O$ and CO on a pure (a) and Cu-doped (b) $Fe_3O_4$(111) surface. Here, CO binds much more strongly to the Cu dopant than to the $Fe_{tet1}$ cations. Reprinted from ref. [542], with permission from Elsevier.

### 5.2.4 Manganese, Titanium, Zirconium

These three metals are grouped together here as there are no surface-science studies on $Fe_3O_4$ other than the study of Bliem et al. [260], already mentioned in the Ni and Co sections above. The adsorption of Mn was found to be similar to Co (mixed $Mn_1$ adatoms and incorporation) in room temperature STM experiments [260], but no corresponding spectroscopic data nor DFT+U calculations were presented. In $MnFe_2O_4$, $Mn^{2+}$ cations occupy octahedral sites, much like $Co^{2+}$ in $CoFe_2O_4$. Studies on $MnFe_2O_4$ nanomaterial suggest that catalytic activity for the Fenton reaction (where Fe dissociates hydrogen peroxide forming free radicals [543-545]) and heavy metal sorption are enhanced by Mn incorporation [524; 546; 547].

Ti and Zr fill the vacancies in the SCV reconstruction immediately on deposition at room temperature (see Figure 98), with little occupation of the adatom site. DFT+U calculations find the adatom site to be unstable. The incorporation in the subsurface octahedral site as $Ti^{4+}$ was found to be the energetically most favourable option with a binding energy of 8.29 eV, which is high because it is referenced against an isolated Ti atom. Spectroscopy data agree with the formation of $Ti^{4+}$ and indicate reduction of Fe to $Fe^{2+}$ in the surface region. $Ti^{4+}$ is known to occupy octahedral sites in solid solution with $Fe_3O_4$ [548], but there are no reports of a corresponding $ZrFe_2O_4$ bulk phase.

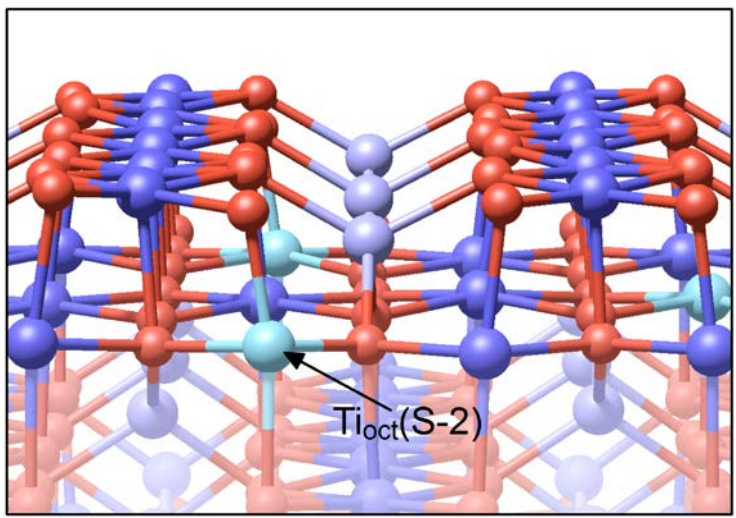

Figure 98: Ti occupies subsurface octahedral sites within a bulk terminated spinel structure following deposition on the SCV terminated $Fe_3O_4$(100) surface at room temperature [260]. In the model $Ti^{4+}$ atoms are cyan, $Fe_{oct}$ are dark blue, $Fe_{tet}$ are light blue, and oxygen atoms red. Figure adapted from ref. [260].

Recently, $TiFe_2O_4$(100) thin films have been grown epitaxially on $Fe_3O_4$(100) by PLD [549; 550]. Droubay et al. [549] performed a systematic study of thin films ranging in stoichiometry from $Fe_3O_4$(100) to $TiFe_2O_4$(100), and found that only pure $Fe_3O_4$(100) exhibited a ($\sqrt{2}\times\sqrt{2}$)R45° reconstruction in RHEED, and that all Ti-containing films exhibit a bulk-terminated (and therefore polar) spinel surface. XPS was performed after a transfer through air and the surface was found to be oxidised, but significant $Fe^{2+}$ signal was recovered after heating to 623 K in UHV. The authors propose that Ti incorporation is linked to an increased number of $Fe_{oct}$ vacancies near the surface under oxidising conditions, citing the work of Dieckmann [127], who showed such defects are predominant in titanomagnetites.

### 5.2.5 Vanadium

The adsorption and oxidation of V on α-$Fe_2O_3$(0001) has been systematically studied in both experiment [356; 551] and theory [342; 552] by Bedzyk and co-workers, motivated to study supported $V_2O_5$ model catalysts. XPS studies [356] found that V adsorbed as $V^{3+}$ up to 2/3 ML coverage, with a concomitant reduction of surface Fe to $Fe^{2+}$. DFT calculations suggest this $V^{3+}$ is adsorbed at threefold hollow sites with a binding energy of 6.2 eV [342]. On the $O_3$-Fe-Fe-termination, the most stable position is in a bulk continuation site, with a binding energy of 9.9 eV. For higher V coverages, only metallic V was observed in XPS. Exposure of the V films to atomic oxygen resulted in the formation of $V_2O_5$, and the reoxidation of the surface Fe [342]. In subsequent XSW studies [551], the authors demonstrated a clear link between the oxidation state of the V, and its position with respect to the iron-oxide substrate. Hydration (through exposure to air) also oxidises the V to $V^{5+}$, although it can be reduced again by annealing at 773 K in $H_2$ [553]. Essentially, this body of work clearly demonstrates that the reduction and oxidation of the V is directly related to the structure of the model catalyst, and that the process is fully reversible depending on the environment (see Figure 99).

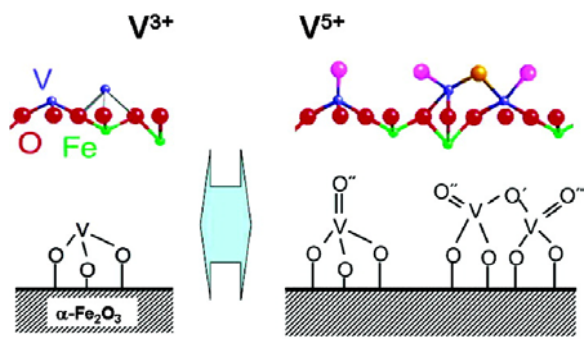

FIGURE 99: Schematic showing the oxidation of V on the α-Fe$_2$O$_3$(0001) surface to form V$_2$O$_5$. Here Fe atoms are green, oxygen red, vanadium blue and adsorbed oxygen pink. Notice that V initially occupies bulk-continuation site of the corundum structure, and that the subsequent reaction with oxygen is site-specific. Reprinted with permission from ref. [551]. Copyright 2007 American Chemical Society.

### 5.2.6 Tungsten (W)

The adsorption and oxidation of W on α-Fe$_2$O$_3$(0001) has been studied by Bedzyk and co-workers [554; 555] utilizing similar approach and methods to that used for their V studies described in the previous section. Again, a natural single crystal was annealed in O$_2$, but here 0.3 ML W was deposited by ALD. Using XSW and XPS, the authors demonstrate that the W cation changes location with respect to the substrate as it changes from the 6+ to 5+ oxidation state, and propose that redox-induced cation dynamics are explained by models that account for W incorporation at the interface in Fe sites. Interestingly, while W clearly prefers a threefold coordinated site at the surface, whether this site is the bulk continuation "A" site or a hollow site above a subsurface Fe cation ("B" site) depends on the oxygen adsorption. A preference for occupation of the A site (0.8 eV) only occurs with coadsorption of one oxygen atom (which makes the W atom fourfold coordinated), while both zero and three oxygen atoms cause a switch to the B site. Much like the studies of V, these results illustrate that the physical and electronic structure of a mixed oxide interface go hand in hand and these are directly related to the environment.

### 5.2.7 Chromium

Given that Cr is the main promoter is in the industrial iron oxide WGS catalyst it is surprising there have been no direct studies of Cr adsorption on iron-oxide surfaces. Henderson studied Cr(CO)$_6$ species adsorbed on α-Fe$_2$O$_3$(0001), and no thermal decomposition was observed. However, exposure to an oxygen plasma led to the formation of a disordered chromium oxide, which diffused into the bulk above 700 K [426].

The bulk [556; 557] [558] and surface [426; 559; 560] properties of 500 Å thick films of (Fe$_{1-x}$Cr$_x$)$_2$O$_3$ thin films deposited on α-Al$_2$O$_3$(0001) substrates by oxygen-plasma-assisted molecular beam epitaxy have been studied recently with a view to engineering the optical band gap for photocatalysis. Cr is found to reduce the optical band gap of α-Fe$_2$O$_3$, which reaches a minimum as the Cr cation fraction increases to 50%. The lowest-energy transitions in the ternary oxide system involve electron excitation from occupied Cr 3*d* orbitals to unoccupied Fe 3*d* orbitals, and they result in a measurable photocurrent. Excitingly, the onset of α-$Fe_2O_3$ photoconductivity can be reduced by nearly 0.5 eV (to 1.60 eV) through addition of Cr.

At the surface, Henderson and Engelhard [559] have shown that light $Ar^+$ sputtering and annealing the ternary oxide film leads to a $Fe_3O_4$(111)-like termination exhibiting (2×2) spots in LEED. Using SIMS, the Fe:Cr ratio in the surface was found to be ~1.9:1, which is less than the 3:1 ratio in the bulk of the as-grown film. XPS spectra suggest that the Cr segregates to the surface as the result of annealing, and further showed a significant proportion of $Fe^{2+}$, consistent with the supposed surface reduction. TPD was performed for the probe molecules $O_2$, $CO_2$, $H_2O$, and NO (Figure 100). In general, the strongly bound sites were attributed to $Fe^{2+}$ and $Fe^{3+}$ sites, with little evidence of binding at $Cr^{3+}$ sites. Water TPD resembles the results acquired from $Fe_3O_4$(111). Through coadsorption studies and an exhaustive comparison to the available literature, it was possible to tentatively assign several of the observed TPD peaks to the sites shown in Table 8. The assignments were hampered by a distinct lack of TPD studies for similar molecules on clean iron oxide surfaces.

Table 8: TPD states associated with each probe molecule with suspected surface sites on (Fe,Cr)$_3$O$_4$(111). Values in parentheses are minor peaks, and the question marks indicate uncertainty in assignment. TPD data shown in Figure 97. Table reproduced from ref. [559].

| adsorbate | $Fe^{2+}$ sites | $Fe^{3+}$ sites | $Cr^{3+}$ sites |
|---|---|---|---|
| $O_2$ | 230 K | (100 K) | ? |
| $CO_2$ | (185 K) | 275 K | ? |
| NO | 370 K | 225 K | ? |
| $H_2O$ | ? | ? | ? |

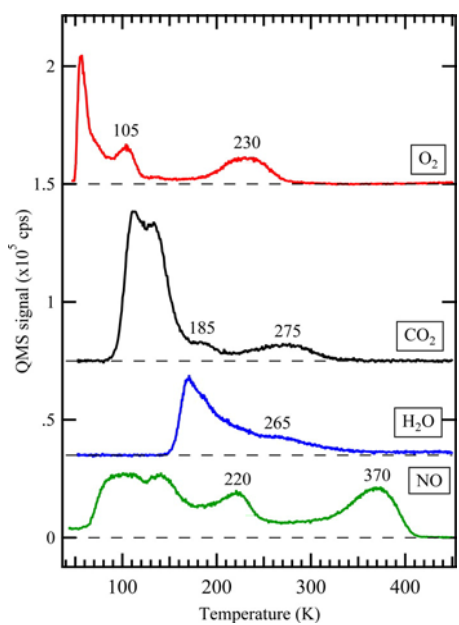

Figure 100: TPD spectra for saturation coverage of $O_2$, $CO_2$, $H_2O$, and NO on the mixed $Cr_{1-x}Fe_{2+x}O_4$(111) termination of a $(Fe_{1-x}Cr_x)_2O_3$ thin film. Peak assignments are given in Table 8. Figure

reproduced from. Reprinted with permission from ref. [559]. Copyright 2014 American Chemical Society.

Henderson [561] also studied NO oxidation on the same surface, again using TPD. Unfortunately, irradiation of adsorbed species with 460 nm light resulted predominantly in photodesorption, rather than photooxidation, which is somewhat disappointing as it limits the extent of possible surface photoreactions. In the absence of photons, no thermal decomposition was detected for adsorbed NO, whereas ~10% of the adsorbed $O_2$ dissociated at $Fe^{2+}$ sites. NO was found to react with preadsorbed $O_2$ to produce surface nitrate, which subsequently decomposed in TPD at 425 K. $Fe^{3+}$ and $Cr^{3+}$ sites were concluded not to participate in NO oxidation. Photoirradiation of $O_2$ adsorbed on the reduced $Cr_{1-x}Fe_{2+x}O_4(111)$ surface found that $O_2$ molecules in the 230 K state preferentially photodesorbed irrespective of the wavelength of light employed [560]. Approximately 10% of adsorbed $O_2$ irreversibly photodissociated, irrespective of wavelength, with the resulting fragments blocking access to both $Fe^{3+}$ and $Fe^{2+}$ sites for subsequent $O_2$ adsorption.

## 5.3   Alkali Metals (Li, Na, K, Rb and Cs), Ca, and Mg

There are several surface science studies of alkali metals at $Fe_3O_4$ surfaces, primarily because alkalis are well known promoters for catalytic reactions. Additionally, alkalis are a common contaminant that segregates during the preparation of natural single crystals [215; 216] and Mg has been observed to diffuse from the MgO substrate through an $Fe_3O_4(100)$ thin film prepared above 670 K [223; 562].

The group of Igor Shvets studied the segregation of K and Ca from natural single crystals [215; 216] and found a series of ordered reconstructions (1×1, 1×2, 1×3 and 1×4) to occur as a function of annealing time at 990 K. STM images of the (1×2) reconstruction (see Figure 98) reveal bright protrusions on the $Fe_{oct}$ row, which probably results from incorporation of Ca in the $Fe_{oct}$ sites. When the sample was annealed for very long times a completely new surface structure was obtained that exhibited a (1×4) periodicity. Strangely, the images bear a striking resemblance to published images of the $Fe_3O_4(110)$ surface (see Figure 49). From AES data, the authors determined that the O/Fe ratio is decreased at the (1×4), in addition to the presence of K and Ca contamination, and proposed that a $Ca_{1-x}Fe_{2+x}O_4$ surface phase is formed in which Ca atoms occupy interstitial sites in the $Fe_{oct}$ lattice (see Figure 101).

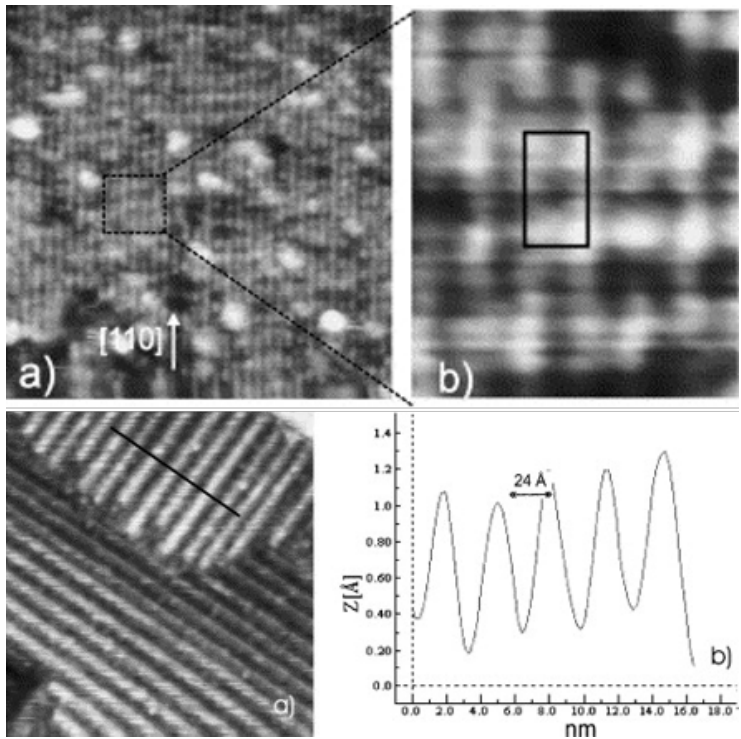

Figure 101: (Top) STM images (200x200 nm$^2$ (left) and 52x48 nm$^2$ (right)) of a Ca-induced (1×2) reconstruction at the (100) surface of a natural Fe$_3$O$_4$ single crystal. The Fe$_{oct}$ rows are clearly visible, with a 6×12 Å superstructure formed by bright protrusions centred on the Fe$_{oct}$ row. (Bottom) STM image of the (1x4) reconstruction obtained after extensive UHV annealing (left), together with a line profile showing the corrugation of the surface (right). Reprinted from ref. [215], with permission from Elsevier.

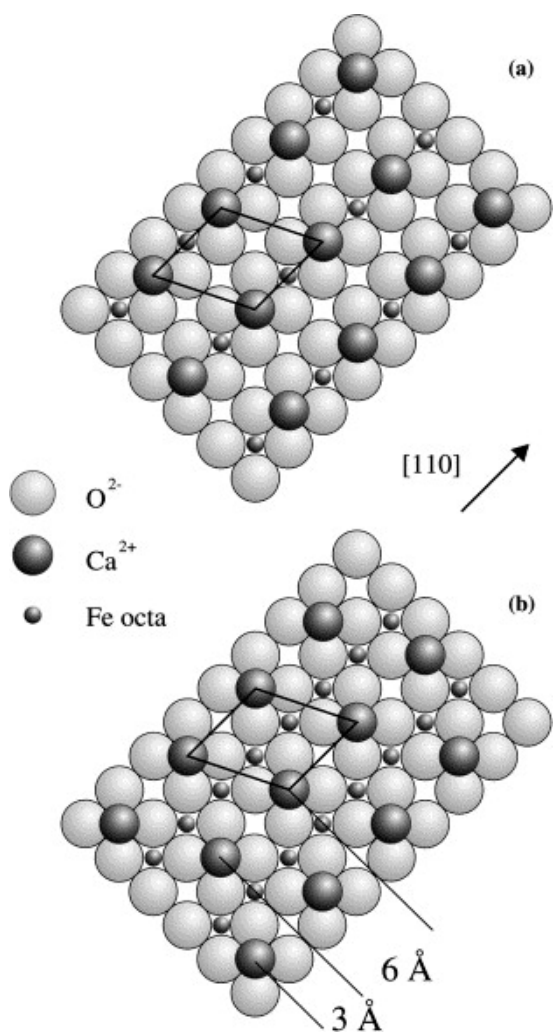

Figure 102: Model for the Fe$_3$O$_4$(100)-(1x4) structure proposed by Shvets and co-workers based on atomically resolved STM images. In (a), Ca$^{2+}$ cations occupy vacant sites in the Fe$_{oct}$ row, while in B they occupy interstitial sites between the rows. Reprinted from ref. [216], with permission from Elsevier.

Anderson et al. [228] found that annealing μm thick films of Fe$_3$O$_4$(100) grown on MgO at 800 K resulted in a surface containing Mg. They recognised that Mg can exist in a solid solution in the Fe$_3$O$_4$(100) lattice, and noted that the Mg-segregated surface exhibited a (1×1) LEED pattern. This result is interesting because it suggests the vacancies in the SCV reconstruction of Fe$_3$O$_4$(100) can be filled from below, not just when metal is evaporated to the surface, and that the SCV reconstruction is not the most favourable reconstruction for Mg-doped Fe$_3$O$_4$(100) samples. In the other examples of lifting discussed above (Ni, Co, Mn, Ti), the SCV reconstruction was always recovered by high temperature annealing. When the sample was annealed in O$_2$, a surface with stoichiometry close to MgFe$_2$O$_4$ emerged that exhibited a (1×4) reconstruction in LEED, and extended rows separated by 24 Å in STM images. The authors proposed a model for such a reconstruction based on an Fe$_{tet}$ terminated surface with every 4$^{th}$ row removed in the [110] direction. It is important to note that the images resemble the (1×4) reconstruction observed by Shvets and co-workers on a natural single crystal (Figure 101), so it seems that all such metals can induce the same reconstruction.

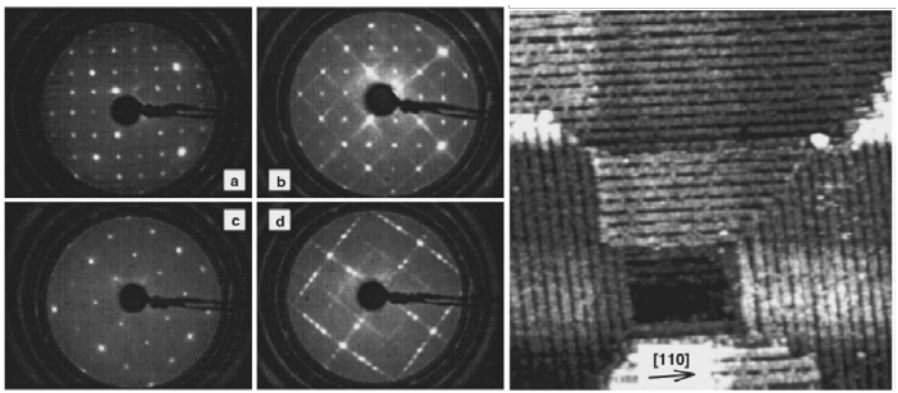

Figure 103: LEED patterns (64 eV) obtained for $Fe_3O_4$(100) thin films grown on MgO(100). (a) A ($\sqrt{2}\times\sqrt{2}$)R45° pattern is obtained after annealing at 660 K in $2\times10^{-6}$ mbar $O_2$, (b) and annealing at 820 K, (c) but prolonged annealing at 880 K results in a (1×1) periodicity. (d) Annealing in oxygen at 890 K leads to a (1×4) reconstruction. (right) STM image of the (1×4) reconstructed phase. Reprinted figure with permission from ref. [228]. Copyright 2005 by the American Physical Society.

Gao et al. [223] discussed the possible diffusion mechanisms for $Mg^{2+}$ within $Fe_3O_4$ and concluded that a vacancy diffusion must dominate over interstitial diffusion. This requires that the films have many $Fe_{oct}$ vacancies, either through non-stoichiometry, or through Frenkel pairs. Since Mg diffusion is essentially controlled by Fe diffusion in such a mechanism, the authors used Fe diffusion data from ref. [563] (diffusion coefficient between $10^{-14}$ and $10^{-12}$ $cm^{-2}$/s at 773 K) to calculate that Mg can traverse a 500 Å film in anywhere between 600 and 6 seconds.

The deposition of K on $Fe_3O_4$(111) thin films has been studied by Shaikhutdinov et al [564] and Joseph et al. [565], in part motivated to shed light on the results of Muhler et al [462; 463] regarding the active phase of the K-promoted iron oxide catalyst used for ethylbenzene dehydrogenation. In the former study, increasing amounts of K were deposited and (4×4), (2×2) and (1×1) structures observed after annealing at 6-700 °C in $10^{-6}$ mbar of $O_2$. Although some K desorbed, K was detected at least 10 layers into the sample by AES following sputtering of the surface. On the basis of the K content derived from AES, formation of a $K_2O/K_2Fe_{22}O_{34}/Fe_3O_4$ interface with K content decreasing from the surface inwards was proposed. Joseph et al performed thermodynamic calculations and XPS and TPD experiments following K deposition at 200 K. The evolution of surface phases was studied with sequential annealing, and a well ordered $KFe_xO_y$ surface was found after annealing at 970 K. Although some K was lost to the substrate, the surface was proposed to consist of 4 K atoms per (2×2) unit cell.

A systematic study of the interaction of alkali metals with the $Fe_3O_4$(111) surface was performed using DFT by Yang et at. [566]. All metals bind preferentially at the same threefold hollow site (coordinated to surface oxygen) donating charge to the surface. However, very low barriers were found for diffusion between threefold hollow sites, so diffusion is expected to be facile at room temperature. The calculated adsorption energy was found to decrease from Li through K, Rb, Cs and Na.

## 5.4  Group IV Elements (C, Si, Ge, Sn)

The adsorption of C, Si, Ge and Sn on Fe$_3$O$_4$(100) has been studied by DFT+U calculations at a coverage of 4 atoms per unit cell. Since the calculations were performed for the DBT structure, there are 4 O$_{surface}$ atoms without a subsurface Fe$_{tet}$ neighbour per unit cell. The considered elements bind to these O atoms (those with a subsurface Fe$_{tet}$ neighbour were not favoured). The adsorption energies E$_{ads}$, tilt angle away from the surface normal Θ, and bond lengths d$_{M-O}$ of these atoms is listed in Table 8. Interestingly, the C atom has a bond length similar to that of C=O, binds very strongly to the surface, and is predicted to restore half-metallicity to the surface. This does not occur for Si, Ge and Sn.

Table 8: Adsorption energies E$_{ads}$, tilt angle away from the surface normal Θ, and bond lengths d$_{M-O}$ for H, C, Si, Ge and Sn atoms adsorbed on the Fe3O4(100) surface as calculated by DFT+*U* calculations.

|  | H | C | Si | Ge | Sn |
| --- | --- | --- | --- | --- | --- |
| E$_{ads}$ (eV) | 3.096 | 5.509 | 4.575 | 3.629 | 3.144 |
| d$_{M-O}$ (Å) | 0.970 | 1.210 | 1.648 | 1.851 | 2.078 |
| Θ (°) | 56.3 | 18.1 | 1.4 | 1.3 | 3.3 |

Since there was no experimental evidence for the adsorption geometry of such atoms this author's group deposited Sn on the Fe$_3$O$_4$(100) surface at low coverage at room temperature and scanned the resulting surface with STM [567]. We find that Sn adsorbs in the "not blocked" site of the SCV reconstruction as Sn$_1$ adatoms, somewhat unsurprisingly given the experience with other metals. There appears to be little evidence for incorporation in the surface at room temperature, but SnFe$_2$O$_4$ is a well-known inverse spinel with Sn$^{4+}$ occupying octahedral sites [523] so Sn most likely diffuses into the bulk at higher temperature. This is not so surprising given the enthalpy of formation of SnO$_2$ is similar to that of Fe, Ni and Co.

## 5.5 Boron (B)

The adsorption of Boron was recently studied at the Fe$_3$O$_4$(100) surface via DFT and DFT+U calculations [94], and is the latest adsorbate of a series of adsorbates that Pratt and co-workers [90-94; 99] predict to recover half-metallicity at the surface. The calculations suggest that B binds strongly to O atoms without a subsurface Fe$_{tet}$ neighbour (B.E. = -5.76 eV), adsorbing with a tilt angle of 59.7° from the surface normal in the direction of the bulk continuation Fe$_{tet}$ site. This work is notable because it is the first of these studies to consider the SCV structure, and the authors report that the adsorption is not significantly affected (other than there is half the amount of O without subsurface Fe$_{tet}$ neighbour). However, an adsorption mode akin to that observed for Ni, Co, Mn, Ti, and Zr [260], i.e. adatom adsorption at the bulk continuation site and incorporation in subsurface vacancies was not considered.

## 6. Summary - What has been learned?

The iron oxides are a fascinating, but complex class of materials to study using the surface science approach. The importance of these materials in the environment and their widespread use in technology makes them exciting candidates for extensive study, and the progress to date suggests that the systems, and the differences between them can ultimately be unravelled.

The ultimate aim is of course to determine the geometric and electronic structure of a given surface, to understand why it forms, and then correlate this information with the surfaces' interaction with the gas phase, liquids and the interfaces to other solids. The first and most crucial aspect of this jigsaw puzzle, determination of the surface structure, has been achieved for $Fe_3O_4(100)$, and for the most part, for $Fe_3O_4(111)$. In the former case, one termination dominates over the range of chemical potential utilized in UHV, and has been observed by several research groups employing different preparation methods. If exceedingly reducing conditions are used in preparation (many sputter/ UHV anneal cycles) or thin film growth on an Fe surface (which leads to an Fe-rich film), Fe-rich terminations can also be stabilized. In either case however, simply annealing the sample at 900 K in $10^{-7}$-$10^{-5}$ mbar $O_2$ for a short time restores the $Fe_{oct}$-O termination [35; 68; 211; 212; 227]. All the available evidence suggests the $(√2×√2)R45°$ surface is terminated by the SCV structure [35], and through LEED *IV*, SXRD and DFT+U calculations, the location of the atoms are known with high precision. The history of this surface clearly shows us that selecting candidate structures based on auto-compensation and nominal bulk charges is too restrictive. Indeed, an analysis of the electronic structure to emerge from DFT+U calculations suggests that polarity compensation is finally achieved by two O atoms per unit cell taking a -1 oxidation state, which was not (and probably could not have been) guessed a-priori. One of the aims in writing this review was to provide reliable reference data for the widespread techniques STM, LEED, and XPS, such that the different terminations can be easily recognised, and to provide confidence in prepared surfaces to be used as the basis for exciting experiments.

In the case of $Fe_3O_4(111)$, the surface is understood with a similar level of detail, but there are continuing issues with preparation because two terminations ($Fe_{tet1}$ and $Fe_{oct2}$) exist close in energy in the relevant chemical potential regime. Thus, until a method is developed to prepare one, or both, of these terminations in monophase, care must be taken to determine which terminations are present, and in which proportion. The $Fe_3O_4(110)$ surface has been shown to exhibit a 1-dimensional reconstruction following sputter/anneal cycles in UHV, but as yet theoretical calculations have only assessed the stability of bulk-like terminations.

The α-$Fe_2O_3(0001)$ surface appears a bit of a minefield at present, because several different terminations have been observed and there is little in the way of clear trends or a clear method to reproducibly prepare a monophase termination other than the $Fe_3O_4(111)$ termination, which is prepared in reducing conditions. This surface appears to be identical to that described above (note, the $Fe_{tet1}$ vs. $Fe_{oct2}$ issue most likely still applies). At least 3 terminations of an α-$Fe_2O_3(0001)$-like nature have evidence in their favour (half-metal terminated, oxygen terminated and Ferryl terminated). The bi-phase surface, which exists between α-$Fe_2O_3(0001)$ and $Fe_3O_4(111)$ terminations has become somewhat controversial because the previously (seemingly widely accepted) model of $Fe_{1-x}O$/α-$Fe_2O_3(0001)$ islands has been challenged by a simple, intuitive explanation i.e., that the outermost two layers are reduced to an $Fe_3O_4(111)$-like structure, while the underlying material remains α-$Fe_2O_3(0001)$. An unambiguous determination of the structure requires modelling of a particularly long-range structure.

In this author's opinion, the most important lesson learned in the study of iron oxide surface to date is that the surface behaviour has its roots in the defect chemistry of the bulk compounds, and the ease with which the structures can be interchanged. Iron oxides maintain a close packed anion lattice over the entire range from $Fe_{1-x}O$ to $Fe_2O_3$, and can be highly defective close to the phase boundaries. Consequently, it is not surprising that these surfaces, where reduction and oxidation ultimately take place, are based on these different phases, and that typical defects are related to excess or missing cations. Similar behaviour can be expected from oxides with similar bulk properties, for example those of Ni, Mn, and Co.

# 7 Future Directions

## 7.1 Structure Determination

As discussed throughout this review, knowledge of the structure of a surface is fundamental to a successful surface science investigation. The emergence of nc-AFM to rival (and even surpass) the resolution offered by STM offers an exciting new way to study metal oxide surfaces [568; 569], particularly insulating single crystals. Barth and Reichling [570] imaged the α-Fe2O3(0001) surface already in 2001 Lauritsen and coworkers [571] have pioneered work in this area for spinel oxides with studies of $MgAl_2O_4$(100) surface, which they propose to be stabilized by anti-site defects [572]. The first STM/AFM images of the $Fe_3O_4$(100) surface shown in Figure 29 reveal the large undulations observed in STM to be predominantly of electronic origin. The well-known rutile $TiO_2$(110) surface has also attracted much attention as a model system to test the capability of nc-AFM on metal oxides [573-576]. It has been shown that both the surface Ti and O sublattices can be imaged depending on the tip termination, and the common defects identified. However, the complexity of interactions involved in nc-AFM make it challenging to simulate the images theoretically. The simultaneous acquisition of STM and AFM images is helpful in this regard [577] because known features can be identified in STM, allowing rapid assignment in AFM. This provides a solid basis to develop the image simulation methods. Looking to new materials, utilizing both STM and AFM channels should allow quicker and more reliable exclusion of candidate models and easier identification of intrinsic surface defects.

Ultimately though, scanning probe studies are useful to develop models for surface structure, but are just part of the story. It is crucial that quantitative structural methods are used to test the models and determine the structure of surfaces precisely. Previously it was assumed that LEED *IV* R-factors might be intrinsically limited by the ionic nature of the bonding in metal oxides, or by an inevitably high defect concentration, but recent LEED *IV* studies achieved excellent agreement ($R_P \approx 0.12$) even for complex metal oxide surfaces such as $Fe_3O_4$(100) [35], $Co_3O_4$(111) [578], and $V_2O_3$(0001) [353; 354]. Crucially, in each case the measured surface was imaged both before and after the LEED experiment to ensure the *IV* curves were acquired from a homogeneous clean surface, and that the electrons did not damage the surface. Going forward, the primary explanation for poor LEED *IV* R-factors (i.e. $R_P \approx 0.3$ and above) must be that the surface was poorly prepared, or that the structural model is incorrect.

The insulating nature of the α-$Fe_2O_3$(0001) surface is problematic for LEED *IV*, but not a problem for SXRD. Consequently, if a recipe to produce a monophase termination can be developed, either using high $O_2$ pressures or through a wet chemical approach, there is every chance that the structure can be determined to a high degree of precision. Perhaps even greater opportunities lie in wait with the α-$Fe_2O_3$(012) surface, the most common surface on nano-hematite, and reported to be easy to prepare under UHV conditions [366]. Again, further progress with this surface requires confirmation of the structural model, which likely requires SPM studies followed by quantitative structural determination.

## 7.2 Relation to Other Oxides

A final question regarding structure is whether the terminations and reconstructions observed for $Fe_3O_4$ are relevant for other metal oxides. In the introduction to this review, it was pointed out that the bulk structure and defect chemistry of iron oxides is akin to those of Co, Mn, Ni. While there is little in the way of surface science studies, there are a few studies to which comparison can be made. In particular, the $Co_3O_4$(111) surface has been shown to exhibit a nominally polar $Co_{tet1}$ termination stabilized through strong relaxations (as proposed for $Fe_3O_4$(111)). This study, performed on 40 Å thick films on Ir(100), achieved a LEED *IV* R-factor of 0.124 [578]. It would be fascinating to see if the $Co_3O_4$(100) surface exhibits a similar SCV reconstruction to $Fe_3O_4$(100). Given the similarity in the lattice parameter of the two compounds, it might be possible to grow the insulating $Co_3O_4$(100) film epitaxially on $Fe_3O_4$(100) and take advantage of the conductivity of $Fe_3O_4$. Another point of comparison are the extensive studies of ultrathin Ni, Co, and Mn oxide films performed by the Netzer group (e.g. [579-582]), among others. These oxides, although not representative of bulk compounds, are in keeping with the ideas developed here because many of the structures are based on a close-packed O-lattice in which different cation arrangements arrange differently depending on the stoichiometry.

## 7.3 Mixed/Doped Oxides

A clear theme of this review is that many metals from the periodic table readily incorporate within the spinel or corundum lattice forming mixed oxides with Fe. There is little work on ternary oxides in the metal-oxide surface science literature, save for some work on perovskites, particularly $SrTiO_3$ (e.g. [583-586]. As mentioned above, a few studies have been performed on the $MgAl_2O_4$. Well-ordered (and well characterised) ternary oxide surface phases have already been observed for 3*d* transition metals (Ni, Co, Mn, Ti [260]) at the $Fe_3O_4$(100) surface, and alkalis (K, Ca, Mg…) that segregate also from ordered structures. Interestingly, even metals with no known bulk ferrite phase, such as Zr [260], can incorporate in the vacancies of the SCV structure. This clearly gives an opportunity to study the effect of doping on the properties of $Fe_3O_4$ surfaces. At present there is a clear desire to better understand how metal oxides are modified by doping, for example, Nilius and Freund [587] have already performed pioneering work showing how Mo dopants donate electrons to MgO and CaO, which can be transferred to adsorbed clusters and molecules. Urgent areas of study in regard to iron oxides are to develop a Cr-free $Fe_3O_4$-based WGS catalyst, or better understand how doping α-$Fe_2O_3$(0001) can reduce the overpotential required for PEC water splitting [167; 556; 559; 561; 588-590].

More generally, the huge range of metal combinations that take the spinel structure combined with their high stability has seen a resurgence in tailoring these compounds for energy related applications. The Centre for Inverse Design, an energy frontier research centre funded by the US DOE, in particular, has focussed strongly on this topic and identified several interesting ternary spinel compounds as candidate materials for p-type transparent conducting oxides. Other than $Fe_3O_4$, very little is known about the surface properties of any spinel compound. Given the lack of availability of such compounds as single crystals, and the similarity of the anion lattice amongst spinels, it may be interesting to grow and study new materials as thin films utilizing $Fe_3O_4$ as a lattice matched (and conductive) substrate.

## 7.4 Single Atom Catalysis

The emerging field of single atom catalysis [481; 482; 488; 591] is based on the exciting, but controversial [592] idea that chemical reactions can be catalysed by single atoms supported on an inexpensive metal oxide. A catalyst based on single-atom active sites promises efficient usage of expensive precious metals. At present, despite the vast improvements in aberration corrected TEM, the field is hampered somewhat by an inability to accurately characterise a catalyst based on single atoms. Knowledge of the environment of the active site is crucial to interpret experimental data, and to provide a basis for DFT-based calculations to model the reactivity. Surface science has traditionally provided the mechanistic insights to understand heterogeneous catalysis, but systems in which single atoms remain stable on a well-characterised surface at reaction temperatures are rare. In this review, it is demonstrated that the $Fe_3O_4$(100) surface stabilizes Au, Pt, Pd and Ag adatoms [35; 259; 261; 262; 402] as high as 700 K, which makes it an excellent candidate system with which to shed light on the fundamental mechanisms underlying an exciting and rapidly growing area.

## 7.5　Hydroxylated Surfaces and Realistic Atmospheres

The so-called pressure gap is one of the most well-known issues in surface science, and restricts the ability to confidently transfer the lessons learned in the vacuum chamber out into the real world. In this review the vast majority of adsorption studies have focussed on clean iron oxide surfaces in ultrahigh vacuum. Ultimately, this level of control is vital for the reliable interpretation of experimental data. However, one has to wonder if the results bear any relation to the surface in a particular application, which mostly occur in ambient or even liquid environments. The trend to develop "high-pressure" variants of surface science techniques is allowing experiments to move toward the application, whilst maintaining a handle on the number of variables. The first high pressure experiments performed on iron oxide surfaces have focussed exclusively on the interaction with water [67; 413]. The results suggest that complete hydroxylation will occur when a clean sample is removed from the vacuum chamber. Indeed it seems likely that a mixed mode layer of OH and water is formed in many cases. Whereas LEED cannot be performed in a high pressure environment, structure determination in a high pressure environment can be performed using SXRD, for example at the ID03 surface diffraction beamline at the ESRF. However, in such experiments trace contaminants from the gas bottle and/or desorption of species from the chamber walls can lead to unintended coadsorption. Kendelewicz and co-workers [67], for example, report that carbon contamination occurs with water exposure already for XPS studies performed in the mbar range.

Clearly the properties of the hydroxylated iron oxide surfaces are at least as interesting as clean iron oxide surfaces. Fortunately, partially and fully hydroxylated surfaces can be formed relatively easily in UHV by cracking $H_2$ molecules onto the surface, and can be detected in O1$s$ XPS spectra. Where the presence of OH reduces surface Fe to $Fe^{2+}$, it can be even detected in STM. This is true on $Fe_3O_4$(100), where the OH structure is known from DFT calculations. It would be interesting to study the adsorption of atomic H on $Fe_3O_4$(111) with STM, and determine if OH species can be imaged there. Ideally, with a well characterized hydroxylated surface one could study how OH groups modify the adsorption of molecules and metals. There are already interesting hints at strong interactions. For example, mobile Pd-CO species become immobilized on encountering a surface OH group on $Fe_3O_4$(001) at room temperature. This suggests that Pd deposited on a pre-hydroxylated surface could be impervious to CO induced agglomeration into nanoparticles. However, the structure of the Pd-H species, and whether CO remains bound to Pd, are presently unknown.

More generally, there is much evidence from $TiO_2$(110) that OH groups can modify the properties of the surfaces [436; 593], and photocatalytic methods have recently been developed to completely hydroxylate the surface [594]. For example, it is known that OH groups donate electrons to the oxide lattice in much the same way as $V_O$s [595], and that this affects the binding strength of coadsorbed atoms and molecules, and can have a direct impact on surface chemistry [596]. Moreover, it has been shown that the nucleation of Au clusters occurs homogeneously on the terraces on a $TiO_2$(110) surface with $V_O$s, but preferentially occurs at step edges in the presence of OH [597].

Very little is known about the atomic-scale structure of iron-oxide surfaces in outside the vacuum chamber. However, Sterrer and coworkers have been performing pioneering work in this area [598-601], and have published STM images of an FeO(111) thin film on Pt(111) in air, in liquid water, and following water evaporation (Figure 104). Using a combination of spectroscopies and theoretical calculations it was shown that the FeO(111) surface is essentially inert in pure water up to millibar pressures, but that water/$O_2$ mixtures lead to significant restructuring of the film and concomitant hydroxylation. In other work, the formation of Pd nanoparticles from solution was studied using $Fe_3O_4$(111) thin film samples transferred out of UHV and contacted with $PdCl_2$ containing precursor solution. After removal of the solution the single crystal surface was washed with ultra-pure water, and dried at room temperature under a He flow. It was concluded that differences in the Pd nucleation behaviour observed between UHV and solution are primarily due to differences in the metal oxide surface (i.e. roughening and hydroxylation) that occur on exposure to the solution. In this authors opinion, these experiments highlight an exciting approach to study well-characterized samples interfaced with realistic environments.

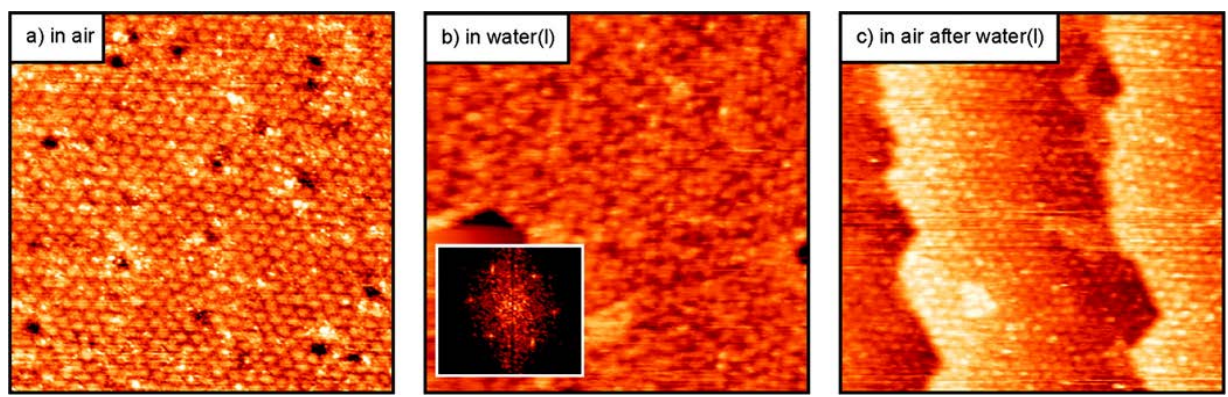

Figure 104: STM images (60 nm × 60 nm) of FeO(111)/Pt(111) prepared in UHV and imaged in (a) air, (b) deionized water, and (c) air again after removing water . The inset in b shows an FFT of the image revealing the hexagonal array of protrusions on the surface in liquid water. A Wandelt-type ECSTM was used for these measurements [602]. Reprinted with permission from ref. [598]. Copyright 2011 American Chemical Society.

Several important applications (e.g. PEC water splitting and groundwater decontamination) are based in this environment. Recent theoretical work on the water/$TiO_2$(110) system suggests that the presence of a second layer of water at the interface already causes rearrangement of the contact layer [603], so it seems likely that experiments on hydroxylated surfaces in UHV may be instructive, but will not be sufficient. Of course, the first question that must be addressed is if the surface structure remains the same in the presence of the liquid, and if similar defects remain. While most surface science techniques are not possible in the liquid environment, STM can be operated in water if the tip is coated by an insulator. Indeed, electrochemical STM is a well-established technique for the study of metal surfaces in the liquid environment, and there are many studies of molecular self-assembly e.g. [604]). These experiments are mostly performed on samples cleaned ex-situ (e.g. clean and flat Au surfaces can be prepared simply and easily by flame annealing), but as yet there are no such methods developed for metal oxides. The technical challenge is thus to prepare a well-characterised oxide surface in UHV, and expose it to the liquid phase in a well-defined way without exposure to air. For structural determination in liquid, SXRD can in principle be used with a suitably modified experimental setup.

## 7.6 Biological Coating Molecules

In recent years many studies have focused on the synthesis and characterization of $Fe_3O_4$ nanoparticles coated with organics such as peptides and amino acids (e.g. [605]. Such systems are interesting with a view to applications in drug delivery, hyperthermic cancer treatment and MRI, but also in the removal of heavy metals from groundwater [606]. The molecules serve to prevent agglomeration of the particles, but also provide a route to functionalisation. A recent study [607] comparing a variety of molecules (l-alanine, l-cysteine, l-glutamic acid, glycine, l-histidine, l-lysine, and l-serine) in solution at pH6 determined that the carboxyl group was involved in binding to the iron oxide surface, and that molecules with a polar side chain packed most densely. To date there has been very little surface science work investigating the bonding of organics on iron oxides. However, in a recent paper, Aschauer and Selloni [608] have recently begun to study this problem from a theoretical viewpoint, performing density functional theory calculations of polyvinyl alcohol (PVA), polyethylene glycol (PEG) and glycine, as well as water on magnetite $Fe_3O_4$(110). The authors suggest that the results serve as a basis for future molecular dynamics calculations to investigate the relative stability of adsorbed coatings in an aqueous environment. Experimentally, there is significant scope to study the adsorption of organic molecules in UHV, and with the current attempts to extend surface science to the liquid phase, this topic seems like a good bet to take advantage of new methods as they emerge.

## 7.7 Realistic Materials

The vast majority of applications of iron oxides utilize powders or nanoparticles, and not macroscopic single crystals. Developments with aberration-corrected TEM have revolutionized our ability to characterize such species, and over recent years an astounding level of control has been developed in the synthesis of nanomaterials. Iron and iron oxide nanoparticles of all kinds of different shapes and sizes including nanorods [609], nanocubes [610-612], and tetrapods [613] can be created by decomposing precursors such as iron pentacarbonyl in a ternary surfactant mixture under mild thermal conditions. In Figure 101, for example, the addition of oleic acid, sodium oleate, and tetraoctylammonium bromide modified the growth rate of different facets such that cubes and octaherda were produced. Exactly how the surfactant species bind to the different facets causing these differences in growth is not well understood, but such tight control of the morphology allows to tune the properties of the material toward applications.

One big question is if the lessons learned on single crystals are relevant when dimensions are reduced to the nanoscale. Even though the study of the surface of such materials is in its infancy there are hints that novel effects occur. For example, Pratt et al. [614] recently showed that strain at the surface of iron oxide nanocubes leads to increased diffusion of cations, potentially explaining enhanced oxidation rates that have been observed for Fe nanoparticles. It would be fascinating to see if cuboid $Fe_3O_4$ nanoparticles exposing (001) facets exhibit the SCV reconstruction, and whether similar Fe rich terminations appear in reducing conditions. Such an experiment could in principle be performed by depositing nanoparticles on an inert substrate *ex-situ*, or *in-situ* using an electrospray technique [615; 616], but the particles will be coated with the surfactant utilized to control the shape during growth and or/solvent. Methods to remove such molecules effectively in UHV without affecting the particles will have to be developed. If this proves possible, methods exist that can be used to study the surfaces of small particles including scanning probe techniques (provided one is fortunate enough that the tip lands atop a nanoparticle), selected area LEED (using a high-resolution LEEM instrument) and photoemission using the latest generation PEEM instruments.

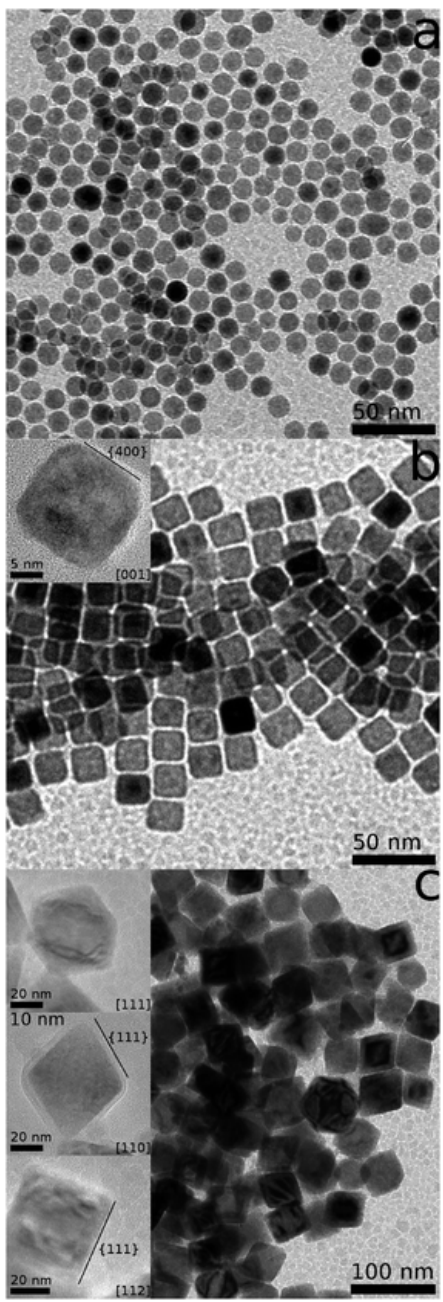

Figure 105: TEM images of iron oxide nanoparticles with different shapes prepared by iron oleate decomposition in the presence of different additives: oleic acid (a), sodium oleate (b), and tetraoctylammonium bromide (c). The inset in (b) is a high resolution image of a core-shell nanocube.

The insets in (c) show different orientations of octahedral nanoparticles. Reproduced from ref. [610] with permission from the PCCP Owner Societies.

# 8 Acknowledgements

First and foremost I would like to thank Prof. Ulrike Diebold for encouraging me to write this review, for continuous support and encouragement throughout its realisation, and for critically reading the manuscript. I am also indebted to Dr. Juan de la Figuera, Prof. Michael Schmid, Dr. Peter Jacobson, and (soon to be Dr.) Roland Bliem for many hours of insightful discussions about iron oxides, and their fascinating surfaces. I would like to thank the many colleagues who have provided high-quality versions of figures, and responded quickly to inquiries about details of their work. I acknowledge the Austrian Science Fund for funding my research on iron oxide surfaces (project P 24925-N20 and START prize Y 847-N20), and must express my gratitude to my students Dr. Zbynek Novotny, Roland Bliem, Oscar Gamba, Jiri Pavelec, and Jan Hulva for their hard work and dedication to our project. Finally, I want to thank my wife, Elena Parkinson, for patiently putting up with many late nights during the final stages of writing, and in particular for painstakingly formatting the titles of all 616 references prior to submission.